\documentclass[oneside, 11pt]{book}
\usepackage{fullpage}
\usepackage[T1]{fontenc}
\usepackage[utf8]{inputenc}
\usepackage{xcolor}
\usepackage[colorlinks=true]{hyperref}
\definecolor{red}{HTML}{F88379}
\hypersetup{
    linkcolor=blue
    ,citecolor=orange
    ,filecolor=purple
    ,urlcolor=orange
    ,menucolor=purple
    ,runcolor=purple
}
\usepackage{amsmath, amssymb, amsthm, amsfonts, graphicx, color, subcaption, enumerate, bm, array, mathtools}
\allowdisplaybreaks[4]

\usepackage[capitalise,nameinlink]{cleveref}
\usepackage{xparse}

\usepackage[colorinlistoftodos,prependcaption,textsize=tiny]{todonotes}

\usepackage{multicol, multirow}
\usepackage{thmtools, thm-restate}
\usepackage{ragged2e}
\usepackage{lineno}
\usepackage{framed}
\usepackage[framemethod=tikz]{mdframed}

\usepackage[linesnumbered, ruled, vlined]{algorithm2e}

\usepackage{subfiles}
\graphicspath{{graphics/}}

\crefname{claim}{Claim}{Claims}
\crefname{property}{Property}{Properties}
\crefname{algocf}{Algorithm}{Algorithms}
\Crefname{algocf}{Algorithm}{Algorithms}

\usepackage[style=trad-alpha,natbib=true,maxcitenames=4]{biblatex}
\usepackage{dirtytalk}

\makeatletter
\g@addto@macro\bfseries{\boldmath}
\makeatother

\newtheorem{theorem}{Theorem}
\newtheorem{lemma}{Lemma}[section]
\newtheorem{claim}[lemma]{Claim}

\newtheorem{corollary}{Corollary}

\newtheorem{proposition}[lemma]{Proposition}

\theoremstyle{definition}
\newtheorem{definition}[lemma]{Definition}

\theoremstyle{remark}
\newtheorem{remark}[lemma]{Remark}
\newtheorem*{remark*}{Remark}

\renewcommand{\epsilon}{\varepsilon}

\newcommand{\m}[1]{\mathcal{#1}}

\renewcommand{\emptyset}{\varnothing}

\newcommand{\F}{{\cal F}}

\renewcommand{\P}{{\cal P}}
\newcommand{\Q}{{\cal Q}}

\newcommand{\kftrs}{\textsc{$k$-FTRS}}
\newcommand{\dist}{\mathrm{dist}}

\newcommand{\ftrs}{\textsc{FTRS}}

\newcommand{\freq}{\textsc{freq}}

\newcommand{\cost}{\operatorname{cost}}
\newcommand{\fair}{\text{fair clustering}}
\newcommand{\nph}{\textbf{NP}\text{-hard}}

\newcommand{\optcorr}{\m{F}^*_{\text{corr}}}
\newcommand{\optcon}{\m{F}^*_{\text{con}}}
\newcommand{\optcl}[1]{\m{F}^*_{#1}}
\newcommand{\fcc}{\texttt{fairfyCC}}
\newcommand{\n}{\nonumber}
\newcommand{\inpset}{\m{C}}
\newcommand{\inpsetd}{\m{C}=\{C_1,\ldots,C_m\} }
\newcommand{\inpcl}[1]{C_{#1}}
\newcommand{\out}{\m{T}}
\newcommand{\outd}{\m{T} = \{ T_1, T_2, \ldots, T_\zeta\}}
\newcommand{\outcl}[1]{T_{#1}}
\newcommand{\cut}{\textsc{cut}}
\newcommand{\nc}{\textsc{newcut}}
\newcommand{\merge}{\textsc{merge}}
\newcommand{\blue}[1]{\textit{blue}\left(#1\right)}
\newcommand{\red}[1]{\textit{red}\left(#1\right)}
\newcommand{\bal}{\texttt{pdc}}
\newcommand{\mopdef}{\bal}
\newcommand{\mopqdef}{\bal}
\newcommand{\surp}[1]{\textit{s}{(#1)}}
\newcommand{\defi}[1]{\textit{d}(#1)}
\newcommand{\ccostf}[1]{\kappa (#1)}
\newcommand{\mcostf}[1]{\mu (#1)}
\newcommand{\algog}{\texttt{AlgoforGeneral}}
\newcommand{\algoc}{\texttt{AlgoforCut}}
\newcommand{\algom}{\texttt{AlgoforMerge}}
\newcommand{\outg}{\m{T}^{cm}}
\NewDocumentCommand{\gencl}{o}{%
  \IfNoValueTF{#1}{D}{D_{#1}}%
}
\newcommand{\pcard}{m}
\newcommand{\pcl}[1]{E_{#1}}
\newcommand{\ccost}{\textit{cut-cost}}
\newcommand{\mcost}{\textit{merge-cost}}
\newcommand{\outcut}{\m{T}^{c}}
\newcommand{\mop}{\m{T}^*}
\newcommand{\mopd}{\m{T}^* = \{ T_1^*, T_2^*, \ldots, T_\psi^*\}}
\newcommand{\mopcl}[1]{T^*_{#1}}
\newcommand{\opt}{\textsc{OPT}}

\newcommand{\abs}[1]{\left\lvert#1\right\rvert}
\newcommand{\cuta}{\textsc{cut-algo}}
\newcommand{\mergea}{\textsc{merge-algo}}
\newcommand{\costone}[1]{\textit{cost}_1(#1)}
\newcommand{\costtwo}[1]{\textit{cost}_2(#1)}
\newcommand{\costthree}[1]{\textit{cost}_3(#1)}
\newcommand{\costfour}[1]{\textit{cost}_4(#1)}
\newcommand{\cuto}{\textsc{cut-OPT}}
\newcommand{\mergeo}{\textsc{merge-OPT}}
\newcommand{\pcls}{\mathcal{E}}

\newcommand{\iprime}{i^{\prime}}
\newcommand{\equ}{\Leftrightarrow}
\newcommand{\clof}{\textsc{closest $p$-fair}}
\newcommand{\npc}{\textbf{NP}\text{-complete}}
\newcommand{\thrp}{\textsc{3partition}}
\newcommand{\costf}[1]{\textit{cost} (#1)}
\newcommand{\ratio}{p}
\newcommand{\fairmop}{\m{F}^{*}}

\newcommand{\fairset}{\m{F}}
\newcommand{\tb}{\textsc{BlueHeavy}}
\newcommand{\tr}{\textsc{RedHeavy}}
\newcommand{\algmf}{\texttt{MakeClustersFair}}
\newcommand{\smp}{s} 
\newcommand{\dmp}{d} 

\newcommand{\ceilenv}[1]{\left\lceil #1 \right\rceil}
\DeclarePairedDelimiter\ceil{\lceil}{\rceil}
\DeclarePairedDelimiter\floor{\lfloor}{\rfloor}
\newcommand{\fptwo}{\texttt{fairpower-of-two}}
\newcommand{\outfptwo}{\m{F}_{\text{fpt}}}
\newcommand{\greedymerge}{\texttt{multi-gm}}
\newcommand{\Sn}{T_a^k}
\newcommand{\Sx}{S_j^k}

\newcommand{\card}[1]{\left\lvert #1 \right\rvert}
\newcommand{\pay}[1]{\operatorname{pay}(#1)}
\newcommand{\gencls}{\m{D}}
\newcommand{\mcostone}{\operatorname{exp}_1}
\newcommand{\mcosttwo}{\operatorname{exp}_2}
\newcommand{\mcostthree}{\operatorname{exp}_3}
\newcommand{\mcostfour}{\operatorname{exp}_4}
\newcommand{\mcostfive}{\operatorname{exp}_5}
\newcommand{\mcostsix}{\operatorname{exp}_6}
\newcommand{\mcostseven}{\operatorname{exp}_7}
\newcommand{\mcosteight}{\operatorname{exp}_8}
\newcommand{\mcostnine}{\operatorname{exp}_9}
\newcommand{\fmulti}{\texttt{make-pdc-fair}}
\newcommand{\fgen}{\texttt{fair-general}}
\newcommand{\outmpf}{\m{F}_{\text{mpf}}}
\newcommand{\fequi}{\texttt{fair-equi}}
\newcommand{\cb}{\texttt{CB}}
\newcommand{\mb}{\texttt{MB}}
\newcommand{\pdca}{\texttt{create-pdc}}
\newcommand{\cutcase}{\texttt{Cut-Case}(c_j)}
\newcommand{\mergecase}{\texttt{Merge-Case}(c_j)}
\newcommand{\setofclusterings}{\mathtt{I}}
\newcommand{\fc}{\texttt{find-candidates}}
\DeclareMathOperator*{\argmin}{arg\,min}
\newcommand{\canset}{\widetilde {\m{F}}}
\NewDocumentCommand{\clsfitting}{g}{%
  \IfNoValueTF{#1}
    {\texttt{ClusterFitting}}%
    {\texttt{ClusterFitting}(#1)}%
}
\NewDocumentCommand{\concls}{g}{%
  \IfNoValueTF{#1}
    {\cls{C}}%
    {\cls{C}_{#1}}%
} 

\newcommand{\fairtriple}{T_{i,j,k}}

\newcommand{\approxFactorFairCorCls}{\gamma}
\newcommand{\tripleconcls}{T}
\newcommand{\cls}[1]{\m{#1}}
\newcommand{\approxFactorClsFair}{\alpha}
\newcommand{\optconcls}{\optcon}
\newcommand{\inpconclss}{\setofclusterings}
\newcommand{\obj}[1]{\mathrm{Obj}\left( #1 \right)}
\newcommand{\optconval}{\mathrm{OPT}}
\newcommand{\avgconval}{\mathrm{AVG}}
\newcommand{\unalignedSet}[1]{I_{#1}}
\newcommand{\pconstream}{\texttt{pairwise streaming model}}
\newcommand{\runtimeClsFair}[1]{t_{2}(#1)}
\newcommand{\runtimeFairCorCls}[1]{t_{1}(#1)}
\newcommand{\errCoreset}{\epsilon}
\NewDocumentCommand{\fairconcls}{g}{%
  \IfNoValueTF{#1}
    {\m{F}}%
    {\m{F}_{\m{C}_#1}}%
} 
\newcommand{\neigh}[1]{\mathtt{N}_{#1}}
\newcommand{\farNeigh}[1]{\mathtt{F}_{#1}}
\newcommand{\stalgo}{\texttt{algo-fair-con-stream}}
\newcommand{\costcor}[1]{\operatorname{cost}(#1)}
\newcommand{\approxFactorCorCls}{\delta}

\date{}

\newenvironment{abstract}{
    \cleardoublepage
    \thispagestyle{plain}
    \null\vfill
    \begin{center}
        {\bfseries\Large Abstract\par}
    \end{center}
    \itshape
}{
    \vfill\null
}

\newtheorem{question}{Question}[chapter]

\begin{document}


\begin{titlepage}
    \centering
    \vspace*{2cm}

    {\bfseries\Large Creating Robust and Fair Graph Structures for Connectivity and Clustering \par}

    \vspace{1.5cm}

    {\large by \par}

    \vspace{0.3cm}

    {\bfseries\Large KUSHAGRA CHATTERJEE \par}

    \vspace{0.2cm}

    {\large A0243436H \par}

    \vspace{1.5cm}

    {\bfseries A THESIS SUBMITTED FOR THE DEGREE OF \par}
    \vspace{0.2cm}

    {\bfseries DOCTOR OF PHILOSOPHY \par}

    \vspace{0.2cm}

    {\large in \par}

    \vspace{0.2cm}

    {\bfseries COMPUTER SCIENCE \par}

    \vspace{0.2cm}

    {\large to the \par}

    \vspace{0.2cm}

    {\bfseries GRADUATE DIVISION \par}

    \vspace{0.2cm}

    {\large of the \par}

    \vspace{0.2cm}

    {\bfseries NATIONAL UNIVERSITY OF SINGAPORE \par}

    \vspace{1cm}

    {\large 2025 \par}

    \vfill

    \large
    Advisor: Dr. Diptarka Chakraborty \\[1 cm]
    Examiner 1: Dr. Chang Yi-Jun \\
    Examiner 2: Dr. Chen Yu

\end{titlepage}

\cleardoublepage
\thispagestyle{plain}

\begin{center}
    {\bfseries\Large Acknowledgement}
\end{center}

\vspace{1em}

I would like to express my deepest gratitude to my PhD advisor, \textbf{Prof.~Diptarka Chakraborty}, for his constant guidance, encouragement, and patience throughout my doctoral studies. I have been extremely fortunate to have an advisor who guided me through every aspect of the research process, including how to read research papers critically, how to identify and approach meaningful research problems, how to develop and analyze solutions rigorously, and how to write and present research clearly and effectively. His insights, rigor, and exceptionally high standards have shaped not only this thesis but also my overall development as a researcher. I am truly grateful to have had him as my advisor.

\vspace{1em}

I am sincerely thankful to my collaborators \textbf{Prof.~Debarati Das}, \textbf{Tien-Long Nguyen}, \textbf{Romina Nobahari}, and \textbf{Keerti Choudhary} for their invaluable intellectual contributions. Working with them has been a rewarding experience marked by numerous insightful discussions and scrutiny of ideas. They provided critical feedback while solving problems, helped identify mistakes in my solutions, and consistently encouraged me to complete the writing—especially during intense periods near submission deadlines. Without their support, persistence, and collaboration, completing this thesis would not have been possible.

\vspace{1em}

I would also like to thank \textbf{Prof.~Prajakta Nimbhorkar}, with whom I collaborated during different stages of my PhD. Beyond our academic interactions, she has been a consistent source of encouragement and support throughout my doctoral journey. Her thoughtful advice and reassuring presence were invaluable during challenging periods, and I am grateful for her generosity and kindness.

\vspace{1em}

I would like to acknowledge the academic environment and research culture at my institution, which provided an intellectually stimulating and supportive setting for my doctoral work. The seminars, discussions, and interactions with faculty members and peers contributed significantly to my growth as a researcher. In particular, I would like to mention \textbf{Prof.~Yi-jun Chang}, whose interactions and discussions helped me substantially improve my research capabilities. In addition, as part of the Graduate Assistantship Programme (GAP) requirements mandated by the university, I had the opportunity to serve as a teaching assistant for the courses CS2040S, CS5230, and CS4234 under the supervision of \textbf{Prof.~Seth Gilbert}, \textbf{Prof.~Arnab Bhattacharyya}, and my advisor, \textbf{Prof.~Diptarka Chakraborty}. This teaching experience was highly enriching and allowed me to deepen my understanding of fundamental concepts while significantly improving my communication and presentation skills as a researcher.

\vspace{1em}

Finally, and most importantly, I would like to express my deepest gratitude to my parents and my grandmother for their constant emotional support, unconditional love, patience, and encouragement throughout these years. Their unwavering belief in me and their sacrifices have been a continual source of strength and motivation during my PhD journey. I also wish to acknowledge my personal faith, and I am grateful for the strength and guidance I draw from my belief in Ma Durga.

\newpage
\thispagestyle{plain}
\begin{center}
    {\bfseries\LARGE Representative Papers}
\end{center}
\vspace{1cm}

This thesis includes contributions from the following set of papers:

\vspace{0.8cm}

\noindent
Chakraborty, D., \textbf{Chatterjee, K.}, \& Choudhary, K. 
Pairwise Reachability Oracle and Preserver under Failures. 
\textit{ICALP 2022} \cite{ICALPchakraborty2021pairwise}.

\vspace{0.5cm}

\noindent
Chakraborty, D., \textbf{Chatterjee, K.}, Das, D., Nguyen, T. L., \& Nobahari, R.
Towards Fair Representation: Clustering and Consensus
\textit{COLT 2025} \cite{chakraborty2025towards}.

\vspace{0.5cm}

\noindent
Chakraborty, D., \textbf{Chatterjee, K.},  Das, D., \& Nguyen, T. L. 
Generalizing Fair Clustering to Multiple Groups: Algorithms and Applications 
\textit{AAAI 2026}(Oral Presentation)  \cite{AAAIchakraborty2026generalizing}.

\vspace{0.5cm}

\noindent
Chakraborty, D., \textbf{Chatterjee, K.},  Das, D., \& Nguyen, T. L. 
A Generic Framework for Fair Consensus Clustering in Streams  
\textit{AAMAS 2026} \cite{AAMASchakraborty2026generic}.

\vfill

\begin{abstract}
Graph algorithms are fundamental to large-scale applications such as navigation systems including Google Maps, Apple Maps, and Waze. Two central challenges in these systems are robustness to failures and fairness in clustering outcomes. This thesis addresses both challenges by developing new algorithmic techniques for (i) fault-tolerant graph structures and (ii) fair clustering, unifying them under the broader goal of designing large-scale applications that are efficient, resilient, and socially responsible.

In the first part of the thesis, we study fault-tolerant reachability preservers. Given a directed graph $G=(V,E)$ on $n$ vertices and a set of pairs $\mathcal{P} \subseteq V \times V$, a reachability preserver is a sparse subgraph that preserves the existence of paths between all pairs in $\mathcal{P}$ even after $k$ failures for some integer parameter $k$. We provide the first non-trivial constructions of dual fault-tolerant pairwise reachability preservers, resilient to two edge or vertex failures. In particular, we present a construction of size $O(n^{4/3}|\mathcal{P}|^{1/3})$, significantly improving upon prior work that was restricted to the single-fault setting.

In the second part of the thesis, we investigate the design of clustering algorithms that are not only representative of the underlying data but also fair with respect to sensitive attributes. Classical clustering methods often fail to provide proportional representation for marginalized groups, thereby amplifying biases present in the data. To address this issue, we study fair variants of clustering problems, where each cluster must approximately reflect the global distribution of protected groups. Our first contribution in this direction is the study of fair consensus clustering for two groups, for which we provide the first constant-factor approximation algorithms that integrate fairness into consensus clustering tasks. To achieve this, we introduce a new framework called closest fair clustering, and show that the problem becomes NP-hard when the two groups are of unequal size.

We then generalize closest fair clustering to arbitrary numbers of groups and develop near-linear-time approximation algorithms that efficiently handle multi-group settings. Building on this framework, we extend our results to obtain improved guarantees for unweighted fair correlation clustering and provide the first approximation algorithms for weighted fair correlation clustering when the weights satisfy probability constraints, as well as for multi-group fair consensus clustering.

Finally, we address the scalability of fair consensus clustering by initiating the study of fair consensus clustering in the streaming model. In this setting, input clusterings arrive sequentially, and memory is limited, making it infeasible to store all inputs. We design the first constant-factor approximation algorithm for fair consensus clustering in the streaming model, which processes the stream while storing only a logarithmic number of input clusterings.

Taken together, these contributions advance the broader vision of designing graph algorithms that make large-scale applications both robust and fair. Fault-tolerant reachability preservers ensure that critical connectivity information is maintained even under multiple failures, while fair clustering algorithms guarantee balanced representation across sensitive groups. By addressing the dual challenges of robustness and fairness, this thesis contributes to the foundations of building graph-based systems that are not only efficient in theory but also trustworthy, inclusive, and dependable in practice.
\end{abstract}

\newpage
\bigskip
\tableofcontents
\bigskip
\thispagestyle{empty}



\newpage
\pagenumbering{arabic}

\chapter{Introduction}

Modern large-scale applications such as Google Maps, Apple Maps, and Waze serve millions of users daily, providing critical services such as navigation, routing, and location-based recommendations. At their core, these systems rely heavily on graph algorithms and clustering methods to deliver fast, accurate, and personalized results. Two central challenges naturally arise in this setting: ensuring \emph{robustness to failures} and guaranteeing \emph{fair representation}. Addressing these challenges requires principled algorithmic techniques, motivating the study of both \emph{fault-tolerant graph structures} and \emph{fair clustering}. In this thesis, we pursue both directions: on the one hand, we study \emph{reachability preservers} as a fundamental tool for fault-tolerant graph sparsification, and on the other hand, we study \emph{fair clustering} to design clustering algorithms that are both accurate and representative. 

Navigation platforms such as Google Maps, Apple Maps and Waze rely on algorithms that solve shortest-path queries at scale under stringent real-time constraints. To achieve fast query response, these systems often rely on precomputation-based architectures such as \emph{Contraction Hierarchies} \cite{GeisbergerDSS08} and \emph{Customizable Route Planning} \cite{delling2017customizable}. These methods are highly efficient but typically return a single ``best'' route, often the shortest path. In practice, however, users may seek multiple alternatives---for example, avoiding highways, toll roads, or unsafe streets. This motivates research into computing \emph{alternative routes} \cite{AbrahamDGW13}. Supporting such customizable objectives directly on the full road network is computationally prohibitive, leading to the development of algorithms that extract \emph{smaller subgraphs} while preserving essential connectivity. For example, work on \emph{subgraph extraction for routing} \cite{ahmadian2024extracting} and \emph{efficient subgraphs for customizable route planning} \cite{delling2017customizable} demonstrates that compact subgraphs enable flexible routing with reduced overhead.

This need for smaller yet faithful subgraphs necessitates the broader study of \emph{graph sparsifiers}. Graph sparsification is the process of approximating a large graph with a sparse subgraph that preserves certain properties of the original. Over the last few decades, multiple notions of sparsification have been introduced, each tailored to a different structural or algorithmic objective. For example, \emph{graph spanners} preserve approximate distances between all pairs of vertices, and have been extensively studied starting with Awerbuch \cite{Awerbuch85}, Peleg and Schäffer \cite{PelegShaffer89}, and Alon et al. \cite{AlonKarpPelegWest91}, culminating in optimal constructions such as those of Thorup and Zwick \cite{ThorupZwick05} and Elkin and Peleg \cite{ElkinPeleg05}. A related notion, \emph{emulators}, preserves approximate distances by allowing additional edges \cite{BaswanaSen07}. Another classical line of work is \emph{cut sparsifiers}, which preserve the capacity of all cuts approximately, beginning with the seminal work of Benczúr and Karger \cite{BenczurKarger96}. More recently, \emph{spectral sparsifiers}, introduced by Spielman and Srivastava \cite{SpielmanSrivastava11}, preserve the quadratic form of the Laplacian matrix, leading to powerful applications in numerical linear algebra and optimization. Each of these sparsifiers captures a different structural aspect of the input graph, and together they form a central toolkit of modern graph algorithms.

Among these sparsification primitives, one of the simplest and most natural objects is the \emph{reachability preserver}. Given a graph $G$ and a set of pairs $\mathcal{P}$ of vertices, a reachability preserver is a sparse subgraph $H$ that preserves the existence of a path between each pair $(s,t) \in \mathcal{P}$ if and only if one exists in $G$. Reachability preservers thus capture one of the most fundamental binary relations in a graph---whether a path exists between two vertices---and serve as a baseline abstraction for path-preserving sparsification. They arise naturally in applications where connectivity, rather than distance or capacity, is the key requirement. For example, in network reliability and fault-tolerant design, ensuring that communication remains possible between designated end vertices is often more important than minimizing distances. Similarly, in road networks, the ability to guarantee that certain critical routes remain present even under subgraph extraction is essential for supporting customizable routing at scale.

In this thesis, we take this perspective further by studying \emph{fault-tolerant reachability preservers}, which remain valid even after the failure of edges or vertices. Fault tolerance is crucial in real-world settings where networks may be subject to disruptions due to accidents, congestion, or hardware failures. While classical sparsification notions largely assume static networks, fault-tolerant variants aim to guarantee structural preservation under adversarial deletions. As such, fault-tolerant reachability preservers unify two key ideas: sparsification for efficiency and robustness for reliability.

In addition to routing, clustering is another core primitive underlying the functionality of large-scale platforms like Google Maps, Apple Maps, and Waze. Clustering is used to group nearby places of interest, such as restaurants, tourist attractions, or services, and to present representative subsets to the user. From a theoretical perspective, clustering is one of the most extensively studied problems in computer science and operations research, with variants such as $k$-center, $k$-median, and $k$-means receiving decades of attention. In the setting of graph metrics, the distance between two vertices is given by the length of the shortest path connecting them. Clustering problems such as $k$-center, $k$-median, and $k$-means have been extensively studied under this formulation, where clusters are formed based on shortest-path distances in the graph. Classic results include the 2-approximation greedy algorithm for $k$-center by Gonzalez \cite{Gonzalez85}, the local search framework for $k$-median by Arya et al. \cite{Arya98}, and LP-rounding approaches for facility location and clustering by Charikar et al. \cite{Charikar99} and Jain and Vazirani \cite{JainVazirani01}. 

Another influential formulation is \emph{correlation clustering}, introduced by Bansal, Blum, and Chawla \cite{BansalBlumChawla04}, where the input is a complete graph with similarity/dissimilarity labels, and the goal is to partition vertices to minimize disagreements. Correlation clustering has since become a central abstraction for unsupervised learning and data integration tasks, with applications ranging from community detection to document classification and entity resolution \cite{charikar2005clustering, Ailon2008JACM, Chawla2018JACM}. Early works established both the algorithmic and hardness foundations of the problem. 
Bansal et al. \cite{Bansal2004} showed that correlation clustering is NP-hard even on complete unweighted graphs, and provided the first constant-factor approximation algorithms for both the minimization and maximization variants. 
Charikar, Guruswami, and Kumar \cite{charikar2005clustering} subsequently developed an LP-based rounding approach that achieved a $3$-approximation for the minimize-disagreement version and an $O(1)$-approximation for the maximize-agreement variant. 
This work set the stage for a long line of research on improving approximation guarantees and extending correlation clustering to richer settings. Ailon, Charikar, and Newman \cite{Ailon2008JACM} proposed the celebrated \emph{pivot algorithm}, which provided a simple and elegant randomized $2.5$-approximation, establishing a new benchmark for practical and theoretical performance. 
Further progress was made by Chawla et al. \cite{Chawla2018JACM}, who refined LP-based techniques to obtain a $2.06$-approximation for the weighted version, which remains a milestone in correlation clustering optimization. 
Their work also established improved hardness of approximation, showing that under the Unique Games Conjecture (UGC), no polynomial-time algorithm can achieve an approximation factor better than $1.36$. In parallel, several works have extended correlation clustering beyond the standard offline and centralized setting. 
The problem has been explored in the massively parallel computation (MPC) model by Behnezhad et al. \cite{behnezhad2022almost} and Cohen-Addad et al. \cite{CohenAddad2021MPC, CohenAddad2024Combinatorial}, where the current state-of-the-art achieves a $1.876$-approximation in a constant number of rounds under sublinear memory. 
Cohen-Addad, Lattanzi, Mitrović, and Parotsidis \cite{CohenAddad2022Online} further studied the online setting, obtaining an $O(\log n)$-competitive algorithm and initiating a new line of work in online clustering. 

Most recently, Cao et al. \cite{Cao2025Dynamic} obtained the current state-of-the-art $1.437$-approximation via a dynamic-programming framework combined with refined rounding analysis.

A fair amount of attention has also been devoted to the weighted version of the correlation clustering problem. In weighted correlation clustering, each edge $(u,v)$ in the graph $G$ is assigned two weights: a positive weight $w^+(u,v)$ and a negative weight $w^-(u,v)$. The objective is to partition the vertices to minimize the total negative weight within clusters, together with the total positive weight across clusters.

Charikar, Guruswami, and Wirth \cite{charikar2005clustering} presented a factor-$4$ approximation algorithm for weighted correlation clustering and showed that the problem is APX-hard. Subsequently, Demaine, Emanuel, Fiat, and Immorlica \cite{demaine2006correlation} studied the problem in general graphs, obtaining an $O(\log n)$-approximation. They further showed that any significant improvement in this approximation factor would imply improved approximations for the notoriously difficult minimum multicut problem.

Later, Ailon, Charikar, and Newman \cite{Ailon2008JACM} introduced an LP-based formulation that achieves a $2.5$-approximation when the edge weights satisfy probability constraints, that is, for every edge $(u,v)$,
\[
w^+(u,v) + w^-(u,v) = 1.
\]
They further obtained a $2$-approximation when the weights satisfy both the probability constraints and triangle inequalities, namely, for every triple $(u,v,w)$,
\[
w^+(u,v) \leq w^+(u,w) + w^+(w,v)
\quad \text{and} \quad
w^-(u,v) \leq w^-(u,w) + w^-(w,v).
\]

More recently, Chawla et al. \cite{Chawla2018JACM} improved the approximation factor to $1.5$ under the same probability and triangle inequality assumptions.

A related and equally important problem is \emph{consensus clustering}, where the goal is to aggregate multiple input clusterings into a single representative clustering \cite{Dahlhaus1997, Voevodski2012, Bhowmick2020}. Consensus clustering is widely used in large-scale applications where multiple models or data sources provide conflicting partitions, such as combining outputs of different machine learning pipelines or integrating user-generated labels. The consensus clustering problem also has a myriad of applications in different domains, such as gene integration in bioinformatics \cite{filkov2004integrating, filkov2004heterogeneous}, data mining \cite{topchy2003combining}, community detection \cite{lancichinetti2012consensus}. The problem (both respect to median and center objective) is not only known to be \texttt{NP}-hard \cite{kvrivanek1986np, swamy2004correlation}, but also known to be \texttt{APX}-hard, i.e., unlikely to possess any $(1+\epsilon)$-approximation (for any $\epsilon >0$) algorithm even when there are only three input clusterings \cite{BonizzoniVDJ08}. While several heuristics have been considered to generate reasonable solutions (e.g., \cite{goder2008consensus, monti2003consensus, wu2014k}), so far, we only know of an $11/7$-approximation algorithm for the median objective \cite{Ailon2008JACM}, and a better than 2-approximation for the center objective \cite{DK25}. 

Both correlation clustering and consensus clustering are valued because they make minimal assumptions on the input and capture the need for robust clustering under noisy or heterogeneous information, a property essential for real-world platforms operating at scale. 

While these formulations capture structural or objective-based requirements, they do not address issues of \emph{fairness}. In practical recommender systems, clustering purely by similarities or aggregating multiple clusterings can produce biased outcomes, for instance by overrepresenting certain types of businesses (e.g., large fast-food chains) while underrepresenting others (e.g., minority-owned restaurants or niche cultural attractions). This imbalance reflects a broader concern in machine learning: algorithmic decision processes often exhibit disparate impact when fairness constraints are ignored. In clustering, the pioneering work of Chierichetti et al. \cite{Chierichetti2017} introduced the notion of \emph{fair clustering}, requiring that the representation of sensitive groups within each cluster reflect their overall proportions in the dataset. Subsequent work expanded this framework: Bera et al. \cite{Bera2019} and Huang et al. \cite{huang2019coresets} developed improved algorithms for fair clustering, while Backurs et al. \cite{Backurs2019} studied scalable approaches, Chen et al. \cite{Chen2019} introduced proportionally fair clustering, and Ahmadian et al. \cite{Ahmadian2020} addressed correlation clustering under fairness constraints. To the best of our knowledge, fairness has not previously been studied in either weighted correlation clustering or consensus clustering. We initiate the study of fairness in both settings.

We also study the problem of fair consensus clustering in the streaming setting. Large-scale applications operate in highly dynamic environments, where data is generated continuously, and decisions must be updated in real time. These systems routinely aggregate information from multiple agents or sources—such as user trajectories, local preferences, sensor data, and third-party inputs—each of which can be viewed as inducing a clustering or partition of the underlying data. Storing and repeatedly processing all such inputs in their entirety is often infeasible due to scale, latency, and memory constraints.

This operational reality naturally motivates the study of space-efficient algorithms in the streaming model, where inputs arrive sequentially, and the algorithm must process them on the fly while retaining only a small subset or a compact sketch of the data. At the end of the stream, all downstream computation is performed using this sketch alone. In such applications, maintaining fairness guarantees is particularly challenging: fairness constraints must be enforced without the ability to store the full input. This raises the fundamental question of whether fairness can be preserved efficiently in the streaming setting, achieving sublinear—ideally logarithmic—space per reported output while retaining strong approximation guarantees. Although multiple passes over the data stream may sometimes be permitted, single-pass algorithms are strongly preferred, as additional passes are often prohibitively expensive in real-world systems \cite{Muthukrishnan05, BabcockEtAl02}.

In the commonly studied insertion-only streaming model for clustering \cite{CharikarOCallaghanPanigrahy03, rosman2014coresets, braverman2019streaming, schmidt2019fair}, data points from an underlying metric space arrive one by one, and deletions are disallowed. A central algorithmic paradigm in this setting is the construction of coresets or other succinct summaries that preserve the clustering structure, enabling approximate solutions for objectives such as $k$-median, $k$-means, and $k$-center with provable guarantees. Seminal work introduced merge-and-reduce frameworks that allow such summaries to be maintained using polylogarithmic space and update time \cite{bentley1980decomposable, HarPeledMazumdar04, FeldmanLangberg11, braverman2019streaming}. More recently, coreset constructions have also been developed for clustering problems under fairness constraints \cite{huang2019coresets, schmidt2019fair, chhaya2022coresets, braverman2022power, xiong2024fair}. However, to the best of our knowledge, none of these results imply sublinear-space streaming algorithms with non-trivial approximation guarantees for the fair consensus clustering problem.

In the context of location-based recommenders, fairness in correlation and consensus clustering is particularly important. Platforms such as Google Maps, Apple Maps, or Waze directly influence user behavior and business visibility through their recommendations. Correlation clustering naturally models the setting where user ratings or co-visit patterns provide noisy similarity/dissimilarity information between businesses, while consensus clustering models the integration of multiple heterogeneous signals (e.g., customer reviews, geographic proximity, and social network influence) into a unified recommendation. Ensuring that the resulting clusters are not only structurally meaningful but also representative of the underlying diversity of businesses and communities is key to building trust and avoiding bias. Fair correlation and consensus clustering thus provide principled frameworks to achieve this balance, ensuring that recommendations remain inclusive, diverse, and socially responsible while maintaining algorithmic efficiency. 

The two themes outlined above---graph sparsification and fair clustering---may at first seem orthogonal, but they address complementary challenges faced by large-scale systems. On one side, fault-tolerant reachability preservers ensure that critical connectivity is maintained even under failures, providing the backbone of reliability for navigation and routing. On the other side, fair clustering ensures that recommendations built on graph metrics are representative and inclusive, mitigating the risk of bias and unequal visibility. Both are essential for the next generation of graph-based systems: efficiency alone is not enough without robustness, and accuracy alone is not enough without fairness. By unifying these perspectives, this thesis contributes to the theoretical foundations of building systems that are not only scalable and accurate but also dependable, equitable, and socially responsible.

\section{Our Contribution}\label{sec:our-contribution}

In this section, we state our contributions.

\section*{Reachability Preservers}

In the context of graph sparsification, {\em reachability preserver} (or {\em reachability subgraph})
for a directed graph $G$ and a set $\m{P}$ of vertex-pairs is a sparse subgraph $H$ with as few edges as possible so that for any pair $(s,t) \in \m{P}$ there is a path from $s$ to $t$ in $H$ if and only if there is such a path in $G$. In the standard static setting (with no failure), this object has been studied widely \cite{CE06, Bodwin17, AB18}. We study these objects in the presence of edge/vertex failures.

Let us formally define fault-tolerant reachability subgraph ({\ftrs}) for a set of vertex-pairs.
\begin{definition}[$\ftrs$]
Let $\m{P}\subseteq V\times V$ be any set of pairs of vertices. A subgraph $H$ of $G$ is said to be a \emph{$k$-Fault-Tolerant 
Reachability Subgraph} of $G$ for $\m{P}$, denoted as $\kftrs(G,\m{P})$, if for any pair $(s,t)\in \m{P}$ and for any subset 
$F\subseteq E$ of at most $k$ edges, $t$ is reachable from $s$ in $G \setminus F$ if and only if $t$ is reachable from 
$s$ in $H \setminus F$.
\label{definition:FTRS}
\end{definition}

For the particular case of single-source, i.e., $\m{P}=\{s\}\times V$, 
Baswana, Choudhary, and Roditty \cite{BCR16} provided a polynomial-time algorithm that, given any $n$-vertex directed graph, constructs an $O(2^k n)$-sized $\kftrs$. As an immediate corollary, we get a $\kftrs$ of size $O(2^kn|\m{P}|)$ (by applying the algorithm of \cite{BCR16} to find subgraph for each source vertex in pairs of $\m{P}$ and then taking the union of all these subgraphs). For the general setting of arbitrary pairs, the only previously known non-trivial result was for single failure \cite{chakraborty2020new}, wherein the authors gave an upper bound of $O(n+\min(|\m{P}|\sqrt{n},~n\sqrt {|\m{P}|}))$ edges. It was left open whether for dual or more failures whether keeping fewer than $O(n|\m{P}|)$ edges sufficient to preserve the pairwise reachability. In particular, does any $n$-vertex graph and a set $\m{P}$ of vertex-pairs always admit a $\kftrs$ of size $o(2^k n |\m{P}|)$?

In this work, we answer the above question affirmatively. For dual failures, we provide an upper bound of $O(n^{4/3} |\m{P}|^{1/3})$ edges on the structure of 2-$\ftrs(G,\m{P})$.

\begin{restatable}[Upper Bound on 2-$\ftrs$]{theorem}{upreachability}
\label{thm:dual-ftrs}
For any directed graph $G=(V,E)$ with $n$ vertices and a set $\m{P} \subseteq V \times V$ of vertex-pairs, there exists a $2$-$\ftrs(G,\m{P})$ having at most $O(n^{4/3} |\m{P}|^{1/3})$ edges. Furthermore, we can find such a subgraph in polynomial time.
\end{restatable}

Clearly, for $\m{P}$ of size $\omega(\sqrt{n})$, the above result breaks below the $O(n |\m{P}|)$ bound.

\section*{Fair Clustering}

Given a set of vertices $V$, clustering is defined as a partitioning of $V$ in a metric space into clusters $C_1, \ldots, C_w$ with the goal of minimizing certain objective functions specific to the application. We call a set of clusters $\m{C} = \{C_1, \ldots, C_w\}$ a clustering.
Each vertex can be viewed as an individual with certain protected attributes, which can be represented by assigning a color to each vertex. As mentioned before, Chierichetti et al  \cite{Chierichetti2017} pioneered the concept of \emph{fair clustering}, where each vertex in the dataset is colored either red or blue. The goal was to partition the data while maintaining balance in each cluster -- the ratio of blue to red vertices in each cluster reflects the overall ratio in the entire set. Later, R\"osner and Schmidt \cite{rosner2018privacy} studied fair clustering for multiple colors. Formally, fair clustering can be defined in the following way.
\begin{definition}[Fair clustering]
    Given a set of vertices $V$, suppose it is divided into $|\chi|$ disjoint groups. Each group $i$ is represented by a color $c_i$. A clustering $\m{F}= \{F_1, \ldots, F_{\iota}\}$ on $V$ is said to be fair if for each cluster $F_j \in \m{F}$ we have
    \[
        c_1(F_j):\cdots:c_{|\chi|}(F_j) = c_1(V):\cdots:c_{|\chi|}(V)
    \]
    where $c_i(S)$ denotes the number of vertices of color $c_i$ in $S$ where $S \subseteq V$.
\end{definition}

The above notion of fairness is called \emph{proportional fairness}. In this thesis, we study fairness for two types of clusterings: $(i)$ \emph{Weighted Correlation Clustering with probability constraints} and $(ii)$ \emph{Consensus Clustering}. Let us define them. 

In weighted correlation clustering with probability constraints, we are given a complete undirected graph $G=(V,E)$ in which each edge $(u,v) \in E$ is assigned two weights: a positive weight $w^+(u,v)$ and a negative weight $w^-(u,v)$, satisfying
\[
w^+(u,v) + w^-(u,v) = 1.
\]
If, for every edge $(u,v)$, the weights satisfy $w^+(u,v), w^-(u,v) \in \{0,1\}$, then the instance is referred to as an \emph{unweighted correlation clustering} instance.

For an arbitrary clustering $\mathcal{C}$, we write $u \sim_{\mathcal{C}} v$ to denote that vertices $u$ and $v$ belong to the same cluster under $\mathcal{C}$, and $u \nsim_{\mathcal{C}} v$ to denote that they belong to different clusters.

The cost of a clustering $\mathcal{C}$ is defined as
\[
\cost(\mathcal{C}) = \sum_{u \sim_{\mathcal{C}} v} w^-(u,v) \;+\; \sum_{u \nsim_{\mathcal{C}} v} w^+(u,v).
\]
The objective of weighted correlation clustering is to find a clustering that minimizes $\cost(\mathcal{C})$. When the clustering $\mathcal{C}$ is additionally required to satisfy fairness constraints, the problem is referred to as \emph{weighted fair correlation clustering}.

\textbf{$\gamma$-approximate weighted fair correlation clustering}: We call a clustering $\m{F}$, a $\gamma$-approximate weighted fair correlation clustering if 

\[\cost(\m{F}) \leq \gamma \cost(\optcorr)\] 
here $\optcorr$ is an optimal weighted fair correlation clustering. For the unweighted correlation clustering instance, that is, when for each edge $(u,v)$ we have $w^+(u,v), w^-(u,v) \in \{0,1\}$, we refer $\m{F}$ as \emph{$\gamma$-approximate unweighted fair correlation clustering}.

To define consensus clustering, we need to define the distance between two clusterings $\m{C}$ and $\m{C}'$ on a vertex set $V$. The distance is measured by the number of pairs $(u,v)$ that are together in $\m{C}$ but separated by $\m{C}'$ and the number of pairs $(u,v)$ that are separated by $\m{C}$ but together in $\m{C}'$. More specifically,
\begin{align*}
\dist(\m{C}, \m{C}') = 
\big| \big\{ \{u,v\} \mid\,& u,v \in V,\; 
[u \sim_{\m{C}} v \land u \not\sim_{\m{C}'} v] \\
&\lor [u \not\sim_{\m{C}} v \land u \sim_{\m{C}'} v] \big\} \big|
\end{align*}
where $u \sim_{\m{C}} v$ denotes whether both $u$ and $v$ belong to the same cluster in $\m{C}$ or not.
 
In Consensus clustering, we are given $m$ clusterings $\m{C}_1, \ldots, \m{C}_m$ on a vertex set $V$ and our goal is to find a clustering $\m{C}$ which minimizes the following objective
\[
        \left( \sum_{i = 1}^m \dist(\m{C}_i, \m{C})^\ell\right)^{1/\ell}
\]
for any integer $\ell \in \mathbb{Z}^+$. In addition to the above, when we want the clustering to be fair, we call that problem as \emph{fair consensus clustering} problem. 

\textbf{$\beta$-approximate consensus clustering}: We call a clustering $\m{F}$, $\beta$-approximate fair consensus clustering if
\[
     \left( \sum_{i = 1}^m \dist(\m{C}_i, \m{F})^\ell\right)^{1/\ell} \leq  \beta \left( \sum_{i = 1}^m \dist(\m{C}_i, \optcon)^\ell\right)^{1/\ell}
\]
here $\optcon$ is an optimal  fair consensus clustering.

To solve the above problems of fair correlation clustering and fair consensus clustering, we introduce a new problem called \emph{closest fair clustering}. In closest fair clustering, our goal is to find a fair clustering $\m{F}$ on $V$ which is closest to a given clustering $\m{C}$. More specifically, given a clustering $\m{C}$, we call a clustering $\m{F}$ a closest fair clustering to $\m{C}$ if $\m{F}$ is fair and
    \[
        \dist(\m{C}, \m{F}) \leq \dist(\m{C}, \m{F}')
    \]
    for all fair clusterings $\m{F}'$.

\textbf{$\alpha$-close $\fair$}: A clustering $\m{F}$ is called $\alpha$-close $\fair$ to a given clustering $\m{C}$ if
\[
    \dist(\m{C}, \m{F}) \leq \alpha \, \, \dist(\m{C}, \optcl{\m{C}})
\]
where $\optcl{\m{C}}$ is one of the closest (optimal) fair clustering to $\m{C}$.

\paragraph{Contributions for Weighted Correlation clustering and Consensus clustering:} We prove that, if there exists an algorithm to find an $\alpha$-close $\fair$, then we can find an $(\alpha + \rho + \alpha \rho)$ - approximate weighted fair correlation clustering and an $(\alpha + 2)$-approximate fair consensus clustering where $\rho$ is the state-of-the-art approximation factor for (unfair) weighted correlation clustering problem. Thus, we prove the following theorems

\begin{restatable}{theorem}{clfairtocor}
\label{thm:closest-fair-to-correlation}
Suppose we have an algorithm to find an $\alpha$-close $\fair$ to any given unfair clustering $\m{D}$, then given a correlation clustering instance $G$ , there exists an algorithm to find an $(\alpha + \rho + \alpha \rho)$ - approximate weighted fair correlation clustering under probability constraints, where $\rho$ is the state-of-the-art approximation factor for the weighted correlation clustering problem under probability constraints.
\end{restatable}

\begin{restatable}{theorem}{clfairtocon}
\label{thm:closest-fair-to-consensus}
Suppose we have an algorithm to find an $\alpha$-close $\fair$ to any given unfair clustering $\m{D}$, then given $m$ input clusterings, $\m{C}_1, \ldots, \m{C}_m$ , there exists an algorithm to find an $(\alpha + 2)$-approximate fair consensus clustering.
\end{restatable}

Theorems~\ref{thm:closest-fair-to-correlation} and~\ref{thm:closest-fair-to-consensus} establish a direct relation of weighted fair correlation clustering and fair consensus clustering with the closest fair clustering problem. 
In particular, any approximation guarantee for closest fair clustering can be transformed into approximation guarantees for the other two problems. 
Therefore, the central algorithmic challenge is to design and analyze algorithms for the closest fair clustering problem, as progress in this setting immediately yields corresponding advances for weighted fair correlation clustering and fair consensus clustering.

\paragraph{Contributions for Closest fair clustering (two colors):}We begin by analyzing the case where the vertex set is partitioned into two disjoint color groups, blue and red, such that the ratio of blue to red vertices in $V$ is $p:1$. 
For this setting, we develop an efficient algorithm that achieves a constant-factor approximation guarantee. 
Specifically, our algorithm constructs a clustering that is within a factor of $17$ of the optimal fair clustering with respect to the input clustering~$\m{C}$.

\begin{restatable}{theorem}{upperboundtwocolors}
\label{thm:closest-fair-p:1}
    Let the vertex set $V$ be partitioned into two disjoint groups, blue and red, where the ratio of blue to red vertices in $V$ is $p:1$ for some integer $p > 1$. Then there exists a polynomial-time algorithm that computes a $17$-close fair clustering with respect to a given input clustering~$\m{C}$.
\end{restatable}

To complement the positive result in Theorem~\ref{thm:closest-fair-p:1}, we show a hardness result for the closest fair clustering problem. More specifically, we show that finding a closest fair clustering to given input clustering is NP-hard in the $p:1$ case.

\begin{restatable}{theorem}{nphard}
\label{thm:closest-fair-np-hard}
The closest fair clustering problem is $\nph$ when the vertex set $V$ is divided into two disjoint groups, blue and red, and the number of blue vertices is $p$ times the number of red vertices, for some integer $p > 1$.
\end{restatable}

\paragraph{Contributions for Closest fair clustering (more than two colors):} We now extend our study to the multi-color setting, where the vertex set $V$ is partitioned into $|\chi| \geq 2$ disjoint color groups $c_1,\ldots,c_{|\chi|}$. 
For a subset $S \subseteq V$, let $c_i(S)$ denote the number of vertices of color $c_i$ in $S$. 
The complexity of the closest fair clustering problem naturally increases with the number of groups, as fairness must be enforced simultaneously across all colors. 
We analyze two important cases: (i) the balanced case, where all colors appear in equal proportion, and (ii) the general case, where color proportions may be arbitrary.

In the balanced case, where 
\[
1:1:\cdots:1 = c_1(V) : c_2(V) : \cdots : c_{|\chi|}(V)
\] 
we design an efficient algorithm that achieves a polylogarithmic approximation guarantee. 
In particular, our algorithm computes an $O(|\chi|^{1.6}\log^{2.81}|\chi|)$-close fair clustering with respect to the input clustering~$\m{C}$ in general for any number of colors and $O(|\chi|^{1.6})$-close fair clustering when the number of colors is a power of $2$.

\begin{restatable}{theorem}{equalmulti}
\label{thm:closest-fair-multicolor-balanced}
    Let $\chi$ be a set of colors and let $V$ be partitioned into disjoint color classes indexed by $\chi$, such that each color class has the same size and $|\chi| > 2$, then there exists a polynomial-time algorithm that computes an $O(|\chi|^{1.6}\log^{2.81}|\chi|)$-close fair clustering with respect to a given input clustering~$\m{C}$. Moreover, when $|\chi|$ is a power of two, the algorithm computes a $O(|\chi|^{1.6})$-close $\fair$.
\end{restatable}

We next consider the general case, where the ratio of group sizes is arbitrary:
\[
p_1 : p_2 : \cdots : p_{|\chi|} = c_1(V) : c_2(V) : \cdots : c_{|\chi|}(V).
\]
Here, we interpret \(p_1 : p_2 : \cdots : p_{|\chi|}\) as the ratio \(c_1(V) : c_2(V) : \cdots : c_{|\chi|}(V)\) reduced to its simplest form, so that each \(p_i\) is an integer. For this setting, we provide a general algorithm that computes an \(O(|\chi|^{3.81})\)-close fair clustering with respect to the input clustering \(\mathcal{C}\).

\begin{restatable}{theorem}{genmulti}
\label{thm:closest-fair-multicolor-general}
There is a polynomial time algorithm that, given an arbitrary clustering $\mathcal{C}$ over a vertex set $V$ where each vertex $v \in V$ has a color in $\chi = \{c_1, \ldots, c_k\}$, finds an $O(|\chi|^{3.81})$-close fair clustering to the clustering $\m{C}$.
\end{restatable}

\paragraph{Corollaries for unweighted fair correlation clustering:} Let us now state the corollaries we obtain for the unweighted fair correlation clustering problem. 

A series of works \cite{charikar2005clustering, Ailon2008JACM, Chawla2018JACM, 
CohenAddad2021MPC, CohenAddad2024Combinatorial, CohenAddad2022Online, Cao2025Dynamic} 
have established constant-factor approximation algorithms for the \emph{standard unweighted correlation clustering} problem, i.e., in the setting where no fairness constraints are imposed on the clustering. Recall in the unweighted correlation clustering problem, for each edge $(u,v)$ we have $w^+(u,v), w^-(u,v) \in \{0,1\}$.
These results represent significant progress toward understanding the approximability of unweighted correlation clustering. 
The current best-known approximation guarantee is due to Cao et al. \cite{Cao2025Dynamic}, 
who achieved an approximation factor of $1.437$ through a refined dynamic programming framework and combinatorial analysis.

\begin{theorem}[\cite{Cao2025Dynamic}]\label{thm:cc-constant}
For the standard unweighted correlation clustering problem (without fairness constraints), 
there exists a polynomial-time algorithm that computes a $1.437$-approximate solution.
\end{theorem}

\noindent By combining \cref{thm:closest-fair-to-correlation} with \cref{thm:closest-fair-multicolor-balanced} and \cref{thm:cc-constant}, we obtain the following corollary.  
 
\begin{corollary}\label{cor:correlation-balanced}
Let $\chi$ be a set of colors and let $V$ be partitioned into color classes indexed by $\chi$, such that $|\chi| \geq 2$ and the color groups are of equal size, 
then there exists a polynomial-time algorithm that computes an $O(|\chi|^{1.6}\log^{2.81}|\chi|)$-approximate unweighted fair correlation clustering.
\end{corollary}

The above corollary improves the state-of-the-art approximation factor of $O(|\chi|^2)$ by Ahmadian et al \cite{Ahmadian2020} and by Ahmadi et al \cite{ahmadi2020fair} for the unweighted fair correlation clustering problem.
\vspace{1em}

\noindent By combining \cref{thm:closest-fair-to-correlation} with \cref{thm:closest-fair-multicolor-general} and \cref{thm:cc-constant}, we obtain the following corollary. 

\begin{corollary}\label{cor:correlation-general}
There is a polynomial time algorithm that, given an arbitrary clustering $\mathcal{C}$ over a vertex set $V$ where each vertex $v \in V$ has a color in $\chi = \{c_1, \ldots, c_k\}$, finds an $O(|\chi|^{3.81})$-approximate unweighted fair correlation clustering.    
\end{corollary}

The above corollary eliminates the dependence on the $M = \max_{j \in [\chi]}p_j$ that appeared in the previous $O(M^2 |\chi|^2)$ bound of \cite{Ahmadian2020, ahmadi2020fair}, for unweighted fair correlation clustering where $M$ can be as large as a polynomial in $|V|$.

\paragraph{Corollaries for weighted fair correlation clustering under probability constraints:} Let us now state the corollaries related to weighted fair correlation clustering problem. By combining \cref{thm:closest-fair-to-correlation} with \cref{thm:closest-fair-p:1}, 
we obtain the following corollary. 

\begin{corollary}\label{cor:correlation-p:1}
If the vertex set $V$ is partitioned into two disjoint groups, blue and red, where the ratio of blue to red vertices in $V$ is $p:1$ for some integer $p > 1$, 
then there exists a polynomial-time algorithm that computes a $(17 + \rho + 17\rho)$-approximate unweighted fair correlation clustering under probability constraints, 
where $\rho$ is the state-of-the-art approximation factor for the weighted correlation clustering problem under probability constraints.
\end{corollary}

For the weighted correlation clustering problem under probability constraints, the state-of-the-art approximation factor is by Ailon et al \cite{Ailon2008JACM}, which provides a $2.5$ approximation. In addition to the probability constraints, if the weights satisfy the triangle inequality constraints, then the state-of-the-art approximation factor is by Chawla et al \cite{Chawla2018JACM}, which provides a $1.5$ approximation. Recall, the triangle inequality constraint needs the weights to satisfy for every triple $(u,v,w)$,
\[
w^+(u,v) \leq w^+(u,w) + w^+(w,v)
\quad \text{and} \quad
w^-(u,v) \leq w^-(u,w) + w^-(w,v).
\]

\begin{theorem}[\cite{Ailon2008JACM}]\label{thm:wcc-constant}
For the standard weighted correlation clustering problem under probability constraints (without fairness constraints), 
there exists a polynomial-time algorithm that computes a $2.5$-approximate solution.
\end{theorem}

\begin{theorem}[\cite{Chawla2018JACM}]\label{thm:wcc-pt-constant}
For the standard weighted correlation clustering problem under both probability and triangle inequality constraints (without fairness constraints), 
there exists a polynomial-time algorithm that computes a $1.5$-approximate solution.
\end{theorem}

\noindent By combining \cref{thm:closest-fair-to-correlation} with \cref{thm:closest-fair-multicolor-balanced} and \cref{thm:wcc-constant}, we obtain the following corollary.  
 
\begin{corollary}\label{cor:weighted-correlation-balanced}
Let $\chi$ be a set of colors and let $V$ be partitioned into disjoint color classes indexed by $\chi$, such that $|\chi| \geq 2$ and the color groups are of equal size, 
then there exists a polynomial-time algorithm that computes an $O(|\chi|^{1.6}\log^{2.81}|\chi|)$-approximate weighted fair correlation clustering under probability constraints.
\end{corollary}

\noindent By combining \cref{thm:closest-fair-to-correlation} with \cref{thm:closest-fair-multicolor-general} and \cref{thm:wcc-constant}, we obtain the following corollary. 

\begin{corollary}\label{cor:weighted-correlation-general}
There is a polynomial time algorithm that, given an arbitrary clustering $\mathcal{C}$ over a vertex set $V$ where each vertex $v \in V$ has a color in $\chi = \{c_1, \ldots, c_k\}$, finds an $O(|\chi|^{3.81})$-approximate weighted fair correlation clustering under probability constraints. 
\end{corollary}

In addition to the probability constraints, if the weights also satisfy the triangle inequality, then \cref{thm:wcc-pt-constant} yields improved constant factors hidden in the \(O\)-notation of the approximation guarantees in \cref{cor:weighted-correlation-balanced} and \cref{cor:weighted-correlation-general}.

We provide the first approximation algorithms for the weighted correlation clustering problem under probability constraints. For brevity, throughout the remainder of this thesis, we refer to weighted correlation clustering with probability constraints simply as \emph{weighted correlation clustering}.

\paragraph{Corollaries for fair consensus clustering:} Similarly, we can state the corollaries for the fair consensus clustering problem. The following corollaries establish the first known approximation guarantees for the fair consensus clustering problem. 

By combining \cref{thm:closest-fair-to-consensus} with \cref{thm:closest-fair-p:1}, we obtain the following corollary.  

\begin{corollary}\label{cor:consensus-p:1}
If the vertex set $V$ is partitioned into two disjoint groups, blue and red, where the ratio of blue to red vertices in $V$ is $p:1$ for some integer $p > 1$, 
then there exists a polynomial-time algorithm that computes a $19$-approximate fair consensus clustering.
\end{corollary}

By combining \cref{thm:closest-fair-to-consensus} with \cref{thm:closest-fair-multicolor-balanced}, we obtain the following corollary.  

\begin{corollary}\label{cor:consensus-balanced}
Let $\chi$ be a set of colors and let $V$ be partitioned into disjoint color classes indexed by $\chi$, such that $|\chi| \geq 2$ and the color groups are of equal size, 
then there exists a polynomial-time algorithm that computes an
$O(|\chi|^{1.6}\log^{2.81}|\chi|)$-approximate fair consensus clustering.
\end{corollary}

By combining \cref{thm:closest-fair-to-consensus} with \cref{thm:closest-fair-multicolor-general}, we obtain the following corollary.  

\begin{corollary}\label{cor:consensus-general}
There is a polynomial time algorithm that, given an arbitrary clustering $\mathcal{C}$ over a vertex set $V$ where each vertex $v \in V$ has a color in $\chi = \{c_1, \ldots, c_k\}$, finds an $O(|\chi|^{3.81})$-approximate fair consensus clustering.  
\end{corollary}

\section*{Fair Consensus Clustering in Streaming Setting}

We study fair consensus clustering in the following streaming model. The input consists of a sequence of triples $(p,j,b)$, where
\begin{itemize}
    \item $p=(u,v)$ is an unordered pair of vertices in $V$,
    \item $j \in [m]$ identifies an input clustering $\mathcal{C}_j$, and
    \item $b \in \{0,1\}$ indicates whether vertices $u$ and $v$ belong to the same cluster in $\mathcal{C}_j$ ($b=0$) or to different clusters ($b=1$).
\end{itemize}

The triples $(p,j,b)$ arrive in an arbitrary order. We refer to this streaming model as $\pconstream$. This model is inspired by prior work on unweighted correlation clustering in the streaming setting, where edges and their labels arrive arbitrarily in a stream \cite{ahn2015correlation, behnezhad2023single, chakrabarty2023single}.


To the best of our knowledge, there has been no prior work explicitly addressing even the standard (unfair) consensus clustering problem in the streaming setting. While it is plausible that techniques based on coresets in streaming models \cite{rosman2014coresets, braverman2016new} could yield approximation algorithms for standard consensus clustering, no such results have been established. In contrast, we provide the first approximation algorithms for \emph{fair} consensus clustering in the streaming setting.

Our main result for this model is summarized in the following theorem.

\begin{restatable}{theorem}{streaming}
\label{thm:explicit.fair.consensus.clustering.randomized}
Suppose there exists an approximation algorithm for fair correlation clustering with running time $\runtimeFairCorCls{n}$, and assume oracle access to an $\alpha$-approximate closest fair clustering algorithm. Then, there exists a $(\approxFactorClsFair + 1.995)$-approximation algorithm for fair consensus clustering in the $\pconstream$ that uses $O(n \log m)$ space and has query time
\[
O\!\left(n^{2}\log^{4} m + \runtimeFairCorCls{n}\log^{3} m\right).
\]
\end{restatable}

\paragraph{Corollaries for fair consensus clustering in streaming setting:} The above \cref{thm:explicit.fair.consensus.clustering.randomized} assumes oracle access to an $\alpha$-approximate closest fair clustering algorithm, which we have already established in previous sections. Consequently, we obtain the following corollaries for fair consensus clustering in the $\pconstream$.

By combining \cref{thm:explicit.fair.consensus.clustering.randomized} with \cref{thm:closest-fair-p:1}, we obtain the following corollary.  

\begin{corollary}\label{cor:stream-consensus-p:1}
If the vertex set $V$ is partitioned into two disjoint groups, blue and red, where the ratio of blue to red vertices in $V$ is $p:1$ for some integer $p > 1$, 
Then, there exists a $18.995$-approximation algorithm for fair consensus clustering in the $\pconstream$ using $O(n \log m)$ space.
\end{corollary}

By combining \cref{thm:explicit.fair.consensus.clustering.randomized} with \cref{thm:closest-fair-multicolor-balanced}, we obtain the following corollary.  

\begin{corollary}\label{cor:stream-consensus-balanced}
Let $\chi$ be a set of colors and let $V$ be partitioned into disjoint color classes indexed by $\chi$, such that $|\chi| \geq 2$ and the color groups are of equal size,
then, there exists a $O(|\chi|^{1.6}\log^{2.81}|\chi|)$-approximation algorithm for fair consensus clustering in the $\pconstream$ using $O(n \log m)$ space.
\end{corollary}

By combining \cref{thm:explicit.fair.consensus.clustering.randomized} with \cref{thm:closest-fair-multicolor-general}, we obtain the following corollary. 

\begin{corollary}\label{cor:stream-consensus-general}
There is a polynomial time algorithm that, given an arbitrary clustering $\mathcal{C}$ over a vertex set $V$ where each vertex $v \in V$ has a color in $\chi = \{c_1, \ldots, c_k\}$, finds an $O(|\chi|^{3.81})$-approximate fair consensus clustering in the $\pconstream$ using $O(n \log m)$ space. 
\end{corollary}

\section{Related Works}
In this section, we write the related works for both fault-tolerant reachability preservers and fair clustering.

\section*{Related Works of Reachability Preservers}

A simple version of a reachability preserver is when there is a single source vertex $s$, and we would like to preserve reachability from $s$ to all other vertices. As mentioned previously, Baswana, Choudhary and Roditty \cite{BCR16} provided an efficient construction of a $k$-fault-tolerant single-source reachability preserver of size $O(2^kn)$. Further, they showed that this upper bound on the size of a preserver is tight up to some constant factor. For the standard static setting (with no faulty edges), a much better bound is known. We know that even to preserve all the pairwise distances, not just reachability, there is a subgraph of size $O\big(n+\min(n^{2/3}|\m{P}|, n\sqrt{|\m{P}|})~\big)$ \cite{CE06, Bodwin17}. Later Abboud and Bodwin \cite{AB18} showed that for any directed graph $G=(V,E)$ given a set $S$ of source vertices and a pair-set $\m{P}\subseteq S \times V$ we can construct a pairwise reachability preserver of size $O\big(n+\min(\sqrt{n |\m{P}| |S|}, (n|\m{P}|)^{2/3})~\big)$. It is further shown that for any integer $d \ge 2$ there is an infinite family of $n$-vertex graphs and vertex-pair sets $\m{P}$ for which any pairwise reachability preserver must be of size $\Omega\big(n^{2/(d+1)}|\m{P}|^{(d-1)/d}\big)$. Note, for undirected graphs, storing spanning forests is sufficient to preserve pairwise reachability information, and thus we can always get a linear size reachability preserver for undirected graphs. We would like to emphasize that all our results in this paper hold for directed graphs.

By \cite{BCR16} we immediately get an oracle of size $O(2^k n)$ for $k$ edge (or vertex) failures with query time $O(2^k n)$. For just dual failures, we have an $O(n)$ size oracle with $O(1)$ query time due to \cite{Choudhary16}. 

For undirected graphs, the optimal bound of $O(kn)$ edges for $k$-fault-tolerant connectivity preserver directly follows from $k$-edge (vertex) connectivity certificate constructions provided by Nagamochi and Ibaraki \cite{NagamochiI:92}. For connectivity oracle, P\v{a}tra\c{s}cu and Thorup \cite{PatrascuT:07} presented a data structure of $O(m)$ size that can handle any $k$ edge failures in $O(k\log^2n\log\log n)$ time to subsequently answer connectivity queries between any two vertices in $O(\log\log n)$ time. For small values of $k$, Duan and Pettie \cite{DuanP:10} improved the update time of \cite{PatrascuT:07} to $O(k^2\log\log n)$ by presenting a data structure of $\widetilde O(m)$ size. For handling vertex failures, Duan and Pettie \cite{DuanP:17} provided a data structure of $O(mk\log n)$ size with $O(k^3 \log^3n)$ update time and $O(k)$ query time. Other closely related problems that have been studied in the fault-tolerant model include computing distance preservers \cite{DTCR08, PP13, Parter15}, depth-first-search tree \cite{BCCK19}, spanners \cite{CLPR09, DK11}, approximate distance preservers \cite{BK13, PP14, BGLP16}, approximate distance oracles \cite{DP09, CLPR10}, compact routing schemes \cite{CLPR10, Chechik13}.

\paragraph{Work done after our work.} 
More recently, Bodwin and Le \cite{BodwinLe2025} studied the problem of \emph{online reachability preservers}, where demand pairs arrive one at a time and the algorithm must irrevocably add edges to the preserver before seeing the next pair. They showed that in this setting, one can construct preservers of size 
\[
|E(H)| \leq O\big(n^{0.72}|\m{P}|^{0.56} + n^{0.6}|\m{P}|^{0.7} + n\big),
\]
polynomially improving the previous best known online upper bound of $O(\min\{n|\m{P}|^{1/2}, n^{1/2}|\m{P}|\}+n)$ (implicit in Coppersmith and Elkin \cite{CE06}). They also studied the \emph{source-restricted online setting}, where all demand pairs have one endvertex in a set $S$, and achieved preservers of size $O((n|\m{P}||S|)^{1/2}+n)$, matching the offline upper bounds in this regime. Beyond upper bounds, their work connected online preservers to the competitive ratio of Online Directed Steiner Forest, improving the state-of-the-art ratio from $O(n^{2/3+\varepsilon})$ by Grigorescu et al \cite{grigorescu2021online} to $O(n^{3/5+\varepsilon})$.

In parallel, other recent developments have further refined the trade-offs between preserver size and adaptivity in the online setting, including results on non-adaptive constructions where all paths must be fixed in advance. These advances highlight that, unlike the offline model where tight extremal bounds are better understood, the online setting still presents significant gaps and open directions.

\section*{Related Works of Fair Clustering}

The study of fair clustering was initiated by Chierichetti et al. \cite{Chierichetti2017}, who introduced the \emph{fairlet} framework for $k$-center/$k$-median/$k$-means clustering with two colors, later extended to multiple colors by R\"osner and Schmidt \cite{rosner2018privacy}. Their proportional-fairness model requires that the demographic distribution within each cluster reflect the global proportions of the dataset. Building on this foundation, Ahmadian et al. \cite{Ahmadian2020} proposed an alternative model where, for each color $c_i$, the number of $c_i$ vertices in any cluster is bounded by a parameter $\sigma_i|F|$, offering finer control of demographic representation. 

 A number of subsequent works explored complementary formulations of fairness. Huang et al. \cite{huang2019coresets} introduced coreset constructions that preserve both cost and fairness constraints, enabling scalable fair $k$-median/$k$-means clustering. Backurs et al. \cite{Backurs2019} developed scalable approximation algorithms that extend fairlet ideas to near-linear running time. Chen et al. \cite{Chen2019} studied \emph{proportionally fair clustering}, requiring each cluster to approximately reflect global proportions, and provided bicriteria approximations. Bera et al. \cite{Bera2019} handled overlapping sensitive groups, ensuring fairness constraints are satisfied simultaneously for multiple (possibly intersecting) populations. Esmaeili et al. \cite{esmaeili2020probabilistic} proposed a group-oblivious probabilistic model, relaxing hard constraints to fairness-in-expectation, useful when group memberships are uncertain.

Beyond geometric clustering, fairness has also been studied in \emph{unweighted correlation clustering}. Ahmadian et al. \cite{Ahmadian2020} initiated this direction by incorporating fairness into the minimize-disagreements objective, followed by alternative formulations \cite{ahmadi2020fair} and improved guarantees \cite{ahmadian2023improved}. Finally, fairness has been extended to rank aggregation, which can be viewed as 1-clustering over permutations. Wei et al. \cite{wei22} introduced proportionate fairness in aggregating rankings, and Chakraborty et al. \cite{chakraborty2022} developed constant-factor approximations for fair rank aggregation.

\section*{Related Works of Fair Clustering in Streaming setting}

Fair clustering was formalized by Chierichetti et al.\cite{Chierichetti2017}, with subsequent work focusing on improving scalability and approximation guarantees under fairness constraints.
A major bottleneck in early approaches was the cost of computing \emph{fairlet decompositions}, which limited applicability to large datasets.
Backurs et al. \cite{backurs2019scalable} addressed this issue by proposing a near-linear time algorithm for fair \(k\)-median clustering.
Their approach leverages probabilistic embeddings into hierarchically well-separated trees (HSTs) to compute \((r,b)\)-fairlet decompositions efficiently, supporting general balance parameters and enabling scalability to massive datasets.

Moving beyond offline settings, Bera et al. \cite{bera2022fair} initiated the study of \emph{fair \(k\)-center clustering} in distributed and streaming models.
They presented the first scalable algorithms in the MapReduce (MPC) and streaming settings, including a constant-round MPC algorithm and a two-pass streaming algorithm with constant-factor approximation guarantees.
A key technical contribution is a compact weighted summary that significantly reduces the size of the fairness LP while preserving solution quality, enabling practical performance on large-scale data.

Lin et al. \cite{lin2024streaming} further advanced fair \(k\)-center clustering in the streaming model.
They proposed a two-pass streaming algorithm achieving a \((3+\varepsilon)\)-approximation for the two-group case using \(O(k \log n)\) memory, matching the best-known offline guarantees.
They also developed a one-pass streaming algorithm with a \((7+\varepsilon)\)-approximation and extended their framework to handle an arbitrary number of demographic groups without increasing asymptotic space complexity.

Subsequently, Guo et al. \cite{guo2025improved} improved the state of the art by designing a one-pass streaming algorithm with a tight approximation ratio of \(5\) using \(O(k \log n)\) memory.
Their approach is based on \(\lambda\)-independent center sets and a refined post-streaming selection procedure.
Additionally, they showed improved guarantees for semi-structured data streams and demonstrated that their algorithm can be adapted to the offline setting to recover a \(3\)-approximation, matching the best known bounds.

While the above works focus primarily on insertion-only streams, Cohen-Addad et al. \cite{cohen2025fair} studied fair clustering in the more challenging \emph{sliding window model}, motivated by recency constraints and data-retention policies.
They established a strong impossibility result, showing that any algorithm achieving a finite multiplicative approximation under exact proportional fairness requires linear space.
To overcome this barrier, they proposed relaxing fairness constraints by a \((1\pm\varepsilon)\) factor and designed a sublinear-space algorithm based on assignment-preserving coresets, demonstrating a fundamental separation between insertion-only and sliding-window streaming models for fair clustering.

An important technique in streaming algorithms for clustering is the construction of coresets. A \emph{coreset} is a small, weighted subset of the input data that approximately preserves the value of an optimization objective for all feasible solutions, thereby enabling efficient algorithms on massive datasets.
Coresets are particularly valuable in streaming and distributed settings due to their composability and mergeability properties.
In the context of fair clustering, however, classical coreset constructions often fail to preserve fairness constraints, motivating the development of \emph{fair coresets} that simultaneously approximate clustering cost while respecting group-based or individual fairness requirements.

Schmidt et al. \cite{schmidt2019fair} initiated the study of fair coresets by introducing new coreset definitions for fair \(k\)-means that restore composability and admit insertion-only streaming algorithms.
Huang et al. \cite{huang2019coresets} extended this line of work by constructing \(\varepsilon\)-coresets for fair \(k\)-median and \(k\)-means under multiple, possibly overlapping sensitive groups, with sizes independent of the dataset.
Chhaya et al. \cite{chhaya2022coresets} further generalized fair coresets to data-dependent constraints, providing the first coreset frameworks for fair regression and individually fair clustering.
More recently, Mahabadi and Trajanovski \cite{mahabadi2023core} developed constant-factor composable coresets for fair diversity maximization under multiple distance-based objectives, demonstrating both strong theoretical guarantees and significant empirical speedups.

\section{Technical Overview}

In this section, we provide a technical overview of both reachability preservers and fair clustering.

\section*{Reachability Preservers}
\begin{figure}[!ht]
\centering
\includegraphics[width=200pt,height=300pt,keepaspectratio]{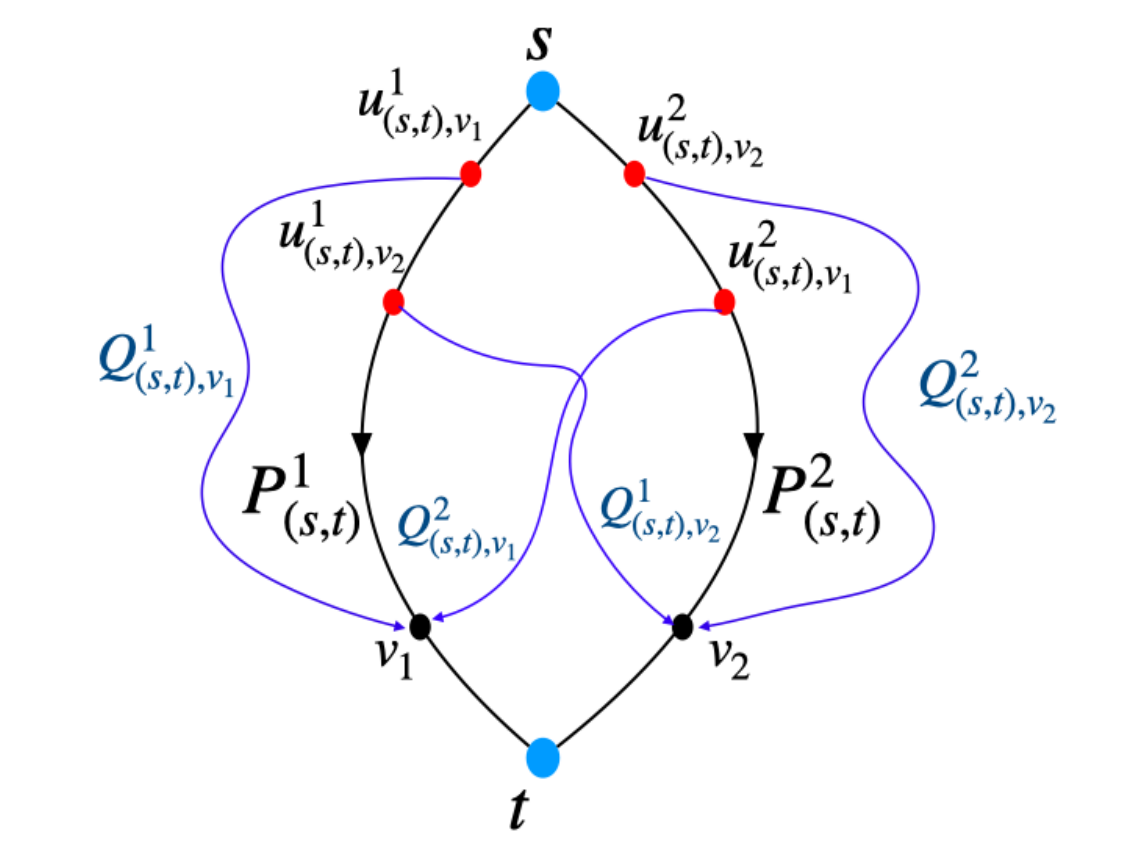}
\caption{$H_{(s,t)}$=2-{\ftrs} for a single pair $(s,t)$. Two black paths are the outer strands and the purple paths are the coupling paths between them.}
\label{fig:single-pair-ftrs}
\end{figure}

Here, we provide a high-level idea for the construction of 2-{\ftrs}. Let us start with a simple construction of a linear (in the number of vertices) sized 2-{\ftrs} for a single pair. Recall, we already know of such a subgraph of $O(n)$ size by \cite{BCR16}. However, this new alternate construction will shed more light on the specific structure of a 2-{\ftrs}, which will play a pivotal role in our constructions of 2-{\ftrs} for multiple pairs. Given a directed graph $G$ and a vertex-pair $(s,t)$, we construct a subgraph $H_{(s,t)}$ as follows: First, consider two "maximally disjoint" $s-t$ paths $P_{(s,t)}^1$ and $P_{(s,t)}^2$ (paths that meet only at the cut-edges and cut-vertices). We refer to these two paths as \emph{outer strands}. Next, we add several \emph{coupling paths} between these two outer strands, which are edge-disjoint from the outer strands. For each vertex $v$ on the outer strands, we check for the "earliest" vertex on the strand $P_{(s,t)}^1$ (and $P_{(s,t)}^2$), from which there is a path $Q_{(s,t),v}^1$ (and $Q_{(s,t),v}^2$) to $v$ that is edge-disjoint with both the outer strands. We refer to these paths $Q_{(s,t),v}^i$  
as coupling paths. Roughly speaking, two outer strands together with the coupling paths constitute the subgraph $H_{(s,t)}$ (see Figure~\ref{fig:single-pair-ftrs}). The actual construction is slightly different. Let us first briefly discuss why the above subgraph is a 2-{\ftrs} for the pair $(s,t)$. Then we will comment on the issue with the above simple construction and how we overcome that.

Consider any two failure edges $f_1$, $f_2$. W.l.o.g. assume, they do not form an $s-t$ cut-set; otherwise, after the failure there won't be any $s-t$ path. Now if both $f_1,f_2$ lie on one of the two outer strands (i.e., either on $P_{(s,t)}^1$ or $P_{(s,t)}^2$), then since these two strands are maximally disjoint, one of them will survive after the failures. So, let $f_1$, $f_2$ lie on the strand $P_{(s,t)}^1$, $P_{(s,t)}^2$ respectively. Then consider the subpaths of $P_{(s,t)}^1$, $P_{(s,t)}^2$ above $f_1$, $f_2$, and the subpaths of $P_{(s,t)}^1$, $P_{(s,t)}^2$ below $f_1$, $f_2$. Since by assumption $f_1, f_2$ does not form an $s-t$ cut-set, there must be a coupling path (edge-disjoint with $P_{(s,t)}^1$, $P_{(s,t)}^2$) from one of the top subpaths to one of the bottom subpaths in $G$. Since $H_{(s,t)}$ consists of all the coupling paths, we get a surviving path in $H_{(s,t)}\setminus \{f_1,f_2\}$. This shows that $H_{(s,t)}$ is a 2-$\ftrs(G,(s,t))$. Moreover, one may observe from the above argument that, after failure of any two edges, one of the surviving paths in $H_{(s,t)}$ must be of the following form: It first follows one of the outer strand from $s$ to some vertex $u$, then takes a coupling path till some vertex $v$ on one of the outer strands, and finally follows the corresponding outer strand from $v$ to $t$. We refer to such a path as \emph{nice path}. The existence of such nice paths helps us in proving the correctness of our 2-{\ftrs} construction for multiple pairs in the subsequent sections.

As we mentioned earlier, our actual construction is slightly different. The main issue with the above simple construction is that the constituted subgraph could be of size $\omega(n)$ (which is worse than \cite{BCR16}) after adding all the coupling paths. To mitigate this issue, instead of adding all the coupling paths, we only add the "essential" coupling paths. (See \cref{sec:reachability-preservers} for details.) It allows us to achieve $O(n)$ size bound without affecting the correctness of 2-{\ftrs}. The guarantee of the existence of nice paths also remains unaffected. Of course, the correctness argument will become slightly more intricate.

Next, we use the above construction of a 2-{\ftrs} of a single pair to study the dual fault-tolerant reachability preserver. Our input is a directed graph $G=(V,E)$ with $n$ vertices, and a vertex-pair set $\m{P} \subseteq V \times V$.

Until now, we only know of $O(n |\m{P}|)$ size 2-{\ftrs} for any $n$-vertex graph $G$ and pair set $\m{P}$. In this work, we provide a deterministic polynomial time construction of a pairwise 2-{\ftrs} of size $O(n^{4/3} |\m{P}|^{1/3})$.  For simplicity, below, we briefly describe a construction that provides a slightly weaker bound, in particular, $O(n \sqrt{|\m{P}|} \log n)$. In the actual construction, we get rid of the $\log n$ factor by constructing a preserver with slack and then using the result of \cite{bodwin2020note}.

To get a sparse 2-{\ftrs} for a vertex-pair set $\m{P}$, we perform two-step sparsification. First, we apply our alternate construction of 2-{\ftrs} for each of the pairs of $\m{P}$. Then take a union of all of these subgraphs to get a $O(n|\m{P}|)$ size intermediate subgraph $H_{inter}$, which is clearly a 2-{\ftrs} for $\m{P}$. Next, we further sparsify this intermediate subgraph. Then we consider the top and bottom $\Theta(n^{2/3}|\m{P}|^{-1/3})$ portions of the outer strands ($P_{(s,t)}^1$, $P_{(s,t)}^2$) of $H_{s,t}$, for each $(s,t)\in \m{P}$. Next, we construct a greedy hitting set containing $O(n^{1/3}|\m{P}|^{1/3} \log n)$ vertices that intersects all these subpaths. We compute linear-sized single-source and single-destination 2-{\ftrs} for each of these vertices in the hitting set using \cite{BCR16}. Let $H_1$ be the union of all these single-source and single-destination 2-{\ftrs}. So, $H_1$ is of size $O(n^{4/3} |\m{P}|^{1/3} \log n)$.

Consider any two failure edges $f_1,f_2$ and a pair $(s,t)\in \m{P}$. If there is a surviving path in $G\setminus \{f_1,f_2\}$, we know that there is a \emph{nice} $s-t$ path in $H_{(s,t)}\setminus \{f_1,f_2\}$. Using an argument similar to that in the oracle construction, if that nice path follows the top or bottom $\Theta(n^{2/3}|P|^{-1/3})$ portion of a outer strand, then there is also an $s-t$ path in $H_1 \setminus \{f_1,f_2\}$. So now on, it suffices to look into the case when the surviving nice path in $H_{(s,t)}\setminus \{f_1,f_2\}$ passes though a coupling path $Q_{(s,t),v}^i$, for some vertex $v$ lying on the bottom $\Theta(n^{2/3}|P|^{-1/3})$ length subpath of a outer strand, that starts from some vertex $u$ lying on the top $\Theta(n^{2/3}|P|^{-1/3})$ length subpath of a outer strand.

If we could include the top and bottom $\Theta(n^{2/3}|P|^{-1/3})$ portion of the outer strands, and all the ``essential'' coupling path $Q_{(s,t),v}^i$'s with endvertices lying on the top and bottom $\Theta(n^{2/3}|P|^{-1/3})$ portions of the outer strands, we will be done. We indeed consider a union of the top and bottom $\Theta(n^{2/3}|P|^{-1/3})$ portion of the outer strands  (for all $(s,t) \in \m{P}$), and let us denote that by $H_2$. So, $H_2$ is of size $O(n^{2/3}|P|^{2/3})$. Unfortunately, we do not have a guarantee on the length of the coupling paths. Thus, if we include all the required coupling paths, we cannot argue about the sparsity of the final subgraph. (This portion of the construction differs significantly from that of our oracle construction.) We consider the subgraph obtained by taking a union of all the "essential" coupling paths with endvertices lying on the top and bottom $\Theta(n^{2/3}|P|^{-1/3})$ portions of the outer strands. (Let us denote this union by $B$.) Then we sparsify this subgraph further. For that purpose, we first isolate all the "high frequency" vertices (iteratively) and remove all the coupling paths containing them. Since total number of coupling paths in this subgraph is only $\Theta(n^{2/3}|P|^{2/3})$, we end up with "a few" ($O(n^{1/3}|P|^{1/3})$) high frequency vertices. Now, observe, in the remaining subgraph (denoted as $H_4$), degree of each vertex is ``small'' (at most $O(n^{1/3}|P|^{1/3})$). Next, for each of the high-frequency vertices, compute linear-sized single-source and single-destination 2-{\ftrs}, and take a union of them to form a subgraph $H_3$. The union of $H_1,H_2,H_3$ and $H_4$ constitute the final subgraph.

It is not difficult to see that a surviving nice $s-t$ path in $H_{(s,t)}\setminus \{f_1,f_2\}$ either passes through one of the high frequency vertices, in which case we get an $s-t$ path in $H_3$; or is included in $H_2 \cup H_4$. Thus, the union of all $H_1, H_2, H_3$ and $H_4$ will be a 2-{\ftrs} for $\m{P}$. It is worth mentioning that the correctness proof works only because of a guarantee of the existence of the nice path. Note, a nice path follows at most one coupling path as a subpath. Thus either that nice path follows the top or bottom  $\Theta(n^{2/3}|P|^{-1/3})$ portion of the outer strands, or the coupling sub-path is part of $B$, which we further sparsify to get $H_3$ and $H_4$. Without the guarantee of a nice path, there could be many coupling sub-paths in a surviving $s-t$ path after failures. Endvertices of these coupling sub-paths may not lie on the top or bottom $\Theta(n^{2/3}|P|^{-1/3})$ portion of the outer strands. As a result, we miss them in $B$. As a consequence, $H_3 \cup H_4$ could not capture those coupling sub-paths.

Observe, each of the $H_i$'s is of size at most $O(n^{4/3}|\m{P}|^{1/3} \log n)$ (the $\log n$ factor is only there for $H_1$). So the total size is also $O(n^{4/3}|P|^{1/3} \log n)$.

\section*{Fair Clustering}

Let us first discuss the algorithms to find approximate weighted fair correlation clustering and fair consensus clustering, given that we have an algorithm to find an $\alpha$-close fair clustering to an unfair clustering $\m{C}$ given as input. 

\paragraph{Weighted Correlation Clustering.}
Suppose we are given two algorithmic primitives:  
(i) an $\alpha$-approximate closest fair clustering algorithm, and  
(ii) a $\rho$-approximation algorithm for standard weighted correlation clustering.  
We show that combining these two primitives yields a fair weighted correlation clustering algorithm with approximation factor $(\alpha + \rho + \alpha\rho)$.

We construct a composed algorithm, denoted by $\fcc$. Given a weighted correlation clustering instance $G$, we first apply the $\rho$-approximation algorithm to obtain a clustering $\mathcal{D}$ whose cost is within a factor $\rho$ of the optimum, but which is not necessarily fair. We then invoke the closest fair clustering algorithm on $\mathcal{D}$ to obtain a fair clustering $\mathcal{F}$ that is $\alpha$-close to $\mathcal{D}$.

We now show that $\mathcal{F}$ is a $(\alpha + \rho + \alpha\rho)$-approximation to the optimal weighted fair correlation clustering. Formally, we prove that
\[
\cost(\mathcal{F}) \leq (\alpha + \rho + \alpha\rho)\,\cost(\optcorr),
\]
where $\optcorr$ denotes an optimal (fair) weighted correlation clustering.

To establish this bound, we rely on the following two inequalities, which hold for any two clusterings $\mathcal{M}$ and $\mathcal{K}$ defined on the same vertex set $V$.

\begin{itemize}
    \item \emph{Inequality 1:} $\cost(\mathcal{M}) \leq \dist(\mathcal{M},\mathcal{K}) + \cost(\mathcal{K})$.
    \item \emph{Inequality 2:} $\dist(\mathcal{M},\mathcal{K}) \leq \cost(\mathcal{M}) + \cost(\mathcal{K})$.
\end{itemize}

Both inequalities follow directly from the definitions of the cost function $\cost(\cdot)$ and the distance measure $\dist(\cdot,\cdot)$.

We now complete the proof using these inequalities.

\begin{enumerate}
    \item By Inequality~1, we have
    \[
    \cost(\mathcal{F}) \leq \cost(\mathcal{D}) + \dist(\mathcal{D},\mathcal{F}).
    \]
    \item Since $\mathcal{D}$ is a $\rho$-approximation to the optimal clustering $\optcorr$, it follows that
    \[
    \cost(\mathcal{D}) \leq \rho\,\cost(\optcorr).
    \]
    \item Because $\mathcal{F}$ is $\alpha$-close to $\mathcal{D}$, we obtain
    \[
    \dist(\mathcal{D},\mathcal{F}) \leq \alpha\,\dist(\mathcal{D},\optcorr).
    \]
    \item Applying Inequality~2 to the term $\dist(\mathcal{D},\optcorr)$ and substituting the above bounds yields
    \[
    \cost(\mathcal{F}) \leq (\alpha + \rho + \alpha\rho)\,\cost(\optcorr).
    \]
\end{enumerate}

Hence, the resulting fair clustering $\mathcal{F}$ achieves a $(\alpha + \rho + \alpha\rho)$-approximation to the optimal weighted fair correlation clustering.

 \paragraph{Consensus Clustering:} Next, we propose an algorithm to find $(\alpha + 2)$-approximate \emph{Fair Consensus Clustering}, given that we have an algorithm to find an $\alpha$-close fair clustering to a given unfair clustering $\m{C}$. Formally, given $m$ input clusterings $\m{C}_1, \ldots, \m{C}_m$ our goal is to find a fair clustering $\m{F}$ such that
 \[
     \left( \sum_{i = 1}^n \dist(\m{C}_i, \m{F})^\ell\right)^{1/\ell} \leq  (\alpha + 2) \left( \sum_{i = 1}^n \dist(\m{C}_i, \optcon)^\ell\right)^{1/\ell}
\]
Here $\optcon$ is an optimal fair consensus clustering.

We begin by computing, for each \(\m{C}_i\), an \(\alpha\)-close fair clustering \(\m{F}_i\) using the closest fair clustering algorithm. Then, we output the fair clustering \(\m{F}_k\) that minimizes the overall \(\ell\)-mean objective.  
Next, we show that this approach guarantees a \((\alpha + 2)\)-approximation for \emph{Fair Consensus Clustering}.

Let \(\m{F}^*\) be an optimal fair consensus clustering, and let \(\m{C}_{i^*}\) be the input clustering closest to \(\m{F}^*\). Using our \(\alpha\)-close fair clustering algorithm, we obtain a fair clustering \(\m{F}_{i^*}\) such that $\dist(\m{C}_{i^*}, \m{F}_{i^*}) \leq \alpha \dist(\m{C}_{i^*}, \optcon)$.
 By applying the triangle inequality, we get $\dist(\m{F}_{i^*}, \optcon) \leq (\alpha + 1) \dist(\m{C}_{i^*}, \optcon)$.
Now, for any other input clustering \(\m{C}_j\), we have  

\[
\dist(\m{C}_j, \m{F}_{i^*}) \leq \dist(\m{C}_j, \optcon) + \dist(\m{F}_{i^*}, \optcon) \leq (\alpha + 2) \dist(\m{C}_j, \optcon),
\]  

since by assumption, \(\m{C}_{i^*}\) is the input closest to \(\optcon\), ensuring \(\dist(\m{C}_{i^*}, \optcon) \leq \dist(\m{C}_j, \optcon)\).  

Thus, we conclude that there exists an input clustering \(\m{C}_{i^*}\) whose corresponding fair clustering \(\m{F}_{i^*}\) achieves a \((\alpha + 2)\)-approximation. Since our algorithm selects the fair clustering $\m{F}_k$ that minimizes the overall objective, it guarantees an overall approximation of \((\alpha + 2)\).  

\paragraph{Closest fair clustering (two colors):} Let us now describe the technical overview of our approach when 
the vertex set $ V $ is divided into two disjoint color groups $\{\text{red, blue}\}$. For a cluster $ C_i $, let $ \blue{C_i} $ denote the blue vertices in $ C_i $, and $ \red{C_i} $ denote the red vertices in $ C_i $.  
Given a clustering $\mathcal{C}$, with the ratio between the number of blue and red vertices being $p:1$ for some integer $p\ge 1$,  we call $\mathcal{C}$ a \emph{fair clustering} if, for each $ i \in [w] $, the ratio between $|\blue{C_i}|$ and $|\red{C_i}|$ is exactly $ p:1 $. Next, we discuss how efficiently we can transform a given clustering $ \mathcal{C} $ into its nearest fair clustering under various integer values of $ p $. 

The algorithm proceeds in two steps:

\begin{enumerate}
    \item \textbf{Creating a $p$-divisible clustering:} Given an input clustering $\mathcal{C} = \{C_1, C_2, \dots, C_w\}$, we transform it into a new clustering $\mathcal{T} = \{T_1, T_2, \dots, T_\psi\}$ such that each cluster $T_i$ contains a number of blue vertices that is a multiple of $ p $. We call such a clustering $\m{T}$ a $p$-divisible clustering. We use the abbreviation $\bal$ for $p$-divisible clustering.

    \item \textbf{Making-Fair Step}: We further process the clusters $ T_1, T_2, \dots, T_\psi $ to obtain a fair clustering $\mathcal{F} = \{F_1, F_2, \dots, F_\iota\}$.
\end{enumerate}

Let $\mathcal{F}^*$ denote an optimal fair clustering closest to the input clustering $\mathcal{C}$, with an associated distance of $\text{OPT}$. By definition, $\mathcal{F}^*$ is also a $\bal$. 

If we propose an $\delta$-approximation algorithm for the balancing step, then the distance between $\mathcal{C}$ and $\mathcal{T}$ is at most $\delta \cdot \text{OPT}$. Furthermore, since the distance between $\mathcal{F}^*$ and $\mathcal{T}$ is at most $(1+\delta) \cdot \text{OPT}$ (by the triangle inequality), getting a $\alpha$-approximation algorithm for the making-fair step ensures that the distance between $\mathcal{T}$ and $\mathcal{F}$ is at most $(1+\delta) \alpha \cdot \text{OPT}$. 

Thus, the total distance between the input clustering $\mathcal{C}$ and the final output fair clustering $\mathcal{F}$ is at most $(\delta + \alpha + \delta \alpha) \cdot \text{OPT}$, leading to an overall approximation ratio of $(\delta + \alpha + \delta \alpha)$. This guarantees a constant-factor approximation when $\delta$ and $\alpha$ are constants.

\vspace{2em}

\noindent \textbf{$3.5$-approximation for creating $\bal$:} Let us provide an high level idea of the algorithm to create a $\bal$ $\m{T}$.

For every cluster $C_i$, we define the \emph{surplus} of $C_i$ as $(|\blue{C_i}| \mod p)$ and its deficit as $(p-(|\blue{C_i}| \mod p))$. To balance each cluster, we either need to remove surplus vertices or add the required number of deficit vertices from another cluster. For each cluster, we define the \emph{cut cost} as the cost of removing surplus vertices and the \emph{merge cost} as the cost of adding deficit vertices. 

Next, we define our cut and merge strategy. We start by classifying clusters into two categories. Call a cluster \emph{merge cluster} if its surplus is at least $p/2$; i.e., its merge cost is smaller than its cut cost. Otherwise, call it a \emph{cut cluster}. For a merge cluster $C_i$, it is more efficient to add $(p-(|\blue{C_i}| \mod p))$ vertices (its deficit), whereas, for a cut cluster $C_i$, it is optimal to remove $(|\blue{C_i}| \mod p)$ vertices (its surplus). A first-line approach is to remove surplus vertices from cut clusters and merge them into merge clusters, ensuring that the number of blue vertices in each cluster becomes a multiple of $p$. We start with this approach, referred to as $\algog$. However, we cannot guarantee that the total surplus across all clusters matches the total deficit. As a result, after this step, we may be left with only cut clusters or only merge clusters.

\begin{figure}[t]
\centering
    \includegraphics[width=11cm]{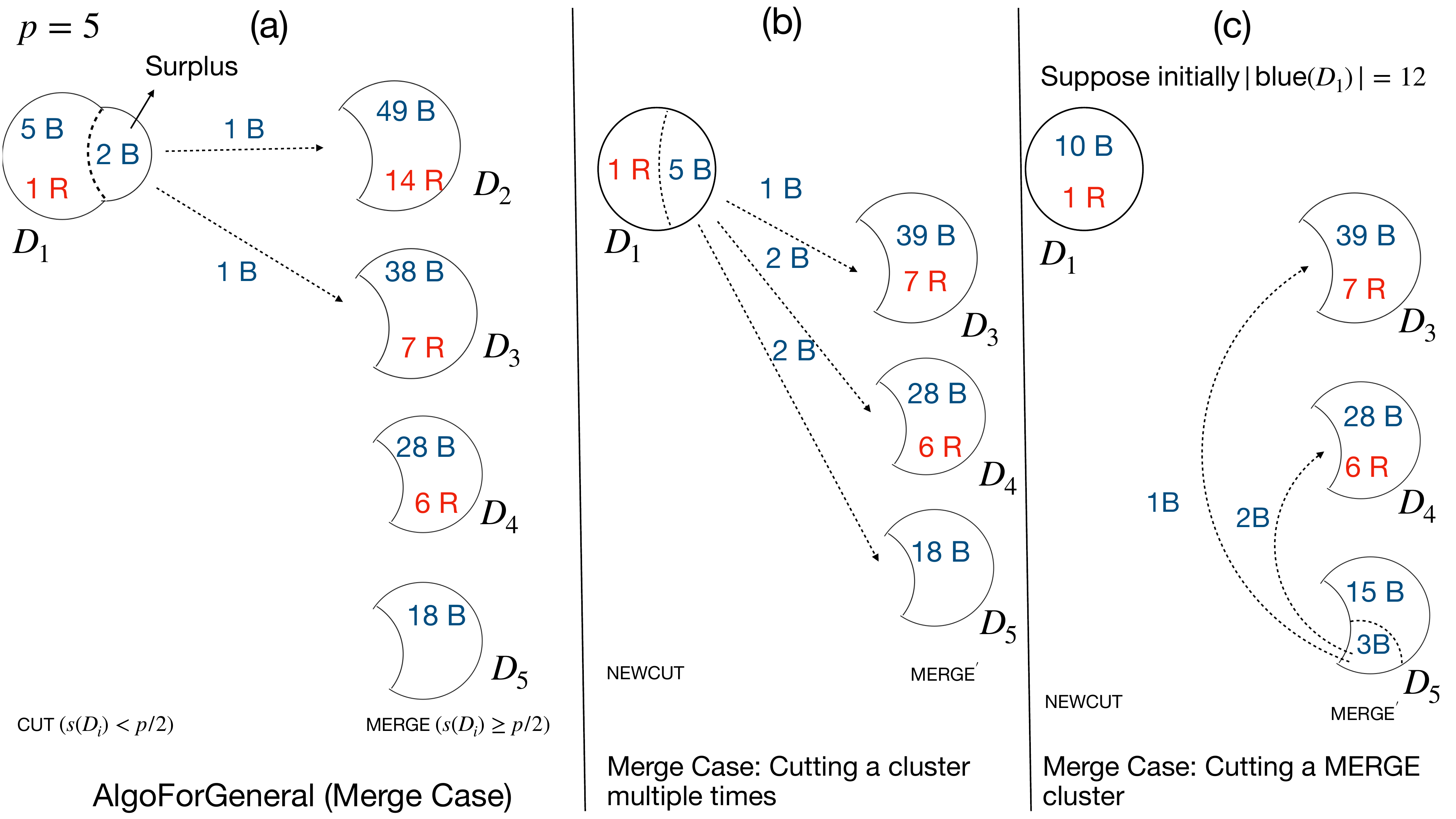}
    \caption{\label{fig:merge-p}Visualization of the scenario when only merge clusters remain after executing AlgoGeneral, which converts (a) to (b). Now, the total deficit $W=5$, and thus, depending on the $cut cost- merge cost$, either an already cut cluster needs to be cut further (depicted in (b)), or a single merge cluster needs to be cut (depicted in (c)). In each cluster, \textcolor{blue}{$xB$} (resp., \textcolor{red}{$yR$}) denotes that the number of \textcolor{blue}{blue} (resp., \textcolor{red}{red}) vertices is $x$ (resp., $y$).
    }
\end{figure}

If only merge clusters remain, then we call our subroutine $\algom$. Here, since we cannot satisfy all merge requests, we need to cut from some merge clusters to provide the necessary vertices. This is challenging, as cutting is now costlier than merging for the remaining unbalanced clusters. The key intuition here is that if cutting is necessary, we need a nontrivial strategy to optimize the cost. To achieve this, we sort the merge clusters in non-increasing order of $\ccost- \mcost$, ensuring optimal cutting decisions. An important observation is that the total deficit $W$ across all clusters is a multiple of $p$. This follows from the existence of a fair clustering, which guarantees that there exists a way to process the input set of clusters such that the number of blue vertices in each cluster is a multiple of $p$. Additionally, if we cut $k$ vertices from a cluster $C_i$ to fulfill the $k$ deficit of another cluster $C_j$, we effectively balance a total deficit of $p$ vertices. This follows intuitively because if $k$ is the surplus of $C_i$, then its deficit is $p-k$, ensuring that the combined deficit of $C_i$ and $C_j$ sums up to $p$. Since the total deficit is $W$, we repeat this operation exactly $W/p$ times, which is crucial for bounding the overall approximation cost. (See~\cref{fig:merge-p}.) 

\begin{figure}[t]
\centering
    \includegraphics[width=11cm]{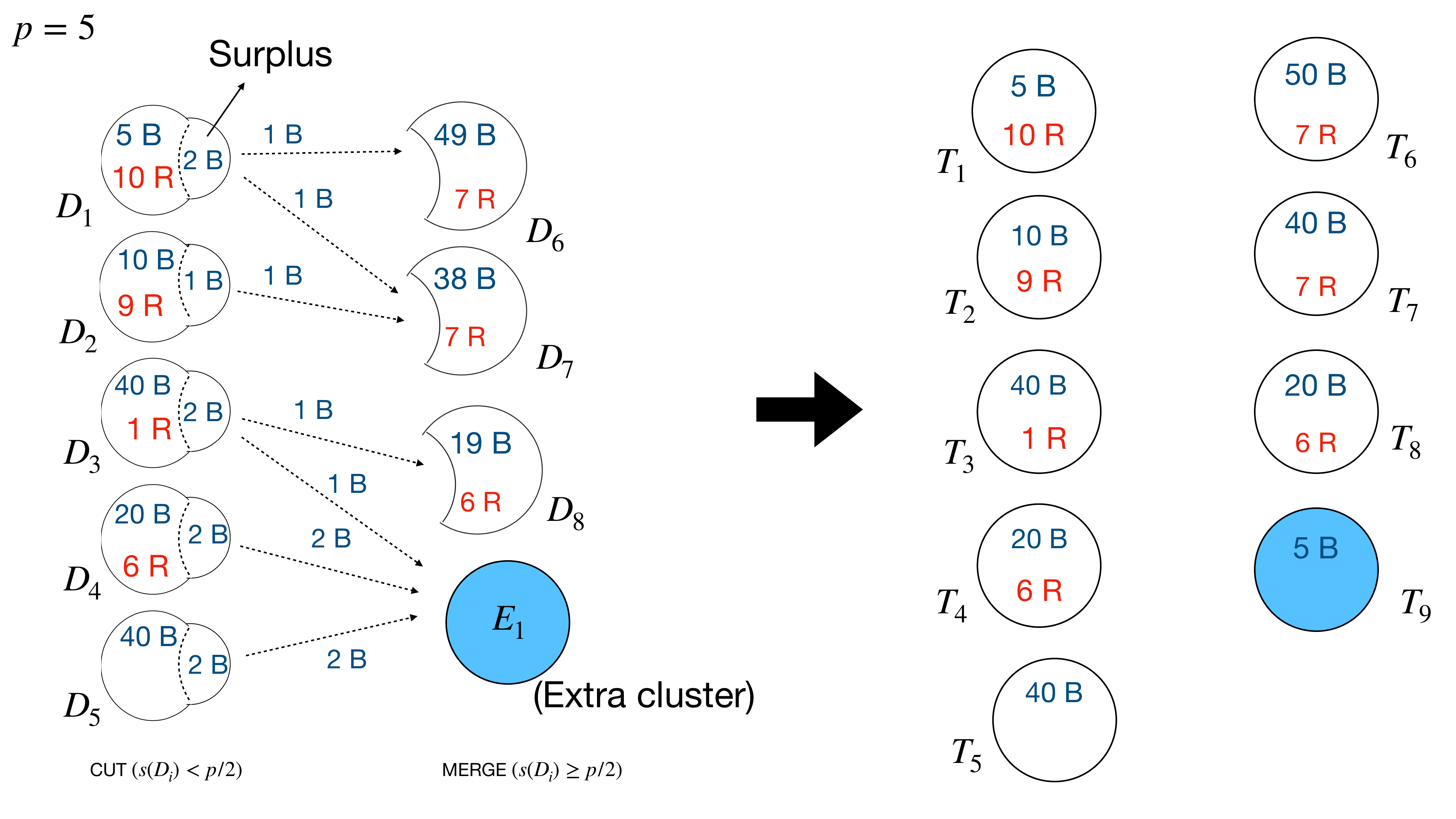}
    \caption{\label{fig:cut-p}Visualization of the scenario when only cut clusters remain after executing AlgoGeneral. The extra surplus vertices of clusters $D_3, D_4, D_5$ form an extra cluster $E_1$ of size $p$, containing only blue vertices. In each cluster, \textcolor{blue}{$xB$} (resp., \textcolor{red}{$yR$}) denotes that the number of \textcolor{blue}{blue} (resp., \textcolor{red}{red}) vertices is $x$ (resp., $y$).
    }
\end{figure}

Next, we consider the simpler case where, after the initial cut-merge processing, we are left only with cut clusters, each having a surplus of size $<p/2$, in this case, we call our subroutine $\algoc$. Here, we simply remove these surplus vertices from each cluster and combine them to form clusters of size exactly $p$, containing only blue vertices. A key question is why forming clusters of size $p$ is the optimal choice rather than larger clusters. The reason is that ensuring size $p$ may require partitioning surplus vertices from specific clusters, incurring additional costs. For example, consider the case where there are three cut clusters with surpluses of $p/3$, $p/3$, and $p/2-1$. Following our strategy to form a cluster of size $p$, we must partition at least one existing cluster. One might wonder whether it is more efficient to combine even more clusters and create a cluster of size $2p$ instead of $p$. However, through careful analysis, we show that accommodating clusters of size $>p$ results in higher merge costs than the partition cost. Thus, forming clusters of size exactly $p$ is the optimal strategy for combining surplus vertices efficiently. (See~\cref{fig:cut-p}.) 

\vspace{2mm}

\noindent \textbf{Making a $\bal$ Fair:}
Given a $\bal$ $\mathcal{T} = \{T_1, T_2, \dots, T_\psi\}$ here we discuss the step to convert it to a fair clustering $\mathcal{F} = \{F_1, F_2, \dots, F_\iota\}$. For this, we present a $ 3$-approximation algorithm $\algmf$. 

 We define a cluster $ T_i $ as $\tr$ if it has surplus red vertices, meaning $ |\blue{\outcl{i}}| < p |\red{\outcl{i}}| $.  
Let $ \surp{T_i} $ denote the set of surplus red vertices in $ T_i $, such that $ |\surp{T_i}| = |\red{\outcl{i}}| - (1/p) |\blue{\outcl{i}}| $ and $\surp{T_i} \subseteq \red{T_i}$.
Otherwise, we define $ T_i $ as $\tb$ if it has a deficit of red vertices, meaning $ |\blue{\outcl{i}}| > p |\red{\outcl{i}}| $.  
Let $ \defi{T_i} $ represent the set of  deficit red vertices deficit in $ T_i $, given by $ |\defi{T_i}| = (1/p) |\blue{\outcl{i}}| - |\red{\outcl{i}}| $ and $\defi{T_i} \subseteq V \setminus T_i$.

Next, we discuss how to eliminate surplus red vertices and redistribute them to address the deficits. To achieve this, we simply remove the surplus vertices from clusters in $ \tr $ and reassign them to clusters in $ \tb $ to compensate for the deficit.  
A crucial observation is that the existence of a fair clustering guarantees that the total surplus from all clusters in $ \tr $ matches the total deficit across all clusters in $ \tb $, thereby making the redistribution process well-defined.
We now proceed to analyze the approximation.

We first claim that, since $p > 1$ removing surplus vertices and redistributing them to fulfill deficits is indeed optimal.  
A key subtlety we need to address is the case where a specific surplus $ \surp{T_i} $ is distributed across multiple clusters to fulfill their deficits. This redistribution involves splitting $ \surp{T_i} $, which incurs an additional cost. However, even in an optimal fair clustering, $ \surp{T_i} $ must be transformed into a fair cluster, requiring the merging of $ p|\surp{T_i}| $ blue vertices with $ \surp{T_i} $. 
The cost of this merging is $ p|\surp{T_i}|^2 $, while the maximum splitting cost of $ \surp{T_i} $ is at most $ |\surp{T_i}|^2/2 $. Since $ p > 1 $, the splitting cost of $ \surp{T_i} $ remains bounded by its merging cost, ensuring the overall $3$-approximation guarantee. In the case where multiple surplus clusters contribute to fulfilling a deficit $ \defi{T_i} $, our algorithm incurs an intra-cluster cost of at most $ |\defi{T_i}|^2/2 $. Since $ p > 1 $, a similar argument as before ensures that, despite this additional cost, we still remain within the claimed approximation guarantee.

\paragraph{Closest Fair Clustering (more than two colors):} Let us now present a technical overview of our approach for constructing a closest fair clustering to a given input clustering $\m{C}$ when there are more than two colored groups. Our contributions can be broadly divided into three parts:
\begin{enumerate}[(i)]
\item When the vertex set $V$ is partitioned into disjoint color groups of equal size, and the total number of colors $|\chi|$ is a power of two.
\item When the vertex set $V$ is partitioned into disjoint color groups of equal size, but the number of colors $|\chi|$ is arbitrary.
\item When the vertex set $V$ is partitioned into disjoint color groups of arbitrary sizes.
\end{enumerate}

Let us now describe the common theme of our algorithms for the above three cases.
Given an input clustering $\m{C}$, we construct a fair clustering $\m{F}$
through a sequence of intermediate clusterings
\[
    \m{C} = \m{J}^0 \rightarrow \m{J}^1 \rightarrow \m{J}^2 \rightarrow \cdots \m{J}^{t - 1} \rightarrow \m{J}^t = \m{F}.
\]
At each intermediate step $\m{J}^i$, the algorithm progressively enforces
fairness across subsets of colors by satisfying a \emph{locally fair condition}.
Specifically, in each iteration, the color set is partitioned into smaller
blocks, and within each block, the algorithm balances color proportions using
a surplus–deficit adjustment procedure similar to the two-color case. As the
process proceeds, these locally fair clusterings become increasingly aligned
with the global fairness constraints, ultimately yielding a fair clustering
$\m{F}$ that closely approximates the input $\m{C}$. Now, suppose at each step we get an approximation factor of $\nu$ that is,
\begin{align}
    \dist(\m{J}^{k-1}, \m{J}^k) \leq \nu \cdot \dist(\m{J}^{k-1}, \m{F}^*) \label{eq:multi-color-one}
\end{align}

Then overall, we get an approximation factor of $((\nu + 1)^t - 1)$ that is

\begin{align}
    \dist(\m{C}, \m{F}) \le ((\nu + 1)^t - 1)\dist(\m{C}, \m{F}^*)\label{eq:multi-color-two}
\end{align}

We prove it using mathematical induction. It is easy to see that \cref{eq:multi-color-two} holds after the $1$st iteration. That is
\[
    \dist(\m{C}, \m{J}^1) \leq \nu \cdot \dist(\m{C}, \m{F}^*)
\]
just by putting $k = 1$ in \cref{eq:multi-color-one}. Note $\m{J}^0 = \m{C}$.

Now, as an induction hypothesis, suppose \cref{eq:multi-color-two} holds for $t = k - 1$ that is

\begin{align}
    \dist(\m{C}, \m{J}^{k - 1}) \le ((\nu + 1)^{k - 1} - 1)\dist(\m{C}, \m{F}^*)\label{eq:multi-color-three}
\end{align}
Note, $\m{J}^t = \m{F}$.

Now, we prove it for $t = k$.

    \begin{align}
        \dist(\m{J}^{k - 1}, \m{J}^k) &\leq \nu \, \dist(\m{J}^{k - 1}, \m{F}^*) \n \\
        &\leq \nu\, (\dist(\m{J}^{k - 1}, \m{C}) + \dist(\m{C}, \m{F}^*)) \n \\ &\text{(triangle inequality)} \nonumber \\
        &\leq \nu\, ((\nu + 1)^{k - 1} - 1)\dist(\m{C}, \m{F}^*) + \nu \, \dist(\m{C}, \m{F}^*) \n \\ &\text{(by IH)} \nonumber \\
        &= \nu \, (\nu + 1)^{k - 1} \dist(\m{C}, \m{F}^*) \label{eq:multi-color-four}
    \end{align}

    Now, we have
    \begin{align}
        \dist(\m{C}, \m{J}^k) &\leq \dist(\m{C}, \m{J}^{k - 1}) + \dist(\m{J}^{k - 1}, \m{J}^k) \n \\
        &\text{(triangle inequality)} \nonumber \\
        &\leq ((\nu + 1)^{k - 1} - 1) \dist(\m{C}, \m{F}^*) + \nu \cdot (\nu + 1)^{k - 1} \dist(\m{C}, \m{F}^*) \nonumber \\
        &\text{(by IH and \cref{eq:multi-color-four})} \nonumber \\
        &= ((\nu + 1)^k - 1) \dist(\m{C}, \m{F}^*) \n
    \end{align}

    Now, let us describe the high-level ideas of the algorithms for each of the above cases.

    \paragraph{Equal Representation with Power-of-Two Colors.}
When the vertex set $V$ is partitioned into disjoint color groups of equal size,
and the total number of colors $|\chi|$ is a power of two, we design an
algorithm called $\fptwo$. The algorithm $\fptwo$ constructs a globally fair
clustering $\outfptwo$ by progressively enforcing fairness across larger groups
of colors in a hierarchical manner. Starting from an input clustering
$\m{C}$ defined over a vertex set $V$, whose points are colored from a color
set $\chi = \{c_1, c_2, \ldots, c_{d}\}$, the algorithm proceeds in
$\log |\chi|$ iterations:
\[
\m{C} = \m{N}^0 \rightarrow \m{N}^1 \rightarrow \cdots \rightarrow
\m{N}^{\log |\chi|} = \outfptwo,
\]
where $\m{N}^i$ denotes the clustering obtained after the $i$th iteration.

In each iteration $i$, the color set $\chi$ is partitioned into contiguous
\emph{blocks} of size $2^i$ as
\[
B_j^i = \{ c_{(j-1)\cdot 2^i + 1}, c_{(j-1)\cdot 2^i + 2}, \ldots,
c_{j\cdot 2^i} \},
\quad \text{for } j = 1, \ldots, |\chi|/2^i.
\]
Hence, iteration $i$ enforces local fairness within each block $B_j^i$, while later
iterations progressively extend this fairness across larger unions of colors.

At the beginning of iteration $i$, the clustering $\m{N}^{i-1}$ is already locally
fair with respect to all blocks of size $2^{i-1}$. The objective of the current
iteration is to produce a new clustering $\m{N}^i$ such that, for every cluster
$N_a^i \in \m{N}^i$ and for every block $B_j^i$, all colors within $B_j^i$
appear equally in $N_a^i$. Formally, for all $x, y \in \{1, \ldots, 2^i\}$ with
$x \neq y$,
\[
c_{(j-1)\cdot 2^i + x}(N_a^i)
= c_{(j-1)\cdot 2^i + y}(N_a^i),
\]
where $c_r(N_a^i)$ denotes the number of points of color $c_r$ in cluster
$N_a^i$.

To achieve this, the algorithm first identifies the \emph{surplus} between each
pair of disjoint consecutive blocks within every cluster. For a given cluster $N_a^i$
and adjacent blocks $B_j^i$ and $B_{j+1}^i$ for odd $j$, the surplus $T_a^j$ is defined as
the excess number of points in the larger block relative to the smaller one.
Formally, if $|B_j^i(N_a^i)| > |B_{j+1}^i(N_a^i)|$, then $T_a^j$ is a subset of
$B_j^i(N_a^i)$ of size $|B_j^i(N_a^i)| - |B_{j+1}^i(N_a^i)|$, chosen so that
each color in $B_j^i$ contributes equally to $T_a^j$. The definition is
symmetric when $|B_{j+1}^i(N_a^i)| > |B_j^i(N_a^i)|$, here $B(S)$ for $S \subseteq V$ and $B \subseteq \chi$ denotes the subset of vertices in $S$ that has a color in $B$.

After identifying these surpluses, the algorithm temporarily removes them from
their respective clusters and collects them into two multisets, $S_j$ and
$S_{j+1}$, corresponding to the adjacent blocks $B_j^i$ and $B_{j+1}^i$.
These two multisets are then passed to the $\texttt{multi-GM}$ subroutine.

\begin{figure}[t]
\centering
    \includegraphics[width=11cm]{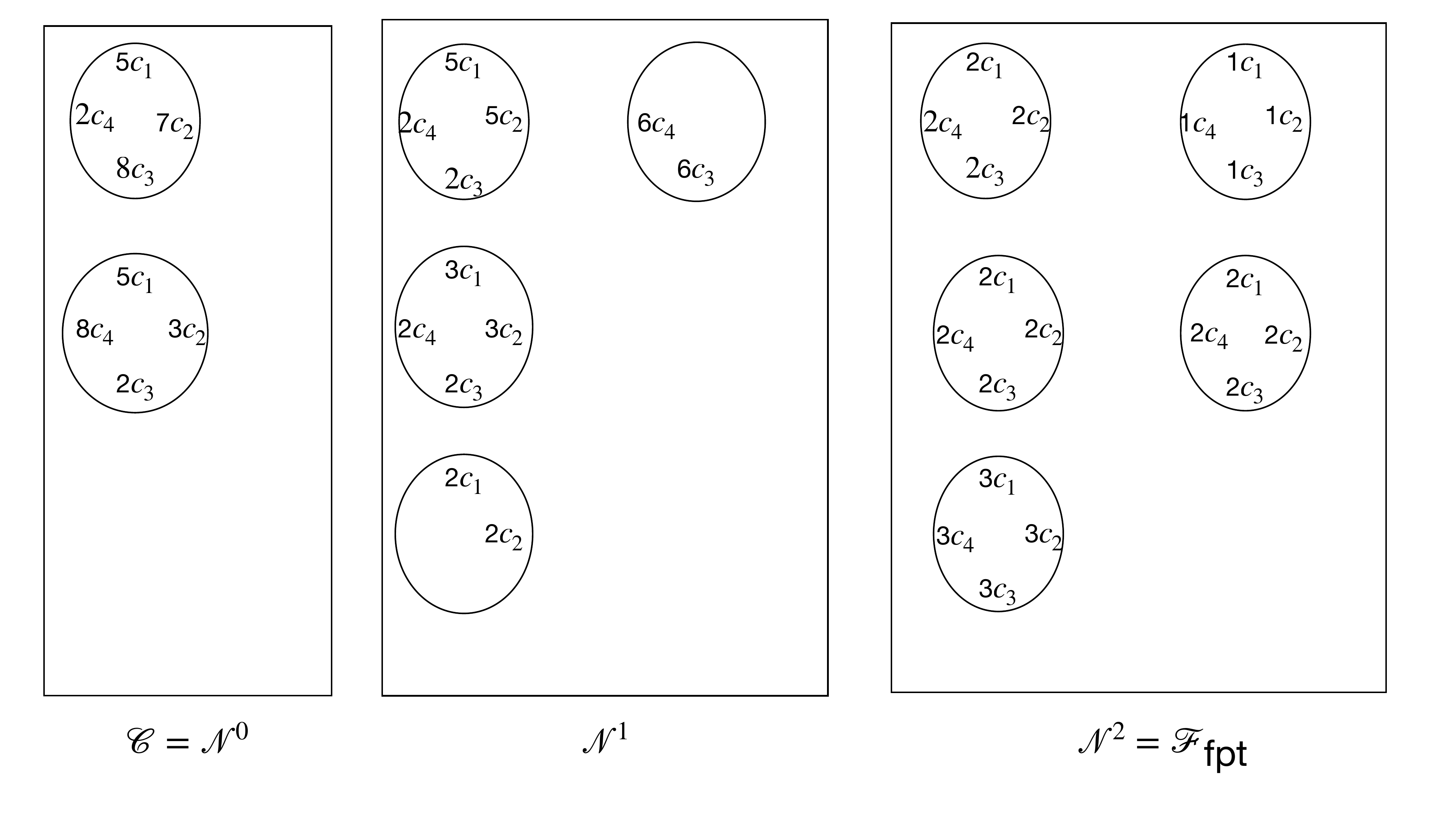}
    \caption{\label{fig:fptwo}Visualization of the $\fptwo$ algorithm with four colors $\{c_1, c_2, c_3, c_4\}$ where $nc_i$ denotes that the number vertices of color $c_i$ is $n$. In the $1$st iteration the color blocks are $B^1_1 = \{c_1, c_2\}$ and $B^1_2 = \{c_3, c_4\}$. So, in $\m{N}^1$ we make number of vertices of color $c_1$ equal to number of vertices of color $c_2$ and number of vertices of color $c_3$ equal to number of vertices of color $c_4$. In the $2$nd iteration we have $B^2_1 = \{c_1, c_2, c_3, c_4\}$, thus all color are equally represented in $\m{N}^2$. 
    }
\end{figure}

The $\texttt{multi-GM}$ subroutine is responsible for merging subsets from
$S_j$ and $S_{j+1}$ to create new clusters that are locally fair with respect
to the union of the two blocks $B_j^i \cup B_{j+1}^i$. It proceeds greedily by
pairing subsets of comparable size from the two collections, trimming from the
larger subset when necessary, and merging them into balanced clusters. Each
resulting cluster thus contains an equal number of points from every color in
$B_j^i \cup B_{j+1}^i$. The newly formed fair clusters are then inserted into
$\m{N}^i$, replacing the previously unbalanced  (see \cref{fig:fptwo}).

Consequently, each iteration enforces fairness locally across pairs of adjacent
color blocks while maintaining at most a factor-$2$ deviation from the previous
clustering. After $\log |\chi|$ iterations, all colors lie within a single
block, and every cluster satisfies global equal representation among all
colors. The cumulative approximation factor across all iterations is therefore
bounded by $O(3^{\log |\chi|}) = O(|\chi|^{1.6})$.

At a high level, the intuition behind the $2$-approximation at each step is as follows. 
For every pair of adjacent color blocks $B^{i}_{2k-1}$ and $B^{i}_{2k}$, 
the algorithm first removes the surplus $T^a_{2k-1}$ from a cluster $N_a^{i-1}$. 
This removal incurs a cost of $|T^a_{2k-1}| \cdot |N_a^{i-1}|$. 
In the subsequent step, within the $\greedymerge$ subroutine, 
these $|T^a_{2k-1}|$ surplus points are merged with an equal number of points taken from other clusters. 
Hence, the total cost incurred by our algorithm for balancing 
the color blocks $(B^{i}_{2k-1}, B^{i}_{2k})$ in the cluster $N_a^{i-1}$ is at most
\[
|T^a_{2k-1}| \left( |N_a^{i-1}| - |T^a_{2k-1}| \right) + |T^a_{2k-1}|^2 
= |T^a_{2k-1}| \cdot |N_a^{i-1}|.
\]

To see that the optimal solution $\m{F}^*$ must incur a comparable cost, 
suppose that in $\m{F}^*$, the cluster $N_a^{i-1}$ is partitioned into 
subsets $X_1, X_2, \ldots, X_s$. 
(The case $s = 1$ corresponds to $N_a^{i-1}$ remaining intact in $\m{F}^*$.)  
For the two color blocks $B^{i}_{2k-1}$ and $B^{i}_{2k}$, 
we define surpluses $S^k_j \subseteq X_j$ in such a way that
\[
\sum_{j=1}^s |S^k_j| = |T^a_{2k-1}|.
\]
The optimal solution must pay a cost of at least
\[
\frac{1}{2} \sum_{j=1}^s 
\Big( |S^k_j| \cdot |N_a^{i-1} \setminus X_j| + |S^k_j| \cdot |X_j| \Big)
= \frac{1}{2} |T^a_{2k-1}| \cdot |N_a^{i-1}|.
\]
The first term, $|S^k_j| \cdot |N_a^{i-1} \setminus X_j|$, 
accounts for pairs $(u,v)$ that were in the same cluster $N_a^{i-1}$ in the input 
but are separated in $\m{F}^*$. 
The second term, $|S^k_j| \cdot |X_j|$, 
captures the merging operation required in $\m{F}^*$ 
to restore local fairness within each subcluster $X_j$. 
The multiplicative factor $\tfrac{1}{2}$ corrects for double counting, 
since a pair $(u,v)$ with $u \in S^k_j$ and $v \in S^k_\ell$ ($j \neq \ell$) 
would otherwise be counted twice in the summation. 
To ensure every unordered pair is counted exactly once, 
we charge each vertex a cost of $\tfrac{1}{2}$.

\medskip
\noindent
Thus, even in the optimal solution $\m{F}^*$, 
the minimum possible cost is at least $\tfrac{1}{2}|T^a_{2k-1}| \cdot |N_a^{i-1}|$, 
while our algorithm pays $|T^a_{2k-1}| \cdot |N_a^{i-1}|$. 
Hence, the cost of our algorithm is at most twice that of $\m{F}^*$, 
establishing the $2$-approximation guarantee for each step.

\paragraph{Equal Representation with Arbitrary Number of Colors.}
When the vertex set $V$ is partitioned into disjoint color groups of equal size,
but the number of colors $|\chi|$ is arbitrary (not necessarily a power of two),
we design an algorithm called $\fequi$. The algorithm $\fequi$ builds upon two
subroutines: $\fptwo$ and $\fmulti$. The subroutine $\fptwo$ is already described. Let us first describe the high-level idea of $\fmulti$ and then present the
construction of $\fequi$.

\paragraph{High-Level Idea of $\fmulti$.}
The subroutine $\fmulti$ takes as input a clustering
$\mathcal{I} = \{ I_1, I_2, \ldots, I_{\eta} \}$ over a vertex set $V$ whose
points are colored from a color set
$\zeta = \{ z_1, z_2, \ldots, z_r \}$. The total number of points of each color
satisfies a prescribed global proportion
\[
z_1(V) : z_2(V) : \ldots : z_r(V) = p_1 : p_2 : \ldots : p_r,
\]
where, without loss of generality, $p_1 > p_2 > \cdots > p_r$. Furthermore, each
cluster $I_i$ in $\mathcal{I}$ is assumed to be \emph{$p$-divisible}—that is,
the number of points of each color $z_j$ in $I_i$ is an integer multiple of
$p_j$.

The goal of $\fmulti$ is to transform $\mathcal{I}$ into a clustering
$\outmpf$ such that every cluster
$F \in \outmpf$ satisfies
\[
z_1(F) : z_2(F) : \ldots : z_r(F) = p_1 : p_2 : \ldots : p_r.
\]
The algorithm proceeds in $\lceil \log_2 r \rceil$ iterations, progressively
aligning the local color proportions toward the global target ratio. Starting
from $\mathcal{F}^0 := \mathcal{I}$, it maintains a sequence of refined
clusterings
\[
\mathcal{I} = \mathcal{F}^0 \rightarrow \mathcal{F}^1 \rightarrow \cdots
\rightarrow \mathcal{F}^{\lceil \log_2 r \rceil} = \outmpf.
\]
In each iteration $t$, $\fmulti$ merges adjacent \emph{color blocks} in a
hierarchical fashion. Initially, each color forms a singleton block
$B^0_j = \{ z_j \}$; at every subsequent iteration $t \ge 1$, consecutive
blocks are merged in pairs:
\[
B^t_i = B^{t-1}_{2i - 1} \cup B^{t-1}_{2i}, \qquad
\text{for } i = 1, 2, \ldots,
\left\lfloor \frac{m_{t-1}}{2} \right\rfloor.
\]
If the number of blocks $m_{t-1}$ is odd, the final block remains unpaired and
is carried forward unchanged. This process continues until all colors are
contained in a single block at iteration
$T = \lceil \log_2 r \rceil$.

At every iteration $t$, the algorithm ensures that all colors within each block
$B^t_i$ appear in every cluster in proportion to their prescribed global ratios.
To achieve this, $\fmulti$ compares the \emph{scaling factors} of two adjacent
blocks $B^{t-1}_{2i-1}$ and $B^{t-1}_{2i}$ within each cluster and tries to equalize the scaling factor. To understand how $\fmulti$ equalizes scaling factors, recall that after iteration $(t-1)$,
every cluster $F^{t-1}$ satisfies the correct color ratios within each block
$B^{t-1}_{2i-1}$ and $B^{t-1}_{2i}$ separately. In particular, for
\[
B^{t-1}_{2i-1} = \{ z_{a_1}, \ldots, z_{a_s} \}
\quad \text{and} \quad
B^{t-1}_{2i} = \{ z_{b_1}, \ldots, z_{b_u} \},
\]
we can write the color counts of $z_{a_j}$ and $z_{b_k}$ for $j \in [1,s]$ and $k \in [1,u]$ respectively inside $F^{t-1}$ as
\[
z_{a_j}(F^{t-1}) = p_{a_j} \cdot x, \quad
z_{b_k}(F^{t-1}) = p_{b_k} \cdot y,
\]
where the integers $x$ and $y$ are the \emph{scaling factors} of the two
blocks. The values $x$ and $y$ indicate how many times the global ratio vector
$(p_{a_1}, \ldots, p_{a_s})$ or $(p_{b_1}, \ldots, p_{b_u})$ is realized inside
the cluster.

When merging these two sub-blocks into a single block
$B^t_i = B^{t-1}_{2i-1} \cup B^{t-1}_{2i}$, our goal is to make the entire set
of colors jointly proportional to their global ratios
$(p_{a_1}, \ldots, p_{a_s}, p_{b_1}, \ldots, p_{b_u})$. This is possible only if
both sub-blocks share the same scaling factor, i.e., $x = y$. If $x \neq y$,
then one group of colors is overrepresented relative to the other.

To correct this imbalance, $\fmulti$ performs a controlled adjustment. Suppose
$x < y$—that is, the sub-block $B^{t-1}_{2i}$ currently represents the
global proportions at a larger scale. The algorithm reduces its contribution by
removing $p_{b_k} \cdot (y - x)$ points of each color $z_{b_k}$ uniformly across
the cluster. If $x > y$ then the sub-block $B^{t-1}_{2i}$  under represents the
global proportions. The algorithm increases its contribution by adding $p_{b_k} \cdot (x - y)$ points of each color
$z_{b_k}$ to match the total mass. After this operation, both
sub-blocks contribute exactly the same total scaling factor $x' = y' = \min(x,
y)$, ensuring that within the combined block $B^t_i$, we have
\[
z_{a_1}(F^t) : \ldots : z_{a_s}(F^t) : z_{b_1}(F^t) : \ldots : z_{b_u}(F^t)
= p_{a_1} : \ldots : p_{a_s} : p_{b_1} : \ldots : p_{b_u}.
\]

\begin{figure}[t]
\centering
    \includegraphics[width=11cm]{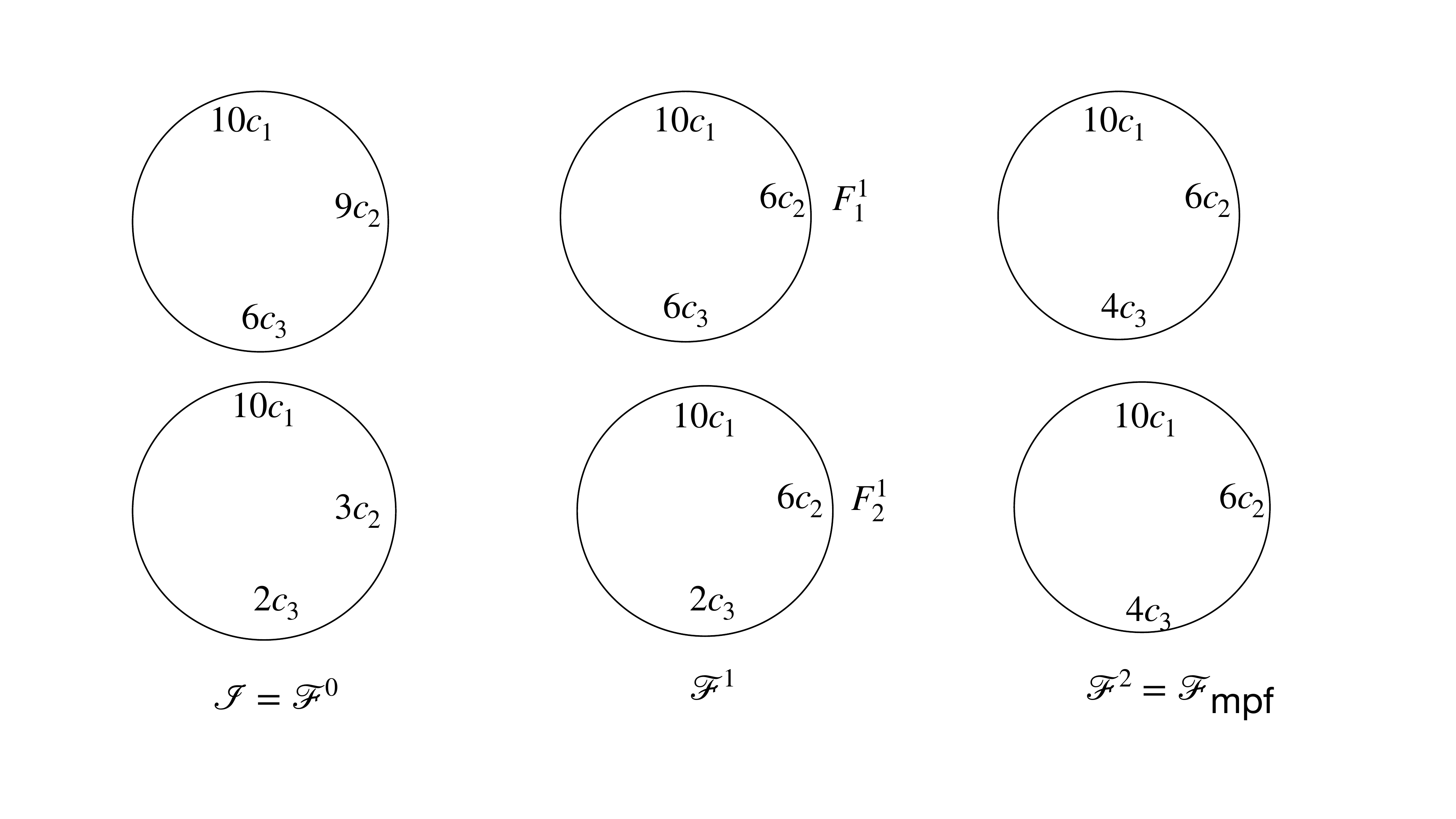}
    \caption{
Visualization of the $\fmulti$ algorithm for three colors $\{c_1, c_2, c_3\}$, 
with $20$, $12$, and $8$ vertices respectively, corresponding to a global ratio of $5\!:\!3\!:\!2$. 
In the first iteration, the color blocks are $B^1_1 = \{c_1, c_2\}$ and $B^1_2 = \{c_3\}$. 
Hence, in the intermediate clustering $\m{F}^1$, each cluster maintains the local ratio 
$c_1:c_2 = 5\!:\!3$. 
The scaling factor of block $B^1_1$ in cluster $F^1_1$ is $x = 2$, 
while that of $B^1_2$ is $y = 3$. 
To align these scaling factors, the algorithm cuts $2(y - x) = 2$ vertices of color $c_3$ 
from $F^1_1$ and merges them into $F^1_2$, 
yielding the final fair clustering $\m{F}^2 = \outmpf$.}
\label{fig:make_pdc_fair}
\end{figure}

(see \cref{fig:make_pdc_fair})

At each step, the algorithm $\fmulti$ incurs an approximation of $6$, thus giving an overall approximation of $7^{\log |\zeta|} = |\zeta|^{2.81}$.

Since in this setting $p_{a_j} \geq p_{b_k}$ for all $j \in [s]$ and $k \in [u]$, 
the high-level idea behind the $6$-approximation at each step of the algorithm $\fmulti$ 
closely mirrors the reasoning behind the $3$-approximation guarantee of the algorithm $\algmf$ 
in the two-color case. 
Recall that $\algmf$ is the algorithm that transforms a $\bal$ into a fair clustering 
when the vertex set is divided into two color groups. 
The additional multiplicative factor of $2$ in the approximation bound of $\fmulti$ 
arises to account for overcounting in the optimal solution, 
since in the multi-color setting each block may contain multiple colors, 
and interactions between these colors can lead to double counting of pairwise contributions.

\paragraph{High-Level Idea of $\fequi$.}
The algorithm $\fequi$ takes as input a clustering $\m{C}$ whose vertices are
colored from $\chi = \{ c_1, c_2, \ldots, c_d \}$, where the colors are
distributed equally but $|\chi|$ is not a power of two. The key idea is to
combine the structural fairness guarantees of $\fptwo$ with the proportional
balancing capability of $\fmulti$.

The algorithm proceeds as follows. First, we partition the color set $\chi$ into
$\log |\chi|$ disjoint groups
$G_1, G_2, \ldots, G_{\log |\chi|}$ such that the size of each group $G_j$ is a
power of two. This is done greedily according to the binary representation of
$|\chi|$: for each index $j \in [\log \lceil |\chi| \rceil]$, if the
corresponding bit in the binary representation of $|\chi|$ is $1$, a group of
size $2^{j-1}$ is created.

Next, $\fequi$ applies $\fptwo$ independently within each color group $G_j$ to
obtain an intermediate clustering
$\m{I} = \{ I_1, I_2, \ldots, I_\eta \}$ satisfying that for every
$I_i \in \m{I}$ and every pair of colors $c_a, c_b \in G_\ell$,
\[
c_a(I_i) = c_b(I_i).
\]
Thus, $\fptwo$ ensures equal representation within each power-of-two-sized
color group.

We then treat each group $G_\ell$ as a single \emph{meta-color} $z_\ell \in \zeta$ and
invoke the $\fmulti$ subroutine on the intermediate clustering $\m{I}$. The
clustering $\m{I}$ serves as a $p$-divisible clustering because for each
cluster $I_i \in \m{I}$, the count $|G_\ell(I_i)|$ is divisible by $|G_\ell|$, where $G_\ell(I_i)$ denotes the subset of vertices of $I_i$ that has a color from the group $G_\ell$.
Applying $\fmulti$ produces a final fair clustering
$\m{F} = \{ F_1, \ldots, F_\iota \}$ such that for every cluster $F_j \in
\m{F}$,
\[
|G_1(F_j)| : |G_2(F_j)| : \ldots : |G_{\log |\chi|}(F_j)| =
|G_1| : |G_2| : \ldots : |G_{\log |\chi|}|.
\]
Since $\fptwo$ already enforces equality within each group $G_\ell$, the
proportional balancing across groups performed by $\fmulti$ ensures that in the
final clustering $\m{F}$, every color $c_u, c_v \in \chi$ satisfies
$c_u(F_j) = c_v(F_j)$. Thus, $\fequi$ outputs a globally fair clustering where
all colors are equally represented in every cluster.

As established earlier, the algorithm $\fptwo$ achieves an approximation factor of $O(|\chi|^{1.6})$. 
In the algorithm $\fequi$, the number of meta-colors generated is at most $\log |\chi|$, 
and consequently, the subroutine $\fmulti$ contributes an additional approximation factor of 
$O(\log^{2.81} |\chi|)$. 
By applying the triangle inequality to combine the errors introduced by the two stages, 
the overall approximation factor of $\fequi$ becomes $O(|\chi|^{1.6} \log^{2.81} |\chi|)$.

   \paragraph{Arbitrary Representation.}
 When the vertex set $V$ is divided into disjoint color groups of arbitrary sizes, 
we design an algorithm $\fgen$ to construct a fair clustering that reflects the global color distribution. 
Let the vertices be colored from the set 
$\chi = \{ c_1, c_2, \ldots, c_{d} \}$, and suppose their total counts satisfy the global ratio
\[
c_1(V) : c_2(V) : \cdots : c_{|\chi|}(V) = p_1 : p_2 : \cdots : p_{|\chi|}.
\]
To achieve fairness under these non-uniform proportions, we adopt a 
two-step approach analogous to the two-color case.

In the first step, we construct a \emph{$p$-divisible clustering} ($\bal$)
$\m{M} = \{ M_1, M_2, \ldots, M_t \}$ using our algorithm $\pdca$. Recall in a $\bal$ $\m{M}$, for every cluster $M \in \m{M}$ 
and every color $c_j \in \chi$, the number of points of color $c_j$ in $M$ 
is divisible by $p_j$.

In the second step, we transform the $p$-divisible clustering $\m{M}$ 
into a \emph{fair clustering} $\m{F}$ using $\fmulti$ such that for every cluster $F \in \m{F}$,
\[
c_1(F) : c_2(F) : \cdots : c_{|\chi|}(F) = p_1 : p_2 : \cdots : p_{|\chi|}.
\]

\paragraph{High level idea of $\pdca$:} To construct a $\bal$, we proceed iteratively over the colors in $\chi$. 
For each color $c_j$, we apply the algorithm $\algog$ introduced in the two-color case. 
At iteration $j$, the algorithm $\algog$ is applied to the current clustering to ensure that, 
within every cluster, the number of vertices of color $c_j$ becomes a multiple of $p_j$. 
Thus, after processing all colors sequentially, in the resulting clustering for all colors $c_j$ the number of vertices of color $c_j$ is divisible by $p_j$. 

Recall that in the two-color setting, the algorithm $\algog$ takes as input a clustering whose vertices are divided into two color groups—say blue and red—such that the global ratio of blue to red points is $p : 1$. 
The output of $\algog$ is a refined clustering where, in every cluster, the number of blue vertices is a multiple of $p$. 
By applying this procedure iteratively to each color $c_j$, 
we generalize the same principle to the multi-color setting, 
thereby obtaining a $\bal$.

As discussed earlier, the algorithm $\algog$ achieves an approximation factor of $3.5$ in the two-color setting. 
When $\algog$ is applied iteratively across all colors in $\chi$, 
these approximation factors accumulate, resulting in an overall approximation factor of $O(|\chi|)$ for the algorithm $\pdca$. 
It is worth noting that if one were to apply a direct analysis using the triangle inequality, 
a constant-factor approximation $\nu$ at each step would compound multiplicatively, yielding an overall factor of $\nu^{|\chi|}$. 
To avoid this exponential blow-up, we adopt a refined analytical approach 
inspired by Chakraborty et al. \cite{chakraborty2025towards}, 
originally developed for the two-color setting where the blue-to-red ratio is an arbitrary $p:q$. 
This sophisticated analysis allows us to bound the cumulative approximation additively rather than multiplicatively, 
thereby ensuring an $O(|\chi|)$ guarantee for $\pdca$.

\paragraph{High-Level Idea of $\fgen$:} 
Given an input clustering $\m{C}$, the algorithm $\fgen$ proceeds in two stages. 
First, it applies $\pdca$ to construct a $\bal$ clustering $\m{M}$, ensuring that within each cluster, 
the number of points of color $c_j$ is divisible by $p_j$. 
Subsequently, it invokes $\fmulti$ on $\m{M}$ to produce the final fair clustering $\m{F}$. 

The algorithm $\pdca$ achieves an approximation factor of $O(|\chi|)$, 
while $\fmulti$ contributes an additional factor of $O(|\chi|^{2.81})$. 
Combining these guarantees, $\fgen$ attains an overall approximation factor of 
$O(|\chi|^{3.81})$.

\section*{Fair Consensus Clustering in the Streaming Setting}

In fair consensus clustering, the input consists of a collection of clusterings
$\setofclusterings = \{\mathcal{C}_1, \ldots, \mathcal{C}_m\}$ defined on a common vertex set.
The objective is to compute a fair clustering $\mathcal{F}$ such that
\[
\left( \sum_{i=1}^m \dist(\mathcal{C}_i, \mathcal{F})^\ell \right)^{1/\ell}
\leq
(\alpha + 1.995)
\left( \sum_{i=1}^m \dist(\mathcal{C}_i, \optcon)^\ell \right)^{1/\ell},
\]
where $\optcon$ denotes an optimal fair consensus clustering, and $\alpha,\ell$ are fixed parameters.
We assume oracle access to an $\alpha$-approximate closest fair clustering algorithm.
Our goal is to compute such a clustering $\mathcal{F}$ using only $O(n \log m)$ space and
$O\!\left(n^2 \log^4 m + \runtimeFairCorCls{n}\log^3 m\right)$ time, where
$\runtimeFairCorCls{n}$ denotes the running time of an unweighted fair correlation clustering algorithm.

We work in the streaming model $\pconstream$ described in \cref{sec:our-contribution}.
We now describe our streaming algorithm, denoted by $\stalgo$.

\paragraph{High-Level Idea of $\stalgo$:} Recall that in the \texttt{pairwise consensus} \texttt{streaming model}, each stream element is a triple $(p,j,b)$ that encodes information about the $j$-th clustering $\mathcal{C}_j \in \setofclusterings$.
Prior to processing the stream, we uniformly sample $4g\log m$ indices
$j_1,\ldots,j_{4g\log m}$ from $[m]$, for a suitable constant $g>1$, and store only the information corresponding to the sampled clusterings
$\mathcal{C}_{j_1},\ldots,\mathcal{C}_{j_{4g\log m}}$.

Using these sampled clusterings, we construct a candidate set of fair clusterings, denoted by $\canset$.
The key guarantee is that, with high probability, $\canset$ contains at least one fair clustering
$\mathcal{F}$ that achieves an $(\alpha+1.995)$-approximation to the optimal fair consensus clustering.
Accordingly, we output the clustering
\[
\mathcal{F}
=
\argmin_{\mathcal{F}' \in \canset}
\left( \sum_{i=1}^m \dist(\mathcal{C}_i, \mathcal{F}')^\ell \right)^{1/\ell}.
\]

A key technical challenge in our setting is evaluating the objective value of a candidate clustering $\mathcal{F}' \in \canset$. In the streaming model, computing its distance to all input clusterings $\mathcal{C}_i \in \setofclusterings$ is infeasible, since only $O(\log m)$ clusterings can be stored at any time. To overcome this limitation, we leverage a result due to Indyk~\cite{indyk1999sublinear, indyk2001high}. Specifically, if we uniformly sample a subset $\mathcal{W} \subseteq \setofclusterings$ of size $\Omega(\varepsilon^{-2}\log m)$, for any $0 < \varepsilon < 1$, then with probability at least $1 - 1/m$, the following guarantee holds:
\[
\left( \sum_{i = 1}^m \dist(\mathcal{C}_i, \mathcal{F}_{\mathrm{best}, \mathcal{W}})^\ell \right)^{1/\ell}
\;\leq\;
(1+\varepsilon)
\left( \sum_{i=1}^m \dist(\mathcal{C}_i, \mathcal{F}_{\mathrm{best}, \setofclusterings})^\ell \right)^{1/\ell}.
\]
Here, $\mathcal{F}_{\mathrm{best}, \mathcal{W}}$ denotes the fair clustering in $\canset$ that minimizes the objective when evaluated only on the sampled set $\mathcal{W}$, while $\mathcal{F}_{\mathrm{best}, \setofclusterings}$ denotes the fair clustering in $\canset$ that minimizes the objective when evaluated over the full set $\setofclusterings$.

We therefore independently sample an additional $64\varepsilon^{-2}\log m$ clusterings from
$\setofclusterings$ to form the set $\mathcal{W}$, and use $\mathcal{W}$ to approximately evaluate
the objective function for all candidates in $\canset$.
Since we store only $O(\log m)$ clusterings in total, and each clustering requires $O(n)$ space,
the overall space complexity of the algorithm is $O(n \log m)$. To analyze the approximation factor and the time complexity of the algorithm, we need to look at the construction of $\canset$. 

\paragraph{Construction of the Candidate Set $\canset$:}
After sampling $4g\log m$ input clusterings, for each sampled clustering
$\mathcal{C}_k$ (where $k \in [4g\log m]$), we compute and store its
$\alpha$-close fair clustering $\mathcal{F}_{\mathcal{C}_k}$ in the candidate set $\canset$.
In addition, for every triple $(\mathcal{C}_x, \mathcal{C}_y, \mathcal{C}_z)$ of sampled clusterings,
we perform the following steps.

\begin{enumerate}[(i)]
    \item We construct an unweighted correlation clustering instance $G_{x,y,z}$ from the three
    clusterings $\mathcal{C}_x$, $\mathcal{C}_y$, and $\mathcal{C}_z$.  
    For each unordered vertex pair $(u,v)$, the edge $(u,v)$ in $G_{x,y,z}$ is assigned a
    positive label if $u$ and $v$ are placed in the same cluster in at least two of the three
    clusterings; otherwise, the edge is assigned a negative label.
    
    \item We compute an unweighted fair correlation clustering $T_{x,y,z}$ on the graph
    $G_{x,y,z}$ using a known algorithm for unweighted fair correlation clustering.
    
    \item We add the clustering $T_{x,y,z}$ to the candidate set $\canset$.
\end{enumerate}

We refer to the above procedure as the \emph{cluster fitting} algorithm, denoted by $\clsfitting$.
We now provide intuition for why the $\clsfitting$ algorithm is effective.

Let $\optcon$ denote an optimal fair consensus clustering. For each input clustering
$\mathcal{C}_i \in \setofclusterings$, let $\unalignedSet{i}$ denote the set of unordered
vertex pairs $(u,v) \in V \times V$ such that $u$ and $v$ are clustered together
(respectively, separated) in $\optcon$, but are separated (respectively, clustered together)
in $\mathcal{C}_i$. By definition,
\[
\dist(\mathcal{C}_i, \optcon) = |\unalignedSet{i}|.
\]

We further define
\[
\opt = \left( \sum_{i=1}^m \dist(\mathcal{C}_i, \optcon)^\ell \right)^{1/\ell}
\quad \text{and} \quad
\avgconval = \frac{\opt}{m}.
\]

It can be shown that if at least $m/g$ input clusterings are close to the optimal solution
$\optcon$, that is, if there exists a set of at least $m/g$ indices such that
\[
|\unalignedSet{i}| \leq (1-\beta)\avgconval
\]
for some parameter $\beta > 1$, then with high probability our sampling procedure selects
one such clustering $\m{C}_i$. In this case, the corresponding $\alpha$-close fair clustering
$\mathcal{F}_{\mathcal{C}_i}$ yields an $(\alpha+1.995)$-approximation to the optimal fair
consensus clustering.

The more challenging case arises when fewer than $m/g$ clusterings are close to $\optcon$.
In this regime, the $\clsfitting$ algorithm becomes crucial. Specifically, consider triples
$(\mathcal{C}_x, \mathcal{C}_y, \mathcal{C}_z)$ for which each of
$|\unalignedSet{x}|$, $|\unalignedSet{y}|$, and $|\unalignedSet{z}|$ is large—exceeding
$(1-\beta)\avgconval$—but whose pairwise intersections are bounded. That is, for all distinct
$r,s \in \{x,y,z\}$,
\[
|\unalignedSet{r} \cap \unalignedSet{s}| \leq \alpha\,\avgconval.
\]
For such triples, the fair clustering $T_{x,y,z}$ produced by the $\clsfitting$ algorithm
is close to the optimal solution $\optcon$, and consequently achieves an
$(\alpha+1.995)$-approximation. The intuition underlying why the clustering $T_{x,y,z}$ is close to $\optcon$ is as follows.
Consider a vertex pair $(u,v) \in \unalignedSet{r} \cap \unalignedSet{s}$ for some
distinct indices $r,s \in \{x,y,z\}$.
Without loss of generality, assume that $u$ and $v$ are clustered together in
$\mathcal{C}_r$ and $\mathcal{C}_s$.
Since $(u,v) \in \unalignedSet{r} \cap \unalignedSet{s}$, it follows that $u$ and $v$
are separated in the optimal fair consensus clustering $\optcon$.

By construction of the correlation clustering instance $G_{x,y,z}$, the edge $(u,v)$
receives a positive label whenever $u$ and $v$ are clustered together in at least two
of the three clusterings $\mathcal{C}_x$, $\mathcal{C}_y$, and $\mathcal{C}_z$.
Therefore, for the pair $(u,v)$ under consideration, the edge $(u,v)$ is labeled positive
in $G_{x,y,z}$.
Consequently, in the unweighted fair correlation clustering $T_{x,y,z}$ computed from
$G_{x,y,z}$, vertices $u$ and $v$ are placed in the same cluster.

When comparing $T_{x,y,z}$ to the optimal clustering $\optcon$, the clustering
$T_{x,y,z}$ incurs a cost for the pair $(u,v)$, since $u$ and $v$ are clustered together
in $T_{x,y,z}$ but separated in $\optcon$.
Thus, every pair belonging to the intersection $\unalignedSet{r} \cap \unalignedSet{s}$
necessarily contributes to the disagreement between $T_{x,y,z}$ and $\optcon$. Therefore, if the total number of such pairs is bounded, the overall disagreement between
$T_{x,y,z}$ and $\optcon$ is small, implying that $T_{x,y,z}$ is close to $\optcon$.

We show that, with high probability, our sampling procedure selects clusterings
$\mathcal{C}_x$, $\mathcal{C}_y$, and $\mathcal{C}_z$ whose corresponding pairwise
intersections are bounded. Consequently, with high probability, the algorithm
achieves an approximation factor of $(\alpha + 1.995)$.

We now analyze the running time of the algorithm. The candidate set $\canset$
contains $O(\log^3 m)$ clusterings. The dominating cost to construct these candidates is by applying the
$\clsfitting$ procedure to all triples of sampled clusterings which requires
$O\!\left(\runtimeFairCorCls{n}\log^3 m\right)$ time.

To select the best clustering from $\canset$, we evaluate the objective function
for each candidate. This requires $O(\log^3 m)$ evaluations. In each evaluation,
we compute the distance between the candidate clustering and each clustering in
the sampled set $\mathcal{W}$. Computing the distance between two clusterings
takes $O(n^2)$ time, and since $|\mathcal{W}| = O(\log m)$, the total time per
evaluation is $O(n^2 \log m)$. Therefore, the total time required for all
evaluations is $O(n^2 \log^4 m)$.

Combining these bounds, the overall running time of the algorithm is
\[
O\!\left(n^2 \log^4 m + \runtimeFairCorCls{n}\log^3 m\right).
\]

\section{Organization}

The remainder of this thesis is organized as follows.
In \cref{sec:reachability-preservers}, we present our results on fault-tolerant reachability preservers.
The \cref{sec:closest-fair-two-colors} studies the closest fair clustering problem when the vertex set $V$ is partitioned into two disjoint color groups, while \cref{sec:closest-fair-with-more-than-two-colors} extends these results to settings with more than two color groups.
In \cref{sec:implications-to-fair-correlation-and-fair-consensus-clustering}, we explore the implications of our closest fair clustering framework for fair correlation clustering and fair consensus clustering.
The ~\cref{sec:streaming} presents our algorithmic results for fair consensus clustering in the streaming setting.
Finally, \cref{sec:future-directions} outlines several open problems and directions for future research.

\chapter{Reachability Preservers}\label{sec:reachability-preservers}

In this chapter, we present a construction of a dual fault-tolerant reachability preserver (2-$\ftrs$) for a given directed graph $G = (V, E)$ and a vertex-pair set $\P \subseteq V \times V$. Specifically, we establish \cref{thm:dual-ftrs}, which we restate below.

\upreachability*

To prove this result, we first describe an alternate construction of a $2$-$\ftrs$ for a single vertex pair. We then extend this construction to handle multiple pairs, thereby obtaining the general result.

\section{An (Alternate) Construction of 2-{\ftrs} for a Single Pair}
\label{section:single-pair}
Given any directed graph $G$ with $n$ vertices, for any pair of vertices $(s,t)$, we already get an $O(n)$ size 2-$\ftrs(G,(s,t))$ from~\cite{BCR16}. Let us state the result of \cite{BCR16} formally here.

\begin{theorem}[\cite{BCR16}]
\label{theorem:ftrs}
For any directed graph $G=(V,E)$, a designated source vertex $s\in V$, and an integer $k\geq 1$, there exists a (sparse) subgraph $H$ of $G$ which is a $\kftrs(G,\{s\}\times V)$ and contains at most $2^k n$ edges. Moreover, such a subgraph is computable in $O(2^kmn)$ time, where $n$ and $m$ are respectively the number of vertices and edges in graph $G$.
\end{theorem}

However, now we study the structure of a 2-$\ftrs(G,(s,t))$ more closely and provide a slightly different construction which will be useful for our construction of 2-$\ftrs$ for multiple pairs. Let us first define a notation to denote the ordering of vertices on a specific path. For any path $P$ and two vertices $u,v$ on $P$, we will write $u <_P v$ if and only if $u$ appears before $v$ on $P$.

Suppose we are given a directed graph $G$ and a pair $(s,t)$. Let us consider two $s-t$ paths $P_{(s,t)}^1$ and $P_{(s,t)}^2$, that intersect only at $s-t$ cut-edges and $s-t$ cut-vertices. We call these two paths \emph{outer strands}. Next, we define coupling paths between these two outer strands. For each vertex on the outer strands, let us first define two \emph{coupling points} of it (one on each outer strand). Consider a vertex $v \in V(P_{(s,t)}^1 \cup P_{(s,t)}^2)$ and $i \in \{1,2\}$ where for a subgraph $H$ of $G$, $V(H)$ represents the set of vertices in the graph $H$. Then consider the vertex $u_{(s,t),v}^i\in P_{(s,t)}^i$ (if exists) for which there is a $u_{(s,t),v}^i-v$ path that is edge-disjoint with both the outer strands, and there is no vertex $u'\in P_{(s,t)}^i$ such that $u' <_{P_{(s,t)}^i} u_{(s,t),v}^i$ and there is a $u'-v$ path that is edge-disjoint with both the outer strands. We refer to $u_{(s,t),v}^1, u_{(s,t),v}^2$ as \emph{coupling points} of $v$. For each $v$ on the outer strands and each coupling point $u_{(s,t),v}^i$ of it, consider any arbitrary $u_{(s,t),v}^i-v$ path (denoted as $Q_{(s,t),v}^i$) of $G$, that is edge-disjoint with both the outer strands. We refer to the path $Q_{(s,t),v}^i$ as a \emph{coupling path} to $v$. Note, for each vertex $v$ on the outer strands, there are at most two coupling points of it (one on each outer strand) and from each coupling point there is a coupling path to $v$. We take two outer strands and some of the coupling paths to build a subgraph $H_{(s,t)}$. More specifically, for a vertex $v \in V(P_{(s,t)}^j)$, if for all the vertices below $v$ on that strand, their coupling points on a strand $P_{(s,t)}^i$ is below the coupling point of $v$ on that strand, then only add the coupling path from the strand $P_{(s,t)}^i$ to $v$. Later, we prove that keeping these "essential" coupling paths is sufficient. 

Below we provide a formal description of the procedure to construct $H_{(s,t)}$. Given $G$ and a pair $(s,t)$, we build a subgraph $H_{(s,t)}$ as follows:
\begin{enumerate}
\item $V(H_{(s,t)})=V(G)$.
\item Add all the edges on the paths $P_{(s,t)}^1$, $P_{(s,t)}^2$ (two outer strands) to $H_{(s,t)}$.
\item For all $i,j \in \{1,2\}$ and $v \in V(P_{(s,t)}^j)$
\begin{enumerate}
\item If for all $v' \in V(P_{(s,t)}^j[v-t])$, $u_{(s,t),v}^i <_{P_{(s,t)}^i} u_{(s,t),v'}^i$, then take the coupling path $Q_{(s,t),v}^i$, and add all its edges to $H_{(s,t)}$. Here. $P_{(s,t)}^j[v-t]$ denotes the subpath of $P_{(s,t)}^j$ from $v$ to $t$. 
\end{enumerate}
\end{enumerate}
Readers may refer to Figure~\ref{fig:single-pair-ftrs} for an example subgraph $H_{(s,t)}$. Note, the above construction procedure of the subgraph $H_{(s,t)}$ runs in polynomial time. 

Before arguing that $H_{(s,t)}$ is a 2-$\ftrs(G,(s,t))$, let us introduce a notion of \emph{nice path}. We call an $s-t$ path in $H_{(s,t)}$ \emph{nice} if it first follows a outer strand till some vertex $u$, then take a coupling path to some vertex $u'$, and finally follows the outer strand on which $u'$ lies till $t$. To state it formally, suppose for two paths $P$ and $Q$, $P \circ Q$ denotes the concatenation of two paths $P$ and $Q$. Now, an $s-t$ path in $H_{(s,t)}$ is nice if and only if it is of the form $P_{(s,t)}^i[s-u] \circ Q_{(s,t),u'}^{i} \circ P_{(s,t)}^j[u'-t]$, for some $i,j \in \{1,2\}$, where subpath $Q_{(s,t),u'}^{i}$ starts from $u$ and is edge-disjoint with both the outer strands $P_{(s,t)}^1$, $P_{(s,t)}^2$. ($Q_{(s,t),u'}^{i}$ could be empty.)

The following claim shows that $H_{(s,t)}$ is a 2-$\ftrs(G,(s,t))$, and also will be useful in the next subsection while arguing about the correctness of our pairwise 2-{\ftrs} for multiple pairs construction.
\begin{claim}
\label{clm:special-path}
For any two edges $f_1,f_2$, if there is an $s-t$ path in $G \setminus \{f_1,f_2\}$, then there must be a nice $s-t$ path in $H_{(s,t)} \setminus \{f_1,f_2\}$.
\end{claim}
\begin{proof}
First of all, one of $f_1,f_2$ must be on $P_{(s,t)}^1$ and another on $P_{(s,t)}^2$. Otherwise, since $P_{(s,t)}^1$ and $P_{(s,t)}^2$ intersect only at $s-t$ cut-edges and $s-t$ cut-vertices, at least one of $P_{(s,t)}^1$, $P_{(s,t)}^2$ remains intact in $H_{(s,t)} \setminus \{f_1,f_2\}$, and thus there is nothing to prove. So from now on without loss of generality, assume, $f_1=(x_1,y_1)$ lies on $P_{(s,t)}^1$ and $f_2=(x_2,y_2)$ lies on $P_{(s,t)}^2$. 

Let $P$ be an $s-t$ path in $G \setminus \{f_1,f_2\}$. Observe, $P$ must intersect either $P_{(s,t)}^1[y_1-t]$ or $P_{(s,t)}^2[y_2-t]$. Let $z'$ be the first vertex on $P$, that is also in $V(P_{(s,t)}^1[y_1-t] \cup P_{(s,t)}^2[y_2-t])$ (such a vertex exists because $P$, $P_{(s,t)}^1$ and $P_{(s,t)}^2$ all end at $t$). Let $z'$ lie on $P_{(s,t)}^j$, for some $j\in \{1,2\}$. Now consider the last vertex before $z'$ on $P$, that is also in $V(P_{(s,t)}^1 \cup P_{(s,t)}^2)$ (such a vertex exists because $P$, $P_{(s,t)}^1$ and $P_{(s,t)}^2$ all start from $s$), and let $z$ be that vertex which is on $P_{(s,t)}^i$, for some $i \in \{1,2\}$. ($i$ could be equal to $j$.) Observe, $z$ must lie on the subpath $P_{(s,t)}^i[s-x_i]$. Furthermore, $P[z-z']$ is edge-disjoint with $P_{(s,t)}^1$ and $P_{(s,t)}^2$. Hence, by the construction of $H_{(s,t)}$, there must be a path $Q_{(s,t),u'}^i$ for some $u'\in V(P_{(s,t)}^j[z'-t])$ which starts from a vertex $u<_{P_{(s,t)}^i} z<_{P_{(s,t)}^i} x_i$ ($u$ could be equal to $z$). Hence, $P_{(s,t)}^i[s-u]$ remains intact in $H_{(s,t)} \setminus \{f_1,f_2\}$. Note, $u' \in V(P_{(s,t)}^j[y_j-t]$ (because $u'\in V(P_{(s,t)}^j[z'-t])$). Thus the subpath $P_{(s,t)}^j[u'-t]$ also remains intact in $H_{(s,t)} \setminus \{f_1,f_2\}$. Lastly, since $Q_{(s,t),u'}^i$ is edge-disjoint with $P_{(s,t)}^1$ and $P_{(s,t)}^2$, it also remains intact in $H_{(s,t)} \setminus \{f_1,f_2\}$. Hence, we get an $s-t$ path $P_{(s,t)}^i[s-u] \circ Q_{(s,t),u'}^{i} \circ P_{(s,t)}^j[u'-t]$ in $H_{(s,t)} \setminus \{f_1,f_2\}$.
\end{proof}

Next, let us use the above claim to conclude that $H_{(s,t)}$ is a 2-$\ftrs(G,(s,t))$. Furthermore, we also argue that the size of $H_{(s,t)}$ is $O(n)$.

\begin{lemma}
\label{lem:single-pair-ftrs}
$H_{(s,t)}$ is a 2-$\ftrs(G,(s,t))$ and contains $O(n)$ edges.
\end{lemma} 
\begin{proof}
Since $H_{(s,t)}$ is a subgraph of $G$, from Claim~\ref{clm:special-path}, we conclude that for any two edges $f_1,f_2$, there is an $s-t$ path in $G \setminus \{f_1,f_2\}$ if and only if there is an $s-t$ path in $H_{(s,t)} \setminus \{f_1,f_2\}$. So, $H_{(s,t)}$ is a 2-$\ftrs(G,(s,t))$.

Let us now compute the size of $H_{(s,t)}$. First, we would like to claim that for each vertex $v \in V(H_{(s,t)}) \setminus V(P_{(s,t)}^1 \cup P_{(s,t)}^2)$, in-degree of $v$ in $H_{(s,t)}$ is at most 4. If not, then there exists a vertex $v \in V(H_{(s,t)}) \setminus V(P_{(s,t)}^1 \cup P_{(s,t)}^2)$ with at least five incoming edges $e_1,e_2,e_3,e_4,e_5$. Suppose during the construction of $H_{(s,t)}$, for each $r \in [5]$, $e_r$ has been included while adding a path $Q_{(s,t),v_r}^{i_r}$, where $v_r \in V(P_{(s,t)}^{j_r})$. Since for each $r \in [5]$, $i_r,j_r \in \{1,2\}$, by the pigeonhole principle, there must exist $r\ne r'\in [5]$ such that $i_r=i_{r'}(=i)$ and $j_r=j_{r'}(=j)$. Without loss of generality, assume, $v_r <_{P_{s,t}^{j}} v_{r'}$. Since $Q_{(s,t),v_r}^{i}$ and $Q_{(s,t),v_{r'}}^{i}$ intersect at $v$, either $u_{(s,t),v_{r'}}^{i}=u_{(s,t),v_r}^{i}$ or $u_{(s,t),v_{r'}}^{i} <_{P_{(s,t)}^i} u_{(s,t),v_r}^{i}$. Hence, by the Step 3(a) of the construction procedure of $H_{(s,t)}$, $Q_{(s,t),v_r}^{i_r}$ would not be added, which leads to a contradiction.

Observe, by the construction of $H_{(s,t)}$, in-degree of each $v \in V(P_{(s,t)}^1 \cup P_{(s,t)}^2)$ is also at most 4. Hence, $H_{(s,t)}$ has total $O(n)$ edges.
\end{proof}

\section{Dual Fault-tolerant Pairwise Reachability Preserver}
\label{section:dual-ftrs}
In this section, we provide the construction of $2$-FTRS for multiple pairs, and thus prove \cref{thm:dual-ftrs}. To prove \cref{thm:dual-ftrs} we provide a construction of \emph{2-$\ftrs$ with slack}. Let us define it below.

\paragraph*{Sparsifiers with Slack. }In case of pairwise FTRS we want to construct an subgraph that provides correct reachability information for all the given vertex-pairs even after edge failures. This is the most standard definition of pairwise FTRS. Next, we define a variant which we refer to as \emph{pairwise FTRS with slack}. Interested readers may find more works on sparsifiers with slack in~\cite{chan2006spanners, konjevod2007compact, dinitz2007compact, bodwin2020note}. Given a graph $G$ and a pair set $\P$, a subgraph $H$ is a $\kftrs$ with slack if there is $\P' \subseteq \P$, $|\P'| = \Omega(|\P|)$ such that $H$ is a $\kftrs$ for the pair set $\P'$.

The following lemma from~\cite{bodwin2020note} shows that to construct a $\kftrs$ for a pair set, it suffices to construct a $\kftrs$ with slack.

\begin{lemma}[~\cite{bodwin2020note}]
\label{lem:ftrs-slack}
Let us consider constants $\alpha,\beta,\gamma >0$ and a parameter $p^*$. Consider a graph $G(V,E)$ with $n$ vertices. If
\begin{itemize}
\item There is a $\kftrs$ of size $O(n^{\alpha})$ for any vertex-pair set $\P$ of size at most $p^*$, and
\item There is a $\kftrs$ with slack of size $O(n^{\beta} |\P|^{\gamma})$ for any vertex-pair set $\P$ of size at least $p^*$,
\end{itemize}
then there is a $\kftrs(G,\P)$ of size at most $O(n^{\alpha} + n^{\beta} |\P|^{\gamma})$. Moreover, if the $\kftrs$ for at most $p^*$ pairs and the $\kftrs$ with slack for at least $p^*$ pairs can be computed in polynomial time, then the $\kftrs(G,\P)$ can also be computed in polynomial time.
\end{lemma}

Thus, due to \cref{lem:ftrs-slack} to prove \cref{thm:dual-ftrs} it suffices to construct $2$-FTRS with slack. To construct that, we need the concept of \emph{fractional hitting set}.

\paragraph*{Fractional Hitting Set. }Let $U:=\{1,2,\cdots,n\}$ be the universe. A set $S \subseteq U$ is said to be a \emph{hitting set} of a family of subsets $S_1,S_2,\cdots,S_m \subseteq U$ if and only if for all $i\in [m]$, $S\cap S_i \ne \emptyset$ (i.e., $S$ intersects all the subsets $S_i$). Given a family of subsets, the problem of finding the minimum sized hitting set is referred to the \emph{hitting set problem}, and is known to be NP-complete. However, there is a simple greedy algorithm that given any family of subsets $S_1,\cdots,S_m$ each of size at least $k$, finds a hitting set of size at most $O(\frac{n}{k} \log n)$.

In this thesis, we consider a variant of the hitting set problem, which we call \emph{fractional hitting set}. A set $S \subseteq U$ is said to be a hitting set with slack of a family of subsets $S_1,\cdots,S_m$ if and only if $|\{i \in [m] \mid S\cap S_i \ne \emptyset\}| = \Omega(m)$ (i.e., $S$ has a non-empty intersection with at least a constant fraction of the input subsets). It is not difficult to see that the above mentioned greedy hitting set algorithm finds a fractional hitting set of size at most $O(n/k)$.

\begin{theorem}
\label{thm:frac-hitting-set}
There is a deterministic algorithm that given a family of subsets $S_1,\cdots,S_m \subseteq [n]$ each of size at least $k$, finds a fractional hitting set of size at most $O(n/k)$ in time $O(n + mk + \frac{n}{k} \log n)$.
\end{theorem}
\begin{proof}
W.l.o.g. assume, each $S_i$ is of size exactly $k$; otherwise we can just consider any $k$ elements from each $S_i$. Let us now consider a slightly modified version of the standard greedy algorithm for the hitting set problem. Initialize $\F:=\{S_1,\cdots,S_m\}$, and $S=\emptyset$. For each $v\in [n]$, compute the number of subsets in $\F$ that contains $v$, i.e., $c(v):=|\{S_i \in \F \mid v \in S_i\}|$. Let $v_{max}$ be a maximizer of $c(v)$. Include $v_{max}$ in $S$, i.e, $S \gets S \cup \{v_{max}\}$. Remove all the subsets that contains $v_{max}$ from $\F$, i.e., $\F \gets \F \setminus \{S_i \mid v_{max} \in S_i\}$. Continue this process $R=4n/k$ times, and output $S$. (Note, instead to continuing the process until $\F=\emptyset$ as in the standard greedy algorithm for the hitting set problem, we terminate after a fixed number of steps.)

Clearly, $|S|=R=O(n/k)$. We would like to claim that $S$ is a fractional hitting set. For that purpose, it suffices to show that at the end of the algorithm $|\F| \le m/10$. Let $n_j$ denote the size of $\F$ after the $j$-th element is added in $S$. $n_0=m$, and we need to show that $n_R \le m/10$. Consider the step when the $j$-th element has been added in $S$. Observe, $\sum_{v \in [n]\setminus H} c(v) = n_{j-1} k$, and thus by a simple averaging, an element $v_{max}^j$ that maximizes $c(v)$ must satisfy $c(v_{max}^j) \ge \frac{n_{j-1}k}{n - j +1}$. Thus,
$$n_j \le \Big(1-\frac{k}{n-j+1}\Big)n_{j-1} \le m \prod_{r=0}^{j-1}\Big(1-\frac{k}{n-r}\Big) < m(1-k/n)^j \le me^{-kj/n}.$$
Then for $R=4n/k$, $n_R < m/10$.

The running time of the modified algorithm is bounded by the standard greedy algorithm for the hitting set problem. For the sake of completeness, let us briefly comment on the running time analysis. At the beginning we can compute the values of $c(v)$'s, and then store them using any standard heap data structure (e.g.~\cite{brodal2012strict}). Then at each step, we extract the max value, and then we update the values of the others. This gives the total time to be $O(n + mk + \frac{n}{k}\log n)$.
\end{proof}

\begin{theorem}
\label{thm:dual-ftrs-slack}
For any directed graph $G=(V,E)$ with $n$ vertices and a set $\P \subseteq V \times V$ of vertex-pairs, there exists a 2-$\ftrs$ with slack having at most $O(n^{4/3} |\P|^{1/3})$ edges. Furthermore, we can find such a subgraph in polynomial time.
\end{theorem}

Now we are ready to describe the construction of $2$-FTRS with slack. We will use the notations $H_{(s,t)}, P_{(s,t)}^i, Q_{(s,t),v}^j$ defined in Section~\ref{section:single-pair}. Apart from them, we need a few more notations. 

\begin{itemize}
    \item $\freq_B(v)$ denotes the frequency of a vertex in a set of paths $B$, more elaborately, for any set $B$ of paths in $G$, for any vertex $v \in V$, $\freq_B(v)$ denotes the number of paths in $B$ that contains $v$.

    \item $P[L]$~:~ The subpath of the path $P$ containing the first $L$ vertices of $P$.
    
    \item $P[-L]$~:~ The subpath of the path $P$ containing the last $L$ vertices of $P$.
\end{itemize}

\textbf{Procedure to Construct a 2-$\ftrs$ with slack for the input $(G,\P)$:}
\begin{enumerate}
\item For each $(s,t)\in \P$, construct $H_{(s,t)}=2\text{-}\ftrs(G,(s,t))$ using the construction provided in Section~\ref{section:single-pair}. (Recall, $H_{(s,t)}$ contains two $s-t$ paths $P_{(s,t)}^1$ and $P_{(s,t)}^2$, that intersect only at $s-t$ cut-edges and $s-t$ cut-vertices.) Define $H_{inter}=\bigcup_{(s,t)\in \P}H_{(s,t)}$.
\item Consider the following family of subsets $\F:=\{V(P^i_{(s,t)}[L]), V(P^i_{(s,t)}[-L]) \mid i\in \{1,2\},(s,t) \in \P\}$, where $L=n^{2/3}|P|^{-1/3}$. Construct a fractional hitting set $S$ of $\F$ (by Theorem~\ref{thm:frac-hitting-set}). Let $\Q:=\{(s,t)\in \P \mid \forall_{i\in \{1,2\}} V(P^i_{(s,t)}[L])\cap S \ne \emptyset \text{ and } V(P^i_{(s,t)}[-L]) \cap S \ne \emptyset\}$.
\item For each $v \in S$, construct a 2-$\ftrs(G,\{v\}\times V)$ and 2-$\ftrs(G,V\times \{v\})$, and take union of all these subgraphs to get a subgraph $H_1$.
\item For each $(s,t) \in \Q$ and $i \in \{1,2\}$, consider the subpaths $P_{(s,t)}^i[L]$ and $P_{(s,t)}^i[-L]$. Let $H_2$ be the graph $\bigcup_{(s,t)\in \P, i\in \{1,2\}}(P_{(s,t)}^i[L] \cup P_{(s,t)}^i[-L])$.
\item Let $B$ be the set of all the paths $Q_{(s,t),v}^i$ (defined in Section~\ref{section:single-pair}) for $(s,t)\in \Q$, $v \in V(\cup_{j \in \{1,2\}}P_{(s,t)}^j[-L])$ and $i \in \{1,2\}$.
\item Initialize an empty set $W$.
\item While $\exists v \in V$ such that $\freq_B(v) \ge \sqrt{L|\Q|}$
\begin{enumerate}
\item Add $v$ to $W$ and remove all the paths that contains $v$ from $B$.
\end{enumerate}
\item For each $w \in W$, construct a 2-$\ftrs(G,\{w\}\times V)$ and 2-$\ftrs(G,V\times \{w\})$, and take union of all these subgraphs to get a subgraph $H_3$.
\item Let $H_4$ be the subgraph obtained by taking the union of all the (remaining) paths in $B$.
\item Return the subgraph $H=H_1\cup H_2 \cup H_3 \cup H_4$.
\end{enumerate}

\paragraph*{Size of the subgraph $H$. }By Theorem~\ref{thm:frac-hitting-set}, $|S|=O(n/L)=O((n|\P|)^{1/3})$. We use Theorem~\ref{theorem:ftrs} to build a 2-$\ftrs(G, \{v\} \times V)$ and 2-$\ftrs(G, V \times \{v\})$. So the size of $H_1$ is $O((n|\P|)^{1/3} \cdot n)$. By the construction (Step 4), $H_2$ is of size at most $O(L \cdot |\P|)$.  

Observe, while defining $B$ (Step 5) before the start of the while loop, we add at most $O(L)$ paths for each $(s,t) \in \Q$. So, $|B|= O(L |\Q|)$. At each iteration of the while loop (Step 7(a)), we remove at least $\sqrt{L|\Q|}$ paths from $B$ and add a new vertex in $W$. At the end of the while loop, for each vertex $v \in V \setminus W$, $\freq_B(v) < \sqrt{L |\Q|}$. So, at the end of the while loop $|W| \le O\Big(\frac{L |\Q|}{\sqrt{L|Q|}}\Big)=O(\sqrt{L|\Q|})$. Again, we use Theorem~\ref{theorem:ftrs} to build a 2-$\ftrs(G, \{w\} \times V)$ and 2-$\ftrs(G, V \times \{w\})$ for each $w \in W$. Thus the size of $H_3$ is at most $O(n \sqrt{L|\Q|})$.

Since $H_4$ is the union of all the leftover paths in $B$ after the while loop, in-degree of each vertex in $H_4$ is at most $\sqrt{L|\Q|}$. So the size of $H_4$ is $O(n \sqrt{L|\Q|})$. Hence, we conclude that the size of the final subgraph $H$ is $O((n|\P|)^{1/3} \cdot n + L \cdot |\Q| + n \sqrt{L|\Q|}) = O(n^{4/3} |P|^{1/3})$, for $L=n^{2/3}|P|^{-1/3}$ (note, $|\Q| \le |\P|\le n^2$).

\paragraph*{Correctness of pairwise 2-{\ftrs} with slack. } 
Next, we would like to claim that the subgraph $H$ is a 2-$\ftrs$ with slack for $(G,\P)$. First, we show that the constructed data structure is a 2-$\ftrs(G,\Q)$. Second, we show that $|\Q| = \Omega(|\P|)$. The second part is quite straightforward. By the definition of fractional hitting set, $S$ intersects all but $1/10$ fraction of the subsets of $\F$. Hence, $|\Q| \ge \frac{3}{5} |\P|$. So it only remains to show the first part.
\begin{lemma}
\label{lem:ftrs-correctness}
The subgraph $H$ is a 2-$\ftrs(G,\Q)$.
\end{lemma}
\begin{proof}
Consider any arbitrary pair $(s,t) \in \Q$. It is immediate from the definition of a 2-$\ftrs$ that any 2-$\ftrs(H_{(s,t)},(s,t))$ is also a 2-$\ftrs(G,(s,t))$. So to prove Lemma~\ref{lem:ftrs-correctness}, it suffices to show that for all $(s,t) \in \Q$, $H$ is a 2-$\ftrs(H_{(s,t)},(s,t))$. From now on we will focus on proving this latter claim.

Suppose there is an $s-t$ path in $H_{(s,t)} \setminus \{f_1,f_2\}$. Observe, by Claim~\ref{clm:special-path}, there must be a nice $s-t$ path $R$ in $H_{(s,t)} \setminus \{f_1,f_2\}$. By the definition of nice path, $R$ would be of the form $P_{(s,t)}^i[s-u] \circ Q_{(s,t),u'}^{i} \circ P_{(s,t)}^j[u'-t]$, for some $i,j \in \{1,2\}$, where the subpath $Q_{(s,t),u'}^{i}$ starts from the vertex $u$ and is edge-disjoint with $P_{(s,t)}^1$ and $P_{(s,t)}^2$. ($Q_{(s,t),u'}^{i}$ could be empty; in that case, $R=P_{(s,t)}^i$ for some $i \in \{1,2\}$.) By the definition of $\Q$, $P_{(s,t)}^i[L]\cap S \ne \emptyset$ and $P_{(s,t)}^j[-L]\cap S \ne \emptyset$. Now, if $u$ does not lie on $P_{(s,t)}^i[L]$, i.e., the length of the subpath $P_{(s,t)}^i[s-u]$ is at least $L$, then $R$ must pass through a vertex $v \in P_{(s,t)}^i[L] \cap S$. Similarly, if $u'$ does not lie on $P_{(s,t)}^j[-L]$, then $R$ must pass through a vertex $v \in P_{(s,t)}^j[-L] \cap S$. In both the scenarios, for some $v \in S$, there is an $s-v$ and $v-t$ path in $H_{(s,t)} \setminus \{f_1,f_2\}$ (and thus also in $G \setminus \{f_1,f_2\}$). This means, by the construction of $H_1$ (Step 3), there is an $s-v$ and $v-t$ path in $H_1 \setminus \{f_1,f_2\}$ as well.

So let us now focus on the case when $u$ lies on $P_{(s,t)}^i[L]$ and $u'$ lies on $P_{(s,t)}^j[-L]$. By the construction (Step 4), $H_2$ contains all the edges of $P_{(s,t)}^i[s-u]$ and $P_{(s,t)}^j[u'-t]$. By the construction of the set $B$ (Step 5), $Q_{(s,t),u'}^{i}$ belongs to the set $B$. Then either during the execution of the while loop (Step 7), $Q_{(s,t),u'}^{i}$ has been removed due to inclusion of some vertex $w \in V(Q_{(s,t),u'}^{i})$ in the set $W$; or $Q_{(s,t),u'}^{i}$ is part of $B$ after the termination of the while loop. For the first case, there is an $s-w$ and $w-t$ path in $H_{(s,t)} \setminus \{f_1,f_2\}$ (and so in $G\setminus \{f_1,f_2\}$), and thus also in $H_3 \setminus \{f_1,f_2\}$ by the construction of $H_3$ (Step 8). In the second case, $Q_{(s,t),u'}^{i}$ is included in $H_4$ (Step 9), and hence $R=P_{(s,t)}^i[s-u] \circ Q_{(s,t),u'}^{i} \circ P_{(s,t)}^j[u'-t]$ also exists in $H_2 \cup H_4$ (and thus in $H\setminus \{f_1,f_2\}$). This concludes the proof.
\end{proof}

Let us now combine Theorem~\ref{thm:dual-ftrs-slack} with Lemma~\ref{lem:ftrs-slack} to finish the proof of Theorem~\ref{thm:dual-ftrs}.

\begin{proof}[Proof of Theorem~\ref{thm:dual-ftrs}]
For single pair, by~\cite{BCR16}, we have a 2-$\ftrs$ of size $O(n)$ that can be computed in polynomial time. By Theorem~\ref{thm:dual-ftrs-slack}, we have a 2-$\ftrs$ with slack of size $O(n^{4/3} |\P|^{1/3})$. Theorem now directly follows from Lemma~\ref{lem:ftrs-slack}.
\end{proof}

\chapter{Closest Fair Clustering with Two Colors}\label{sec:closest-fair-two-colors}

In this chapter, we present an algorithm to construct a fair clustering $\m{F}$ which is $17$-close to an input clustering $\m{C}$, when the irreducible ratio of blue to red vertices is \( p:1 \) for some integer \( p > 1 \). More specifically, we prove \cref{thm:closest-fair-p:1}, which we restate below.

\upperboundtwocolors*

Our approach proceeds in two stages. 
In the first stage, given an input clustering \( \mathcal{C} \), we compute a \emph{$p$-divisible clustering} \( \mathcal{T} \). 
In the second stage, we take \( \mathcal{T} \) as input and produce a fair clustering \( \mathcal{F} \). 

\begin{definition}[$p$-Divisible Clustering]\label{def:pdivisible}
Let \( V \) be a set of vertices colored from the set \( \chi = \{ c_1, \ldots, c_r \} \), where the color proportions in \( V \) satisfy 
\[
p_1 : p_2 : \cdots : p_r = c_1(V) : c_2(V) : \cdots : c_r(V)
\]
and \( c_i(V) \) denotes the number of vertices of color \( c_i \) in \( V \).  
A clustering \( \mathcal{T} = \{ T_1, \ldots, T_m \} \) of \( V \) is called a \emph{$p$-divisible clustering} if, for every cluster \( T_i \in \mathcal{T} \) and each color \( c_j \in \chi \), the count \( c_j(T_i) \) of vertices of color \( c_j \) in \( T_i \) is divisible by \( p_j \); that is, \( p_j \mid c_j(T_i) \).
\end{definition}

In the special case of two colors—blue and red—where the global ratio of blue to red vertices is \( p:1 \), a $p$-divisible clustering \( \mathcal{T} \) ensures that, for every cluster \( T_i \in \mathcal{T} \), the number of blue vertices \( |\blue{T_i}| \) is a multiple of \( p \).  
We use the abbreviation \( \bal \) to denote a $p$-divisible clustering.  

A clustering \( \mathcal{T} \) is said to be \emph{$\delta$-close} $\bal$ to an arbitrary unfair clustering \( \mathcal{D} \) if
\[
\dist(\mathcal{D}, \mathcal{T}) \le \delta \cdot \dist(\mathcal{D}, \mathcal{T}^*),
\]
where \( \mathcal{T}^* \) denotes the closest $\bal$ to \( \mathcal{D} \).

Finally, we complement our upper bound result by showing that the problem becomes NP-hard when the ratio between the number of blue and red vertices is \( p:q \) with \( p \neq q \).

\section{Creating a p-divisible clustering.}

In this section, given a clustering $\inpsetd$ of a set of red-blue colored vertices, where the irreducible ratio of blue to red vertices is $ p $ for some integer $ p > 1 $, we construct a $\bal$ $\outd$. Our main result for this section is the following:

\begin{lemma} \label{thm:main-multiple-of-p}
    There exists a polynomial time algorithm that, given any clustering $\inpset$ on $n$ colored vertices with each vertex having a color either red or blue and a non-negative integer $p$, finds its $3.5$-close $\mopdef$.
\end{lemma}


We provide the algorithm to find a $3.5$-close $\mopdef$ in \cref{subsec:algorithms} and provide the analysis of the algorithm in \cref{subsec:analysis}

\subsection{Details of the Algorithm} \label{subsec:algorithms} 
Before describing our algorithm, let us first introduce a few notations that we use later.

For any input cluster $C_i$, if its number of blue vertices is not a multiple of $p$, then we define its surplus as the minimum number of blue vertices whose removal makes the number of blue vertices a multiple of $p$, and we define its deficit as the minimum number of blue vertices whose addition makes the number of blue vertices to be a multiple of $p$. Formally, we define them as follows.

\begin{itemize}
    \item Surplus of $C_i$, $\surp{C_i}$: For any cluster $C_i \subseteq V$, $\surp{C_i} = |\blue{C_i}| \mod p$.
    \item Deficit of $C_i$, $\defi{C_i}$: For any cluster $C_i \subseteq V$ if $p \nmid |\blue{C_i}|$ then $\defi{C_i} = p - \surp{C_i}$ otherwise $\defi{C_i} = 0$.
\end{itemize}

We call the operation of taking some vertices, $S$ (say) from a cluster $\inpcl{i} $, that is $\inpcl{i} = \inpcl{i} \setminus S$ as \emph{cut} and the operation of merging some vertices $M$ (say) to a cluster $\inpcl{i}$ that is $\inpcl{i} = \inpcl{i} \cup M$ as \emph{merge}. Let us define some more terms that we need to describe our algorithms.

\begin{itemize}
    \item Cut cost, $\ccostf{\inpcl{i}}$: For any cluster $\inpcl{i} \in \inpset$, $\ccostf{\inpcl{i}} = (\surp{\inpcl{i}} \cdot (|\inpcl{i}| - \surp{\inpcl{i}})$.

    \item Merge cost, $\mcostf{\inpcl{i}}$: For any cluster $\inpcl{i} \in \inpset$, $\mcostf{\inpcl{i}} = d(\inpcl{i}) \cdot |\inpcl{i}|$.

\end{itemize}

\paragraph{Description of the algorithm $\algog$.}
Given a clustering, $\inpset$, we partition the clusters $\inpcl{i} \in \inpset$ into two types based on the $\surp{\inpcl{i} }$ value of the cluster $\inpcl{i} $. If $\surp{\inpcl{i} } \leq p/2$ we call these clusters as $\cut$ clusters and if $\surp{\inpcl{i} } > p/2$ we call these clusters as $\merge$ clusters. Thus

\begin{itemize}
    \item $\cut = \{\inpcl{i} \mid \surp{\inpcl{i}} \leq p/2\}$.
    \item $\merge = \{\inpcl{i} \mid \surp{\inpcl{i}} > p/2\}$.
\end{itemize}

We first sort the clusters $\inpcl{i} \in \merge$ based on the value of $(\ccostf{\inpcl{i}} - \mcostf{\inpcl{i}})$ in \emph{non-increasing} order. Our algorithm $\algog$ cuts a set of $\surp{\inpcl{i} }$ blue vertices from each cluster $\inpcl{i}  \in \cut$ and merges them to clusters in $\merge$ sequentially starting from the beginning in the sorted list. In each cluster $\inpcl{j} \in \merge$, we merge a set of $\defi{\inpcl{j}}$ blue vertices. Now, we have two cases: after a certain vertex of time, the remaining set of clusters in $\cut$ becomes empty, and we are only left with some clusters in $\merge$; in this case, we call our subroutine $\algom$. To remove any ambiguity, we refer to the remaining clusters in $\merge$ at this vertex as $\merge'$, and the set of clusters created from $\cut$ after cutting the surplus vertices $s(C_i)$ from each cluster $C_i \in \cut$ be $\nc$. 

Another case is when the remaining set of clusters in $\merge$ becomes empty, and we are only left with some clusters in $\cut$. In this case, we call our subroutine $\algoc$. Again, To remove any ambiguity, we refer to the remaining set of clusters in $\cut$ as $\cut'$.

\paragraph{Description of the Subroutine $\algom$.}

The subroutine $\algom$ takes as input two sets of clusters: $\nc$ and $\merge'$. 
Here, $\nc$ denotes the set of clusters that have already become $p$-divisible after the initial phase of $\algog$, and $\merge'$ denotes the remaining clusters from $\merge$ whose number of blue vertices is not yet a multiple of $p$. 

We begin by computing the total deficit of the clusters in $\merge'$, defined as
\[
W = \sum_{C_j \in \merge'} \defi{C_j}.
\]
Intuitively, $W$ represents the total number of blue vertices that must be added (through merging) across all clusters in $\merge'$ to make them $p$-divisible.

For every cluster $C_k \in \nc \cup \merge'$, we partition the set of blue vertices $\blue{C_k}$ into
\[
\frac{|\blue{C_k}| - \surp{C_k}}{p}
\]
subsets of size $p$, and one remaining subset of size $\surp{C_k}$. 
We number the subset of size $\surp{C_k}$ as the \emph{$0$th subset}, and label all other subsets of size $p$ sequentially starting from $1$. 
We denote the $z$th subset of cluster $C_k$ by $W_{m,z}$.

Each subset $W_{m,z}$ is associated with a cutting cost defined as follows:
\begin{align}
\kappa_0(C_k) &= \surp{C_k} \cdot (|C_k| - \surp{C_k}), && \text{(cost for cutting the $0$th subset)}, \\
\kappa_z(C_k) &= p \cdot (|C_k| - (zp + \surp{C_k})), && \text{(cost for cutting the $z$th subset, $z \ge 1$)}. \notag
\end{align}

The subroutine $\algom$ is invoked in the situation where only clusters in $\merge$ remain non–$p$-divisible. 
Ideally, we would prefer to perform merge operations for all clusters in $\merge$, since for these clusters we have $\surp{C_i} > p/2$, which implies $\mcostf{C_i} < \ccostf{C_i}$. 
However, in order to make every cluster $C_i \in \merge$ satisfy the divisibility condition $p \mid |\blue{C_i}|$, we must also perform cut operations on some clusters to supply the required blue vertices. 
Therefore, we choose to cut from clusters $C_i$ that have the smallest difference $(\ccostf{C_i} - \mcostf{C_i})$, ensuring that the incurred cutting cost is not disproportionately large relative to their merging benefit.

After the main algorithm $\algog$ terminates, all clusters in $\nc$ are already $p$-divisible. 
Nevertheless, it may still be more cost-effective to cut subsets of size $p$ from some of these clusters in $\nc$ than from certain clusters in $\merge'$. 
This means that the algorithm may cut from the same cluster multiple times if it minimizes the total cost.

At each iteration of $\algom$, among all clusters $C_\ell \in \nc \cup \merge'$, the algorithm selects a cluster $C_k$ with the minimum cost associated with its least uncut subset $W_{m,v_m}$, where $v_m$ denotes the smallest index corresponding to a subset of size $p$ not yet cut from $C_k$. 
The cost of the subset $W_{m,v_m}$ is defined as:
\begin{align}
\textit{cost}(W_{m,v_m}) &=
\begin{cases}
\kappa_{v_m}(C_k), & \text{if } v_m \ge 1,\\
\kappa_0(C_k) - \mcostf{C_k}, & \text{if } v_m = 0.
\end{cases}
\end{align}

If multiple clusters have the same minimum cost, the algorithm breaks ties arbitrarily. 
Let $C_k$ be the selected cluster. 
The algorithm then cuts the subset $W_{m,v_m}$ from $C_k$. 
If $C_k \in \merge'$, it is subsequently removed from $\merge'$ and the subset $W_{m,v_m}$ is merged sequentially into the first available clusters in $\merge'$, which are maintained in non-increasing order of $(\ccostf{C_i} - \mcostf{C_i})$.

The process continues iteratively until the total deficit $W$ becomes zero. 
We later prove that exactly $W/p$ subsets need to be cut to eliminate the entire deficit, thereby ensuring that all clusters become $p$-divisible.


\paragraph{Description of the Subroutine $\algoc$.}

The subroutine $\algoc$ takes as input the set of clusters $\cut'$, which represents the remaining clusters in $\cut$ after the main algorithm $\algog$ terminates.  
For each cluster $\inpcl{i} \in \cut'$, the algorithm removes $\surp{\inpcl{i}}$ blue vertices—corresponding to its surplus—and collects these vertices to form new \emph{auxiliary} clusters, each of size $p$.  
This process is repeated sequentially until, in every cluster $\inpcl{i} \in \cut'$, the number of blue vertices becomes a multiple of $p$.  
As a result, all clusters in the final output satisfy the $p$-divisibility property.

We provide the pseudocode of the algorithms $\algog$, $\algoc$, $\algom$ in \cref{alg:algo-for-general}, \cref{alg:algo-for-cut} and \cref{alg:algo-for-merge} respectively. Let the output of the algorithm $\algog$ be $\out$. 

\paragraph{Runtime analysis of $\algog$.} The expensive step in the $\algog$ is that in the first step of the algorithm, we sort the clusters in $\merge$ which would take $O\left(|\merge| \, \, \log (|\merge|)\right) = O(n \log n)$ time. The algorithm uses a vertex $v \in V$ at most twice, once while cutting the vertex $v$ from its parent cluster $C_i \in \inpset$ (say) and another when merging the vertex $v$ to some cluster $C_j \neq C_i$. Hence, the cutting and merging process of the algorithm can take at most $O(n)$ time (recall, $|V|=n$). 

In the subroutine $\algom$, at each iteration, the algorithm needs to find the cluster that has the minimum cost attached to its least-numbered subset. The find-Min operation can be done in $O(1)$ by implementing a priority queue with a Fibonacci heap. Hence, overall the time complexity of $\algog$ is $O(|\merge| \, \, \log (|\merge|) + |V|) = O(n \log n)$. 


\subsection{Approximation Guarantee for $\algog$} \label{subsec:analysis}

In this section, we prove that $\dist(\inpset, \out) \leq 3.5 \, \dist(\inpset, \mop)$ which is stated formally in \cref{thm:main-multiple-of-p}

Our proof is divided into the following two cases:
\begin{itemize}
    \item \emph{Merge} case: When $\sum_{\inpcl{i} \in \cut} \surp{\inpcl{i}} \leq \sum_{\inpcl{j} \in \merge} \defi{\inpcl{j}}$; and
    \item \emph{Cut} case: When $\sum_{\inpcl{i} \in \cut} \surp{\inpcl{i}} > \sum_{\inpcl{j} \in \merge} \defi{\inpcl{j}}$.
\end{itemize}

In \cref{subsubsec:merge-case}, we argue that in the merge case $\dist(\inpset, \out) \leq 3 \, \dist(\inpset, \mop)$.

\begin{restatable} {lemma} {mainmerge}
    If $\sum_{\inpcl{i} \in \cut} \surp{\inpcl{i}} \leq \sum_{\inpcl{j} \in \merge} \defi{\inpcl{j}}$, the algorithm $\algog$ produces a $\mopdef$ $\out$ such that $\dist(\inpset, \out) \leq 3 \; \dist(\inpset, \mop)$.
    \label{lem:main-merge-case}
\end{restatable}

In \cref{subsec:cut-case}, we argue that in the cut case $\dist(\inpset, \out) \leq 3.5 \, \dist(\inpset, \mop)$.

\begin{restatable} {lemma} {maincut}
    If $\sum_{\inpcl{i} \in \cut} \surp{\inpcl{i}} > \sum_{\inpcl{j} \in \merge} \defi{\inpcl{j}}$, the algorithm $\algog$ produces a $\mopdef$ $\out$ such that $\dist(\inpset, \out) \leq 3.5 \; \dist(\inpset, \mop)$.
    \label{lem:main-cut-case}
\end{restatable}

It is straightforward to see that to conclude the proof of \cref{thm:main-multiple-of-p}, it suffices to show \cref{lem:main-merge-case} and \cref{lem:main-cut-case}. In \cref{subsubsec:structure-of-M*}, we discuss some important properties of $\mop$ that we need to prove \cref{lem:main-merge-case} and \cref{lem:main-cut-case}. In \cref{subsubsec:merge-case} we prove \cref{lem:main-merge-case} and in \cref{subsec:cut-case} we prove \cref{lem:main-cut-case}.

\subsubsection{Structure of $\mop$} \label{subsubsec:structure-of-M*}
To prove \cref{lem:main-merge-case} and \cref{lem:main-cut-case}, we need an important property about the structure of $\mop$. We need to define some notations to state that important property of $\mop$.



Consider any vertex $v \in V$. Suppose in the clustering $\inpset$, $v \in C_i$, for some $i$, and in the clustering $\mop$, $v \in T^*_j$, for some $j$. Note, the number of vertices $u$ such that $u \in \inpcl{i} \setminus \mopcl{j}$ is given by

\begin{align}
\sum\limits_{\mopcl{k} \neq \mopcl{j}} \abs{\inpcl{i} \cap \mopcl{k}} \label{equn:expression-1}
\end{align}

Similarly the number of vertices $u$ such that $u \in \inpcl{i} \setminus \mopcl{j}$ is given by

\begin{align}
\sum\limits_{\inpcl{k} \neq \inpcl{i}} \abs{\inpcl{k} \cap \mopcl{j}} \label{equn:expression-2}
\end{align}

Thus by the definition of the distance function, we have
\begin{align}
\dist(\inpset, \mop) = \sum\limits_{v \in V}1/2 \cdot \left(\sum\limits_{k \mid \mopcl{k} \neq \mopcl{j}} \abs{\inpcl{i} \cap \mopcl{k}} + \sum\limits_{m \mid \inpcl{m} \neq \inpcl{i}} \abs{\inpcl{m} \cap \mopcl{j}}\right) \label{equn:consensus-metric}
\end{align}.

The $1/2$ comes before the expressions \ref{equn:expression-1} and \ref{equn:expression-2} because when a pair $(u,v)$ we charge $1/2$ for the vertex $u$ and $1/2$ for the vertex $v$. $\dist(\inpset, \mop)$ counts the total number of pairs that are together in $\inpset$ but separated by $\mop$ and the pairs that are in different clusters in $\inpset$ but together in $\mop$. 


Let us denote the cost paid by $\mop$ for a vertex $v \in V$ that belongs to $\inpcl{i} \in \inpset$ for some $i$ and also to $\mopcl{j} \in \mop$ for some $j$ by the notation $\opt_v$ where

\[ \opt_v = 1/2 \cdot \left(\sum\limits_{k \mid \mopcl{k} \neq \mopcl{j}} \abs{\inpcl{i} \cap \mopcl{k}} + \sum\limits_{m \mid \inpcl{m} \neq \inpcl{i}} \abs{\inpcl{m} \cap \mopcl{j}} \right)\].

Let us denote the cost paid by $\mop$ for a cluster $\inpcl{i} \in \inpset$ by

\[ \opt_{\inpcl{i}} = \sum_{v \in \inpcl{i}} \opt_v\].

Now, we are ready to state the important property of $\mop$. We show that if for a cluster $\inpcl{i} \in \inpset$, if $\surp{\inpcl{i}} \leq p/2$ then

\[
    \opt_{\inpcl{i}} \geq s(C_i) (|C_i| - s(C_i)) + \frac{1}{2} s(C_i)(p - s(C_i)) 
\]
that is $\mop$ must pay at least the cost $s(C_i) (|C_i| - s(C_i)) + 1/2 s(C_i)(p - s(C_i))$ for the cluster $\inpcl{i}$ if the surplus of $\inpcl{i}$ is less than or equal to $p/2$. 

Again if $\surp{\inpcl{i}} > p/2$ then 

\[
    \opt_{\inpcl{i}} \geq (p - s(C_i)) (|C_i| - s(C_i)) + \frac{1}{2} s(C_i)(p - s(C_i)) 
\] 

that is $\mop$ must pay at least the cost $(p - s(C_i)) (|C_i| - s(C_i)) + \frac{1}{2} s(C_i)(p - s(C_i))$ for the cluster $\inpcl{i}$ if the surplus of $\inpcl{i}$ is greater than $p/2$. 

We state this formally in the following lemma.

\begin{lemma} \label{lem:main-structure-of-M*}
    For a cluster $\inpcl{i} \in \inpset$ we have
    \begin{enumerate}
        \item if $\surp{\inpcl{i}} \leq p/2$ then $\opt_{\inpcl{i}} \geq s(C_i) (|C_i| - s(C_i)) + \frac{1}{2} s(C_i)(p - s(C_i))$.
        \item if $\surp{\inpcl{i}} > p/2$ then $\opt_{\inpcl{i}} \geq (p - s(C_i)) (|C_i| - s(C_i)) + \frac{1}{2} s(C_i)(p - s(C_i)) $.
    \end{enumerate}
\end{lemma}

For proving \cref{lem:main-structure-of-M*}, we need the help of the following propositions.

\begin{proposition}\label{prop:mod-sum-ineq}
    Consider any integer $p \ge 2$, and $\alpha_1,\ldots,\alpha_t \in [p]$, for $t \ge 1$. Let $ s = \sum_{i=1}^t \alpha_i \mod p$. Then
    $\sum_{i=1}^t \alpha_i \left( p - \alpha_i \right) \ge s \left(p-s\right)$.
\end{proposition}

\begin{proof}
    Suppose, $(\alpha_1 + \alpha_2) \mod p = s_1$, hence $(\alpha_1 + \alpha_2) = kp + s_1$ where $k \in \{0,1\}$.

    First, we prove
    \[ 
        \alpha_1(p - \alpha_1) + \alpha_2(p - \alpha_2) \geq s_1(p - s_1)
    \]
    Now,
    \begin{align}
      &\alpha_1(p - \alpha_1) + \alpha_2(p - \alpha_2) \n \\
      = \, \, &\alpha_1p - \alpha_1^2 + \alpha_2p - \alpha_2^2 \n \\
      = \, \, &p(\alpha_1 + \alpha_2) - (\alpha_1 + \alpha_2)^2 + 2\alpha_1 \alpha_2 \n \\
      = \, \, &p(kp + s_1) - (kp + s_1)^2 + 2\alpha_1\alpha_2 && (\text{Replacing} \, \, (\alpha_1 + \alpha_2) \, \, \text{with} \, \, (kp + s_1) )\n \\
      = \, \, &\left(kp + s_1\right)\left(p - \left(kp + s_1\right)\right) + 2 \alpha_1 \alpha_2 \label{equn:prop-equn}
    \end{align}

    Case $1$: When $k = 0$ then \cref{equn:prop-equn} becomes
    \[
        s_1(p - s_1) + 2\alpha_1\alpha_2 \geq s_1(p - s_1) 
    \]

    Case $2$: When $k = 1$ then \cref{equn:prop-equn} becomes

    \begin{align}
        &(p + s_1)(-s_1) + 2\alpha_1\alpha_2 \n \\
        = \, \, &(p - s_1)s_1 + 2(\alpha_1\alpha_2 - s_1p) \n \\
        = \, \, &(p - s_1)s_1 + 2(\left(s_1p + r(p - r - s_1)\right) - s_1p) \n \\
        &(\text{Replacing} \, \, \alpha_1 \, \, \text{with} \, \, (p - r) \, \, \text{and} \, \, \alpha_2 \, \, \text{with} \, \, (s_1 + r) )\n \\
        = \, \, &(p - s_1)s_1 + 2(\left(s_1p + r(p - \alpha_2)\right) - s_1p) \n \\
        \geq \, \, &(p - s_1)s_1 && (\textbf{as} \, \, \alpha_2 \in [p]) \n
    \end{align}

    Hence, we get 

    \begin{align}
        \alpha_1(p - \alpha_1) + \alpha_2(p - \alpha_2) \geq s_1(p - s_1) && (\text{where} \, \, s_1 = (\alpha_1 + \alpha_2) \mod p) \label{equn:prop-main}
    \end{align}

    Now,

    \begin{align}
        \sum_{i=1}^t \alpha_i \left( p - \alpha_i \right) &= \alpha_1(p - \alpha_1) + \alpha_2(p - \alpha_2) + \sum_{i = 3}^t \alpha_i(p - \alpha_i) \n \\
        &\geq s_1(p - s_1) + \alpha_3(p - \alpha_3) + \sum_{i = 4}^t\alpha_i(p - \alpha_i) \n \\ 
        &(\text{using} \, \, \cref{equn:prop-main}, \, \, \text{assume} \, \, s_1 = (\alpha_1 + \alpha_2) \mod p) \n \\
        &\geq s_2(p - s_2) + \alpha_4(p - \alpha_4) + \sum_{i = 5}^t\alpha_i(p - \alpha_i) \n \\
        &(\text{using} \, \, \cref{equn:prop-main}, \, \, \text{assume} \, \, s_2 = (s_1 + \alpha_3) \mod p) \n \\
        & \vdots \n \\
        &\geq s(p - s) && (\text{Proved}) \n
    \end{align}
\end{proof}

\begin{proposition}\label{prop.bound.the.firsterm}
    Given positive integers $ n $, $ p $, and $ x_{1},x_{2}, \dots, x_{t} $ such that $ n = x_{1}+x_{2}+\dots + x_{t} $. Let $ b, b_{1},b_{2}, \dots, b_{t} <p $ be nonnegative integers satisfying $ b\leq n $, $ b_{i} \leq x_{i}$, and $ b = (b_{1} + b_{2} + \dots + b_{t})(\mod{p}) $. Let $ q_{i} = p-b_{i} (\mod{p}) $. Consider $ A = \sum_{i<j}x_{i}x_{j}  + \sum_{i=1}^{t}(x_{i}-b_{i})q_{i} $. Then
    \begin{enumerate}
        \item If $ b\leq p/2 $, then $ A \geq b(n-b) $. \label{enu.first.term.case1}
        \item If $ b>p/2 $, then $ A \geq (p-b)(n-b) $. \label{enu.first.term.case2}
    \end{enumerate}
\end{proposition}
\begin{proof}
    Without loss of generality, assume that $ x_{1}\leq x_{2}\leq \dots \leq x_{t} $.

    (\ref{enu.first.term.case1}) Suppose that $ b\leq p/2 $. We show that $ A \geq b(n-b) $.
    
    As $ x_{t} = \max\{x_{i}\} $, it follows that $ A \geq \sum_{i<j}x_{i}x_{j} = \sum_{i=1}^{t}x_{i}\left( \dfrac{n-x_{i}}{2} \right) \geq n\left( \dfrac{n-x_{t}}{2} \right) $. Additionally, $ A\geq x_{t}(n-x_{t}) + (x_{t}-b_{t})q_{t} $. Let $ B = n \left( \dfrac{n-x_{t}}{2} \right) $, and $ C = x_{t}(n-x_{t}) + (x_{t}-b_{t})q_{t} $. Then $ A\geq B $ and $ A\geq C $. Therefore, it suffices so that $ B\geq b(n-b) $ or $ C\geq b(n-b) $. We proceed by cases.

    \textbf{Case 1}: $ x_{t}\leq b $.
    \begin{itemize}
            \item If $ n\leq x_{t}+b $, then $ C - b(n-b) > x_{t}(n-x_{t}) - b(n-b) = (x_{t}-b)(n-x_{t}-b) \geq 0 $.
            \item If $ n > x_{t} + b $, then
                \begin{align}
                    B - b(n-b) &= \dfrac{1}{2}(n(n-x_{t}) - 2b(n-b)) \nonumber \\
                               &\geq \dfrac{1}{2}(n(n-x_{t})-b(n-b) - b(n-x_{t})) && (\text{as }b\geq x_{t}) \nonumber \\
                               &= \dfrac{1}{2}(n-b)(n-x_{t}-b) \nonumber \\
                               &\geq 0. && (\text{as }n\geq b \text{ and } n > x_{t}+b)\nonumber
                \end{align}
    \end{itemize}

    \textbf{Case 2:} $ x_{t} > b $.
    \begin{itemize}
        \item If $ n \geq x_{t}+b $, then $ C -
            b(n-b) > x_{t}(n-x_{t}) - b(n-b) = (x_{t}-b)(n-x_{t}-b) \geq 0 $.
        \item If $ n<x_{t}+b $, then $ b>n-x_{t} = x_{1} + x_{2} + \dots + x_{t-1} \geq b_{1} + b_{2} + \dots + b_{t-1} $. Note that $ b  = (b_{1} + b_{2} + \dots + b_{t}) (\mod{p}) $. It follows that $ b_{t} = b-(b_{1} + b_{2} + \dots + b_{t-1}) (\mod{p}) = b - (b_{1} + b_{2} + \dots + b_{t-1})$. Hence, $ 0 < b_{t} < b $ and $ q_{t} = p-b_{t} (\mod p) = p - b_{t} > p-b\geq b $, as $ b\leq p/2 $. This leads to
            \begin{align}
                C - b(n-b) &= (x_{t}-b)(n-x_{t}-b) + (x_{t}-b_{t})q_{t} \nonumber \\
                            &> (x_{t}-b)(n-x_{t}-b) + (x_{t}-b)q_{t} && (\text{as }b_{t}<b) \nonumber \\
                            &= (x_{t}-b)(n-x_{t}-b+q_{t}) \nonumber \\
                            &\geq 0. && (\text{as }x_{t}\geq b, n\geq x_{t}, \text{ and }q_{t}>b)\nonumber
            \end{align}
    \end{itemize}
    (\ref{enu.first.term.case2}) Suppose that $ b>p/2 $. We show that $ A \geq (p-b)(n-b) = a(n-b) $, where $ a=p-b\leq p/2 <b $. As before, we have $ A\geq B $ and $ A\geq C $, where $ B= n\left(\dfrac{n-x_{t}}{2}\right) $, and $ C=x_{t}(n-x_{t}) + (x_{t}-b_{t})q_{t} $. Similarly, we proceed by cases.

    \textbf{Case 1:} $ x_{t} \leq a$.
    \begin{itemize}
        \item If $ n\leq x_{t}+a $, then $ C - a(n-b) \geq x_{t}(n-x_{t}) - a(n-a) = (x_{t}-a)(n-x_{t}-a) \geq 0 $.
        \item If $ n>x_{t}+a $, then
            \begin{align}
                B - a(n-b) &\geq \dfrac{1}{2}(n(n-x_{t})-2a(n-a)) \nonumber \\
                           &\geq \dfrac{1}{2}(n(n-x_{t})-a(n-a)-a(n-x_{t})) && (\text{as } a\geq x_{t}) \nonumber \\
                           &= (n-a)(n-x_{t}-a) \nonumber \\
                           &\geq 0. && (\text{as } n\geq a \text{ and }n\geq x_{t}+a) \nonumber.
            \end{align}
    \end{itemize}
    \textbf{Case 2:} $ x_{t}>a $.
    \begin{itemize}
        \item If $ n\geq x_{t}+a $, then $ C-a(n-b)\geq x_{t}(n-x_{t}) - a(n-a) = (x_{t}-a)(n-x_{t}-a)\geq 0 $.
        \item If $ n<x_{t}+a $, then $ a>n-x_{t} = x_{1} + x_{2} + \dots + x_{t-1} \geq b_{1} + b_{2} + \dots + b_{t-1} $. Note that $ b=(b_{1}+b_{2}+\dots + b_{t}) (\mod{p}) $. It follows that $ b_{t} = b-(b_{1}+b_{2}+\dots +b_{t-1}) (\mod{p}) = b-(b_{1}+b_{2}+\dots + b_{t-1}) $. Hence, $ 0<b_{t}<b $, and $ q_{t} = p-b_{t}(\mod{p}) = p-b_{t} $. Therefore, $ q_{t} = p-b+b_{1}+b_{2}+\dots b_{t-1} \geq p-b = a $. If $ b_{t}\leq a $, it follows that
            \begin{align}
                C - a(n-b) &\geq x_{t}(n-x_{t}) + (x_{t}-b_{t})q_{t} - a(n-b) \nonumber \\
                           &\geq (x_{t}-a)(n-x_{t}-a) - (x_{t}-a)q_{t} && (\text{as }b_{t}<a) \nonumber\\
                           &= (x_{t}-a)(n-x_{t}+q_{t}-a) \nonumber \\
                           &\geq 0. &&(\text{as } x_{t} > a,\ n\geq x_{t} \text{ and } q_{t}\geq a) \nonumber
            \end{align}
            It remains to consider $ b_{t}> a $. We have
            \begin{align}
                &\>C - a(n-b) \nonumber \\
                &= \left( x_{t}(n-x_{t})  - b_{t}(n-a) + x_{t}(b_{t}-a)\right) + b_{t}(n-a)-x_{t}(b_{t}-a) + (x_{t}-b_{t})q_{t} - a(n-b) \nonumber \\
                &= (x_{t}-b_{t})(n-x_{t}-a) + (b_{t}-a)(n-x_{t}) + a(b-b_{t}) + (x_{t}-b_{t})q_{t} \nonumber \\
                &\geq (x_{t}-b_{t})(n-x_{t}-a+q_{t}) + (b_{t}-a)(n-x_{t}) + a(b-b_{t}) \nonumber\\
                &\geq 0. \nonumber
            \end{align}
    \end{itemize}
    This concludes the proof.
\end{proof}

Now we are ready to prove \cref{lem:main-structure-of-M*}

\begin{proof}[Proof of \cref{lem:main-structure-of-M*}]
    Suppose in $\mop$, the cluster $\inpcl{i}$ is split into $t$ parts $X_{i,1}, \ldots, X_{i,t}$; more specifically,
    \begin{itemize}
        \item $\inpcl{i} = \bigcup_{j = 1}^t X_{i,j} $, and
        \item $\forall j \neq k \in [t]$, $ X_{i,j} \cap X_{i,k} = \emptyset$.
    \end{itemize}
     For each part $X_{i,j}$, consider a set $S_{i,j} \subseteq X_{i,j}$ that consists of $(\blue{X_{i,j}} \mod p)$ many blue vertices. Now, since $\mop$ is a $\mopdef$ we get for the clusters $X_{i,k}$ for $k \in \{1, \ldots, t\}$ at least $(p - |S_{i,k}|)$ blue vertices from clusters other than $\inpcl{i}$ must be merged. Let, these set of $(p - |S_{i,k}|)$ blue vertices be $S_{i,k}'$. Hence we get,

     \begin{align}
         \opt_{\inpcl{i}} \geq \sum_{j < k}|X_{i,j}||X_{i,k}| + \sum_{k = 1}^t |X_{i,k} \setminus S_{i,k}||S_{i,k}'| + 1/2 \sum_{k = 1}^t |S_{i,k}|  |S_{i,k}'| \label{equn:main-equn}
     \end{align}
      
     \begin{flalign*}
     &\text{Note that} \, \, \sum_{k = 1}^t |S_{i,k}| \mod p = s_i && \text{recall $\surp{\inpcl{i}} = s_i$}
     \end{flalign*}

    Hence, from \cref{prop:mod-sum-ineq} we get,

    \begin{align}
         &1/2 \sum_{k = 1}^t |S_{i,k}|  |S_{i,k}'| \n \\
         = \, \,& 1/2 \sum_{k = 1}^t |S_{i,k}|  (p - |S_{i,k}|) \n \\
         \geq \, \, &1/2 \, \, s_i  \cdot (p - s_i) \label{equn:main-second-term}
    \end{align}

    It suffices to show that
    \begin{align}
        \sum_{j<k}|X_{i,j}||X_{i,k}| + \sum_{k=1}^{t}|X_{i,k}\setminus S_{i,k}||S'_{i,k}| \geq \surp{\inpcl{i}}(|\inpcl{i}| - \surp{\inpcl{i}}),\label{eq.opt.cut.cost.first.term}
    \end{align}
    when $ \surp{\inpcl{i}}\leq p/2 $, and
    \begin{align}
        \sum_{j<k}|X_{i,j}||X_{i,k}| + \sum_{k=1}^{t}|X_{i,k}\setminus S_{i,k}||S'_{i,k}| \geq (p-\surp{\inpcl{i}})(|\inpcl{i}| - \surp{\inpcl{i}}),\label{eq.opt.merge.cost.first.term}
    \end{align}
    when $ \surp{\inpcl{i}} > p/2 $.

    To this end, we let $ |\inpcl{i}| = n,\ \surp{\inpcl{i}} = b<p $, $ |X_{i,j}| = x_{j} $, $ |S_{i,j}| = b_{j}<p $, and $ |S'_{i,j}| = q_{j} $. Then, $ n = x_{1}+x_{2}+\dots x_{t} $, $ b = (b_{1}+b_{2}+ \dots + b_{t}) (\mod{p}) $, $ q_{j} = (p-b_{j}) (\mod{p}) $, and $ x_{j} \geq b_{j} $, for all $ j $. This allows us to apply~\cref{prop.bound.the.firsterm}, which claims the correctness of~\eqref{eq.opt.cut.cost.first.term} and~\eqref{eq.opt.merge.cost.first.term}.
\end{proof}

\subsubsection{Approximation guarantee in the merge case} \label{subsubsec:merge-case}

In this section, we prove that if $\sum_{\inpcl{i} \in \cut} \surp{\inpcl{i}} \leq \sum_{\inpcl{i} \in \merge} \defi{\inpcl{i}}$, that is in the merge case, the distance between input clustering $\inpset$ and the output clustering $\out$ of the algorithm $\algog$, $\dist(\inpset, \out)$ is at most $3 \, \, \dist(\inpset, \mop)$. let us restate it formally below.

\mainmerge*

To prove \cref{lem:main-merge-case}, here we need to recall and also define some costs paid by our algorithm $\algog$. For a cluster $\outcl{k} \in \out$ we define $\pi(\outcl{k})$ as a parent of $\outcl{k}$ iff $\pi(\outcl{k}) \in \inpset$ and $\outcl{k}$ is formed either by cutting some vertices from $\pi(\outcl{k})$ or by merging some vertices to $\pi(\outcl{k})$.

\begin{itemize}
    \item $\cuta$: The set of clusters $C_i \in \inpset$ from where the algorithm has cut some blue vertices.
    \item $\mergea$: The set of clusters $C_i \in \inpset$ to where the algorithm has merged some blue vertices. We have $\mergea = \inpset \setminus \cuta$.
    \item Cost of cutting the $z$th subset of size $p$ from $\inpcl{i}$ where $z \in \{0, \ldots, (|\blue{\inpcl{i}}| - s_i)/p \}$: 
    \begin{itemize}
        \item $\kappa_0(\inpcl{i}) = s_i (|\inpcl{i}| - s_i)$.
        \item For $z \in \{1, \ldots, (|\blue{\inpcl{i}}| - s_i)/p \}$: $\kappa_z(\inpcl{i}) = p \, (|\inpcl{i}| - (s_i + zp))$.
    \end{itemize}
    Let $y_{i,z}$ take the value $1$ if the algorithm cuts the $z$th subset from $C_i$.
    Hence, we define 
    \begin{align}
     \costone{\m{T}} = \sum_{C_i \in \cuta} \sum_{z = 0}^t y_{i,z} \kappa_z(C_i) \label{eq:cost-one-merge} 
    \end{align}
    \item Cost of merging in a cluster $C_j$ which is $\mcostf{C_j}$.
    Hence, we define 
    \begin{align}
        \costtwo{\m{T}} = \sum_{C_j \in \mergea} \defi{C_j} |C_j| \label{eq:cost-two-merge}
    \end{align}
    \item Inter-cluster cost paid by the vertices that belong to a subset of size $p$ of $\inpcl{i}$ that we cut from $\inpcl{i}$ in the algorithm $\algom$:
    \begin{itemize}
        \item (Intercluster cost) $\costthree{\out}$: $\sum\limits_{\outcl{k} \in \out} |(\outcl{k} \setminus \pi(\outcl{k})) \setminus \inpcl{i}|  |(\outcl{k} \setminus \pi(\outcl{k})) \cap \inpcl{i}|$.
    \end{itemize}
    If the algorithm $\algog$ cuts a subset $W_{i,z}$ from a cluster $C_i$ then
    we can upper bound $\costthree{\out}$ by 
    \begin{align}
        \costthree{\out} \leq |W_{i,z}| (p - |W_{i,z}|) \label{eq:cost-three-merge} 
    \end{align}
    \item Intracluster cost within the vertices that belong to a subset of size $p$ of $\inpcl{i}$ that we cut from $\inpcl{i}$ in the algorithm $\algom$:
    \begin{itemize}
        \item (Intracluster cost) $\costfour{\out}$: $\sum\limits_{\substack{\outcl{k}, \outcl{j} \in \out \\ k \neq j}} |(\outcl{k} \setminus \pi(\outcl{k})) \cap \inpcl{i}| |(\outcl{j} \setminus \pi(\outcl{j})) \cap \inpcl{i})|$.
    \end{itemize}
    If the algorithm $\algog$ cuts a subset $W_{i,z}$ from a cluster $C_i$ and it further gets split into parts $W^{(1)}_{i,z}, W^{(2)}_{i,z}, \ldots, W^{(t)}_{i,z}$ then
    we can upper bound $\costfour{\out}$ by 
    \begin{align}
        \costfour{\out} \leq \frac{1}{2}\sum_{j = 1}^t|W^{(k)}_{i,z}| (|W_{i,z}| - |W^{(k)}_{i,z}|) \label{eq:cost-four-merge}
    \end{align}
\end{itemize}


Now we prove the following claim that we need to prove \cref{lem:main-merge-case}.

\begin{claim}\label{clm:merge-case-structure}
Suppose in $\mop$, a cluster $\inpcl{i} \in \inpset$ gets split  into $t$ parts $X_{i,1}, X_{i,2}, \ldots, X_{i,t}$ more specifically,
    \begin{itemize}
        \item $X_{i,j} \subseteq T_{r_j}^*$ for some $T_{r_j}^* \in \mop$.
        \item $\inpcl{i} = \bigcup\limits_{j = 1}^t X_{i,j}$ and
        \item $X_{i,j} \cap X_{i,k} = \emptyset \, \, \forall j \neq k$.
    \end{itemize}
    then
    \begin{enumerate}
        \item Either $\exists j \in [t]$, such that $X_{i,j} = T_{r_j}^*$ (for some $T_{r_j}^* \in \mop)$ and $\red{\inpcl{i}} \subseteq X_{i,j}$ and $|X_{i,\ell}| < p$ $\forall \, \, \ell \neq j$.
        \item or $|X_{i,\ell}| < p$, $\forall \, \, \ell \in [t]$.
    \end{enumerate}
\end{claim}

Now, to prove the previous claim, we need to prove the other claims below. 

\begin{claim}\label{clm:bound-on-part-other-than-Xij}
        Consider a partition $X_{i,j}$ of $\inpcl{i}$. Suppose, $X_{i,j} \subseteq T_{r_j}^*$ (for some $T_{r_j}^* \in \mop$) then $|T_{r_j}^* \setminus X_{i,j}| \leq s_{i,j}$ where $s_{i,j} = |\blue{X_{i,j}}| \mod p$.
\end{claim}

\begin{proof}
    Suppose for a partition $X_{i,j}$, $|T_{r_j}^* \setminus X_{i,j}| > s_{i,j}$. To prove this first we construct $\m{M}$ from $\mop$ such that $\dist(\inpset, \m{M}) < \dist(\inpset, \mop)$. This would contradict the fact that $\mop$ is the closest $\mopdef$ and thus we will conclude $|T_{r_j}^* \setminus X_{i,j}| \leq s_{i,j}$.

    \paragraph{Construction of $\m{M}$ from $\mop$:} $\m{M} = \mop \setminus \{ T_{r_j}^*\} \cup \{ (X_{i,j} \setminus S_{i,j}), (T_{r_j}^* \setminus (X_{i,j} \setminus S_{i,j}))\}$.

    That is $\m{M}$ we remove the set $T_{r_j}^*$ from $\mop$ and add two sets $(X_{i,j} \setminus S_{i,j})$ and $(T_{r_j}^* \setminus (X_{i,j} \setminus S_{i,j}))$ where $S_{i,j} \subseteq \blue{X_{i,j}}$ such that $|S_{i,j}| = s_{i,j}$.

    Note that,

    \begin{align}
        \dist(\inpset, \m{M}) = \dist(\inpset, \mop) + |X_{i,j} \setminus S_{i,j}||S_{i,j}| - |X_{i,j} \setminus S_{i,j}||T_{r_j}^* \setminus X_{i,j}| \n
    \end{align}

    This is because, 
        \begin{itemize}
            \item \textbf{Reason behind the term $|X_{i,j} \setminus S_{i,j}||S_{i,j}|$} : The pairs $(u,v)$ such that $u \in (X_{i,j} \setminus S_{i,j})$ and $v \in S_{i,j}$ are not counted in $\dist(\inpset, \mop)$ because $u$ and $v$ are present in the same clusters $\inpcl{i} \in \inpset$ and $T_{r_j}^* \in \mop$. In $\m{M}$, since, these pairs $u$ and $v$ belongs to different clusters $(X_{i,j} \setminus S_{i,j})$ and $(T_{r_j}^* \setminus (X_{i,j} \setminus S_{i,j}))$ respectively, hence, these pairs are counted in $\dist(\inpset, \m{M})$.

            \item \textbf{Reason behind the term $|X_{i,j} \setminus S_{i,j}||T_{r_j}^* \setminus X_{i,j}|$} : The pairs $(u,v)$ such that $u \in (X_{i,j} \setminus S_{i,j})$ and $v \in T_{r_j}^* \setminus X_{i,j}$ are counted in $\dist(\inpset, \mop)$ because such pairs $u$ and $v$ are present in the different clusters in $\inpset$ but in the same cluster $T_{r_j}^* \in \mop$. In $\m{M}$, since, these pairs $u$ and $v$ belongs to different clusters $(X_{i,j} \setminus S_{i,j})$ and $(T_{r_j}^* \setminus (X_{i,j} \setminus S_{i,j}))$ respectively, hence, these pairs are not counted in $\dist(\inpset, \m{M})$.
        \end{itemize}
    Now,
    \begin{align}
        \dist(\inpset, \m{M}) &= \dist(\inpset, \mop) + |X_{i,j} \setminus S_{i,j}||S_{i,j}| - |X_{i,j} \setminus S_{i,j}||T_{r_j}^* \setminus X_{i,j}| \n \\
        &< \dist(\inpset, \mop) + |X_{i,j} \setminus S_{i,j}| s_{i,j} - |X_{i,j} \setminus S_{i,j}| s_{i,j} && (\textbf{as} |S_{i,j}| = s_{i,j} \, \, \text{and} \, \, |\mop \setminus X_{i,j}| > s_{i,j}). \n \\ 
        &= \dist(\inpset, \mop). \n
    \end{align}
\end{proof}

\begin{claim}\label{clm:at-most-one-partition-size-more-p}
    There exists at most one partition $X_{i,j}$ such that $|X_{i,j}| \geq p$
\end{claim}

\begin{proof}
     Let us assume there exists two partitions $X_{i,j}$ and $X_{i,j'}$ such that $|X_{i,j}| \geq p$, $|X_{i,j'}|\geq p$ and $X_{i,j} \subseteq T_{r_j}^*$ and $X_{i,j'} \subseteq T_{r_{j'}}^*$ for some $T_{r_j}^*$ and $T_{r_j'}^*$ in $\mop$.

      Without loss of generality, assume, $|T_{r_j}^* \setminus X_{i,j}| \leq |T_{r_{j'}}^* \setminus X_{i,j'}|$. Now we construct $\m{M}$ from $\mop$ in such a way that $\dist(\inpset, \m{M}) < \dist(\inpset, \mop)$ which is a contradiction, and thus we will conclude that there exists at most one partition $X_{i,j}$ such that $|X_{i,j}| \geq p$.

      \paragraph{Construction of $\m{M}$ from $\mop$ :}

      \begin{align}
          \m{M} = \mop \setminus \left\{ T_{r_j}^*, T_{r_{j'}}^* \right\} \cup \left\{ \left( T_{r_j}^* \cup (X_{i,j'} \setminus S_{i,j'}) \right), \left(T_{r_{j'}}^* \setminus  (X_{i,j'} \setminus S_{i,j'})\right)\right\} \n
      \end{align}

      That is in $\m{M}$ we remove sets $T_{r_j}^*$ and $T_{r_{j'}}^*$ from $\mop$ and add two sets $\left( T_{r_j}^* \cup (X_{i,j'} \setminus S_{i,j'}) \right)$ and $\left(T_{r_{j'}}^* \setminus  (X_{i,j'} \setminus S_{i,j'})\right)$ where $S_{i,j'} \subseteq \blue{X_{i,j'}}$ such that $|S_{i, j'}| = (\blue{X_{i,j'}} \mod p)$.

      Note that,

      \begin{align}
          \dist(\inpset, \m{M}) = \dist(\inpset, \mop) + |(X_{i,j'} \setminus S_{i,j'})||T_{r_j}^* \setminus X_{i,j}| + |(X_{i,j'} \setminus S_{i,j'})||S_{i,j'}| \n \\
          - |(X_{i,j'} \setminus S_{i,j'})||T_{r_{j'}}^* \setminus X_{i,j'}| - |(X_{i,j'} \setminus S_{i,j'})||X_{i,j}| \n
      \end{align}

      This is because

      \begin{itemize}
          \item \textbf{The reason behind the term $|(X_{i,j'} \setminus S_{i,j'})||T_{r_j}^* \setminus X_{i,j}|$:} The pairs $(u,v)$ such that $u \in (X_{i,j'} \setminus S_{i,j'})$ and $v \in T_{r_j}^* \setminus X_{i,j}$ are not counted in $\dist(\inpset, \mop)$ because $u$ and $v$ are present in different clusters both in $\inpset$ and $\mop$. In $\m{M}$, since, these pairs $u$ and $v$ belong to the same cluster $\left( T_{r_j}^* \cup (X_{i,j'} \setminus S_{i,j'}) \right)$. Hence, these pairs are counted in $\dist(\inpset, \m{M})$.

          \item \textbf{The reason behind the term $|(X_{i,j'} \setminus S_{i,j'})||S_{i,j'}|$:} The pairs $(u,v)$ such that $u \in (X_{i,j'} \setminus S_{i,j'})$ and $v \in S_{i,j'}$ are not counted in $\dist(\inpset, \mop)$ because $u$ and $v$ are present in the same clusters $\inpcl{i} \in \inpset$ and $T_{r_{j'}}^* \in \mop$. In $\m{M}$, since, these pairs $u$ and $v$ belongs to different clusters $\left( T_{r_j}^* \cup (X_{i,j'} \setminus S_{i,j'}) \right)$ and $\left(T_{r_{j'}}^* \setminus  (X_{i,j'} \setminus S_{i,j'})\right)$ respectively. Hence, these pairs are counted in $\dist(\inpset, \m{M})$.

          \item \textbf{The reason behind the term $|(X_{i,j'} \setminus S_{i,j'})||T_{r_{j'}}^* \setminus X_{i,j'}|$:} The pairs $(u,v)$ such that $u \in (X_{i,j'} \setminus S_{i,j'})$ and $v \in T_{r_{j'}}^* \setminus X_{i,j'}$ are counted in $\dist(\inpset, \mop)$ because such pairs $u$ and $v$ are present in the different clusters in $\inpset$ but in the same cluster $T_{r_{j'}}^* \in \mop$. In $\m{M}$, since, these pairs $u$ and $v$ belongs to different clusters $\left( T_{r_j}^* \cup (X_{i,j'} \setminus S_{i,j'}) \right)$ and $\left(T_{r_{j'}}^* \setminus  (X_{i,j'} \setminus S_{i,j'})\right)$ respectively. Hence, these pairs are not counted in $\dist(\inpset, \m{M})$.

          \item \textbf{The reason behind the term $|(X_{i,j'} \setminus S_{i,j'})||X_{i,j}|$:} The pairs $(u,v)$ such that $u \in (X_{i,j'} \setminus S_{i,j'})$ and $v \in X_{i,j}$ are counted in $\dist(\inpset, \mop)$ because such pairs $u$ and $v$ are present in the same cluster $\inpcl{i} \in \inpset$ but in different clusters in  $\mop$. In $\m{M}$, since, these pairs $u$ and $v$ belongs to same cluster $\left( T_{r_j}^* \cup (X_{i,j'} \setminus S_{i,j'}) \right)$. Hence, these pairs are not counted in $\dist(\inpset, \m{M})$.
      \end{itemize}

      Now,
      \begin{align}
          \dist(\inpset, \m{M}) &= \dist(\inpset, \mop) + |(X_{i,j'} \setminus S_{i,j'})||T_{r_j}^* \setminus X_{i,j}| + |(X_{i,j'} \setminus S_{i,j'})||S_{i,j'}|\n \\
          &- |(X_{i,j'} \setminus S_{i,j'})||T_{r_{j'}}^* \setminus X_{i,j'}| - |(X_{i,j'} \setminus S_{i,j'})||X_{i,j}| \n \\
          &= \dist(\inpset, \mop) + |(X_{i,j'} \setminus S_{i,j'})|\left(|T_{r_j}^* \setminus X_{i,j}| - |T_{r_{j'}}^* \setminus X_{i,j'}|\right) + |(X_{i,j'} \setminus S_{i,j'})|\left(|S_{i,j'}| - |X_{i,j}|\right) \n \\
          &< \dist(\inpset, \mop) + |(X_{i,j'} \setminus S_{i,j'})|\left(|T_{r_j}^* \setminus X_{i,j}| - |T_{r_{j'}}^* \setminus X_{i,j'}|\right) + |(X_{i,j'} \setminus S_{i,j'})|\left(|S_{i,j'}| - p\right) \n  \\ 
          &(\textbf{as} \, \, |X_{i,j}| \geq p) \n \\
          &< \dist(\inpset, \mop) \n\\ 
          &(\textbf{as} \, \, |T_{r_j}^* \setminus X_{i,j}| \leq |T_{r_{j'}}^* \setminus X_{i,j'}| \, \, \text{and} \, \, |S_{i,j'}| < p) \n.
      \end{align}   
\end{proof}

Now from \cref{clm:at-most-one-partition-size-more-p}, we get that there exists at most one partition $X_{i,j}$ such that $|X_{i,j}| \geq p$ the following claims are related to the partition $X_{i,j}$. Recall each partition $X_{i,\ell} \subseteq T_{r_\ell}^*$ for all $\ell \in [t]$ where $T_{r_\ell}^* \in \mop$.

\begin{claim}\label{clm:geq-p-minus-si}
     $|\blue{\bigcup\limits_{\ell \neq j}X_{i,\ell}}| \geq (p - s_{i,j})$
\end{claim}

\begin{proof}
   Suppose for contradiction, 
     \[ |\blue{\bigcup_{\ell \neq j} X_{i,\ell}}| < (p - s_{i,j})\]
    Now we prove,
    \[ |\blue{\bigcup_{\ell \neq j} X_{i,\ell}}| < (p - s_{i,j}) \Rightarrow |\blue{\bigcup_{\ell \neq j} X_{i,\ell}}| = s_i - s_{i,j} \]

    \begin{align}
        &\left(  |\blue{X_{i,j}}| + |\blue{\bigcup_{\ell \neq j}X_{i,\ell}}| \right) \mod p = s_i. \n \\
        \Rightarrow & \, \, |\blue{X_{i,j}}| \mod p + |\blue{\bigcup_{\ell \neq j}X_{i,\ell}}| \mod p = s_i. \n \\
        \Rightarrow &\, \, s_{i,j} + |\blue{\bigcup_{\ell \neq j}X_{i,\ell}}| \mod p = s_i. \n \\
        \Rightarrow &\, \, |\blue{\bigcup_{\ell \neq j}X_{i,\ell}}| \mod p = s_i -s_{i,j} \label{equn:bound-on-blue-vertices}
    \end{align}
    Since, $|\blue{\bigcup\limits_{\ell \neq j}}X_{i,\ell}| < (p - s_{i,j})$ and due to \cref{equn:bound-on-blue-vertices} we have
    \[ |\blue{\bigcup\limits_{\ell \neq j} X_{i, \ell}}| = s_i - s_{i,j}\]

    Now we prove if $|\blue{\bigcup_{\ell \neq j}X_{i,\ell}}| = s_i - s_{i,j}$ then there exists a clustering $\m{M}$ such that $\dist(\inpset, \m{M}) < \dist(\inpset, \mop)$ which is a contradiction and thus we conclude our claim which is $|\blue{\bigcup\limits_{\ell \neq j}X_{i,\ell}}| \geq (p - s_i)$.

    \paragraph{Construction of $\m{M}$ from $\mop$ \\\\}

    Note that since  $\blue{T_{r_j}^*} \mod p = 0$ and by \cref{clm:bound-on-part-other-than-Xij} we get $|T_{r_j}^* \setminus X_{i,j}| \leq s_{i,j}$ thus

    \[ |\blue{T_{r_j}^* \setminus X_{i,j}}| = p - s_{i,j} > s_i - s_{i,j}\]

    Let $Y \subseteq \blue{T_{r_j}^* \setminus X_{i,j}}$ such that $|Y| = s_i - s_{i,j}$.

    Here, the construction of $\m{M}$ from $\mop$ is a bit more involved, so first, we provide an informal description of the construction, and then we give the formal description.

    Recall, the cluster $C_i$ is split into $t$ partitions $X_{i,1}, X_{i,2}, \ldots, X_{i,t}$ in $\mop$. Among these partitions $|X_{i,j}| \geq p$ and each of these partitions $X_{i,\ell} \subseteq T^*_{r_\ell}$ for some $T^*_{r_\ell} \in \mop$.

    While constructing $\m{M}$ from $\mop$ we do the following steps

    \begin{enumerate}
        \item Divide $Y$ into $(t - 1)$ disjoint parts $Y_1, Y_2, \ldots, Y_{j - 1},Y_{j + 1}, \ldots, Y_{t}$ such that
        \begin{itemize}
            \item $|Y_\ell| = |\blue{X_{i,\ell}}|$ $\forall \ell \neq j$ and
            \item $\bigcup_{\ell \neq j}Y_\ell = Y$
        \end{itemize}
        \item Remove $t$ clusters $T_{r_\ell}^*$ for all $\ell \in [t]$ from $\mop$.
        \item Add $t$ new clusters to $\m{M}$
        \begin{itemize}
            \item First cluster : $\inpcl{i} \cup (T^*_{r_j} \setminus (X_{i,j} \cup Y))$
            \item Remaining $(t - 1)$ clusters : $(T^*_{r_\ell} \setminus X_{i,\ell}) \cup Y_\ell) \, \,  \forall \ell \neq j$.
        \end{itemize}
        
        We move the red vertices of $X_{i, \ell}$ for all $\ell \neq j$ to the set $T_{r_j}^*$.
        
        The main part of the above construction is that we have divided $Y$ into $(t - 1)$ disjoint parts $Y_1, Y_2, \ldots, Y_{j - 1}, Y_{j + 1}, \ldots, Y_{t}$ such that $|Y_{\ell}| = |\blue{X_{i, \ell}}| \, \, \forall \ell \neq j$. After that, we swap the vertices in $\blue{X_{i,\ell}}$ with the vertices in $Y_\ell$.

    \end{enumerate}

    Formally,

    \[
    \m{M} = \mop \setminus \{T^*_{r_\ell} \mid \ell \neq j \} \cup \left\{\left( C_i \cup \left( T_{r_j}^* \setminus \left( X_{i,j} \cup Y \right) \right) \right),  \left(  \left( T_{r_\ell}^* \setminus X_{i, \ell} \right) \cup Y_\ell \right) \mid \ell \neq j  \right\}
    \]

    where $Y_\ell \subseteq Y$ such that 

    \begin{itemize}
        \item $|Y_\ell| = |\blue{X_{i,\ell}}|$ $\forall \ell \neq j$ and
        \item $Y_{m} \cap Y_{n} = \emptyset$ $\forall m \neq n$ and $m, n \in [t]$.
        \item $\bigcup_{\ell \neq j}Y_\ell = Y$.
    \end{itemize}

    Note that,
    \begin{align}
     \dist(\inpset, \m{M}) = \dist(\inpset, \mop) &+  \left|\bigcup_{\ell \neq j}X_{i,\ell}\right| \left| T^*_{r_j} \setminus (X_{i,j} \cup Y)\right| + |Y| \left| T^*_{r_j} \setminus (X_{i,j} \cup Y)\right| + \sum_{\ell \neq j}|Y_\ell| |T^*_{r_\ell} \setminus X_{i,\ell}| \n \\
     &- \left|\bigcup_{\ell \neq j}X_{i,\ell}\right| \left|X_{i,j}\right| - |Y| |X_{i,j}| - \sum_{\ell \neq j}|\blue{X_{i,\ell}}| |T^*_{r_\ell} \setminus X_{i,\ell}| \n 
     \end{align}

    This is because
    \begin{itemize}
        \item \textbf{Reason behind the term $\left|\bigcup_{\ell \neq j}X_{i,\ell}\right| \left| T^*_{r_j} \setminus (X_{i,j} \cup Y)\right|$}: The pairs $(u,v)$ such that $u \in \bigcup_{\ell \neq j}X_{i,\ell}$ and $v \in T_{r_j}^* \setminus (X_{i,j} \cup Y)$ are not be counted in $\dist(\inpset, \mop)$ because $u$ and $v$ are present in different clusters both in $\inpset$ and $\mop$. In $\m{M}$, since these pairs $u$ and $v$ belong to the same cluster, these pairs are counted in $\dist(\inpset, \m{M})$.
        
        \item \textbf{Reason behind the term $|Y| |T_{r_j}^* \setminus (X_{i,j} \cup Y)|$}: The pairs $(u,v)$ such that $u \in Y$ and $v \in T_{r_j}^* \setminus (X_{i,j} \cup Y)$ may not be counted in $\dist(\inpset, \mop)$ because $u$ and $v$ are present in same clusters both in $\inpset$ and $T_{r_j}^* \in \mop$. In $\m{M}$, since these pairs $u$ and $v$ belong to the different clusters, these pairs are counted in $\dist(\inpset, \m{M})$.
        
        \item \textbf{Reason behind the term $\sum_{\ell \neq j}|Y_\ell| T^*_{r_\ell} \setminus X_{i,\ell}|$}: The pairs $(u,v)$ such that $u \in Y_\ell$ and $v \in T^*_{r_\ell} \setminus X_{i,\ell}$ are not be counted in $\dist(\inpset, \mop)$ because $u$ and $v$ are present in different clusters both in $\inpset$ and $\mop$. In $\m{M}$, since, these pairs $u$ and $v$ belong to the same cluster $T^*_{r_\ell}$ these pairs are counted in $\dist(\inpset, \m{M})$.

        \item \textbf{Reason behind the term $\left|\bigcup_{\ell \neq j}X_{i,\ell}\right| \left|X_{i,j}\right|$}: The pairs $(u,v)$ such that $u \in X_{i,j}$ and $v \in Y$ are counted in $\dist(\inpset, \mop)$ because $u$ and $v$ are present in different clusters $\mop$ but present in the same cluster in $C_i \in \inpset$. In $\m{M}$, since these pairs $u$ and $v$ belong to the same cluster, these pairs are not counted in $\dist(\inpset, \m{M})$.

        \item \textbf{Reason behind the term $|Y||X_{i,j}|$}: The pairs $(u,v)$ such that $u \in Y$ and $v \in X_{i,j}$ are counted in $\dist(\inpset, \mop)$ because $u$ and $v$ are present in same cluster $T_{r_j}^* \in \mop$ but present in different clusters in $\inpset$. In $\m{M}$, since these pairs $u$ and $v$ belong to the different clusters, these pairs are not counted in $\dist(\inpset, \m{M})$.

        \item \textbf{Reason behind the term $\sum_{\ell \neq j}|\blue{X_{i,\ell}}| |T^*_{r_\ell} \setminus X_{i,\ell}|$}: The pairs $(u,v)$ such that $u \in \blue{X_{i,\ell}}$ and $v \in T^*_{r_\ell} \setminus X_{i,\ell}$ are counted in $\dist(\inpset, \mop)$ because $u$ and $v$ are present in same cluster $\mop$ but present in different clusters in $\inpset$. In $\m{M}$, since these pairs $u$ and $v$ belong to the different clusters, these pairs are not counted in $\dist(\inpset, \m{M})$.

    \end{itemize}

    Now,

    \begin{align}
     \dist(\inpset, \m{M}) &= \dist(\inpset, \mop) +  \left|\bigcup_{\ell \neq j}X_{i,\ell}\right| \left| T^*_{r_j} \setminus (X_{i,j} \cup Y)\right| + |Y| \left| T^*_{r_j} \setminus (X_{i,j} \cup Y)\right| + \sum_{\ell \neq j}|Y_\ell| T^*_{r_\ell} \setminus X_{i,\ell}| \n \\
     &- \left|\bigcup_{\ell \neq j}X_{i,\ell}\right| \left|X_{i,j}\right| - |Y| |X_{i,j}| - \sum_{\ell \neq j}|\blue{X_{i,\ell}}| |T^*_{r_\ell} \setminus X_{i,\ell}| \n \\
     &< \dist(\inpset, \mop) \n \\
     &(\textbf{by} \, \, \cref{clm:bound-on-part-other-than-Xij} \, \,  \text{we have}  \, \, T_{r_j}^* \setminus (X_{i,j} \cup Y) = s_{i,j} \, \textbf{and} \, \, |X_{i,j}| > p ) \n
     \end{align}

\end{proof}

\begin{claim}\label{clm:equal-to-one-cluster}
    $X_{i,j} = T_{r_j}^*$ (for some $T_{r_j}^* \in \mop$).
\end{claim}

\begin{proof}
     Suppose for contradiction $X_{i,j} \subset T_{r_j}^*$. at first we prove $\red{T_{r_j}^* \setminus X_{i,j}} = \emptyset$.

     Suppose $\red{T_{r_j}^* \setminus X_{i,j}} \neq \emptyset$ then we construct $\m{M}$ from $\mop$ such that $\dist(\inpset, \m{M}) < \dist(\inpset, \mop)$ which is a contradiction and thus it will conclude $\red{T_{r_j}^* \setminus X_{i,j}} = \emptyset$.

     \paragraph{Construction of $\m{M}$ from $\mop$ \\\\}

     \[ \m{M} = \mop \setminus \{T_j^*\} \cup \left\{ \left( T_j^* \setminus \red{T_{r_j}^* \setminus X_{i,j}} \right), \red{T_{r_j}^* \setminus X_{i,j}} \right\} \]

     That is in $\m{M}$ we remove the set $T_j^*$ from $\mop$ and add sets $\left( T_j^* \setminus \red{T_{r_j}^* \setminus X_{i,j}} \right)$ and $\red{T_{r_j}^* \setminus X_{i,j}}$.

     Note that,
     \begin{align}
         \dist(\inpset, \m{M}) &= \dist(\inpset, \mop) + |\red{T_{r_j}^* \setminus X_{i,j}}| |\blue{T_{r_j}^* \setminus X_{i,j}}| - |\red{T_{r_j}^* \setminus X_{i,j}}| |X_{i,j}| \n
     \end{align}

     This is because

\begin{itemize}
    \item \textbf{Reason behind the term $|\red{T_{r_j}^* \setminus X_{i,j}}| |\blue{T_{r_j}^* \setminus X_{i,j}}|$ :} The pairs $(u,v)$ such that $u \in \red{T_{r_j}^* \setminus X_{i,j}}$ and $v \in \blue{T_{r_j}^* \setminus X_{i,j}}$ are not counted in $\dist(\inpset, \mop)$ because $u$ and $v$ are present in same clusters both in $\inpset$ and $T_{r_j}^* \in \mop$. In $\m{M}$, since, these pairs $u$ and $v$ belong to the different clusters $\red{T_{r_j}^* \setminus X_{i,j}}$ and $\left( T_j^* \setminus \red{T_{r_j}^* \setminus X_{i,j}} \right)$ respectively thus these pairs are counted in $\dist(\inpset, \m{M})$.

    \item \textbf{Reason behind the term $|\red{T_{r_j}^* \setminus X_{i,j}}| |X_{i,j}|$ :} The pairs $(u,v)$ such that $u \in \red{T_{r_j}^* \setminus X_{i,j}}$ and $v \in X_{i,j}$ are counted in $\dist(\inpset, \mop)$ because $u$ and $v$ are present in the same cluster $T_{r_j}^* \in \mop$ but present in different clusters in $\inpset$. In $\m{M}$, since these pairs $u$ and $v$ belong to the different clusters, these pairs are not counted in $\dist(\inpset, \m{M})$.
\end{itemize}

Now, we have proved that $\red{T_{r_j}^* \setminus X_{i,j}} = \emptyset$. We also know that $|\blue{T_{r_j}^* \setminus X_{i,j}}| = (p - s_{i,j})$ thus $|T_{r_j}^* \setminus X_{i,j}| = (p - s_{i,j})$. From \cref{clm:geq-p-minus-si} we get that $|\blue{\bigcup\limits_{\ell \neq j}X_{i,\ell}}| \geq (p - s_{i,j})$. Let $Y \subseteq \blue{\bigcup\limits_{\ell \neq j}X_{i,\ell}}$ such that $|Y| = (p - s_{i,j})$.

To prove $X_{i,j} = T_{r_j}^*$ (for some $T_{r_j}^* \in \mop$) we construct $\m{M}$ from $\mop$ by swapping the vertices present in $Y$ with the vertices in $T_{r_j}^* \setminus X_{i,j}$. After that, we prove $\dist(\inpset, \m{M}) < \dist(\inpset, \mop)$. 

\paragraph{Construction of $\m{M}$ from $\mop$}: Suppose $Y_\ell = Y \cap X_{i,\ell}$ $\forall \ell \neq j$. We divide $T_{r_j}^* \setminus X_{i,j}$ into subsets of small size $Z_1, \ldots, Z_t$ such that $|Z_{\ell}| = |Y_{\ell}|$ $\forall \ell \neq j$. 

$\m{M} = \mop \setminus \left\{ T^*_{r_k} \, \middle\vert \, k \in \left\{1, \ldots, t \right\}\right\} \, \cup \, \left\{ (X_{i,j} \cup Y), (T^*_{r_\ell} \setminus Y_\ell \, \middle\vert \, \forall \ell \neq j)\right\}$.

Note that,

\begin{align}
    \dist(\inpset, \m{M}) < \, \, &\dist(\inpset, \mop) \n \\
    &+ \left(p\sum_{\ell \neq j}|Y_{\ell}|\, \, + p\sum_{\ell \neq j}|Z_{\ell}| + \sum_{\ell_1 < \ell_2} |Z_{\ell_1}| |Z_{\ell_2}|\right) \n \\
    &- \left( |X_{i,j}|\sum_{\ell \neq j}|Z_{\ell}| + |X_{i,j}|\sum_{\ell \neq j}|Y_{\ell}| + \sum_{\ell_1 < \ell_2} |Y_{\ell_1}| |Y_{\ell_2}|\right)
\end{align}

\begin{itemize}
    \item \textbf{Reason behind the terms $p\sum_{\ell \neq j}|Y_{\ell}|$ and $p\sum_{\ell \neq j}|Z_{\ell}|$:} The pairs $(u,v)$ such that $u \in Y_\ell$ and $v \in X_{i,\ell}$ for $\ell \neq j$ are not counted in $\dist(\inpset, \mop)$ because $u$ and $v$ are present in the same clusters both in $\mop$ and $\inpset$. In $\m{M}$, since these pairs $u$ and $v$ belong to the different clusters, these pairs are counted in $\dist(\inpset, \m{M})$. Since, $|X_{i,\ell}| \leq p$ for $\ell \neq j$ thus we have the term $p\sum_{\ell \neq j}|Y_{\ell}|$.

    Again, since The pairs $(u,v)$ such that $u \in Z_\ell$ and $v \in X_{i,\ell}$ for $\ell \neq j$ are not counted in $\dist(\inpset, \mop)$ because $u$ and $v$ are present in different clusters both in $\mop$ and $\inpset$. In $\m{M}$, since, these pairs $u$ and $v$ belong to the same cluster these pairs are counted in $\dist(\inpset, \m{M})$ and thus we have the term $p\sum_{\ell \neq j}|Y_{\ell}|$.

    \item \textbf{Reason behind the term $\sum_{\ell_1 < \ell_2} |Z_{\ell_1}| |Z_{\ell_2}|$ :} The pairs $(u,v)$ such that $u \in Z_{\ell_1}$ and $v \in Z_{\ell_2}$ are not counted in $\dist(\inpset, \mop)$ because $u$ and $v$ are present in the same clusters both in $\mop$ and $\inpset$. In $\m{M}$, since these pairs $u$ and $v$ belong to the different clusters, these pairs are counted in $\dist(\inpset, \m{M})$. 

    \item \textbf{Reason behind the terms $|X_{i,j}|\sum_{\ell \neq j}|Z_{\ell}|$ and $|X_{i,j}|\sum_{\ell \neq j}|Y_{\ell}|$:} The pairs $(u,v)$ such that $u \in X_{i,j}$ and $v \in Z_\ell$ are counted in $\dist(\inpset, \mop)$ because $u$ and $v$ are present in the same cluster $T_{r_j}^* \in \mop$ but present in different clusters in $\inpset$. In $\m{M}$, since these pairs $u$ and $v$ belong to the different clusters, these pairs are not counted in $\dist(\inpset, \m{M})$.

    Similarly, the pairs $(u,v)$ such that $u \in X_{i,j}$ and $v \in Y_\ell$ are counted in $\dist(\inpset, \mop)$ because $u$ and $v$ are present in different clusters in $\mop$ but present in the same cluster in $\inpset$. In $\m{M}$, since these pairs $u$ and $v$ belong to the same cluster, these pairs are not counted in $\dist(\inpset, \m{M})$.

    \item \textbf{Reason behind the term $\sum_{\ell_1 < \ell_2} |Y_{\ell_1}| |Y_{\ell_2}|$ :} The pairs $(u,v)$ such that $u \in Y_{\ell_1}$ and $v \in Y_{\ell_2}$ are counted in $\dist(\inpset, \mop)$ because $u$ and $v$ are present in different clusters in $\mop$ but in same cluster in $\inpset$. In $\m{M}$, since these pairs $u$ and $v$ belong to the same cluster, these pairs are not counted in $\dist(\inpset, \m{M})$. 
\end{itemize}
\end{proof}

\begin{claim}\label{clm:red-subseteq}
    $\red{\inpcl{i}} \subseteq X_{i,j}$.
\end{claim}

\begin{proof}
    Suppose not then $\red{\bigcup_{\ell \neq j} X_{i,\ell}} \neq \emptyset$

    \paragraph{Construction of $\m{M}$ from $\mop$}:

    $\m{M} = \mop  \setminus \left\{ T^*_{r_\ell} \, \, \middle\vert \, \, (\red{C_i} \cap T^*_{r_\ell}) \neq \emptyset\right\}\cup \left\{ \left(T_{r_j}^* \cup \red{\bigcup\limits_{\ell \neq j} X_{i,\ell}} \right), \left( T^*_{r_\ell} \setminus \red{C_i} \, \,  \middle\vert \, \, i \neq \ell \right) \right\}$

    Note that,

    \begin{align}
        \dist(\inpset, \m{M}) &< \dist(\inpset, \mop) + p\left| \red{\bigcup_{\ell \neq j}X_{i,\ell}}\right| - \left|X_{i,j}\right| \left|\red{\bigcup_{\ell \neq j} X_{i,\ell}}\right| \n \\
        &< \dist(\inpset, \mop) && (\textbf{as} \, \, |X_{i,j}| > p)\n
    \end{align}

    \begin{itemize}
        \item \textbf{Reason behind the term $p\left| \red{\bigcup_{\ell \neq j}X_{i,\ell}}\right|$}: The pairs $(u, v)$ such that $u \in \red{\bigcup_{\ell \neq j} X_{i,\ell}}$ and $v \in bigcup_{\ell \neq j} X_{i,\ell}$ are not counted in $\dist(\inpset, \mop)$ because these pairs $u$ and $v$ lie in the same cluster for both $\inpset$ and $\mop$ but since these pairs are present in different clusters in $\m{M}$ they are counted in $\dist{\inpset, \m{M}}$. Now, since we know that for all $\ell \neq j$, $|X_{i,\ell}| \leq p$ hence the total cost paid for these pairs is at most $p \, \,  |\red{\bigcup_{\ell \neq j} X_{i,\ell}}|$.

        \item \textbf{Reason behind the term $\left|X_{i,j}\right| \left|\red{\bigcup_{\ell \neq j} X_{i,\ell}}\right|$}: The pairs $(u,v)$ such that $u \in X_{i,j}$ and $v \in \red{\bigcup_{\ell \neq j} X_{i,\ell}}$ are counted in $\dist(\inpset, \mop)$ because these pairs $u$ and $v$ lie in the different clusters in $\mop$ but same cluster in $\inpset$. In $\m{M}$, since these pairs lie in the same cluster, these pairs are not counted in $\dist{\inpset, \m{M}}$. Now, since in~\cref{clm:equal-to-one-cluster} we proved that $X_{i,j} = T^*_{r_j}$ there is no other cost paid by the algorithm for merging the vertices of $\red{\bigcup_{\ell \neq j} X_{i,\ell}}$ to $X_{i,j}$.
    \end{itemize}
    
\end{proof}

Now, we are ready to prove \cref{clm:merge-case-structure}

\begin{proof}[Proof of \cref{clm:merge-case-structure}]
    From \cref{clm:equal-to-one-cluster} we get that $X_{i,j} = T_{r_j}^*$, from \cref{clm:red-subseteq} we get $\red{\inpcl{i}} \subseteq X_{i,j}$ and from \cref{clm:at-most-one-partition-size-more-p} we get $|X_{i,\ell}| < p$ $\forall \, \, \ell \neq j$. Thus, $(i)$ is true.

    If such a partition $X_{i,j}$ does not exist, then $(ii)$ is true. Hence, either $(i)$ is true or $(ii)$ is true.

\end{proof}

We also need \cref{clm:merge-case-lp-bound} and \cref{clm:cost-three-and-cost-four} to prove \cref{lem:main-merge-case}. 

\begin{claim}\label{clm:merge-case-lp-bound}
    Suppose $y_{i,j}$ takes the value $1$ if we cut the $j$th subset from a cluster $\inpcl{i}$ in the subroutine $\algom$ otherwise $0$. Then

    \begin{align}
        \sum_{\inpcl{i} \in \cuta}\left(\sum_{j = 1}^n y_{i,j} \, \kappa_j(\inpcl{i}) + y_{i,0} \, \kappa_0(\inpcl{i}) - \mcostf{\inpcl{i}}\right) + \sum_{\inpcl{i} \in \inpset} \mcostf{\inpcl{i}} \leq \dist(\inpset, \mop). \n
    \end{align}
\end{claim}

\begin{proof}
    From~\cref{clm:merge-case-structure} we get that if in $\mop$, $C_i$ gets split into $X_{i,1}, \ldots, X_{i,t}$ that is 

    \begin{itemize}
        \item $X_{i,j} \subseteq T^*_k$ (for some $T_k^* \in \mop$) $\forall j \in [t]$.
        \item $X_{i,j_1} \cap X_{i,j_2} = \emptyset$ for $j_1 \neq j_2$ and $j_1, j_2 \in [t]$.
        \item $\bigcup\limits_{j \in [t]} X_{i,j}= C_i$.
    \end{itemize}
    
    then

    \begin{enumerate}
        \item Either $\exists j \in [t]$, such that $X_{i,j} = T_{r_j}^*$ (for some $T_{r_j}^* \in \mop)$ and $\red{\inpcl{i}} \subseteq X_{i,j}$ and $|X_{i,\ell}| < p$ $\forall \, \, \ell \neq j$.
        \item or $|X_{i,\ell}| < p$ $\forall \, \, \ell \in [t]$.
    \end{enumerate}

    This implies 
    
    \[\left|  \bigcup\limits_{\ell \neq j} X_{i,\ell} \right| \mod p = s_i\]

    Hence,

    \[\left|  \bigcup\limits_{\ell \neq j} X_{i,\ell} \right| = ap + s_i  \, \, \text{for some constant $a$.} \]

    Let us partition the set $\blue{C_i}$ into subsets $W_{i,0}, W_{i,1}, \ldots , W_{i,m}$ such that $m = \left( |\blue{C_i}| - s_i \right)/p$ where $|W_{i,0}| = s_i$ and $|W_{i,z}| = p$ for all $z > 0$.

    If $\left|  \bigcup\limits_{\ell \neq j} X_{i,\ell} \right| = ap + s_i$ then we say $\mop$ has cut $(a + 1)$ subsets $W_{i,0}, W_{i,1}, \ldots, W_{i,a}$ from $C_i$.

    Now since $|X_{i, \ell}| < p$ $\forall \ell \neq j$ hence the pair $(u,v)$ such that $u \in W_{i,z_1}$ and $v \in W_{i,z_2}$ for $z_1 \neq z_2$ must be counted in $\dist(\inpset, \mop)$.

    Hence, the cost of 

    \begin{enumerate}
        \item Cutting $W_{i,0}$ from $C_i$ is : $\kappa_0(C_i) = s_i(|C_i| - s_i)$
        \item Cutting $W_{i,z}$ from $C_i$ for $z > 0$ is : $\kappa_z(C_i) = p(|C_i| - (zp + s_i))$.
    \end{enumerate}

    Recall in $\mop$, $C_i$ gets split into $X_{i,1}, \ldots, X_{i,t}$. We say a cluster $C_i \in \inpset$ also belongs to a set $\cuto$ (say) if $t > 1$; otherwise, it belongs to a set $\mergeo$ (say). Informally, $\mop$ has cut some vertices from the clusters $C_i \in \cuto$ and merged some vertices to the clusters $C_i \in \mergeo$. 
    
    $\mop$ must pay the minimum cost for cutting from the clusters that belong to $\cuto$ and merging to the clusters in $\mergeo$. More specifically,

    \begin{align}
        \dist(\inpset, \mop) \geq \text{min} &\sum_{C_i \in \cuto} \sum_{z = 0}^m x_{i,z} \kappa_z (C_i) + \sum_{C_i \in \mergeo} \mu(C_i) \n \\
        &\text{subject to} \n \\
        &x_{i,z} \leq x_{i,z'} \, \, \text{for $z > z'$} \label{equn:constraint-one}\\
        &\sum_{C_i \in \inpset} \sum_{z = 0}^m x_{i,z} = \frac{W_g}{p} \, \, \text{where $W_g = \sum\limits_{C_i \in \merge}\defi{C_i}$} \label{equn:constraint-two}\\
        &x_{i,z} \in \{ 0, 1\} \label{equn:constraint-three}
    \end{align}

    Here, $m = (|\blue{C_i}| - s_i)/p$ and $x_{i,z}$ takes the value $1$ if $\mop$ cuts the subset $W_{i,z}$ from a cluster $C_i$ otherwise $0$. Constraint \cref{equn:constraint-one} denotes that if $\mop$ cuts a subset $W_{i,z}$ then it must cut all subsets $W_{i,z'}$ such that $z' < z$. Constraint \cref{equn:constraint-two} denotes that $\mop$ cuts exactly $W_g/p$ such subsets where $W_g$ is the sum of the deficits of all clusters that belong to $\merge$. Note the difference between the notations $W = \sum_{C_i \in \merge'}\defi{C_i}$ which is used in our subroutine $\algom$ and the notation $W_g = \sum_{C_i \in \merge}\defi{C_i}$. Since here, we are analyzing the algorithm $\algog$, we are using $W_g$ instead of $W$.

    Now, we prove two claims that explain the second constraint.

    \begin{claim}\label{clm:our-algorithm-cuts-w/p}
        $\algom$ cuts exactly $W_g/p$ subsets from the clusters $C_i \in \inpset$.
    \end{claim}

    \begin{proof}
        Suppose the algorithm cuts a subset $W_{i,z}$ from the cluster $C_i$ and merges it to another set of clusters $C_{m_1}, C_{m_2}, \ldots, C_{m_t}$(say).

        When $\algog$ cuts a partition $W_{i,z}$ the deficit of $C_i$ reduces by $(p - |W_{i,z}|)$ that is it gets updated to
        \begin{align}
            \defi{C_i} = \defi{C_i} - (p - |W_{i,z}|) \label{equn:main-claim-merge-case-one}
        \end{align}
        and when it merges this subset to a set of clusters $C_{m_1}, C_{m_2}, \ldots, C_{m_t}$ then the total deficit of clusters other than $C_i$ reduces by $|W_{i,z}|$ that is it gets updated to
        \begin{align}
            \sum_{C_j \neq C_i}\defi{C_j} = \sum_{C_j \neq C_i}\defi{C_j} - |W_{i,z}| \label{equn:main-claim-merge-case-two}
        \end{align}
        Adding \cref{equn:main-claim-merge-case-one} and \cref{equn:main-claim-merge-case-two}, we get that the total deficit decreases exactly by $p$ for cutting one part.
        \begin{align}
            \sum_{C_i \in \nc \cup \merge'}\defi{C_i} = \sum_{C_i \in \nc \cup \merge'}\defi{C_j} - p .
        \end{align}
         Hence, to decrease the deficit $W_g$ to zero, our algorithm cuts exactly $W/p$ subsets.
    \end{proof}

    \begin{claim}\label{clm:mop-cuts-w/p}
        $\mop$ cuts exactly $W_g/p$ subsets from the clusters $C_i \in \inpset$.
    \end{claim}

    \begin{proof}
        In the clustering, $\mop$ for each cluster $T_i \in \mop$, $\blue{T_i}$ is divisible by $p$ but in the clustering $\inpset$ (input of $\algog$) the total deficit is $W_g$. 

        If $\mop$ cuts a partition $W_{i,z}$ from $\inpcl{i}$ then deficit of $C_i$ would reduce by $(p - |W_{i,z}|)$. That is, it gets updated to 
        \begin{align}
            \defi{C_i} = \defi{C_i} - (p - |W_{i,z}|). \label{equn:main-claim-two-merge-case-one}
        \end{align}
        Now, these $|W_{i,z}|$ vertices can reduce the total deficit of other clusters by at most $|W_{i,z}|$. That is, it gets updated to at most
        \begin{align}
            \sum_{C_j \neq C_i}\defi{C_j} \geq \sum_{C_j \neq C_i}\defi{C_j} - |W_{i,z}| .\label{equn:main-claim-two-merge-case-two}
        \end{align}
        Adding \cref{equn:main-claim-two-merge-case-one} and \cref{equn:main-claim-two-merge-case-two}, we get that $\mop$ can decrease the total deficit at most by $p$ by cutting one part.
        \begin{align}
            \sum_{C_i \in \nc \cup \merge'}\defi{C_i} \geq \sum_{C_i \in \nc \cup \merge'}\defi{C_j} - p .
        \end{align}
         Hence, to decrease the deficit $W_g$ to zero, $\mop$ must cut at least $W/p$ subsets.

         Since, by \cref{clm:our-algorithm-cuts-w/p}, we get our algorithm can fulfill the deficit of $W_g$ only by cutting $W_g/p$ subsets. Since $\mop$ is the optimal clustering, it must cut at most $W_g/p$ subsets.

         Thus we get $\mop$ cuts exactly $W_g/p$ subsets.
    \end{proof}

    Now,

    \begin{align}
        &\text{min} \sum_{C_i \in \cuto} \sum_{z = 0}^m x_{i,z} \kappa_z (C_i) + \sum_{C_i \in \mergeo} \mu(C_i) \n \\
        \equ \, \, &\text{min} \sum_{C_i \in \cuto} \sum_{z = 0}^m x_{i,z} \kappa_z (C_i) - \sum_{C_i \in \cuto} \mu(C_i) + \sum_{C_i \in \inpset} \mu(C_i) \n \\
        \equ \, \, &\text{min} \sum_{C_i \in \cuto} \left( \sum_{z = 1}^m x_{i,z} \kappa_z (C_i) + x_{i,0} \kappa_0(C_i) - \mu(C_i) \right) + \sum_{C_i \in \inpset} \mu(C_i) \n 
    \end{align}

    Now, to prove our claim

    \begin{align}
        \sum_{\inpcl{i} \in \cuta}\left(\sum_{j = 1}^n y_{i,z} \, \kappa_j(\inpcl{i}) + y_{i,0} \, \kappa_0(\inpcl{i}) - \mcostf{\inpcl{i}}\right) + \sum_{\inpcl{i} \in \inpset} \mcostf{\inpcl{i}} \leq \dist(\inpset, \mop). \n
    \end{align}

    We need to prove that $x_{i,z} = 1$ iff $y_{i,z} = 1$. 

    \begin{claim}
        $x_{i,z} = 1$ iff $y_{i,z} = 1$.
    \end{claim}

    \begin{proof}
        Let's look at the objective function of the above ILP.

    \begin{align}
        \text{min} \sum_{C_i \in \cuto} \left( \sum_{z = 1}^m x_{i,z} \kappa_z (C_i) + x_{i,0} \kappa_0(C_i) - \mu(C_i) \right) + \sum_{C_i \in \inpset} \mu(C_i) \n
    \end{align}

    and the constraint $1$ of the ILP

    \begin{align}
        x_{i,z} \leq x_{i,z'} \, \, \text{for $z > z'$}\n
    \end{align}

    The above constraint implies that if $\mop$ cuts a cluster $C_i$, it must cut the $0$th subset $W_{i,0}$ of $C_i$. Now, to minimize the above objective function, we can assume the cost assigned to the $0$th subset of $C_i$ is $\ccostf{C_i} - \mcostf{C_i}$, (note $\kappa_0(C_i) = \ccostf{C_i}$).

    and the cost assigned to the $z$th subset of $C_i$ is $\kappa_z(C_i)$.

    More specifically,

    \begin{align}
        &\textit{cost}(W_{i,z}) = \ccostf{C_i} - \mcostf{C_i} \, \, \text{for $z = 0$} \n \\
        &\textit{cost}(W_{i,z}) = \kappa_z(C_i) \, \, \text{for $z \geq 1$}
    \end{align}

    Now, minimizing the above objective function is equivalent to minimizing the total cost of subsets that $\mop$ cuts. 
    
    Since, in \cref{clm:mop-cuts-w/p} we get $\mop$ cuts exactly $W_g/p$ subsets so to solve the above ILP exactly our algorithm needs to cut $W_g/p$ subsets that have the minimum cost. Note that for all clusters $C_i \in \cut$, we have 
    
    \begin{align}
    \textit{cost}(W_{i,0}) &= (\ccostf{C_i} - \mcostf{C_i}) \n \\
    &\leq (\ccostf{C_j} - \mcostf{C_j}) \, \, \forall C_j \in \merge
    \end{align}

    This is because 

    \begin{align}
        &\textit{cost}(W_{j,0}) - \textit{cost}(W_{i,0}) \n \\
        &= s_j(|C_j| - s_j) - (p - s_j) |C_j| - s_i(|C_i| - s_i) + (p - s_i)|C_i| \n \\
        &= (2s_j - p) |C_j| - s_j^2 + (p - 2s_i)|C_i| + s_i^2 \label{equn:expression-claim-one} \\
        &\geq 0 \n
    \end{align}

    Since $s_j = (\blue{C_j} \mod p )\geq p/2$ and $s_i = (\blue{C_i} \mod p) < p/2$ we get that all the terms in the \cref{equn:expression-claim-one} are positive except the term $-s_j^2$. Now, $|C_i|p > s_j^2$ because $p > s_j$, and we can assume $|C_i| \geq p$ because if $C_i$ is less than $p$ then $\ccostf{C_i} = 0$ and we can ignore those clusters because we cut no vertices from a cluster $C_i$ where $|C_i| < p$.
    
    Hence, $\mop$ must cut the subset $W_{i,0}$ from all clusters $C_i \in \cut$.

    Now, in the subroutine $\algom$ at each iteration, we select the subsets $W_{i,z}$ for which $\textit{cost}(W_{i,z})$ is the minimum. From \cref{clm:our-algorithm-cuts-w/p} we get $\algog$ also cuts $W_g/p$ subsets which is same as $\mop$.
    
    Thus, our algorithm also selects the subsets that have the minimum cost and hence $x_{i,j} = 1$ iff $y_{i,j} = 1$. 
    \end{proof}
\end{proof}

Now we prove \cref{clm:cost-three-and-cost-four} where we provide a bound on $\costthree{\out}$ and $\costfour{\out}$ paid by our algorithm $\algog$. In the proof of \cref{clm:cost-three-and-cost-four}, we are going to use the following notations.

Suppose a subset $W_{i,z}$ of $C_i$ gets split into $W^{(1)}_{i,z}, \ldots, W^{(t)}_{i,z}$ and gets merged into clusters $C_{m_1}, \ldots, C_{m_t} \in \merge$. Let $T_{j_1}, \ldots, T_{j_t} \in \out$ be the clusters formed from $C_{m_1}, \ldots, C_{m_t}$ respectively by either merging deficit number of blue vertices to these clusters.

So, the pairs $(u,v)$ such that $u \in W^{(k)}_{i,z}$ and $v \in T_{j_{k}} \setminus (C_{m_{k}} \cup W^{(k)}_{i,z})$ must be counted in $\dist(\inpset, \out)$. Let us denote the number of such pairs by $\costthree{W_{i,z}}$.

Again, the pairs $u \in W^{(k_1)}_{i,z}$ and $v \in W^{(k_2)}_{i,z}$ must be counted in $\dist(\inpset, \out)$. Let us denote the number of such pairs by $\costfour{W_{i,z}}$.

Now, let us assume 

\begin{align*}
    \costthree{\out} = \sum_{C_i \in \inpset} \sum_{z = 0}^{m} \costthree{W_{i,z}}.
\end{align*}

\begin{align*}
    \costfour{\out} = \sum_{C_i \in \inpset}\sum_{z = 0}^m \costfour{W_{i,z}}
\end{align*}

where $m = (|\blue{C_i}| - s_i)/p$. Now, we prove the following claim. To prove the following claim, we need to recall the following definitions

\begin{itemize}
    \item $\mergea$: The set of clusters $C_m \in \inpset$ where the algorithm $\algog$ has merged the deficit amount of vertices to it.
    \item $\cuta$: The set of clusters $C_i \in \inpset$ where the algorithm $\algog$ has cut the surplus amount of vertices from it.
\end{itemize}
    
\begin{claim}\label{clm:cost-three-and-cost-four}
    $\costthree{\out} + \costfour{\out} \leq 2 \, \, \dist(\inpset, \mop)$.
\end{claim}

\begin{proof}
    Case $1$: \textbf{When} $|W_{i,z}| \leq p/2$: In this case, we prove that

    \begin{align*}
        \costthree{W_{i,z}} + \costfour{W_{i,z}} \leq W^{(1)}_{i,z} |C_{m_1}| + \mu(C_{m_2}) + \ldots + \mu(C_{m_{t - 1}}) + W^{(t)}_{i,z} |C_{m_t}|
    \end{align*}

    Thus adding over all clusters $C_i \in \inpset$ and all subsets $W_{i,z} \in C_i$ we get that 

    \begin{align*}
        \sum_{C_i \in \inpset} \sum_{z = 0}^{m} \costthree{W_{i,z}} + \sum_{C_i \in \inpset}\sum_{z = 0}^m \costfour{W_{i,z}} &= \costthree{\out} + \costfour{\out} \n \\ 
        &\leq \sum_{C_m \in \mergea} \mu(C_m) \n \\
        &(\text{Reason provided below)}) \n \\
        &< 2 \, \sum_{C_m \in \mergea}s(C_m) (|C_m| - s(C_m)) + \frac{1}{2} s(C_m)(p - s(C_m)) \n \\
        &< 2 \, \dist(\inpset, \mop) \n \\
        & (\text{from} \, \, \cref{lem:main-structure-of-M*})
    \end{align*}

    Reason: This is because for the clusters $C_{m_k}$ for $k \in [2, (t - 1)]$ the subset $W^{(k)}_{i,z}$ covers the deficit of the clusters $C_{m_k}$ fully that is $T_{j_k} \setminus (C_{m_k} \cup W^{(k)}_{i,z}) = \emptyset$ and for $k = 1$ and $k = t$ we have charged only the merge cost contributed by the parts $W^{(1)}_{i,z}$ and $W^{(t)}_{i,z}$ respectively.

    Thus, now we prove 

    \begin{align*}
        \costthree{W_{i,z}} + \costfour{W_{i,z}} \leq W^{(1)}_{i,z} |C_{m_1}| + \mu(C_{m_2}) + \ldots + \mu(C_{m_{t - 1}}) + W^{(t)}_{i,z} |C_{m_t}|
    \end{align*}

    Recall $\costthree{W_{i,z}}$ is the number of pairs $(u,v)$ such that $u \in W^{(k)}_{i,z}$ and $v \in T_{j_{k}} \setminus (C_{m_{k}} \cup W^{(k)}_{i,z})$. Thus 

    \begin{align}
        \costthree{W_{i,z}} &= \sum_{k = 1}^t\left|W^{(k)}_{i,z}\right| \left|T_{j_{k_1}} \setminus (C_{m_{k_1}} \cup W^{(k)}_{i,z})\right|  \n \\
        &< \sum_{k = 1}^t\dfrac{1}{2} \left( \left|W^{(k)}_{i,z}\right|  \dfrac{p}{2}  \right) && \left( \textbf{as} \, \, \left|T_{j_{k}} \setminus (C_{m_{k}} \cup W^{(k)}_{i,z})\right| < p/2 \right)\n \\
        &\leq \dfrac{1}{2} \sum_{k = 1}^t \left( \left|W^{(k)}_{i,z}\right|  \left|C_{m_k} \right|  \right) \label{equn:equation-one}
    \end{align}

    The $1/2$ in the above expression comes from the fact that when we sum over all the clusters $C_i \in \inpset$ and all subsets $W_{i,z} \in C_i$ we are counting the pairs twice once while considering the subset $W^{(k)}_{i,z}$ and again while considering the subset $T_{j_{k}} \setminus (C_{m_{k}} \cup W^{(k)}_{i,z})$. 

    Recall $\costfour{W_{i,z}}$ is the number of pairs $(u,v)$ such that $u \in W^{(k_1)}_{i,z}$ and $v \in W^{(k_2)}_{i,z}$. Thus

    \begin{align}
        \costfour{W_{i,z}} &= \sum_{\substack{k_1, k_2 = 1 \\ k_1 \neq k_2}}^t\left|W^{(k_1)}_{i,z}\right| \left|W^{(k_2)}_{i,z}\right|  \n \\
        &< \sum_{k = 1}^t \dfrac{1}{2} \left( \left|W^{(k)}_{i,z}\right|  \dfrac{p}{2}  \right) && \left( \textbf{as} \, \, \left|W^{(k_2)}_{i,z}\right| < p/2 \right)\n \\
        &\leq \dfrac{1}{2} \sum_{k = 1}^t \left( \left|W^{(k)}_{i,z}\right|  \left|C_{m_k} \right|  \right) \label{equn:equation-two}
    \end{align}

    The $1/2$ in the above expression comes from the fact that when we sum over all the clusters $C_i \in \inpset$ and all subsets $W_{i,z} \in C_i$ we are counting the pairs twice once while considering the subset $W^{(k_1)}_{i,z}$ and again while considering the subset $W^{(k_2)}_{i,z}$. 

    Thus from \cref{equn:equation-one} and \cref{equn:equation-two} we get that,

    \begin{align}
       \costthree{W_{i,z}} + \costfour{W_{i,z}} &\leq \sum_{k = 1}^t \left( \left|W^{(k)}_{i,z}\right|  \left|C_{m_k} \right|  \right) \n \\
       &= W^{(1)}_{i,z} |C_{m_1}| + \mu(C_{m_2}) + \ldots + \mu(C_{m_{t - 1}}) + W^{(t)}_{i,z} |C_{m_t}| \n
    \end{align}

The reason behind the last expression is as stated before, for the clusters $C_{m_k}$ for $k \in [2, (t - 1)]$ the parts $W^{(k)}_{i,z}$ covers the deficit of the clusters $C_{m_k}$ entirely, that is $T_{j_k} \setminus \left(C_{m_k} \cup W^{(k)}_{i,z}\right) = \emptyset$.

Case $2$: \textbf{When} $|W_{i,z}| > p/2$ : For a cluster $C_m \in \mergea$ the deficit of the cluster $C_m$, $\defi{C_m} < p/2$. This implies the deficit of $C_m$ is either filled by exactly one subset $W_{i,z}$ or by two subsets $W_{i,z} \in C_i$ and $W_{j,z'} \in C_j$.

We divide our proof again into two subcases:

Case $2(a)$: \textbf{When} the deficit of $C_m$ is filled by a subset $W_{i,z} \in C_i$: Let us assume a cluster $\outcl{j} \in \out$ is created from the cluster $\inpcl{k} \in \inpset$ by merging the deficit $\defi{\inpcl{k}} = d_m$ (say) to it. Let, a part of $W_{i,z}$, $W^{(1)}_{i,z}$ (say) be used to fill $d_m$. Hence, $\costthree{W^{(1)}_{i,z}} = 0$ because the deficit is filled up by a subset that belongs to only one cluster $\inpcl{i} \in \inpset$. 
    
Again, $\costfour{W_{i,z}} = (|W^{(1)}_{i,z}|  \, \, (|W_{i,z}| - |W^{(1)}_{i,z}|)\, )$ because when we merge $W^{(1)}_{i,z}$ part of $W_{i,z}$ to $\inpcl{k}$ and the remaining part $(W_{i,z} - W^{(1)}_{i,z})$ is merged to other clusters in $\mergea$. Hence, 
    
    \begin{align}
        \costthree{W^{(1)}_{i,z}} + \costfour{W_{i,z}} &= |W^{(1)}_{i,z}| \, \, (|W_{i,z}| - |W^{(1)}_{i,z}|) \nonumber \\
        &\leq (d_m \cdot (p - d_m)) && (\because |W^{(1)}_{i,z}| = d_m \, \, \text{and} \, \, |W_{i,z}| \leq p) \nonumber \label{equn:algo-pays-case-1} \\
    \end{align}

    From \cref{lem:main-structure-of-M*} we get that for a cluster $C_m \in \mergea$
    
     
     \begin{align}
        \opt_{C_m} &\geq \surp{C_m}(|C_m| - \surp{C_m}) + \frac{1}{2} \surp{C_m}(p - \surp{C_m})\n \\
        &> \frac{1}{2} d_m \, \, (p - d_m) &&(\because |\inpcl{k}| > (p - d_m)) \label{equn:mop-pays}
     \end{align}

     Hence, by \cref{equn:algo-pays-case-1} and \cref{equn:mop-pays} we get, that we can charge the $2\,\,\opt_{C_m}$ for $\costthree{W^{(1)}_{i,z}} + \costfour{W_{i,z}}$.

Case $2(b)$: \textbf{When} deficit of the cluster $\defi{\inpcl{k}} = d_m$ is filled by the subsets of exactly two clusters $W_{i,z} \in C_i$ and $W_{j,z} \in C_j$: Let us assume a cluster $\outcl{j} \in \out$ is created from the cluster $\inpcl{k}$ by merging the deficit $\defi{\inpcl{k}} = d_m$ (say) to it. Let, a part of $W_{i,z}$, $W^{(2)}_{i,z}$ is used to fill the deficit of $\inpcl{k}$. Similarly, let part of $W_{j,z}$, $W^{(1)}_{j,z}$ is used to fill the deficit of $\inpcl{k}$. Hence,

\[
    \costthree{W^{(2)}_{i,z}} + \costthree{W^{(1)}_{j,z}} = |W^{(2)}_{i,z}| \, \, |W^{(1)}_{j,z}|
\]

and 

\[
    \costfour{W_{j,z}} = |W^{(1)}_{j,z}| \, \, (|W_{j,z}| - |W^{(1)}_{j,z}|)
\] 

( note we have already charged the cost $\costfour{W_{i,z}}$ in case $1$ ). 

Hence, 

\[
    \costthree{W^{(2)}_{i,z}} + \costthree{W^{(1)}_{j,z}} + \costfour{W_{j,z}} = (|W^{(1)}_{i,z}| \, \, |W^{(1)}_{j,z}|) + (|W^{(1)}_{j,z}| \, \, (|W_{j,z}| - |W^{(1)}_{j,z}|)).  
\] 

Now, WLOG we can assume $(|W_{j,z}| - |W^{(1)}_{j,z}|) < p/2$
because otherwise there exists a portion $W^{(1')}_{j,z}$ of $(W_{j,z} \setminus W^{(1)}_{j,z})$ that is used to fill the deficit of some cluster fully in $\mergea$ and thus we have charged $\costfour{w_{j,z}}$ in case $(1)$.

    Hence,

    \begin{align}
        \costthree{W^{(2)}_{i,z}} + \costthree{W^{(1)}_{j,z}} + \costfour{W_{j,z}} &= (|W^{(2)}_{i,z}| \, \, |W^{(1)}_{j,z}|) + (|W^{(1)}_{j,z}| \, \, (|W_{j,z}| - |W^{(1)}_{j,z}|)) \nonumber \\
        &\leq |W^{(2)}_{i,z}| \, \, d_m + |W^{(1)}_{j,z}| \, \,  p/2 \n \\ 
        &(\textbf{as} \, \, |W^{(1)}_{j,z}|  < d_m \, \text{and} \,  (|W_{j,z}| - |W^{(1)}_{j,z}|) < p/2) \nonumber \\
        &\leq (|W^{(2)}_{i,z}| + |W^{(1)}_{j,z}|) \, \, (p - d_m) \n \\ 
        &(\textbf{as} \, \, d_m \, \, \text{and} \, \, p/2 \leq (p - d_m)) \nonumber \\
        &= d_m \, \, (p - d_m) \label{equn:algo-pays-case-2} 
    \end{align}

     Hence, by \cref{equn:algo-pays-case-2}, \cref{equn:mop-pays} and \cref{lem:main-structure-of-M*} we get, that we can charge the $2 \, \, \opt_{C_m}$ for $\costthree{W^{(2)}_{i,z}} + \costthree{W^{(1)}_{j,z}} + \costfour{W_{j,z}}$.

     Since,

     \begin{align}
         d_m(p - d_m) &= (p - s(C_m)) s(C_m) \n \\
         &< 2 \, (s(C_m) (|C_m| - s(C_m)) + \frac{1}{2} s(C_m)(p - s(C_m))) \n \\
         &< 2 \opt_{\inpcl{k}} \n
     \end{align}

    Thus, combining the above two cases, we get,

    \begin{align*}
        \sum_{C_i \in \inpset} \sum_{z = 0}^{m} \costthree{W_{i,z}} + \sum_{C_i \in \inpset}\sum_{z = 0}^m \costfour{W_{i,z}} &= \costthree{\out} + \costfour{\out} \n \\ 
        &\leq 2 \, \, \sum_{C_m \in \mergea} \opt_{C_m} \n \\
        &\leq 2 \, \dist(\inpset, \mop) \n \\
        & (\text{from} \, \, \cref{lem:main-structure-of-M*})
    \end{align*}

\end{proof}

Now we are ready to prove \cref{lem:main-merge-case}

\begin{proof}[Proof of \cref{lem:main-merge-case}]
   Suppose $y_{i,j}$ takes the value $1$ if we cut the $j$th subset of size $p$ from a cluster $\inpcl{i}$ in the subroutine $\algom$ otherwise $0$. Hence,

   \begin{align}
       \dist(\m{C},\m{T}) &= \sum_{\inpcl{i} \in \cuta} \sum_{j = 0}^n y_{i,j} \kappa_i(\inpcl{j}) + \sum_{\inpcl{j} \in \mergea} \mcostf{\inpcl{j}} + \costthree{\out} + \costfour{\out}. \n \\
       &= \sum_{\inpcl{i} \in \cuta}\left(\sum_{j = 1}^n y_{i,j} \, \kappa_j(\inpcl{i}) + y_{i,0} \, \kappa_0(\inpcl{i}) - \mcostf{\inpcl{i}}\right) + \sum_{\inpcl{i} \in \inpset} \mcostf{\inpcl{i}} + \costthree{\out} + \costfour{\out}. \n \\
       &\leq \dist(\inpset, \mop) + 2 \, \dist(\inpset, \mop). \n \\
       &= 3\, \, \dist(\inpset, \mop). \n
   \end{align}
   
\end{proof}

\subsubsection{Approximation guarantee in the cut case} \label{subsec:cut-case}

 In this section, we show that, in the cut case, the clustering output by our algorithm $\algog$ has a distance of at most $3.5\opt$ from the input clustering. In particular, we prove the following lemma that we restate below.
\maincut*

To prove \cref{lem:main-cut-case}, we need to define and recall some definitions.


\begin{align}
    \costone{\out}: \sum_{C_i \in \inpset} s_i(|C_i| - s_i) \label{equn:cost-one} 
\end{align}

It is the total cost of cutting the surplus blue vertices from a cluster $C_i \in \inpset$.

\begin{align}
    \costtwo{\out}: \sum_{C_j \in \inpset}(p - s_j) |C_j| \label{equn:cost-two}
\end{align}

It is the total cost of merging the $\defi{C_j}$ number of blue vertices to a cluster $C_j \in \inpset$.

Now, let's recall the following definitions


where for a cluster $\outcl{k} \in \out$ we defined $\pi(\outcl{k})$ as a parent of $\outcl{k}$ iff $\pi(\outcl{k}) \in \inpset$ and $\outcl{k}$ is formed either by cutting some vertices from $\pi(\outcl{k})$ or by merging some vertices to $\pi(\outcl{k})$.

Thus, our algorithm pays the above four types of costs. 

To prove \cref{lem:main-cut-case}, first, we need to prove some claims.

 We first argue (in~\cref{lem.opt.si.times.p.min.si}) that $\mop$ must pay $\sum_{\inpcl{i} \in \inpset} \frac{s_i (p - s_i)}{2}$ in the cut case where $s_i = \surp{\inpcl{i}}$.

\begin{claim}\label{lem.opt.si.times.p.min.si}
    \begin{align}
        \dist(\inpset, \mop) \geq \sum_{\inpcl{i} \in \inpset}\dfrac{s_{i}(p-s_{i})}{2} \nonumber
    \end{align}
\end{claim}

\begin{proof}
    Suppose there is an input cluster $\inpcl{k} \in \inpset$ that is split into $t$ parts (say) $X_{k,1},\ldots,X_{k,t}$.

    We can assume that the cluster $\inpcl{k}$ consists only of blue vertices. If it contains red vertices, the cost paid by $\mop$ for splitting $\inpcl{k}$ into $t$ parts would be more because the red vertices would increase the size of the cluster $\inpcl{k}$. Let us assume that the surplus of $X_{k,i}$ be $s_{k,j}$ and to make the number of blue vertices in each cluster a multiple of $p$ $\mop$ must merge at least $(p - s_{k,j})$ blue vertices to $X_{k,j}$. Hence,

    \begin{align}
        \opt_{C_k} \geq \frac{1}{2} \left(\sum_{j = 1} ^ t s_{k,j} (p - s_{k,j}) \right) \label{equn:one}
    \end{align}

    Now, since $\sum_{j = 1} ^ t s_{k,j} \mod p = s_k$ thus from \cref{prop:mod-sum-ineq} we get,

    \begin{align}
        \frac{1}{2} \left( \sum_{j = 1} ^ t s_{k,j} (p - s_{k,j}) \right) \geq \frac{s_k (p - s_k)}{2} \label{equn:two}
    \end{align}

    Now,

    \begin{align}
        \dist(\inpset, \mop) &= \sum_{C_i \in \inpset} \opt_{C_i} \n \\
        &\geq \sum_{C_i \in \inpset} \frac{s_i (p - s_i)}{2} && (\text{from} \, \cref{equn:one} \,  \text{and} \,  \cref{equn:two}) \n
    \end{align}

    This completes the proof.
\end{proof}

The following lemma is a consequence of~\cref{lem.opt.si.times.p.min.si}
\begin{claim}\label{lem.opt.si.times.p.min.si.breakdown}
    \begin{align}
        \opt \geq \sum_{\inpcl{i} \in \cut\setminus\cut'}\dfrac{1}{2}s_{i}^{2} + \sum_{\inpcl{i} \in \merge}\dfrac{1}{2}d_{i}^{2} + \sum_{\inpcl{i} \in \cut'}\dfrac{s_{i}(p-s_{i})}{2} \nonumber 
    \end{align}
Here, $s_i = \surp{\inpcl{i}}$ and $C_i = \defi{\inpcl{i}}$.
\end{claim}
\begin{proof}
    We have
    \begin{align}
        \sum_{\inpcl{i} \in \inpset}\dfrac{s_{i}(p-s_{i})}{2}  = \sum_{\inpcl{i} \in \cut\setminus\cut'}\dfrac{s_{i}(p-s_{i})}{2}  + \sum_{\inpcl{i} \in \merge}\dfrac{s_{i}(p-s_{i})}{2} + \sum_{\inpcl{i} \in \cut'}\dfrac{s_{i}(p-s_{i})}{2}.\nonumber
    \end{align}
    
    For each $\inpcl{i} \in \cut$, $s_{i} \leq \frac{p}{2}$. Hence $s_{i}(p-s_{i}) \geq s_{i}\frac{p}{2} \geq s_{i}^{2}$.

    For each $\inpcl{i} \in \merge$, $s_{i} \geq \frac{p}{2}$. Hence $s_{i}(p-s_{i}) \geq \frac{p}{2} (p-s_{i}) \geq d_{i}^{2}$  (where $C_i = (p - s_i)$).

    These observations imply that
    \begin{align}
        \sum_{\inpcl{i} \in \cut\setminus\cut'}\dfrac{1}{2}s_{i}^{2} + \sum_{\inpcl{i} \in \merge}\dfrac{1}{2}d_{i}^{2} + \sum_{\inpcl{i} \in \cut'}\dfrac{s_{i}(p-s_{i})}{2} \leq \sum_{\inpcl{i} \in \inpset}\dfrac{s_{i}(p-s_{i})}{2}  \leq \opt,\nonumber
    \end{align}
    where the last inequality is from~\cref{lem.opt.si.times.p.min.si}.

    This completes the proof.
\end{proof}

\begin{claim}\label{lem:cost-one-plus-two-cut-case}
    If $\sum_{\inpcl{i} \in \cut} \surp{\inpcl{i}} > \sum_{\inpcl{j} \in \merge} \defi{\inpcl{j}}$ then $\costone{\out} + \costtwo{\out} \leq 2 \, \opt$.
\end{claim}

\begin{proof}
   By the definition of $\costone{\out}$, we have $\costone{\out} = \sum_{\inpcl{i} \in \cuta}\ccostf{\inpcl{i}}$. Now, if $\sum_{\inpcl{i} \in \cut} \surp{\inpcl{i}} > \sum_{\inpcl{j} \in \merge} \defi{\inpcl{j}}$ that is in the cut case we have $\cuta = \cut$. Recall that $\cuta$ is the set of clusters $\inpcl{i} \in \inpset$ from which we cut some vertices at any vertex during the execution of our algorithm $\algog$. Since in the cut case, we only cut from the clusters $\inpcl{i} \in \cut$ that is the clusters $\inpcl{i}$ for which $\surp{\inpcl{i}} \leq p/2$ we have $\cuta = \cut$. Hence, in the cut case, we have, 
   \begin{align}
      \costone{\out} = \sum_{\inpcl{i} \in \cuta}\ccostf{\inpcl{i}} = \sum_{\inpcl{i} \in \cut}\ccostf{\inpcl{i}}. \label{equn:cut-case-first}
   \end{align}
    
    Similarly, in the cut case, we also have $\mergea = \merge$ thus we have

    \begin{align}
    \costtwo{\out} = \sum_{\inpcl{i} \in \mergea}\mcostf{\inpcl{i}} = \sum_{\inpcl{i} \in \merge}\mcostf{\inpcl{i}}. \label{equn:cut-case-second}
    \end{align}
    Hence, from \cref{equn:cut-case-first} and \cref{equn:cut-case-second} we have

    \begin{align}
        \costone{\out} + \costtwo{\out} = \sum_{\inpcl{i} \in \cut}\ccostf{\inpcl{i}} + \sum_{\inpcl{i} \in \merge}\mcostf{\inpcl{i}}. \label{equn:cut-case-third}
    \end{align}

    Now, from \cref{lem:main-structure-of-M*} we get for all clusters $\inpcl{i} \in \cut$ we have 
    
    \begin{align}
        \opt_{\inpcl{i}} &\geq s(C_i)(|C_i| - s(C_i)) + \frac{1}{2} s(C_i) (p - s(C_i)) \n \\
        &> \ccostf{C_i} \n
    \end{align} 
    and for all clusters $\inpcl{i} \in \merge$ we have 
    
    \begin{align}
        \opt_{\inpcl{i}} &\geq (p - s(C_i)) (|C_i| - s(C_i)) + \frac{1}{2} s(C_i) (p - s(C_i)) \n \\
        &\geq 2 \, \mu(C_i) \n
    \end{align} 
    Thus we have

    \begin{align}
        \dist(\inpset, \mop) &= \sum_{\inpcl{i} \in \cut} \opt_{\inpcl{i}} + \sum_{\inpcl{i} \in \merge} \opt_{\inpcl{i}} \nonumber \\
        &\geq \sum_{\inpcl{i} \in \cut} \ccostf{\inpcl{i}} + 2 \sum_{\inpcl{i} \in \merge} \mcostf{\inpcl{i}} .\label{equn:cut-case-fourth}
    \end{align}

    Now, from \cref{equn:cut-case-fourth} and \cref{equn:cut-case-third} we get

    \begin{align}
        \costone{\out} + \costtwo{\out} &\leq 2 \, \, \dist(\inpset, \mop) \n \\
        &\leq 2 \, \, \opt.\n
    \end{align}
\end{proof}

\begin{claim} \label{lem:cost 3 + 4 cut case}
    $\costthree{\out} + \costfour{\out} \leq 1.5 \, \opt$.
\end{claim}
\begin{proof}

    We introduce some notations.

    For each cluster $ \inpcl{i}\in \cut $, denote by $ X_{i,{1}}, X_{i,{2}}, \dots, X_{i,{\ell_{i}}} $ the partition of $ s_{i} $ blue vertices that are split from $ \inpcl{i} $, and each of these parts are merged to different clusters $ \inpcl{j} $'s in $ \merge $, for $ k=1,2,\dots, \ell_{j} $.

     Let $\pcls$ be the set of clusters of size $p$ formed by $\algoc$. Let $\pcl{j}$ be its $j$th cluster. Represents $ \pcl{j} = Y_{j,1} \cup Y_{j,2} \cup \dots \cup Y_{j,y_{j}} $ where each $ Y_{j,k} $ is a set of blue vertices coming from a cluster $ \inpcl{j_{k}}\in \cut $.

    We prove this claim by showing the correctness of the following claims.
    \begin{claim}\label{clm:cost 4}
        $ \costfour{\out} \leq \sum_{\inpcl{i} \in \cut\setminus\cut'}\dfrac{1}{2}s_{i}^{2} + \sum_{\inpcl{i} \in \cut'}{\left( \sum_{1\leq j< k\leq \ell_{i}} |X_{i,j}||X_{i,k}| \right)}. $
    \end{claim}
    \begin{claim}\label{clm:cost 3} 
    $ \costthree{\out} \leq \sum_{\inpcl{i}\in \merge}\dfrac{1}{2}d_{i}^{2} + \dfrac{1}{2}\sum_{\pcl{j}\in \pcls}\left(\sum_{k=1}^{y_{j}}|Y_{j,k}|(p-|Y_{j,k}|)\right). $
    \end{claim}
    \begin{claim}\label{clm:three half cost}
        $$ \sum_{\inpcl{i} \in \cut'}{\left( \sum_{1\leq j< k\leq \ell_{i}} |X_{i,j}||X_{i,k}| \right)} +  \dfrac{1}{2}\sum_{\pcl{j}\in \pcls}\left(\sum_{k=1}^{y_{j}}|Y_{j,k}|(p-|Y_{j,k}|)\right) \leq \dfrac{3}{2}\sum_{\inpcl{i}\in \cut'}\dfrac{s_{i}(p-s_{i})}{2}. $$
    \end{claim}
    Given the correctness of the claims above, the lemma can be proved as follows.
    {\small
    \begin{align}
        &\> \costthree{\out} + \costfour{\out} \nonumber \\
        &\leq \sum_{\inpcl{i} \in \cut\setminus\cut'}\dfrac{1}{2}s_{i}^{2} + \sum_{\inpcl{i} \in \merge}\dfrac{1}{2}d_{i}^{2} \nonumber\\
        &\> + \sum_{\inpcl{i} \in \cut'}\left( \sum_{1\leq j<k \leq \ell_{i}} |X_{i,j}||X_{i,k}| \right) + \dfrac{1}{2}\sum_{\pcl{j}\in \pcls}\left(\sum_{k=1}^{y_{j}}|Y_{j,k}|(p-|Y_{j,k}|)\right) && (\text{by~\cref{clm:cost 4} and~\cref{clm:cost 3}}) \nonumber \\
        &\leq \sum_{\inpcl{i} \in \cut\setminus\cut'}\dfrac{1}{2}s_{i}^{2} + \sum_{\inpcl{i} \in \merge}\dfrac{1}{2}d_{i}^{2} + \dfrac{3}{2}\sum_{\inpcl{i}\in \cut'}\dfrac{s_{i}(p-s_{i})}{2} && (\text{by~\cref{clm:three half cost}})\nonumber \\
        &\leq \dfrac{3}{2} \left(\sum_{\inpcl{i} \in \cut\setminus\cut'}\dfrac{1}{2}s_{i}^{2} + \sum_{\inpcl{i} \in \merge}\dfrac{1}{2}d_{i}^{2} + \sum_{\inpcl{i}\in \cut'}\dfrac{s_{i}(p-s_{i})}{2} \right) \nonumber \\
        &\leq \dfrac{3}{2}\opt. && (\text{by~\cref{lem.opt.si.times.p.min.si.breakdown}}) \nonumber
    \end{align}
}
    We now proceed to the proof of each of the above claims to finish the proof:

    \begin{proof}[Proof for~\cref{clm:cost 4}]
        By definition,
        \begin{align}
            \costfour{\out}  &= \sum_{\inpcl{i} \in \cut}{\left( \sum_{1\leq j< k\leq \ell_{i}} |X_{i,j,}||X_{i,k}| \right)} \nonumber\\
            &= \sum_{\inpcl{i} \in \cut \setminus \cut'}{\left( \sum_{1\leq j< k\leq \ell_{i}} |X_{i,j}||X_{i,k}| \right)} + \sum_{\inpcl{i} \in \cut'}{\left( \sum_{1\leq j< k\leq \ell_{i}} |X_{i,j}||X_{i,k}| \right)}\nonumber            
        \end{align}
        Note that $ \sum_{1\leq j \leq \ell_{i}}|X_{i,j}| = s_{i} $, therefore $\sum_{1\leq j<k\leq \ell_{i}} |X_{i,j}| |X_{i,k}| \leq \frac{1}{2}\left( \sum_{1\leq j\leq \ell_{i}} |X_{i,j}| \right)^{2} = \frac{1}{2}s_{i}^{2}$, it follows that
        \begin{align}
            \costfour{\out} \leq \sum_{\inpcl{i} \in \cut\setminus\cut'}\dfrac{1}{2}s_{i}^{2} + \sum_{\inpcl{i} \in \cut'}{\left( \sum_{1\leq j< k\leq \ell_{i}} |X_{i,j}| |X_{i,k}| \right)}. \nonumber 
        \end{align}
    \end{proof}

    \begin{proof}[Proof for~\cref{clm:cost 3}]
        For each cluster $\inpcl{i} \in \merge$, $\algog$ brings $d_{i}$ blue vertices from some other clusters $\inpcl{\iprime} \in \cut$ into $\inpcl{i}$. Denote by $ Y_{i,1}, Y_{i,2}, \dots, Y_{i,y_{i}} $ the partition of these $ d_{i} $ blue vertices such that in the following order, $ \algog $ merges into $ \inpcl{i} $: $ Y_{i,1} $ from $ \inpcl{i_{1}} $, $ Y_{i,2} $ from $ \inpcl{i_{2}} $, and so on, $ Y_{i,y_{i}} $ from $ \inpcl{i_{y_{i}}} $. This implies
        \begin{align}
        d_{i} = |Y_{i,1}| + |Y_{i,2}| + \dots + |Y_{i,y_{i}}| \label{eq.defi of Ci},
        \end{align}
        and the cost contributed to $\costthree{\out}$ by this cluster $\inpcl{i}$ is
        \begin{align}
            \dfrac{1}{2} \left( \sum_{j=1}^{y_{i}} |Y_{i,j}|(d_{i} - |Y_{i,j}|) \right) &\leq \dfrac{1}{2y_{i}}d_{i} (y_{i} - 1)d_{i}\label{eq.cost 4.chebyshev ineq}\\
            &\leq \dfrac{1}{2}d_{i}^{2}. \label{eq.cost 4. bound}
        \end{align}

        To obtain~\eqref{eq.cost 4.chebyshev ineq}, we use Chebyshev inequality ($\frac{a_{1}b_{1}+a_{2}b_{2}+\dots a_{n}b_{n}}{n} \leq \frac{a_{1}+a_{2}+\dots a_{n}}{n}\frac{b_{1}+b_{2}+\dots b_{n}}{n}$, for $a_{1}\leq a_{2}\leq \dots \leq a_{n}, b_{1}\geq b_{2}\geq \dots \geq b_{n}$), combining with~\eqref{eq.defi of Ci}.

        Similarly, for each cluster of size $ p $, $ \pcl{j} \in \pcls $, we can assume $\pcl{j}$ is formed by $ Y_{j,1} $ from $ \inpcl{j_{1}} $, $ Y_{j,2} $ from $ \inpcl{j_{2}} $, and so on, $ Y_{j,{y_{j}}} $ from $ \inpcl{j_{y_{j}}} $. Therefore, the cost contributed to $\costthree{\out}$ by $P_{j}$ is 
        \begin{align}
            \dfrac{1}{2}\sum_{k=1}^{y_{j}}|Y_{j,k}|(p-|Y_{j,k}|) \label{eq.cost 3 contributed by Pj}
        \end{align}
        From~\eqref{eq.cost 4. bound} and~\eqref{eq.cost 3 contributed by Pj}, we have 
    $ \costthree{\out} \leq \sum_{\inpcl{i}\in \merge}\dfrac{1}{2}d_{i}^{2} + \dfrac{1}{2}\sum_{\pcl{j}\in \pcls}\left(\sum_{k=1}^{y_{j}}|Y_{j,k}|(p-|Y_{j,k}|)\right). $
    \end{proof}

    \begin{proof}[Proof for~\cref{clm:three half cost}]
        Let $$ A = \sum_{\inpcl{i} \in \cut'}{\left( \sum_{1\leq j< k\leq \ell_{i}} |X_{i,j}||X_{i,k}| \right)} +  \dfrac{1}{2}\sum_{\pcl{j}\in \pcls}\left(\sum_{k=1}^{y_{j}}|Y_{j,k}|(p-|Y_{j,k}|)\right),$$
        and $ A' = \sum_{\inpcl{i}\in \cut'}\dfrac{s_{i}(p-s_{i})}{2} $. We need to show that $ A \leq \dfrac{3}{2}A' $.
        
        In each cluster $\inpcl{i} \in \cut'$, $ \algoc $ splits $s_{i}$ blue vertices from it, and these $s_{i}$ blue vertices are split again into at most two parts. Therefore, we can write $ \sum_{1\leq j< k\leq \ell_{i}} |X_{i,j}||X_{i,k}|  = |X_{i,1}|(s_{i}-|X_{i,1}|) = \alpha_{i,1}(s_{i} - \alpha_{i,1}) $, where $ 0<\alpha_{i,1} = |X_{i,1}| \leq s_{i} $.

        Each cluster $ \pcl{j}\in \pcls $ is formed by merging $ Y_{j,1} $ from $ \inpcl{j_{1}} \in \cut' $, $ Y_{j,2} $ from $ \inpcl{j_{2}} \in \cut' $, and so on, $ Y_{j,{y_{j}}} $ from $ \inpcl{j_{y_{j}}} \in \cut' $, in that order. Therefore, $ |Y_{j,k}| = s_{j_{k}} $ for $ 2\leq k< y_{j} $, and $ |Y_{j,1}| = \alpha_{j_{1}} \leq s_{j_{1}} $, $ |Y_{j,y_{j}}| = \beta_{j_{y_{j}}} \leq s_{j_{y_{j}}} $.

        $ A $ can be rewritten as
        \begin{align}
            A &= \sum_{\inpcl{i} \in \cut'}\alpha_{i,1}(s_{i}-\alpha_{i,1}) \nonumber \\
              &\> + \dfrac{1}{2}\sum_{\pcl{j} \in \pcls}\left( \alpha_{j_{1}}(p-\alpha_{j_{1}}) + \sum_{t=2}^{y_{j}-1}s_{j'_{t}}(p-s_{j'_{t}}) + \beta_{j_{y_{j}}}(p-\beta_{j_{y_{j}}}) \right) \nonumber \\
              &=\sum_{P_{j} \in \pcls} \left( \dfrac{1}{2}   \left( \alpha_{j_{1}}(p-\alpha_{j_{1}}) + \sum_{t=2}^{y_{j}-1}s_{j'_{t}}(p-s_{j'_{t}}) + \beta_{j_{y_{j}}}(p-\beta_{j_{y_{j}}}) \right)\nonumber + \alpha_{j_{1}}(s_{j_{1}} - \alpha_{j_{1}}) \right) \nonumber \\
              &=  \sum_{P_{j} \in \pcls} \Bigl( \dfrac{1}{2}   \left( \alpha_{j_{1}}(p-s_{j_{1}}) + \sum_{t=2}^{y_{j}-1}s_{j'_{t}}(p-s_{j'_{t}}) + \beta_{j_{y_{j}}}(p-s_{j_{y_{j}}}) \right)\nonumber\\
              &\> + 3\alpha_{j_{1}}(s_{j_{1}} - \alpha_{j_{1}}) + \beta_{j_{y_{j}}}(s_{j_{y_{j}}}-\beta_{j_{y_{j}}}) \Bigl) \nonumber \\
              &= \sum_{\inpcl{j}\in \cut'} \dfrac{s_{j}(p-s_{j})}{2} + \sum_{\pcl{j}\in \pcls}\left(3\alpha_{j_{1}}(s_{j_{1}} - \alpha_{j_{1}}) + \beta_{j_{y_{j}}}(s_{j_{y_{j}}}-\beta_{j_{y_{j}}}) \right) \nonumber \\
              &= A' + \sum_{\pcl{j}\in \pcls} \dfrac{1}{2}(3\alpha_{j_{1}}(s_{j_{1}} - \alpha_{j_{1}}) + \beta_{j_{y_{j}}}(s_{j_{y_{j}}}-\beta_{j_{y_{j}}})). \nonumber
        \end{align}

        Since $s_{j} \leq \frac{p}{2}$ for all $\inpcl{j} \in \cut$, and $ p = \alpha_{j_{1}} + s_{j_{2}} + \dots + s_{j_{y_{j}}} + \beta_{j_{y_{j}}} $, it follows that
\begin{align}
    A' &\geq \sum_{\pcl{j} \in \pcls} \dfrac{1}{2} (\alpha_{j_{1}} + s_{j_{2}} + \dots + s_{j_{y_{j}}} + \beta_{j_{y_{j}}})\dfrac{p}{2} & \nonumber \\
       &= |\pcls|\dfrac{p^{2}}{4}. \nonumber 
\end{align}
Moreover, $\alpha_{j_{1}}(s_{j_{1}} - \alpha_{j_{1}}) \leq \dfrac{s_{j_{1}}^{2}}{4} \leq \dfrac{p^{2}}{16}$, and $\beta_{j_{y_{j}}}(s_{j_{y_{j}}}-\beta_{j_{y_{j}}})) \leq \dfrac{s_{j_{y_{j}}}^{2}}{4} \leq \dfrac{p^{2}}{16}$. Combining these facts, we get
\begin{align}
    A &\leq A' + |\pcls|\dfrac{1}{2}\left( 3\dfrac{p^{2}}{16} + \dfrac{p^{2}}{16} \right) \nonumber\\
      &= A' + |\pcls|\dfrac{p^{2}}{8} \leq \dfrac{3}{2}A', \nonumber
\end{align}
which is desired.

\end{proof}

\end{proof}

Now we are ready to prove \cref{lem:main-cut-case}.

\begin{proof}[Proof of \cref{lem:main-cut-case}]
   Let $\out$ be the output of $\algog$ in the cut case. Let $\costone{\out}, \costtwo{\out},$ $\costthree{\out},$ $\costfour{\out}$ be the four costs that are paid by our algorithm when in $\algog$, we are left with the clusters present only in $\cut$.~\cref{lem:main-cut-case} is essentially showing that
\begin{align}
    \costone{\out} + \costtwo{\out} + \costthree{\out} + \costfour{\out} \leq 2 \, \opt + 1.5\opt. \nonumber
\end{align}

We complete the proof by applying~\cref{lem:cost-one-plus-two-cut-case}, which gives 
\begin{align}
   \costone{\out} + \costtwo{\out} \leq 2 \, \opt \nonumber,
\end{align}
and applying~\cref{lem:cost 3 + 4 cut case}, which gives
\begin{align}
    \costthree{\out} + \costfour{\out} \leq 1.5\opt \nonumber.
\end{align} 
\end{proof}

Hence, from \cref{lem:cost-one-plus-two-cut-case} and \cref{lem:cost 3 + 4 cut case} we get that $\mop$ pays at most $3.5 \, \, \opt$ in the cut case.

\section{Making a p-divisible clustering fair.}

Given a $p$-divisible clustering $\out = \{\outcl{1}, \outcl{2}, \ldots, \outcl{\zeta}\}$, where the vertex set $V$ is partitioned into two color groups, blue and red, with the ratio of blue to red vertices being $p\!:\!1$, our goal is to compute a closest fair clustering $\fairmop$ to $\out$. 
Recall that a clustering is said to be \emph{fair} if, in every cluster, the number of blue vertices is exactly $\ratio$ times the number of red vertices. 
Our main result in this section is stated below.

\begin{lemma}\label{thm:make.p.cluster.fair}
Let $p > 1$ be an integer, and let the vertex set $V$ be partitioned into two color groups, blue and red, such that the ratio of blue to red vertices is $p\!:\!1$. 
Then, there exists a linear-time algorithm that, given any $p$-divisible clustering $\mathcal{T}$, computes a $3$-close fair clustering $\mathcal{F}$ to $\m{C}$. More specifically,
\[
\dist(\mathcal{T}, \mathcal{F}) \leq 3 \cdot \dist(\mathcal{T}, \mathcal{F}^*),
\]
 where $\mathcal{F}^*$ denotes a closest fair clustering to $\mathcal{T}$.
\end{lemma}

In~\cref{sec:algo.make.pcluster.fair}, we introduce the algorithm $\algmf$ (\cref{alg:algo-mf}), which takes $ \out $ as its input, and outputs a fair cluster $ \fairset $. We analyze in~\cref{sec:analyze.make.pcluster.fair} that $ \dist(\out, \fairset) \leq 3\dist(\out, \fairmop) $.

\subsection{Details of the Algorithm}\label{sec:algo.make.pcluster.fair}
The high-level idea of $\algmf$ (\cref{alg:algo-mf}) is the following. The $ \mopqdef \ \out $ is partitioned into two sets. Each cluster $ \outcl{i} $ is chosen to be in one of the two sets based on whether $ |\blue{\outcl{i}}| < \ratio |\red{\outcl{i}}| $ or $ |\blue{\outcl{i}}| > \ratio |\red{\outcl{i}}|  $. If it is the former case, $ \outcl{i} $ is placed into the first set $ \tr $, while if it is the latter case, $ \outcl{i} $ is placed into the second set $ \tb $. After that, a set of $ |\red{\outcl{i}}| - \dfrac{|\blue{\outcl{i}}|}{\ratio} $ red vertices are cut from each cluster $ \outcl{i} \in \tr $. These vertices are merged to clusters in the set $ \tb $. Each cluster $ \outcl{j} \in \tb $ are merged with exactly $ \dfrac{|\blue{\outcl{j}|}}{\ratio} - |\red{\outcl{j}}| $ red vertices.

\paragraph{Runtime Analysis of $\algmf$.} First note, each vertex is used at most twice during the execution of the algorithm $\algmf$ -- once when we cut the red vertex $v \in V$ from the cluster in the set $\tr$ and again when we merge that red vertex $v \in V$ to cluster that belongs to the set $\tb$. Hence, the total runtime of the algorithm $\algmf$ would be $O(|V|)$.

\subsection{Approximation Guarantee of $\algmf$}\label{sec:analyze.make.pcluster.fair}
In this section, we aim to prove~\cref{thm:make.p.cluster.fair}.

Consider a cluster $ \outcl{i}\in \out $. Suppose that in $ \fairmop $, $ \outcl{i} $ is spread through clusters $ X_{i,1} \cup B_{i,1} \cup R_{i,1}, X_{i,2} \cup B_{i,2} \cup R_{i,2}, \dots, X_{i,t} \cup B_{i,t} \cup R_{i,t} $, such that $ \outcl{i} = \cup_{j=1}^{t}X_{i,j} $, and $ B_{i,j}, R_{i,j} $ are the set of blue vertices and red vertices from some clusters $ \outcl{k} $ ($ k \neq i $), respectively. Then the distance between the two clusterings $ \fairmop $ and $ \out $ is
\begin{align}
    \dist(\out, \fairmop) &= \sum_{\outcl{i}\in \out} \left(\sum_{j<k} |X_{i,j}||X_{i,k}| + \sum_{j=1}^{t}\dfrac{|X_{i,j}|(|B_{i,j}|+ |R_{i,j}|)}{2} \right) \nonumber \\
                      &= \sum_{\outcl{i}\in \out} \opt_{\outcl{i}}, \nonumber
\end{align}
where $ \opt_{\outcl{i}} =  \sum_{j<k} |X_{i,j}||X_{i,k}| + \sum_{j=1}^{t}\dfrac{|X_{i,j}|(|B_{i,j}|+ |R_{i,j}|)}{2} $.

Prior to proving the correctness of~\cref{thm:make.p.cluster.fair}, we demonstrate some useful lower bounds for $ \opt_{\outcl{i}} $. In particular, let $ \dmp = \frac{|\blue{\outcl{i}}|}{\ratio} - |\red{\outcl{i}}| $, $ \smp = |\red{\outcl{i}}| - \frac{|\blue{\outcl{i}}|}{\ratio} $, then
\begin{itemize}
    \item If $ \outcl{i}\in \tb $, then $ \opt_{\outcl{i}} \geq \dfrac{1}{2}\dmp |\outcl{i}| $;
    \item If $ \outcl{i}\in \tr $, then $ \opt_{\outcl{i}} \geq \dfrac{1}{2}\smp(|\outcl{i}|-s) $;
    \item If $ \outcl{i}\in \tr $, then $ \opt_{\outcl{i}} \geq \dfrac{\smp_{i}^{2}}{2} $.
\end{itemize}
To show this, we leverage the following lemma.

\begin{lemma}\label{lem:make.estimated.extra.integral}
    Given positive numbers $ x_{1}, x_{2}, \dots, x_{t} $ and a positive integer $ S $ such that $ S \leq \sum_{i=1}^{t}x_{i} $. Then for any $ a_{1}, a_{2}, \dots, a_{n} $ such that $ a_{i} \geq 0 $, there exists non-negative integers $ y_{1}, y_{2}, \dots, y_{t} $ satisfying:
    \begin{itemize}
        \item $ 0\leq y_{i} \leq \ceilenv{x_{i}}$, for $ i=1,2,\dots,t $.
        \item $ \sum_{i=1}^{t}y_{i}= S $.
        \item $ \sum_{i=1}^{t}a_{i}y_{i} \leq \sum_{i=1}^{t}a_{i}x_{i} $.
    \end{itemize}
\end{lemma}
\begin{proof}
    Without loss of generality, assume that $ a_{1}\leq a_{2}\leq \dots \leq a_{n} $.

    We will show that the output of the following algorithm is the desired $ y_i $'s.

    \begin{algorithm}[htbp]
        \DontPrintSemicolon
        \caption{Choose $ (y_{1}, y_{2}, \dots, y_{t}) $.}
        \label{alg.make.estimated.extra.integral}
        \SetKwInOut{KwIn}{Input}
        \SetKwInOut{KwOut}{Output}
        $ y_{j} \gets \floor{x_{j}}, \forall j $

        \While{$ \sum_{i=1}^{t}y_{i} > S $}{
            \For{$ i=1 \to t $}{
                \uIf{$ \sum_{i=1}^{t} y_{i} >S $ and $ y_{i}>0 $}{
                    $ y_{i} \gets y_{i} - 1 $ \label{line: decreases}
                }
            }
        }

        \While{$ \sum_{i=1}^{t}y_{i} < S $}{
            \For{$ i=1 \to t $}{
                \uIf{$ \sum_{i=1}^{t} y_{i} <S $ and $ y_{i}< \ceil{x_{i}} $}{
                    $ y_{i} \gets y_{i} + 1 $ \label{line: increases}
                }
            }
        }

        \KwRet{$ (y_{1}, y_{2}, \dots, y_{t}) $}
    \end{algorithm}

    Let $ y_{i} $'s be the output of~\cref{alg.make.estimated.extra.integral}.
    %
    
    $ \bullet $ $ 0\leq y_{i} \leq \ceilenv{x_{i}} $, for $ i=1,2,\dots, t $.

    Observe that $ y_{i} $ is decreased by $ 1 $ (\cref{line: decreases}) only if $ y_{i}>0 $, and is increased by $ 1 $ (\cref{line: increases}) only if $ y_{i}<\ceil{x_{i}} $. Hence, each $ y_{i} $ output by this algorithm satisfies $ 0\leq y_{i} \leq \ceil{x_{i}} $.

    $ \bullet $ $ \sum_{i=1}^{t}y_{i} = S $.

    To see this, we consider $ (y_{1},y_{2}, \dots, y_{t}) $ after the execution of each while loop. Initially, $ y_{i} $ is set to $ \floor{x_{i}} $. If $ \sum_{i=1}^{t}y_{i} > S $, the first while loop gradually decreases $ \sum_{i=1}^{t}y_{i} $ by $ 1 $ by decreasing a nonzero $ y_{i} $ by $ 1 $, until $ \sum_{i=1}^{t}y_{i} = y $. Since $ \min\sum_{i=1}^{t}y_{i}=0 $ and $ S\geq 0 $, there must be an iteration where $ \sum_{i=1}^{t}y_{i} = S $. After this iteration, the first while loop is done. The second while loop will not be executed, and the algorithm will return $ (y_{1}, y_{2}, \dots, y_{t}) $ satisfying $ \sum_{i=1}^{t}y_{i} = S $. 

    On the other hand, if $ \sum_{i=1}^{t}\floor{x_{i}}<y $, the first while loop is not executed. The second while loop gradually increases $ \sum_{i=1}^{t}y_{i} $ by $ 1 $ by increasing a $ y_{i} < \ceil{x_{i}} $ by $ 1 $, until $ \sum_{i=1}^{t}y_{i} $. Since $ S\leq \sum_{i=1}^{t}x_{i}\leq \sum_{i=1}^{t}\ceil{x_{i}} $, there always exists some $ y_{i} (\leq \ceil{x_{i}})$ that can be increased whenever $ \sum_{i=1}^{t}y_{t} < S $. Hence, there must be an iteration where $ \sum_{i=1}^{t}y_{i}=S $. After this iteration, the second while loop is done, and the algorithm outputs $ (y_{1}, y_{2}, \dots, y_{t}) $ such that $ \sum_{i=1}^{t}y_{i} = S $. 
    
    
    $ \bullet $ $ \sum_{i=1}^{t}a_{i}y_{i} \leq \sum_{i=1}^{t}a_{i}x_{i} $.

    We first claim that there exists $ m\leq t $ such that for all $ i< m $, we have $ y_{i}\geq x_{i} $, and for all $ i\geq m $, we have $ y_{i}\leq x_{i} $.

    If $ y_{i} \leq x_{i} $  for all $ j $, then $ m = 1 $ satisfies the claim. If there exists $ y_{i} > x_{i}$, since $ \sum_{i=1}^{t}y_{i} < \sum_{i=1}^{t}x_{i} $, there also exists $ y_{i} < x_{i} $. Let $ m $ be the minimum index such that $ y_{m}<x_{m} $, then $ y_{i} \geq x_{i}$ for all $ j<m $. Now we show that $ y_{i}\leq x_{i} $ for all $ j\geq m $.

    To see this, note that in~\cref{alg.make.estimated.extra.integral}, if the second while loop is not executed, then the algorithm only decreases $ y_{i} $. Thus $ y_{i}\leq \floor{x_{i}} \leq x_{i} $ for all $ j $. The second while loop is executed only if the first while loop is not. In that case, the second while loop starts with $ y_{i} = \floor{x_{i}} $. If $ y_{i} $ is increased (\cref{line: increases}), then $ y_{i} = \ceil{x_{i}} > x_{i} $. Assume to the contrary that there exists $ j>m $ such that $ y_{j} > x_{j} $. This implies that before $ y_{j} $ is increased, $ \sum_{i=1}^{t}y_{i} < S $. This means that in the iteration of the for loop where $ i = m $, we have $ \sum_{i=1}^{t}y_{i}< S $ and by definition of $ m $, $ y_{m}<x_{m} $. It follows that $ y_{m} $ is increased and becomes greater than $ x_{m} $, which is a contradiction since in the output, $ y_{m}<x_{m} $.
    
    This concludes that $ y_{i}\leq x_{i} $, for all $ i\geq m $.
    
    Finally, we show that $ \sum_{i=1}^{t}a_{i}y_{i} \leq \sum_{i=1}^{t}a_{i}x_{i} $. We have
    \begin{align}
        \sum_{i=1}^{t}x_{i} &\geq \sum_{i=1}^{t}y_{i} \nonumber \\
        \Leftrightarrow \sum_{i=m}^{t}(x_{i} - y_{i}) &\geq \sum_{i=1}^{m-1} (y_{i} - x_{i}) \geq 0. \nonumber
    \end{align}
    Moreover, note that $ a_{1}\leq a_{2}\leq \dots \leq a_{t} $, therefore
    \begin{align}
        \sum_{i=m}^{t}(x_{i}-y_{j})a_{i} &\geq \sum_{i=m}^{t}(x_{i}-y_{i})a_{m} \nonumber \\
                                         &\geq \sum_{i=1}^{m-1}(y_{i}-x_{i})a_{m} \nonumber \\
                                         &= \sum_{i=1}^{m-1}(y_{i}-x_{i})a_{i},\nonumber
    \end{align}
    which is equivalent to 
    \begin{align}
        \sum_{i=1}^{t}a_{i}y_{i} \leq \sum_{i=1}^{t}a_{i}x_{i}. \nonumber
    \end{align}
    The proof is complete.
\end{proof}

\begin{lemma}\label{lem:cost.of.opt.when.blue>.p.red}
    For each $ \outcl{i}\in \tb $, let $ \dmp = \dfrac{|\blue{\outcl{i}}|}{\ratio} - |\red{\outcl{i}}| $. Then $ \opt_{\outcl{i}} \geq \dfrac{1}{2}\dmp |\outcl{i}| $.
\end{lemma}
\begin{proof}
    By definition, $\opt_{\outcl{i}} = \sum_{1\leq j< k\leq t}|X_{i,j}||X_{i,k}| + \sum_{j=1}^{t}\dfrac{|X_{i,j}|(|B_{i,j}| + |R_{i,j}|)}{2}. \nonumber$

    Denote by $ \blue{X_{i,j}} $ and $ \red{X_{i,j}} $ the set of blue vertices and the set of red vertices in $ X_{i,j} $, respectively. Then $ X_{i,j} = \blue{X_{i,j}} \cup \red{X_{i,j}} $, and $ |\blue{\outcl{i}}| = \sum_{j=1}^{t}|\blue{X_{i,j}}| $, $ |\red{\outcl{i}}| = \sum_{j=1}^{t}|\red{X_{i,j}}| $, and $ |\blue{X_{i,j}}|, |\red{X_{i,j}}| \geq 0 $.

    Let $ S = \left\{j: |\blue{X_{i,j}}| > \ratio|\red{X_{i,j}}|  \right\} $. Then $ S $ is a nonempty set because otherwise $ |\blue{\outcl{i}}| = \sum_{j=1}^{t}|\blue{X_{i,j}}| \leq  \ratio(\sum_{j=1}^{t}|\red{X_{i,j}}|) = \ratio |\red{\outcl{i}}| $, which contradicts to the assumption that $ \outcl{i}\in \tb $.

    For each $ j\in S $, consider arbitrary $ \dmp_{j} $ blue vertices from $ \blue{X_{i,j}} $ ($ d_{j}\leq |\blue{X_{i,j}}| $). Denote $  \dmp' = \sum_{j\in L}\dmp'_{j}  $. Then
    \begin{align}
        \sum_{1\leq j<k \leq t}|X_{i,j}||X_{i,k}| &\geq \sum_{j\in S} (\dmp'_{j}\sum_{k\notin S} |X_{i,k}|) + \dfrac{1}{2}\sum_{j\in S} (\dmp'_{j}\sum_{k\in S\setminus\{j\}} |X_{i,k}| ) \nonumber \\
                                                  &\geq \dfrac{1}{2}\sum_{j\in S}\dmp'_{j}|\outcl{i}| - \dfrac{1}{2}\sum_{j\in S} \dmp'_{j}|X_{i,j}| \nonumber \\
                                                  &= \dfrac{1}{2}(\sum_{j\in S}\dmp'_{j})|\outcl{i}| - \dfrac{1}{2}\sum_{j\in S} \dmp'_{j}|X_{i,j}|. \label{eq. cost paid for dmpj}
    \end{align}

    Let $ B_{j} = |B_{i,j}|,\ R_{j} = |R_{i,j}| $. Let $ \dmp_{j} = \dfrac{|\blue{X_{i,j}}|}{\ratio} - |\red{X_{i,j}}| $. 

    Since $ X_{i,j}\cup B_{i,j}\cup R_{i,j} $ is a cluster in the fair clustering $ \fairmop $, it means that $ p(|\red{X_{i,j}}| + R_{j}) = q(|\blue{X_{i,j}}|+ B_{j}) $. Therefore, if $ j\in S $, then $ B_{j} + R_{j} \geq \dfrac{|\blue{X_{i,j}}|}{\ratio} - |\red{X_{i,j}}| = \dmp'_{j} >0 $. Thus
    \begin{align}
        \sum_{j\in S}\dfrac{|X_{i,j}|(B_{j} + R_{j})}{2} \geq \sum_{j\in S}\dfrac{|X_{i,j}|\dmp'_{j}}{2}. \label{eq. cost paid for merge Rj} 
    \end{align}
    From~\eqref{eq. cost paid for dmpj} and~\eqref{eq. cost paid for merge Rj}, it follows that
    \begin{align}
        \opt_{\outcl{i}} \geq \dfrac{1}{2} (\sum_{j\in S}\dmp_{j})|\outcl{i}| - \dfrac{1}{2}\sum_{j\in S}\dmp_{j}|X_{i,j}| + \dfrac{1}{2}\sum_{j\in S}\dmp_{j}'|X_{i,j}|. \label{eq.bound.opt.dmp} 
    \end{align}
If we can choose $ \dmp_{j} (\leq |\blue{X_{ij}}|) $ such that  $ \sum_{j\in S}\dmp_{j} = \dmp $, and $ \sum_{j\in S}\dmp_{j}|X_{i,j}| \leq \sum_{j\in S}\dmp'_{j}|X_{i,j}| $, then from~\eqref{eq.bound.opt.dmp}, it implies that $ \opt_{\outcl{i}}\geq \dfrac{1}{2}\dmp|\outcl{i}| $, which concludes the proof.

    To this end, note that  $ \dmp_{j} > 0 $ for all $ j\in S $, and $ \dmp_{j}\leq 0 $ if $ j\notin S $. As a result, $ \dmp = \sum_{j=1}^{t}\dmp_{j} \leq \sum_{j\in S}\dmp_{j} $. Finally, applying~\cref{lem:make.estimated.extra.integral} (by setting $ \dmp_{j} = x_{j},\ |X_{i,j}| = a_{j} $), there exists integers $ \dmp'_{j} $ satisfying:
    \begin{itemize}
        \item $ 0\leq \dmp'_{j} \leq \ceilenv{\dmp_{j}}$, which implies $ \dmp'_{j}\leq |\blue{X_{i,j}}| $ as $ \dmp_{j} \leq \dfrac{|\blue{X_{i,j}|}}{\ratio} $ and $ \ratio >1 $.
        \item $ \sum_{j\in S}\dmp'_{j} = d $.
        \item $ \sum_{j\in S}\dmp'_{j}|X_{i,j}| \leq \sum_{j\in S}\dmp_{j}|X_{i,j}| $.
    \end{itemize}
\end{proof}

%
\begin{lemma}\label{lem:cost.of.opt.when.blue<.p.red}
For each $ \outcl{i}\in \tr $, let $ \smp = |\red{\outcl{i}}| - \dfrac{|\blue{\outcl{i}}|}{\ratio} $. Then $ \opt_{\outcl{i}} \geq \dfrac{1}{2}\smp(|\outcl{i}|-s) $. 
\end{lemma}
\begin{proof}
    By definition, $\opt_{\outcl{i}} = \sum_{1\leq j< k\leq t}|X_{i,j}||X_{i,k}| + \sum_{j=1}^{t}\dfrac{|X_{i,j}|(|B_{i,j}| + |R_{i,j}|)}{2}. \nonumber$

    Denote by $ \blue{X_{i,j}} $ and $ \red{X_{i,j}} $ the set of blue vertices and the set of red vertices in $ X_{i,j} $, respectively. Then $ X_{i,j} = \blue{X_{i,j}} \cup \red{X_{i,j}} $, and $ |\blue{\outcl{i}}| = \sum_{j=1}^{t}|\blue{X_{i,j}}| $, $ |\red{\outcl{i}}| = \sum_{j=1}^{t}|\red{X_{i,j}}| $, and $ |\blue{X_{i,j}}|, |\red{X_{i,j}}| \geq 0 $.

    Let $ L = \left\{j: |\blue{X_{i,j}}| < \ratio|\red{X_{i,j}}|  \right\} $. Then $ L $ is a nonempty set because otherwise $ |\blue{\outcl{i}}| = \sum_{j=1}^{t}|\blue{X_{i,j}}| \geq  \ratio(\sum_{j=1}^{t}|\red{X_{i,j}}|) = \ratio |\red{\outcl{i}}| $, which contradicts to the assumption that $ \outcl{i}\in \tr $.

    For each $ j\in L $, consider arbitrary $ \smp'_{j} $ red vertices from $ \red{X_{i,j}} $. Denote $  \smp' = \sum_{j\in L}\smp'_{j}  $. Then
    \begin{align}
        \sum_{1\leq j<k \leq t} |X_{i,j}||X_{i,k}| &\geq \sum_{j\in L} \left(\smp'_{j} \sum_{k\notin L}|X_{i,k}|\right) + \sum_{j\in L} \left(\smp'_{j} \sum_{k\in L\setminus \{j\} }(|X_{i,k}| - \smp'_{k})\right) \nonumber \\
        &=\sum_{j\in L}\smp'_{j}\left(|\outcl{i}|-\smp'_{j}\right) - \sum_{j\in L} \smp'_{j}\left(|X_{i,j}-\smp'_{j}|\right) \nonumber \\
        &= \smp'\left(|\outcl{i}|-\smp'\right) - \sum_{j\in L} \smp'_{j}\left(|X_{i,j}|-\smp'_{j}\right).\label{eq. cost paid for smpj} 
    \end{align}

    For each $ j $, let $ B_{j} = |B_{i,j}|,\ R_{j} = |R_{i,j}| $. Let $ \smp_{j} = |\red{X_{i,j}}| - \dfrac{|\blue{X_{i,j}}|}{\ratio} $.

    Since $ X_{i,j}\cup B_{i,j}\cup R_{i,j} $ is a cluster in the fair clustering $ \fairmop $, it means that $ p(|\red{X_{i,j}}| + R_{j}) = q(|\blue{X_{i,j}}|+ B_{j}) $. Therefore, if $ j\in L $, then $ B_{j} + R_{j} \geq \ratio |\red{X_{i,j}}| - |\blue{X_{i,j}}| = \ratio\smp_{j} >0 $. Thus 
    \begin{align}
        \sum_{j\in L} \dfrac{|X_{i,j}|(B_{j} + R_{j})}{2} \geq \dfrac{\ratio}{2}\sum_{j\in L}|X_{i,j}|\smp_{j} \geq \dfrac{1}{2}\sum_{j\in L}|X_{i,j}|\smp_{j} .\label{eq. cost paid for merge Bj} 
    \end{align}
    From~\eqref{eq. cost paid for smpj} and~\eqref{eq. cost paid for merge Bj}, it follows that
    \begin{align}
        \opt_{\outcl{i}} &\geq \smp'\left(|\outcl{i}|-\smp'\right) - \sum_{j\in L} \smp_{j}(|X_{i,j}|-\smp_{j}) + \dfrac{1}{2}\sum_{j\in L}\smp'_{j}|X_{i,j}| \nonumber \\
                         &\geq \dfrac{1}{2} \left(\smp'\left(|\outcl{i}|-\smp'\right) - \sum_{j\in L} \smp_{j}(|X_{i,j}|-\smp_{j}) + \sum_{j\in L}\smp'_{j}|X_{i,j}|\right). \label{eq.cost.bound.smp}
    \end{align}

    If we can choose $ \smp_{j}\ (\leq |\red{X_{i,j}}|) $ such that $ \sum_{j\in L}\smp_{j} = \smp $ and $ \sum_{j\in L} \smp_{j}|X_{i,j}| \leq \sum_{j\in L}\smp'_{j}|X_{i,j}| $, then from~\eqref{eq.cost.bound.smp}, it implies that $ \opt_{\outcl{i}}\geq \dfrac{1}{2}\smp(|\outcl{i}| - \smp) $, which completes the proof.

    To this end, note that $ \smp_{j} > 0 $ for all $ j\in L $, and $ \smp_{j}\leq 0 $ for all $ j\notin L $. Consequently, $ \smp = \sum_{j=1}^{t}\smp_{j} \leq \sum_{j\in L}\smp_{j} $. Finally, applying~\cref{lem:make.estimated.extra.integral} (by setting $ \smp_{j} = x_{j}, |X_{i,j}| = a_{j} $), there exists integers $ \smp'_{j} $ satisfying
    \begin{itemize}
        \item $ 0\leq \smp'_{j}\leq \ceilenv{\smp_{j}} < |\red{X_{i,j}}|$.
        \item $ \sum_{j\in S}\smp'_{j} = \smp $.
        \item $ \sum_{j\in L}\smp_{j}|X_{i,j}| \leq \sum_{j\in L}\smp'_{j}|X_{i,j}| $.
    \end{itemize}

\end{proof} 

\begin{lemma}\label{lem:cost.split.the.split.when.blue<.p.red}
    For each cluster $ \outcl{i}\in \tr $, let $ \smp_{i} = |\red{\outcl{i}}| - \dfrac{|\blue{\outcl{i}|}}{\ratio} $. Then $$ \opt_{\outcl{i}} \geq \dfrac{\smp_{i}^{2}}{2}. $$
\end{lemma}
\begin{proof}
    By definition,
    \begin{align}
        \opt_{\outcl{i}} = \sum_{1\leq j< k\leq t}|X_{i,j}||X_{i,k}| + \sum_{j=1}^{t}\dfrac{|X_{i,j}|(|B_{i,j}| + |R_{i,j}|)}{2}. \nonumber
    \end{align}
    Denote by $ r_{j}, b_{j} $ the number of red vertices and blue vertices in $ X_{i,j} $ respectively. Then $ |X_{i,j}| = r_{j} + b_{j} $. Let $ R_{j} = |R_{i,j}| $, and $ B_{j} = |B_{i,j}| $.

    Let $ L = \{j:\ \ratio r_{j} > b_{j} \} $. Then $ L \neq \emptyset $.

    For each $ j\in L $, let $ E_{j} = r_{j} - \dfrac{b_{j}}{\ratio} > 0 $. Since $ X_{i,j}\cup B_{i,j}\cup R_{i,j} $ is a cluster in the fair clustering $ \fairmop $, it means that $ p(r_{j} + R_{j}) = q(b_{j} + B_{j}) $. Therefore, if $ j\in L $, then $ B_{j} \geq \ratio r_{j} - b_{j} >0  $. We have
    \begin{align}
        \opt_{\outcl{i}} &\geq \sum_{j\in L}\dfrac{|X_{i,j}||B_{i,j}|}{2} + \sum_{j<k;\ j,k \in L}|X_{i,j}||X_{i,k}| \nonumber \\
                         &\geq \sum_{j\in L}\dfrac{r_{j}(\ratio r_{j}-b_{j})}{2} + \sum_{j<k;\ j,k \in L} r_{j}r_{k} && (\text{since }|B_{i,j}| \geq \ratio r_{j}-b_{j},\ |X_{i,j}|\geq r_{j})\nonumber \\
                         &\geq \dfrac{1}{2}\left(\sum_{j\in L}  r_{j}(r_{j} - \dfrac{b_{j}}{\ratio}) + 2\sum_{j<k;\ j,k\in L}r_{j}r_{k} \right) && (\text{since }\ratio >1)\nonumber \\
                         &\geq \dfrac{1}{2} \left(\sum_{j\in L} E_{j}^{2} + 2\sum_{j<k; j,k \in L}E_{j}E_{k} \right) && (\text{since } r_{j} \geq r_{j} -\dfrac{b_{j}}{\ratio})\nonumber \\
                         & = \dfrac{(\sum_{j\in L}E_{j})^{2}}{2} .\nonumber
    \end{align}
    Note that, if $ j\notin L $, $ r_{j}-\dfrac{b_{j}}{\ratio}\leq 0 $. Therefore, $ \smp_{i} = |\red{\outcl{i}}| - \dfrac{|\blue{\outcl{i}}|}{\ratio} = \sum_{j=1}^{t}r_{j} - \dfrac{\sum_{j=1}^{t}b_{j}}{\ratio} \leq \sum_{j\in L}(r_{j} - \dfrac{b_{j}}{\ratio}) = \sum_{j\in L}E_{j}$. It follows that $ \opt_{\outcl{i}} \geq \dfrac{(\sum_{j\in L}E_{j})^{2}}{2} \geq \smp_{i}^{2}/2 $.

     This concludes the proof.
\end{proof}

Recall that for each cluster $ \outcl{i}\in \tr $,~\cref{alg:algo-mf} cuts $ |\red{\outcl{i}}| - \dfrac{|\blue{\outcl{i}}|}{\ratio} $ red vertices from it. Each cluster in $ \outcl{i}\in \tb $ will be merged with $ \dfrac{|\blue{\outcl{i}}|}{\ratio} - |\red{\outcl{i}}| $ red vertices, which are previously split from clusters in $ \tr $. Note that, during this process, vertices in $ |\red{\outcl{i}}| - \dfrac{|\blue{\outcl{i}}|}{\ratio} $ red vertices from the same cluster $ \outcl{i}\in \tr $ may be separated from each other to be merged with different clusters in $ \tb $. Also, $ \dfrac{|\blue{\outcl{i}}|}{\ratio} - |\red{\outcl{i}}| $ vertices that are merged to a cluster $ \outcl{i}\in \tb $ may come from different clusters in $ \tr $. After this process, every cluster $ \outcl{i} $ has become fair. Therefore, every red vertex that is cut from $ \tr $ would have been merged to some cluster in $ \tb $.

The distance $ \dist(\out, \fairset) $ is the number of pairs of vertices that are clustered together in one clustering (either $ \out $ or $ \fairset $) and are separated by the other (either $ \fairset $ or $ \out $). We decompose the number of such pairs into four types of costs. In particular, $ \dist(\out, \fairset) = \costone{\out} + \costtwo{\out} + \costthree{\out} + \costfour{\out} $, where each cost is defined as follows.
\begin{itemize}
    \item $\costone{\out}$: the number of pairs $ (u,v) $ in which in $ \out $, $ u,v \in \outcl{i} $ and $ \outcl{i} \in \tr $, while in $ \fairset $, $ u $ is kept in $ \outcl{i} $ and $ v $ is cut from $ \outcl{i} $. Since for each $ \outcl{i} \in \tr $,~\cref{alg:algo-mf} splits $ \smp_{i} = |\red{\outcl{i}}| - \dfrac{|\blue{\outcl{i}}|}{\ratio} $ from it, it follows that
        \begin{align}
            \costone{\out} = \sum_{\outcl{i}\in \tr}\smp_{i}(|\outcl{i}| - \smp_{i}). \label{eq.cost1.make.p.fair}
        \end{align}
    \item $ \costtwo{\out} $: the number of pairs $ (u,v) $ in which in $ \out $, $ u\in \outcl{i} \in \tb $ and $ v\in \outcl{j} \in \tr $, while in $ \fairset $, $ v $ is split from $ \outcl{j} $ and is merged to $ \outcl{i} $, so that $ u $ and $ v $ are in the same cluster. Since for each $ \outcl{i} \in \tb $,~\cref{alg:algo-mf} merges $ \dmp_{i} = \dfrac{|\blue{\outcl{i}}|}{\ratio} - |\red{\outcl{i}}| $ red vertices to it, it follows that
        \begin{align}
            \costtwo{\out} = \sum_{\outcl{i}\in \tb}\dmp_{i}|\outcl{i}|. \label{eq.cost2.make.p.fair}
        \end{align}
    \item $ \costthree{\out} $: the number of pairs $ (u,v) $ in which in $ \out $, $ u\in \outcl{j} $, and $ v\in \outcl{k} $ where $ \outcl{j} $ and $ \outcl{k} $ are distinct clusters in $ \tr $, while in $ \fairset $, $ u $ is split from $ \outcl{j} $, $ v $ is split from $ \outcl{k} $ and they are both merged to a cluster $ \outcl{i}\in \tb $. For each cluster $ \outcl{i}\in \tb $,~\cref{alg:algo-mf}  merges exactly $ \dmp_{i} = \dfrac{|\blue{\outcl{i}}|}{\ratio} - |\red{\outcl{i}}| $ red vertices to it. Suppose that among these vertices, $ \alpha_{i_{1}} $ vertices are brought from a cluster $ \outcl{i_{1}}\in \tr $, $ \alpha_{i_{2}} $ vertices are brought from a cluster $ \outcl{i_{2}}\in \tr $, and so on, $ \alpha_{i_{t}} $ vertices are brought from a cluster $ \outcl{i_{t}}\in \tr $, such that $ \sum_{j=1}^{t}\alpha_{i_{j}} = \dmp_{i} $, and $ \outcl{i_{1}}, \outcl{i_{2}},\dots, \outcl{i_{t}} $ are distinct clusters. It follows that
        \begin{align}
            \costthree{\out} &= \sum_{\outcl{i}\in \tb } (\sum_{1\leq j <k \leq i_{t}} \alpha_{i_{j}}\alpha_{i_{k}}) \nonumber \\
                             &\leq \sum_{\outcl{i}\in \tb}\dfrac{(\sum_{j=1}^{t}\alpha_{i_{j}})^{2}}{2} = \sum_{\outcl{i}\in \tb}\dfrac{\dmp_{i}^{2}}{2}. \label{eq.cost3.make.p.fair}
        \end{align}
    \item $ \costfour{\out} $: the number of pairs $ (u,v) $ in which $ u, v $ are among $ \smp_{i} = |\red{\outcl{i}}| - \dfrac{|\blue{\outcl{i}}|}{\ratio} $ red vertices split from a cluster $ \outcl{i}\in \tr $, but $ u $ and $ v $ are then separated from each other because $ u $ and $ v $ are merged to different clusters $ \outcl{j} $ and $ \outcl{k} $ in $ \tb $. Suppose that among these vertices, $ \alpha_{i_{1}} $ vertices are merged to a cluster $ \outcl{i_{1}}\in \tb $, $ \alpha_{i_{2}} $ vertices are merged to a cluster $ \outcl{i_{2}}\in \tb $, and so on, $ \alpha_{i_{t}} $ vertices are merged to a cluster $ \outcl{i_{t}}\in \tb $, such that $ \sum_{j=1}^{t}\alpha_{i_{j}} = \smp_{i} $, and $ \outcl{i_{1}}, \outcl{i_{2}},\dots, \outcl{i_{t}} $ are distinct clusters. It follows that
        \begin{align}
            \costfour{\out} &= \sum_{\outcl{i}\in \tr } (\sum_{1\leq j <k \leq i_{t}} \alpha_{i_{j}}\alpha_{i_{k}}) \nonumber \\
                             &\leq \sum_{\outcl{i}\in \tr}\dfrac{(\sum_{j=1}^{t}\alpha_{i_{j}})^{2}}{2} = \sum_{\outcl{i}\in \tr}\dfrac{\smp_{i}^{2}}{2}. \label{eq.cost4.make.p.fair}
        \end{align}
\end{itemize}

We are ready to prove~\cref{thm:make.p.cluster.fair}
\begin{proof}[Proof of~\cref{thm:make.p.cluster.fair}]
    Recall that $ \dist(\out, \fairset) = \costone{\out} + \costtwo{\out} + \costthree{\out} + \costfour{\out} $. 

    We show that $ \costone{\out} + \costtwo{\out} \leq \dfrac{1}{2}\dist(\out, \fairmop) $, and $ \costthree{\out} + \costfour{\out} \leq \dist(\out, \fairmop) $.

    Applying~\cref{lem:cost.of.opt.when.blue<.p.red} for each $ \outcl{i}\in \tr $, we have $ \opt_{\outcl{i}} \geq \dfrac{1}{2}\smp_{i}(|\outcl{i}|-\smp_{i}) $. Applying~\cref{lem:cost.of.opt.when.blue>.p.red} for each $ \outcl{i} \in \tb $, we have $ \opt_{\outcl{i}} \geq \dfrac{1}{2}\dmp_{i}|\outcl{i}| $. Consequently, it follows that
    \begin{align}
        \costone{\out} + \costtwo{\out} &= \sum_{\outcl{i}\in \tr}\smp_{i}(|\outcl{i}| - \smp_{i}) + \sum_{\outcl{i}\in \tb}\dmp_{i}|\outcl{i}| && (\text{by~\eqref{eq.cost1.make.p.fair} and~\eqref{eq.cost2.make.p.fair}}) \nonumber \\
                                        &\leq 2\sum_{\outcl{i}\in \tr}\opt_{\outcl{i}} + 2\sum_{\outcl{i}\in \tb}\opt_{\outcl{i}} \nonumber \\
                                        &\leq 2\dist(\out, \fairmop). \nonumber
    \end{align}

    In each $ \outcl{i}\in \tb $, $ |\outcl{i}| \geq |\blue{\outcl{i}}| > \ratio\dmp_{i} $. Therefore, applying~\cref{lem:cost.of.opt.when.blue>.p.red}, it follows that $\opt_{\outcl{i}} \geq \dmp_{i}|\outcl{i}| \geq \ratio \dmp_{i}^{2}$. Using~\eqref{eq.cost3.make.p.fair}, we have $ \costthree{\out} \leq \sum_{\outcl{i}\in \tb}\dfrac{\dmp_{i}^{2}}{2} \leq \sum_{\outcl{i}\in \tb}\dfrac{\opt_{\outcl{i}}}{2\ratio} $.

    For each $ \outcl{i}\in \tr $, applying~\cref{lem:cost.split.the.split.when.blue<.p.red}, we have $ \opt_{\outcl{i}} \geq \dfrac{\smp_{i}^{2}}{2} $. Using~\eqref{eq.cost4.make.p.fair} implies that $ \costfour{\out} \leq \sum_{\outcl{i}\in \tr }\dfrac{\smp_{i}^{2}}{2} \leq \sum_{\outcl{i}\in \tr}\opt_{\outcl{i}} $. Then $ \costthree{\out} + \costfour{\out} $ is upper bounded by
    \begin{align}
        \sum_{\outcl{i}\in \tb}\dfrac{\opt_{\outcl{i}}}{2\ratio} + \sum_{\outcl{i}\in \tr}\opt_{\outcl{i}} \leq \sum_{\outcl{i}\in \out}\opt_{\outcl{i}} = \dist(\out, \fairmop). \nonumber
    \end{align}

     To summarize, we have
     \begin{align}
         \dist(\out, \fairset) &= \costone{\out} + \costtwo{\out} + \costthree{\out} + \costfour{\out}\nonumber \\
                           &\leq 2\dist(\out, \fairmop) + \dist(\out, \fairmop) \nonumber \\
                           &= 3\dist(\out, \fairmop). \nonumber
     \end{align}
     This completes the proof.
\end{proof}


\section{Concluding the Approximation Guarantee}

In this section, we prove \cref{thm:closest-fair-p:1} by combining the results of \cref{thm:make.p.cluster.fair} and \cref{thm:main-multiple-of-p} to establish the overall approximation guarantee. 
Specifically, we show that the clustering $\m{F}$ obtained by first applying $\algog$ to the input clustering $\m{C}$ to obtain $\m{T}$, and then applying $\algmf$ to $\m{T}$, yields a $17$-close fair clustering with respect to $\m{C}$. 
Instead of directly proving \cref{thm:closest-fair-p:1}, we first prove the following lemma, which immediately implies \cref{thm:closest-fair-p:1} by setting the parameters $\delta = 3.5$ and $\alpha = 3$.

\begin{lemma}
    \label{thm:combine-p-blue-fair}
    Consider any $\delta,\alpha \ge 1$. Suppose there is a $t_1(n)$-time algorithm Algorithm-A() that, given any clustering on $n$ points, produces an $\delta$-close $\bal$, and a $t_2(n)$-time algorithm Algorithm-B() that, given any $\bal$ on $n$ points, produces a $\alpha$-close $\fair$. Then there is an $O(t_1(n) + t_2(n) + n)$-time algorithm that, given any clustering on $n$ points, produces an $\left(\delta + \alpha +\delta \alpha \right)$-close $\fair$.
\end{lemma}

\begin{proof}
Consider the following simple algorithm: Given a clustering $\inpset$ as input, first run Algorithm-A() to generate a $\bal$ $\m{T}$, and then run Algorithm-B() with $\m{T}$ as input to generate a $\fair$ $\m{F}$. Finally, output $\m{F}$.

It is straightforward to see that the running time of the above algorithm is $O(t_1(n) + t_2(n) + n)$. We now claim that $\m{F}$ is an $\left(\delta + \alpha +\delta \alpha \right)$-close $\fair$ of the input clustering $\inpset$.

Let $\fairmop$ be an (arbitrary) closest $\fair$ of the input $\inpset$, and $\mop$ be an (arbitrary) closest $\bal$ of the input $\inpset$. Since by the definition, any $\fair$ is also a $\bal$, we get
\begin{equation}
    \label{eq:opt-fair-bal}
    \dist(\inpset, \mop) \le \dist(\inpset,\fairmop).
\end{equation}

Now, by the guarantee provided by Algorithm-A(), we get
\begin{align}
    \label{eq:output-bal}
    \dist(\inpset,\m{T}) &\le \delta \cdot \dist(\inpset, \mop) \nonumber\\
    &\le \delta \cdot \dist(\inpset, \fairmop) &&\text{(By \cref{eq:opt-fair-bal})}.
\end{align}

Next, let $\m{C}^*$ be an (arbitrary) closest $\fair$ of $\m{T}$ (recall, $\m{T}$ is an output of Algorithm-A()). Then, by the guarantee provided by Algorithm-B(), we get
\begin{align}
    \label{eq:outbal-optfair}
    \dist(\m{T},\m{F}) &\le \alpha\cdot \dist(\m{T},\m{C}^*)\nonumber\\
    & \le \alpha \cdot \dist(\m{T},\fairmop) &&\text{(Since $\fairmop$ is a valid $\fair$)}\nonumber\\
    &\le \alpha \cdot \left(\dist(\m{T},\inpset) + \dist(\inpset,\fairmop) \right) &&\text{(By the triangle inequality)}\nonumber\\
    &\le \alpha (\delta + 1)\cdot \dist(\inpset,\fairmop)&&\text{(By \cref{eq:output-bal})}.
\end{align}

Thus, finally, we deduce that

\begin{align*}
    \dist(\inpset,\m{F}) & \le \dist(\inpset,\m{T}) + \dist(\m{T},\m{F}) &&\text{(By the triangle inequality)}\\
    & \le \delta \cdot \dist(\inpset, \fairmop) + \alpha (\delta + 1)\cdot \dist(\inpset,\fairmop) &&\text{(By \cref{eq:output-bal} and \cref{eq:outbal-optfair})}\\
    & = (\delta + \alpha +\delta \alpha) \cdot \dist(\inpset, \fairmop)
\end{align*}
and that concludes the proof.
\end{proof}

\section{Hardness of Closest Fair Clustering Problem}\label{sec:NP-hard}

In this section, we prove that given a clustering $\m{C}$ over a set of $n$ red-blue colored vertices where the ratio between the total number of blue and red vertices is $p$ for some integer $p > 1$, the problem of finding a Closest {\fair} $\m{F}^*$ is \textbf{NP}-hard. More specifically, we prove \cref{thm:closest-fair-np-hard}, which we restate below.

\nphard*

Let us start by defining the corresponding decision version of the problem.

\begin{definition}[$\clof$]
    Consider a positive integer $p$. Given a clustering $\m{C}$ over a set of red-blue colored vertices where the ratio between the total number of blue and red vertices is $p$, and a non-negative integer $\tau$, decide between the following:
    \begin{itemize}
        \item YES: There exists a {\fair} (on input vertex set) $\m{F}$ such that $\dist(\m{C}, \m{F}) \le \tau$;
        \item NO: For every {\fair} (on input vertex set) $\m{F}$, $\dist(\m{C}, \m{F}) > \tau$.
    \end{itemize}
\end{definition}

 We show the following theorem.

\begin{lemma}\label{thm:main}
    For any integer $p \geq 2$, the $\clof$ problem is $\npc$.
\end{lemma}

It is immediate from \cref{thm:main} that \cref{thm:closest-fair-np-hard} follows as a direct consequence. Hence, we focus on establishing \cref{thm:main}.

Observe that the $\clof$ problem clearly belongs to the class \textbf{NP}. To prove its \textbf{NP}-hardness, we present a polynomial-time reduction from the well-known \emph{3-Partition problem} (defined below) to $\clof$. This reduction establishes that $\clof$ is \textbf{NP}-complete.

\begin{definition}[$3$-Partition problem]
    Given a (multi)set of positive integers $S = \{ x_1, \ldots, x_n\}$, decide whether (YES:) there exists a partition of $S$ into $m$ (disjoint) subsets $S_1, S_2, \ldots, S_m \subseteq S$ where $m = n/3$, such that
    \begin{itemize}
        \item For all $i$, $|S_i| = 3$; and
        \item For all $i$, $\sum_{x_j \in S_i}x_j = T$, where $T = \frac{\sum_{x_j \in S}x_j}{n/3}$,
    \end{itemize}
    or (NO:) no such partition exists.
\end{definition}

Note that, by partition of $S$ we mean $S_i \cap S_j = \emptyset$ for all $i \neq j$ and $\bigcup_{i = 1}^m S_i = S$.

We know that the $3$-Partition problem is in fact \emph{strongly $\npc$}~\cite{garey1975complexity}, i.e., the $3$-Partition problem remains $\npc$ even if all the integers $x_i \in S$ are bounded by a polynomial in $n$; more specifically, $\max_{x_i \in S} x_i \le n^c$ for some non-negative constant $c$. Furthermore, the $3$-Partition problem remains strongly $\npc$ even when for all $x_i \in S$, $x_i \in (T/4, T/2)$, where $T= \frac{3}{n} \sum_{x_i \in S} x_i$~\cite{garey1975complexity}. From now on, we refer to this restricted (with a restriction that each $x_i \in (T/4, T/2)$) variant of the $3$-partition problem as the $\thrp$ problem.

\begin{theorem}(\cite{garey1975complexity}) \label{thm:garey-johnson}
    $\thrp$ is strongly $\npc$.
\end{theorem}

We divide our \textbf{NP}-hardness proof into two parts. First, we prove $\clof$ is \textbf{NP}-hard for any integer $p \geq 3$, and then we prove $\clof$ is \textbf{NP}-hard also for $p = 2$ using a slightly different reduction.

\subsection{Hardness of $\clof$ for $p \geq 3$:}
 In this section, we prove the following lemma.

\begin{lemma}\label{lem:main-one}
    For any integer $p \geq 3$, the $\clof$ problem is \textbf{NP}-hard.
\end{lemma}

To prove \cref{lem:main-one}, we show a reduction from $\thrp$ to $\clof$.

\paragraph{Reduction from $\thrp$ to $\clof$ :} Let us consider a $\thrp$ instance $S = \{ x_1, x_2, \ldots, x_n\}$, where for all $i \in [n]$, $x_i \le n^c$, for some non-negative constant $c$. Given the $\thrp$ instance $S = \{ x_1, x_2, \ldots, x_n\}$ we create a $\clof$ instance $( \m{C}, \tau )$ as follows:

\begin{itemize}
    \item $ \m{C} = \{B_1, \ldots, B_{n/3}, R_1,\ldots,R_n \} $, where for each $i \in \{1,\ldots,n/3\}$, $B_i$ is a monochromatic blue cluster (i.e., containing only blue vertices) of size $pT$ and for each $j \in \{1,\ldots, n\}$, $R_j$ is a monochromatic red cluster (i.e., containing only red vertices) of size $x_j$ (i.e., $|R_j| = x_j$);

    \item \[ \tau = \frac{1}{2} \sum_{j = 1}^n x_j \left(T - x_j\right) + \frac{n}{3}pT^2.\]
\end{itemize}

Since each $x_i \le n^c$, the size of the instance $( \m{C}, \tau )$ is polynomial in $n$. Moreover, it is straightforward to see that the reduction runs in polynomial time.

Now, we argue that the above reduction maps a YES instance of the $\thrp$ to a YES instance of the $\clof$.

\begin{lemma}\label{lem:yes-instance}
    If $S$ is a YES instance of the $\thrp$, then $(\m{C}, \tau) $ is also a YES instance of the $ \clof$.
\end{lemma}

\begin{proof}
    Suppose $S$ is a YES instance of the $\thrp$. Then there exists a partition $S_1, S_2, \ldots, S_{n/3} \, \, \text{of} \, \, S = \{ x_1, x_2, \ldots, x_n\}$ such that for all $1\le i \le n/3$, $|S_i| = 3$ and
    \[
        \sum_{x_j \in S_i}x_j = T \, \, \text{where} \, \, T = \frac{\sum_{x_k \in S} x_k}{\frac{n}{3}}.
    \]
    Let, $S_i = \{ x_{i_1}, x_{i_2}, x_{i_3}\}$. By our construction of $(\m{C}, \tau)$, we have $|R_{i_j}| = x_{i_j}$, for $j \in \{1,2,3\}$.
   
    Now, we construct a fair clustering $\m{F}$ by merging each $B_i$ with $R_{i_1},R_{i_2}, R_{i_3}$. More formally,
    
    \[
       \m{F} = \left\{ \left(B_i \cup \bigcup_{j \in \{1,2,3\}}R_{i_j} \right)  \, \,  \middle| \, \, i \in [n/3] \right \} .
    \]
    Note that $\m{F}$ is a fair clustering because for each cluster $F \in \m{F}$ we have $|\blue{F}| = pT$ (since $\blue{F}=B_i$ for some $i \in [n/3]$) and 
    
    \[|\red{F}| = \sum_{j \in \{1,2,3\}}|R_{i_j}| = \sum_{j \in \{1,2,3\}}x_{i_j} = T.\] 
    
    Next, we claim that
    \[
        \dist(\m{C}, \m{F}) = \frac{1}{2} \sum_{j = 1}^n x_j  \left(T - x_j\right) + \frac{n}{3}pT^2 = \tau.
    \]

    \begin{itemize}
        \item \textbf{Reason behind the term $\frac{1}{2} \sum_{j = 1}^n x_j \left(T - x_j \right)$}: In $\m{F}$, each red cluster $R_j$ is merged with $(T - |R_j|)$ red vertices from other clusters. This expression counts the number of such pairs. $1/2$ comes from the fact that we are counting a pair twice, once while considering the cluster $R_j$ and again while considering a cluster $R_{j'}$, for $j' \neq j$, that is merged with $R_j$. (One important thing to note is that since $S$ is a YES instance of the $\thrp$, we do not need to cut any red cluster $R_j$ in $\m{C}$ to form the {\fair} $\m{F}$.)
        \item \textbf{Reason behind the term $(n/3)pT^2$}: To form each cluster of $\m{F}$ we have merged $T$ red vertices with $pT$ blue vertices. We have $n/3$ clusters in $\m{F}$, hence the number of such pairs in total is $(n/3)pT^2$.
    \end{itemize}
    Hence, $(\m{C},\tau) $ is a YES instance of the $ \clof$.
\end{proof}

Now, we argue that our reduction maps a NO instance of the $\thrp$ to a NO instance of the $\clof$.

\begin{lemma}\label{lem:no-instance}
 For $p \geq 3$, if $S $ is a NO instance of the $\thrp$, then $(\m{C}, \tau) $ is also a NO instance of the $ \clof$.
\end{lemma}

\begin{proof}
    To prove $(\m{C}, \tau) $ is a NO instance of the $ \clof$ we consider an (arbitrary) closest {\fair} $\m{F}^*$ to $\m{C}$, and then argue that $\dist(\m{C}, \m{F}^*) > \tau$.

    Let us recall, $S = \{x_1,\ldots, x_n\}$, $\m{C} = \{ B_1,\ldots,B_{n/3}, R_1,\ldots,R_n\}$ where $B_i$'s are monochromatic blue clusters each of size $pT$, where $T = \sum_{x_i \in S} x_i / (n/3)$ and for each $\ell \in [n]$, $R_\ell$ is a monochromatic red cluster of size $x_\ell$.

    To prove $\dist(\m{C}, \m{F}^*) > \tau$ we prove the following claim regarding the structure of $\m{F}^*$ when $p \geq 3$.

     \begin{claim}\label{clm:np-main-claim}
        Consider any integer $p \geq 3$. Then, for all $B_i$, there exists $F_a \in \m{F}^*$ such that $\blue{F_a} = B_i$.
    \end{claim}

    For now, let us assume the above claim and prove \cref{lem:no-instance}. Since, the number of monochromatic blue clusters in $\m{C}$ is $n/3$, assuming \cref{clm:np-main-claim} we get $|\m{F}^*| = n/3$.

    According to the definition of consensus metric, $\dist(\m{C}, \m{F}^*)$ counts total number of pairs $(u,v)$ that are together in $\m{C}$ but separated by $\m{F}^*$ and the number of pairs $(u,v)$ that are in different clusters in $\m{C}$ but together in $\m{F}^*$.

    For a cluster $F_a \in \m{F}^*$, let us define $\costf{F_a}$ to be the number of pairs $u,v$ such that $u,v \in F_a$ but in different clusters in $\m{C}$.

    Formally, we define $\costf{F_a}$ as
    \[
        \costf{F_a} := |\{ (u,v) \mid u,v \in F_a \, \, \text{but} \, \, u \in C_i, v \in C_j, C_i, C_j \in \m{C}, i \neq j \}|.
    \]

    It is easy to verify that 

    \[
        \costf{F_a} = pT^2 + \frac{1}{2}\sum_{\ell = 1}^n|R_{a,\ell}|(T - |R_{a,\ell}|)
    \]
    where $R_{a,\ell} = R_\ell \cap F_a$.

    In the above expression, $pT^2$ counts the number of blue-red pairs in $F_a$, note by assuming \cref{clm:np-main-claim} we have $|\blue{F_a}| = pT$ and since $F_a$ is fair we have $|\red{F_a}| = T$. In the above expression $(1/2 \cdot \sum_{\ell = 1}^n|R_{a,\ell}|(T - |R_{a,\ell}|))$ counts the number of red-red pairs $(u,v) \in F_a$ such that $u \in R_{\ell_1}$ and $v \in R_{\ell_2}$ for $\ell_1 \neq \ell_2$.

    It is straightforward to observe

    \[
        \dist(\m{C}, \m{F}^*) \geq \sum_{F_a \in \m{F}^*} \costf{F_a}.
    \]

    Now we prove 

    \[
        \sum_{F_a \in \m{F}^*} \costf{F_a} > \tau.
    \]

    For that purpose, we use the following claim, the proof of which is provided later.
    \begin{claim}\label{clm:exists-ell-lese-ml}
        For each $\ell \in [n]$, let $m_\ell = \max_{a: F_a \in \m{F}^*}|R_{a,\ell}|$. There exists an $\ell \in [n]$ such that $m_{\ell} < x_{\ell}$.
    \end{claim}

    So, from now on, we assume the above claim and proceed with arguing $\sum_{F_a \in \m{F}^*} \costf{F_a} > \tau$.

    \begin{align}
        \sum_{F_a \in \m{F}^*} \costf{F_a} &= \sum_{F_a \in \m{F}^*} \left( pT^2 + \frac{1}{2}\sum_{\ell = 1}^n|R_{a,\ell}|(T - |R_{a,\ell}|) \right) \n \\
        &= \frac{n}{3}pT^2 + \frac{1}{2}\sum_{F_a \in \m{F}^*}\sum_{\ell = 1}^n|R_{a,\ell}|(T - |R_{a,\ell}|) \n \\
        &\geq \frac{n}{3}pT^2 + \sum_{\ell = 1}^n (T - m_\ell) \sum_{F_a \in \m{F}^*}|R_{a,\ell}| \, \, \, (\text{recall}\, \, m_\ell = \max_a|R_{a,\ell}|) \n \\
        &= \frac{n}{3}pT^2 + \sum_{\ell = 1}^n (T - m_\ell) |R_\ell| \n \\
        &= \frac{n}{3}pT^2 + \sum_{\ell = 1}^n (T - m_\ell) x_\ell \, \, \, (\text{recall}\, \, x_\ell = |R_{\ell}|) \n \\
        &> \frac{n}{3}pT^2 + \sum_{\ell = 1}^n (T - x_\ell) x_\ell \n = \tau \n
    \end{align}

    where the last inequality follows from \cref{clm:exists-ell-lese-ml}. Thus we get that $\dist(\m{C}, \m{F}^*) > \tau$, which implies $(\m{C},\tau) $ is a NO instance of the $ \clof$.
    \end{proof}

    We are now left with proving \cref{clm:np-main-claim} and \cref{clm:exists-ell-lese-ml}. Let us first show \cref{clm:exists-ell-lese-ml}.

    \begin{proof}[Proof of \cref{clm:exists-ell-lese-ml}]
        For the sake of contradiction, suppose for all $\ell$, $m_\ell = x_\ell$. Note $m_\ell \leq x_\ell$ because
        \begin{align}
            &m_\ell = \max_a|R_{a,\ell}|,\, \, \text{and} \n \\
            &\sum_{F_a \in \m{F}^*}|R_{a,\ell}| = x_\ell. \label{equn:np-claim-two} 
        \end{align}

        If $m_\ell = x_\ell$, then it immediately follows from the definition,

        \begin{equation}
            \label{eq:np-sta-one}
            \exists a: |R_{a,\ell}| = x_\ell, \text{ and } \forall_{b \ne a}, |R_{b,\ell}| = 0.
        \end{equation}

        Now we create partitions $S_a$ of $S$ in the following way: For each $F_a \in \m{F}^*$,

        \[
            S_a = \{ x_\ell \mid |R_{a,\ell}| = x_\ell \}.
        \]

        Since by \cref{clm:np-main-claim}, there are $n/3$ clusters $F_a \in \m{F}^*$, the number of such subsets would be $n/3$. Clearly, the above subsets $S_a$'s create a partitioning of $S$, that is

        \begin{itemize}
            \item $S_a \cap S_b = \emptyset$ for $a \neq b$, follows from \cref{eq:np-sta-one};
            \item $\bigcup S_a = S$, follows from \cref{equn:np-claim-two} (that says $R_{a,\ell}$ cannot be empty for all $a$).  
        \end{itemize}

        Since for each $a$, $F_a$ is a fair cluster containing $pT$ blue vertices (by \cref{clm:np-main-claim}) and $R_{a,\ell} = R_\ell \cap F_a$, we have that for all $a$,
        
        \[
            \sum_{x_\ell \in S_a}x_\ell = T.
        \]

        Further, since $x_\ell \in (T/4,T/2)$, we must have $|S_a| = 3$. However, since $S$ is a NO instance of $\thrp$, such a partitioning of $S$ does not exist, leading to a contradiction.

        Hence, there must exist an $\ell$ such that $m_\ell < x_\ell$.
    \end{proof}

    It only remains to prove \cref{clm:np-main-claim}.

    Recall here, our goal is to prove that for every $F_i \in \m{F^*}$, there exists a monochromatic blue cluster $B_j$ such that $\blue{F_i} = B_j$. Before proceeding with the proof, we briefly outline the main steps. First, we show that for any $F_i \in \m{F^*}$, if there exists a monochromatic blue cluster $B_j$ such that $|B_j \cap F_i|$ is a multiple of $p$, then for all $k \neq j$, we must have $B_k \cap F_i = \emptyset$. Next, we prove that for every $F_i \in \m{F^*}$, there exists some $B_j$ such that $|B_j \cap F_i|$ is a multiple of $p$. Combining these two claims, we conclude the proof of \cref{clm:np-main-claim}.

    \paragraph{Proof of \cref{clm:np-main-claim}.} For a cluster $B_i$, suppose it is partitioned into $Y_i^{1}, Y_i^{2}, \ldots, Y^{t}_i$ in $\m{F}^*$ that is
    \begin{itemize}
        \item $Y^{k}_i \subseteq F$ for some $F \in \m{F}^*$ for all $k \in [t]$
        \item For $k \neq \ell \in [t]$ if $Y^{k}_i \subseteq F_a$ and $Y^{\ell}_i \subseteq F_b$ then $a \neq b$ for some $F_a, F_b \in \m{F}^*$.
        \item $Y^{k}_i \cap Y^{\ell}_i = \emptyset$  for  $k \neq \ell$.
        \item $\bigcup_k Y_i^{k} = B_i$.
    \end{itemize}

    Now, we argue that if for a cluster $B_i$, there exists $Y^{k}_i \subseteq F$ for some $F \in \m{F}^*$ such that $p$ divides $|Y^{k}_i|$, then $F$ does not contain blue vertices from any other cluster $B_j$.

    \begin{claim}\label{clm:claim-np}
        Consider any integer $ p > 1$. Suppose for a cluster $B_i$, there exists $Y^{k}_i \subseteq F$ for some $F \in \m{F}^*$ such that $p$ divides $|Y^{k}_i|$, then $\blue{F} = Y^{k}_i$.
    \end{claim}

    \begin{proof}
        For the sake of contradiction, suppose there exists a cluster $F \in \m{F}^*$ such that $Y^{k}_i \subseteq F$ and $p$ divides $|Y^{k}_i|$, but $B_j \cap F \neq \emptyset$ for some $j \neq i$. In this case, we can construct a {\fair} $\m{M}$ such that $\dist(\m{C}, \m{M}) < \dist(\m{C}, \m{F}^*)$ which contradicts the fact that $\m{F}^*$ is a closest {\fair}. 

        \textbf{Construction of $\m{M}$ from $\m{F}^*$}: Let $Q \subseteq \red{F}$ such that $|Q| = |Y^{k}_i|/p$.

        \[
            \m{M} = \m{F}^* \setminus \{ F\} \cup \{ (Y_i^k \cup Q), F \setminus (Y^k_i \cup Q)\}.
        \]

        That is, in $\m{M}$ we remove the cluster $F$ from $\m{F}^*$ and add two clusters $(Y_i^k \cup Q)$ and $(F \setminus (Y^k_i \cup Q))$.

        Since $F$ is a fair cluster, both the clusters $(Y_i^k \cup Q)$ and $F \setminus (Y_i^k \cup Q)$ are fair.

        Let us now argue that $\dist(\m{C},\m{M}) < \dist(\m{C},\m{F}^*)$. Roughly, the above construction only incurs an additional cost for pairs between $Q$ and $\red{F} \setminus Q$, while saving a cost for pairs between $Q$ and $\blue{F} \setminus Y_i^k$ (note, $\blue{F} \setminus Y_i^k = \blue{F} \setminus B_i$). Further, note that $|\blue{F} \setminus B_i| = p |\red{F} \setminus Q|$. 

        Thus,
        \begin{align*}
            \dist(\m{C}, \m{M}) & \le \dist(\m{C}, \m{F}^*) + |Q| |\red{F} \setminus Q| - p|Q| |\red{F} \setminus Q|\\
            &< \dist(\m{C}, \m{F}^*) &&\text{(for $p > 1$)}.
        \end{align*}
    \end{proof}

    Now, we prove that for all $B_i$ and for all $k \in [t]$, $|Y^{k}_i|$ must be a multiple of $p$. To prove this, we first establish the following two claims. 

    \begin{claim}\label{clm:np-small-claim-one}
        Consider any integer $p > 1$. If for a cluster $F \in \m{F}^*$, $|B_i \cap \blue{F}| < p$ for all $B_i$, then $|\blue{F}| \leq p$.
    \end{claim}

    \begin{proof}
        Consider any cluster $F \in \m{F}^*$. For any $B_i$, let $Y_i := B_i \cap \blue{F} $. WLOG assume $|Y_1| \ge |Y_2| \ge \cdots \ge |Y_n|$. No, suppose for all $i \in [n]$, $|Y_i| < p$. 

        For the sake of contradiction, suppose $|\blue{F}| > p$. Then, there must exist an index $x \in [n]$ such that

        \begin{align}
            &\sum_{i = 1}^{x - 1} |Y_i| \leq p < \sum_{i = 1}^x |Y_i|.\n
        \end{align}

        Note $x \geq 2$ because $|Y_i| < p$ for all $i$.

        Let $Y_x' \subseteq Y_x$ such that $\sum_{i = 1}^{x - 1} |Y_i| + |Y_x'| = p$. Now, since $|Y_1| \ge |Y_x| \ge |Y'_x|$, $|Y_x'| \leq p/2$.

        Next, we construct $\m{M}$ from $\m{F}^*$ in the following way

        \[
            \m{M} = \m{F}^* \setminus \{ F \} \cup \left\{ \left(\bigcup_{i = 1}^{x - 1} Y_i \cup Y_x' \cup Q \right), \left(F \setminus \left(\bigcup_{i = 1}^{x - 1} Y_i \cup Y_x' \cup Q \right) \right) \right\}
        \]

        where $Q \subseteq \red{F}$ such that $|Q| = 1$.

        That is, in $\m{M}$ we remove the set $F$ from $\m{F}^*$ and add two clusters $\left(\bigcup_{i = 1}^{x - 1} Y_i \cup Y_x' \cup Q \right)$ and $\left(F \setminus \left(\bigcup_{i = 1}^{x - 1} Y_i \cup Y_x' \cup Q \right) \right)$.

        It is easy to see that $\m{M}$ is a {\fair}. Next,

        \begin{align}
            \dist(\m{C}, \m{M}) &\le \dist(\m{C}, \m{F}^*) + |Y_x'| |Y_x \setminus Y_x'| + |Q||\red{F} \setminus Q| - \left|\bigcup_{i = 1}^{x - 1}Y_i \right| |Y_x \setminus Y_x'| - |Q| \left| \blue{F} \setminus \left( \bigcup_{i = 1}^{x - 1} Y_i \cup Y_x'\right) \right| \n \\
            &\leq \dist(\m{C}, \m{F}^*) + |Y_x'| (|Y_x| - |Y_x'|) + |Q| |\red{F} \setminus Q| - (p - |Y_x'|) (|Y_x| - |Y_x'|)  - |Q|\cdot  p |\red{F} \setminus Q| \n \\
            &(\text{as} \, \, \left| \blue{F} \setminus \left( \bigcup_{i = 1}^{x - 1} Y_i \cup Y_x'\right) \right| \geq p |\red{F} \setminus Q|) \n \\
            &< \dist(\m{C}, \m{F}^*) \, \, \, (\text{as} \, \, |Y_x'| \leq \frac{p}{2}\, \, \text{and}\, \, |\red{F} \setminus Q| < p |\red{F} \setminus Q|\, \, \text{for $p > 1$} ) \n
        \end{align}
        contradicting the fact that $\m{F}^*$ is a closest {\fair} to $\m{C}$.

        \begin{itemize}
            \item \textbf{Reason behind the term $|Y_x'| |Y_x \setminus Y_x'|$}: The pairs $(u,v)$ such that $u \in Y_x'$ and $v \in Y_x \setminus Y_x'$ are not counted in $\dist(\m{C}, \m{F}^*)$ because the pairs $u$ and $v$ may lie in the same cluster both in $B_i \in \m{C}$ and $F \in \m{F}^*$. These pairs are counted in $\dist(\m{C}, \m{M})$ because they lie in different clusters in $\m{M}$. 
            \item \textbf{Reason behind the term $|Q||\red{F} \setminus Q|$}: The pairs $(u,v)$ such that $u \in Q$ and $v \in \red{F} \setminus Q$ may not be counted in $\dist(\m{C}, \m{F}^*)$ because the pairs $u$ and $v$ may lie in the same cluster both in $\m{C}$ and $\m{F}^*$. These pairs are counted in $\dist(\m{C}, \m{M})$ because they lie in different clusters in $\m{M}$. 
            \item \textbf{Reason behind the term $\left|\bigcup_{i = 1}^{x - 1}Y_i \right| |Y_x \setminus Y_x'| $}: First note that, $\bigcup_{i = 1}^{x - 1}Y_i \neq \emptyset$ as $x \geq 2$. The pairs $(u,v)$ such that $u \in \bigcup_{i = 1}^{x - 1}Y_i$ and $v \in Y_x \setminus Y_x'$ are counted in $\dist(\m{C}, \m{F}^*)$ because the pairs $(u, v)$ lie in different clusters in $\m{C}$ but in the same cluster $F \in \m{F}^*$. These pairs are not counted in $\dist(\m{C}, \m{M})$ because they lie in different clusters both in $\m{C}$ and $\m{M}$. 
            \item \textbf{Reason behind the term $|Q| \left| \blue{F} \setminus \left( \bigcup_{i = 1}^{x - 1} Y_i \cup Y_x'\right) \right| $}: The pairs $(u,v)$ such that $u \in Q$ and $v \in \blue{F} \setminus \left( \bigcup_{i = 1}^{x - 1} Y_i \cup Y_x'\right)$ are counted in $\dist(\m{C}, \m{F}^*)$ because the pairs $(u, v)$ lie in different clusters in $\m{C}$ but in the same cluster $F \in \m{F}^*$. These pairs are not counted in $\dist(\m{C}, \m{M})$ because they lie in different clusters both in $\m{C}$ and $\m{M}$. 
        \end{itemize}
    \end{proof}

    \begin{claim}\label{clm:np-small-claim-two}
        Consider any integer $p > 1$. If there exists a cluster $B_i$ such that $|B_i \cap F| \geq p$ for some $F \in \m{F}^*$, then $|\blue{F} \setminus B_i| \leq p - (|B_i \cap F| \mod p)$.
    \end{claim}

    \begin{proof}
        Let $Y_i = B_i \cap F$. Suppose for the sake of contradiction, $|\blue{F} \setminus B_i| > p - (|Y_i| \mod p)$.

        In that case, we construct $\m{M}$ from $\m{F}^*$ in the following way:

        \[
            \m{M} = \m{F}^* \setminus \{ F\} \cup \{ (Y_i \cup W \cup Q), F \setminus (Y_i \cup W \cup Q) \}
        \]
        where $W \subseteq \blue{F} \setminus Y_i$ such that $|W| = p - (|Y_i| \mod p)$ and $Q \subseteq \red{F}$ such that $|Q| = |Y_i \cup W|/p$ (note, $|Y_i \cup W|$ is a multiple of $p$).

        That is, in $\m{M}$ we remove the set $F$ from $\m{F}^*$ and add two clusters $(Y_i \cup W \cup Q)$ and $F \setminus (Y_i \cup W \cup Q)$. By the construction, $\m{M}$ is a {\fair}.

        Observe that $|Y_i|=|B_i \cap F| \ge p > |W|$. Then,

        \begin{align}
            \dist(\m{C}, \m{M}) &\le \dist(\m{C}, \m{F}^*) + |W| |\blue{F} \setminus (Y_i \cup W)| + |Q| |\red{F} \setminus Q| \n \\
            &- |Y_i||\blue{F} \setminus (Y_i \cup W)| - |Q| |\blue{F} \setminus (Y_i \cup W)| \n \\
            &\leq \dist(\m{C}, \m{F}^*) + |W| |\blue{F} \setminus (Y_i \cup W)| + |Q| |\red{F} \setminus Q|\n\\
            &- |Y_i||\blue{F} \setminus (Y_i \cup W)| - |Q| \cdot p |\red{F} \setminus Q| \n \\
            &< \dist(\m{C}, \m{F}^*)  \, \, (\text{since} \, \, p > 1 \, \, \text{and} \, \, |Y_i| > |W|)\n
        \end{align}
        contradicting the fact that $\m{F}^*$ is a closest {\fair} to $\m{C}$.

        \begin{itemize}
            \item \textbf{Reason behind the term $|W| |\blue{F} \setminus (Y_i \cup W)|$}: The pairs $(u,v)$ such that $u \in W$ and $v \in \blue{F} \setminus (Y_i \cup W)$ may not be counted in $\dist(\m{C}, \m{F}^*)$ because they may lie in same clusters both in $\m{C}$ and $\m{F}^*$. They are counted in $\dist(\m{C}, \m{M})$ because they lie in different clusters in $\m{M}$.
            \item \textbf{Reason behind the term $|Q| |\red{F} \setminus Q|$}: The pairs $(u,v)$ such that $u \in Q$ and $v \in \red{F} \setminus Q$ may not be counted in $\dist(\m{C}, \m{F}^*)$ because they may lie in same clusters both in $\m{C}$ and $\m{F}^*$. They are counted in $\dist(\m{C}, \m{M})$ because they lie in different clusters in $\m{M}$.
            \item \textbf{Reason behind the term $|Y_i||\blue{F} \setminus (Y_i \cup W)|$}: The pairs $(u,v)$ such that $u \in Y_i$ and $v \in \blue{F} \setminus (Y_i \cup W)$ are counted in $\dist(\m{C}, \m{F}^*)$ because they lie in different clusters in $\m{C}$ but the same cluster $F \in \m{F}^*$. They are not counted in $\dist(\m{C}, \m{M})$ because they lie in different clusters both in $\m{C}$ and $\m{M}$.
            \item \textbf{Reason behind the term $|Q| |\blue{F} \setminus (Y_i \cup W)| $}: The pairs $(u,v)$ such that $u \in Q$ and $v \in \blue{F} \setminus (Y_i \cup W)$ are counted in $\dist(\m{C}, \m{F}^*)$ because they lie in different clusters in $\m{C}$ but in the same cluster $F \in \m{F}^*$. They are not counted in $\dist(\m{C}, \m{M})$ because they lie in different clusters both in $\m{C}$ and $\m{M}$.
        \end{itemize}
    \end{proof}

    We are now ready to prove that for all $B_i$, for all $k \in [t]$, $|Y^k_i|$ must be a multiple of $p$.

    \begin{claim}\label{clm:np-all-partition-mop}
        Consider any integer $p > 1$. For all $B_i$, for all $k \in [t]$, $|Y^{k}_i|$ is a multiple of $p$.
    \end{claim}

    \begin{proof}
        Suppose there exists $m_1,\ldots,m_z$ such that $|Y_i^{m_w}|$ is not a multiple of $p$ for $w \in [z]$. In this case, we can construct a {\fair} $\m{M}$ such that $\dist(\m{C}, \m{M}) < \dist(\m{C}, \m{F}^*)$ which would contradict $\m{F}^*$ is a closest {\fair} to $\m{C}$. 
        
        Let, $Y_i^{m_w} \subseteq F_w \in \m{F}^*$ for all $w \in [z]$. WLOG, assume $|Y_i^{m_x}| \geq |Y_i^{m_y}|$ if $m_x < m_y$. Thus $Y_i^{m_1}$ is the partition of $B_i$ of largest size.  

        Before providing the construction of $\m{M}$ let us prove there exists a $B_i$ such that $|\blue{F_1}\setminus Y_i^{m_1}| \leq \left|\bigcup_{w=2}^z Y_i^{m_w} \right|$.

        \begin{itemize}
            \item \textbf{Case 1:} $\exists B_i$, $|Y_i^{m_1}| \geq p$: 
            
            In this case we know by \cref{clm:np-small-claim-two} that $|\blue{F_1} \setminus Y^{m_1}_i| \leq p - (|Y_i^{m_1}| \mod p)$.

        Further, note that $\left| \bigcup_{w=2}^z Y_i^{m_w} \right| \geq p - (|Y_i^{m_1}| \mod p)$. Hence, $|\blue{F_1}\setminus Y_i^{m_1}| \leq \left|\bigcup_{w=2}^z Y_i^{m_w} \right|$.

             \item \textbf{Case 2:} For all $B_j$, $|Y_j^{m_1}| < p$:

            Fix any cluster $B_i$. 

            Since for all $B_j$, $|Y_j^{m_1}| < p$, we get that for all $B_j$, $|B_j \cap F_1| < p$. Then by \cref{clm:np-small-claim-one}, we get $|\blue{F_1}| \leq p$.

            Thus, $|\blue{F_1} \setminus Y_i^{m_1}| \leq p - |Y^{m_1}_i|$.

            Further note that, $|\bigcup_{w = 2}^z Y_i^{m_w}| \geq p - |Y_i^{m_1}|$.

            Hence, $|\blue{F_1}\setminus Y_i^{m_1}| \leq \left|\bigcup_{w=2}^z Y_i^{m_w} \right|$.

        \end{itemize}    

        Next, we provide a construction of $\m{M}$ from $\m{F}^*$. Consider a cluster $B_i$ for which $|\blue{F_1}\setminus Y_i^{m_1}| \leq \left|\bigcup_{w=2}^z Y_i^{m_w} \right|$, the existence of which is already argued. In this case, informally, we want to construct a clustering $\m{M}$ by swapping the locations of blue vertices in $\blue{F_1}\setminus Y_i^{m_1}$ with an equal number of blue vertices present in $\bigcup_{w=2}^z Y_i^{m_w}$. We choose an equal number of blue vertices to $|\blue{F_1}\setminus Y_i^{m_1}|$ in $\bigcup_{w=2}^z Y_i^{m_w}$ greedily starting from $Y_i^{m_2}$.

        Formally, we do the following operation: First, we find an index $x$ such that

        \begin{align}
            \sum_{w = 2}^{x - 1} |Y_i^{m_w}| <|\blue{F_1} \setminus Y_i^{m_1}| \leq \sum_{w = 2}^x |Y_i^{m_w}|.\n 
        \end{align}

        Let, $Y_i^{m_x'} \subseteq Y_i^{m_x}$ such that
        \[
           |\blue{F_1} \setminus Y_i^{m_1}| = \sum_{w = 2}^{x-1}|Y_i^{m_w}| + |Y_i^{m_x'}|. \n
        \]

        Then, we divide the vertices in $\blue{F_1} \setminus Y_i^{m_1}$ into $(x - 1)$ disjoint subsets $Z_1, Z_2, \ldots, Z_{x-2},Z_{x-1}$ such that for $1 \leq j \leq (x-2)$ we have $|Z_j| = |Y_i^{m_{j + 1}}|$ and $|Z_{x-1}| = |Y^{m_x'}_i|$.

        \textbf{Construction of $\m{M}$ from $\m{F}^*$}:

        \begin{align}
        \m{M} = &\m{F}^* \setminus \{ F_w \mid 1 \leq w \leq x\} \cup \{ (F_w \setminus Y_i^{m_w}) \cup Z_{w - 1} \mid 2 \leq w \leq (x - 1)\} \n \\
        &\cup \{ (F_x \setminus Y^{m_x'}_i) \cup Z_{x - 1} \} \cup \left\{ F_1 \setminus (\blue{F_1} \setminus Y_i^{m_1}) \cup \left(\bigcup_{w = 2}^{x -1}Y_i^{m_w} \right) \cup Y_i^{m_x'}\right\} \n
        \end{align}

        That is, in $\m{M}$, from each $F_w$ for $2 \leq w \leq x-1 $ we remove its subset $Y^{m_w}_i$ and add the set $Z_{w-1}$ to it. From $F_x$ we remove its subset $Y_i^{m_x'}$ and add the set $Z_{x-1}$ to it. From $F_1$ we remove its subset $(\blue{F_1} \setminus Y^{m_1}_1)$ and add the set $\left(\bigcup_{w = 2}^{x -1}Y_i^{m_w} \right) \cup Y_i^{m_x'}$ to it. It is easy to observe that $\m{M}$ is a {\fair}.

        Now,

        \begin{align}
            \dist(\m{C},\m{M}) &\le \dist(\m{C},\m{F}^*) + |Z_{x - 1}||Y_i^{m_x} \setminus Y_i^{m_x'}| + |Y_i^{m_x'}||Y_i^{m_x} \setminus Y_i^{m_x'}|\n\\
            &+ \left(\sum_{r<w \in [x-2]}|Z_r| |Z_w|  + |Z_{x-1}| \sum_{w \in [x-2]}|Z_w|\right) \n \\
            &- |Z_{x - 1}||Y^{m_1}_i| - |Y_i^{m_x'}||Y^{m_1}_i| - \left( \sum_{r < w \in \{2,\cdots,x - 1\}} |Y_i^{m_r}| |Y_i^{m_w}| + |Y_i^{m_x'}| \sum_{w \in \{2,\cdots,x - 1\}} |Y_i^{m_w}|\right) \n \\
             &< \dist(\m{C}, \m{F}^*) \n 
             \end{align}
             where the last inequality follows since

             \begin{align*}
                 &|Y^{m_1}_i| > (|Y^{m_x}_i \setminus Y^{m_x'}_i|,\; \text{ and}\\
                 &\sum_{r<w \in [x - 2]}|Z_r| |Z_w|  + |Z_{x-1}| \sum_{w \in [x-2]}|Z_w|  =  \sum_{r < w \in \{2,\cdots,x - 1\}} |Y_i^{m_r}| |Y_i^{m_w}| + |Y_i^{m_x'}| \sum_{w \in \{2,\cdots,x - 1\}} |Y_i^{m_w}|.
             \end{align*}
        This contradicts the fact that $\m{F}^*$ is a closest {\fair} to $\m{C}$.

        \begin{itemize} 
            \item \textbf{Reason behind the term $|Z_{x - 1}||Y_i^{m_x} \setminus Y_i^{m_x'}|$:} In $\dist(\m{C}, \m{F}^*)$ the pairs $(u,v)$ such that $u \in Z_{x - 1}$ and $v \in Y_i^{m_x}\setminus Y_i^{m_x'}$ are not counted because they are in different clusters both in $\m{C}$ and $\m{F}^*$ but these pairs are counted in $\dist(\m{C}, \m{M})$ because they are in the same cluster $\{ (F_x \setminus Y^{m_x'}_i) \cup Z_{x - 1} \}$ in $\m{M}$.

            \item \textbf{Reason behind the term $|Y_i^{m_x'}||Y_i^{m_x} \setminus Y_i^{m_x'}|$:} In $\dist(\m{C}, \m{F}^*)$ the pairs $(u,v)$ such that $u \in Y_i^{m_x'}$ and $v \in Y_i^{m_x} \setminus Y_i^{m_x'}$ are not counted because they are in same clusters both in $\m{C}$ and $\m{F}^*$ but these pairs are counted in $\dist(\m{C}, \m{M})$ because they are in different clusters in $\m{M}$.

            \item \textbf{Reason behind the term $\left(\sum_{r<w \in [x - 2]}|Z_r| |Z_w|  + |Z_{x-1}| \sum_{w \in [x-2]}|Z_w|\right)$:} In $\dist(\m{C}, \m{F}^*)$ the pairs $(u,v)$ such that $u \in Z_r$ and $v \in Z_w$ for $r < w$ may not be counted because they may be in same clusters both in $\m{C}$ and $\m{F}^*$ but these pairs are counted in $\dist(\m{C}, \m{M})$ because they are in different clusters in $\m{M}$.

            \item \textbf{Reason behind the term $|Z_{x - 1}||Y^{m_1}_i|$:} In $\dist(\m{C}, \m{F}^*)$ the pairs $(u,v)$ such that $u \in Z_{x - 1}$ and $v \in Y_i^{m_1}$ are counted because they are in different clusters in $\m{C}$ but in the same cluster $F_1$ in $\m{F}^*$ but these pairs are not counted in $\dist(\m{C}, \m{M})$ because they are in different clusters both in $\m{C}$ and $\m{M}$.

            \item \textbf{Reason behind the term $|Y_i^{m_x'}||Y^{m_1}_i|$:} In $\dist(\m{C}, \m{F}^*)$ the pairs $(u,v)$ such that $u \in Y_i^{m_x'}$ and $v \in Y_i^{m_1}$ are counted because they are in different clusters in $\m{F}^*$ but in the same cluster $B_i$ in $\m{C}$ but these pairs are not counted in $\dist(\m{C}, \m{M})$ because they are in the same clusters both in $\m{C}$ and $\m{M}$.

            \item \textbf{Reason behind the term $\left( \sum_{r < w \in \{2,\cdots,x - 1\}} |Y_i^{m_r}| |Y_i^{m_w}| + |Y_i^{m_x'}| \sum_{w \in \{2,\cdots,x - 1\}} |Y_i^{m_w}|\right)$:} In $\dist(\m{C}, \m{F}^*)$ the pairs $(u,v)$ such that $u \in Y_i^{m_r}$ and $v \in Y_i^{m_w}$ for $r < w$ where $r,w \in \{ 2,\ldots, (x-1) \}$ or $u \in Y_i^{m_x'}$ and $v \in Y_i^{m_w}$ where $w \in \{ 2,\ldots, x -1 \}$ are counted because they are in the same cluster $B_i$ in $\m{C}$ but different clusters in $\m{F}^*$. These pairs are not counted in $\dist(\m{C}, \m{M})$ because they are in the same cluster 
            in $\m{M}$.
        \end{itemize}
        
    \end{proof}

    Next, we complete the proof of \cref{clm:np-main-claim}.

    \begin{proof}[Proof of \cref{clm:np-main-claim}]
        Suppose a cluster $B_i$ is split into $t$ parts $Y_i^1, Y_i^2, \ldots, Y_i^t$ in $\m{F}^*$. In this claim, we would like to show that $t = 1$. For the sake of contradiction, assume $t \geq 2$.  By \cref{clm:np-all-partition-mop} we get $p$ divides $|Y_i^k|$, for all $k \in [t]$. 
        
        For each $k \in [t]$, let $Y_i^{k} \subseteq F_{a_k}$ for some $F_{a_k} \in \m{F}^*$. By \cref{clm:claim-np}, we know that $\blue{F_{a_k}} = Y_i^k$. Next, we argue that since $t \geq 2$, we can construct a {\fair} $\m{M}$ from $\m{F}^*$ such that $\dist(\m{C}, \m{M}) < \dist(\m{C}, \m{F}^*)$ contradicting the fact that $\m{F}^*$ is a closest {\fair} to $\m{C}$.

        \textbf{Construction of $\m{M}$:}

        \[
            \m{M} = \m{F}^* \setminus \{ F_{a_k} \in \m{F}^* \mid  k \in [t]\} \cup \left\{ \bigcup_{k \in [t]} F_{a_k} \right\}.
        \]

    That is, while constructing $\m{M}$, we remove all the clusters $F_a$ from $\m{F}^*$ if $Y_i^{k} \subseteq F_a$ for some $k \in [t]$ and add one cluster by merging all these clusters. By the construction, clearly $\m{M}$ is a {\fair}.

    Next, we argue that $\dist(\m{C}, \m{M}) < \dist(\m{C}, \m{F}^*)$ for $ p\ge 3$. 

    \begin{align}
        \dist(\m{C}, \m{M}) &\le \dist(\m{C}, \m{F}^*) + \sum_{k \neq \ell} |\blue{F_{a_k}}| |\red{F_{a_\ell}}| + \sum_{k < \ell} |\red{F_{a_k}}||\red{F_{a_\ell}}| - \sum_{k < \ell} |\blue{F_{a_k}}||\blue{F_{a_\ell}}|\n \\
        &= \dist(\m{C}, \m{F}^*) + \sum_{k \ne \ell} \frac{|Y_i^k||Y_i^\ell|}{p} + \sum_{k < \ell} \frac{|Y_i^k|}{p}\frac{|Y_i^\ell|}{p} - \sum_{k < \ell} |Y_i^k||Y_i^\ell| \n \\
        &\left(\text{as }\forall_k \, \, \blue{F_{a_k}} = Y_i^k, \text{by \cref{clm:claim-np}, and } \, \, |\red{F_{a_k}}| = \frac{|Y_i^k|}{p}\right) \n \\
        &= \dist(\m{C}, \m{F}^*) + 2\sum_{k < \ell} \frac{|Y_i^k||Y_i^\ell|}{p} + \sum_{k < \ell} \frac{|Y_i^k|}{p}\frac{|Y_i^\ell|}{p} - \sum_{k < \ell} |Y_i^k||Y_i^\ell| \n \\
        &< \dist(\m{C}, \m{F}^*) \, \, \, (\text{for} \, \, p \geq 3) .\n
    \end{align}

    \begin{itemize}
        \item \textbf{Reason behind the term $\sum_{k \neq \ell} |\blue{F_{a_k}}| |\red{F_{a_\ell}}|$}: The pairs $(u,v)$ such that $u \in \blue{F_{a_k}}$ and $v \in \red{F_{a_\ell}}$ (for $k \neq \ell$) may not be counted in $\dist(\m{C}, \m{F}^*)$ because $u$ and $v$ may lie in different clusters both in $\m{C}$ and $\m{F}^*$. These pairs are counted in $\dist(\m{C}, \m{M})$ because they lie in the same cluster in $\m{M}$.
        \item \textbf{Reason behind the term $\sum_{k < \ell} |\red{F_{a_k}}||\red{F_{a_\ell}}|$}: The pairs $(u,v)$ such that $u \in \red{F_{a_k}}$ and $v \in \red{F_{a_\ell}}$ (for $k < \ell$) may not be counted in $\dist(\m{C}, \m{F}^*)$ because $u$ and $v$ may lie in different clusters both in $\m{C}$ and $\m{F}^*$. These pairs are counted in $\dist(\m{C}, \m{M})$ because they lie in the same cluster in $\m{M}$.
        \item \textbf{Reason behind the term $\sum_{k < \ell} |\blue{F_{a_k}}||\blue{F_{a_\ell}}|$}: The pairs $(u,v)$ such that $u \in \blue{F_{a_k}}$ and $v \in \blue{F_{a_\ell}}$ (for $k < \ell$) are counted in $\dist(\m{C}, \m{F}^*)$ because $u$ and $v$ lie in the same cluster $B_i \in \m{C}$ but different clusters in $\m{F}^*$. These pairs are not counted in $\dist(\m{C}, \m{M})$ because they lie in the same cluster in $\m{M}$.
    \end{itemize}

\end{proof}

We now conclude the proof of~\cref{lem:main-one}.

\begin{proof}[Proof of \cref{lem:main-one}]
    By~\cref{lem:yes-instance} and~\cref{lem:no-instance} we infer that for any $p \geq 3$, $S $ is a YES instance of the $ \thrp$ if and only if $(\m{C}, \tau) $ is a YES instance of the $ \clof$. It now immediately follows from our reduction that for any $p \geq 3$, $\clof$ is \textbf{NP}-hard because $\thrp$ is strongly $\npc$ (\cref{thm:garey-johnson}).
\end{proof}

\begin{remark}\label{remark:np-hard}
    We would like to remark that the above \textbf{NP}-hardness proof can be extended for any fixed $p/q$ ratio on the input blue to red vertices as long as $p/q > 1+\sqrt{2}$. More specifically, consider any two integers $p,q \ge 2$ such that $p/q > 1+\sqrt{2}$. Then the closest fair clustering problem on red-blue colored vertices where the ratio between the total number of blue and red vertices is $p/q$, is \textbf{NP}-hard. We can show this result by making the following modifications to the reduction: Take monochromatic red clusters $R_i$'s of size $q x_i$, set $\tau = \frac{q^2}{2} \sum_{j = 1}^n x_j \left(T - x_j\right) + \frac{n}{3}pqT^2$; everything else remain the same. The proof argument will be similar. One important vertex to note is that $p/q > 1+\sqrt{2}$ is crucial for the argument used in the proof of \cref{clm:np-main-claim}.
\end{remark}

Now in the following paragraph, we prove that $\clof$ is \textbf{NP}-hard also for $p = 2$.

\subsection{Hardness of $\clof$ for $p = 2$:}

In this section, we prove the following lemma.

\begin{lemma}\label{lem:main-two}
    For $p = 2$, the $\clof$ problem is \textbf{NP}-hard.
\end{lemma}

We again reduce the $\thrp$ problem to the $\clof$ problem for the case when $p = 2$. The reduction follows the same structure as in the case where $p \geq 3$, with the only difference being the choice of the parameter $\tau$. For completeness, we restate the reduction here.

\paragraph{Reduction from $\thrp$ to $\clof$:}  
Consider a $\thrp$ instance $S = \{ x_1, x_2, \ldots, x_n \}$, where for all $i \in [n]$, it holds that $x_i \le n^c$ for some fixed non-negative constant $c$. Given this instance, we construct a corresponding $\clof$ instance $(\mathcal{C}, \tau)$ as follows:

\begin{itemize}
    \item Define the set of clusters as  
    \[
        \mathcal{C} = \{ B_1, \ldots, B_{n/3}, R_1, \ldots, R_n \},
    \]
    where each $B_i$ (for $i \in \{1, \ldots, n/3\}$) is a \emph{monochromatic blue} cluster of size $pT$, and each $R_j$ (for $j \in \{1, \ldots, n\}$) is a \emph{monochromatic red} cluster of size $x_j$, i.e., $|R_j| = x_j$.

    \item Set the value of $\tau$ as  
    \[
        \tau = p\sum_{\ell = 1}^n x_\ell^2 + \frac{p^2}{2} \sum_{\ell = 1}^n x_\ell(T - x_\ell)
    \]
\end{itemize}

Since, for all $i \in [n]$, it holds that $x_i \le n^c$ for some fixed non-negative constant $c$, it is straightforward to see that the above reduction is a polynomial time reduction.

Now, we argue that the above reduction maps a YES instance of the $\thrp$ to a YES instance of the $\clof$.

\begin{lemma}\label{lem:yes-instance-p=2}
    If $S$ is a YES instance of the $\thrp$, then $(\m{C}, \tau) $ is also a YES instance of the $ \clof$.
\end{lemma}

\begin{proof}
    Suppose $S$ is a YES instance of the $\thrp$. Then there exists a partition $S_1, S_2, \ldots, S_{n/3} \, \, \text{of} \, \, S = \{ x_1, x_2, \ldots, x_n\}$ such that for all $1\le i \le n/3$, $|S_i| = 3$ and
    \[
        \sum_{x_\ell \in S_i}x_\ell = T \, \, \text{where} \, \, T = \frac{\sum_{x_\ell \in S} x_\ell}{\frac{n}{3}}.
    \]
    Let, $S_i = \{ x_{i_1}, x_{i_2}, x_{i_3}\}$. By our construction of $(\m{C}, \tau)$, we have $|R_{i_\ell}| = x_{i_\ell}$, for $\ell \in \{1,2,3\}$.
   
    Now, we construct a fair clustering $\m{F}$ by merging $R_{i_\ell}$ with $p|R_{i_\ell}|$ blue vertices in $B_i$ for $\ell \in \{ 1, 2, 3\}$. More formally,
    
    \[
       \m{F} = \left\{ \left(B_{i_\ell} \cup R_{i_\ell} \right)  \, \,  \middle| \, \, i \in [n/3], \ell \in \{1, 2, 3\} \right \} .
    \]
    where $B_{i_\ell} \subseteq B_i$ such that $|B_{i_\ell}| = p|R_{i_\ell}|$ for $\ell \in \{1, 2, 3\}$.
    
    Note that $\m{F}$ is a $\fair$ because for each cluster $F \in \m{F}$ we have $|\blue{F}| = p|\red{F}|$.
    
    Next, we claim that
    \[
        \dist(\m{C}, \m{F}) = p\sum_{\ell = 1}^n x_\ell^2 + \frac{p^2}{2} \sum_{\ell = 1}^n x_\ell(T - x_\ell) = \tau.
    \]

    \begin{itemize}
        \item \textbf{Reason behind the term $p\sum_{\ell = 1}^n x_\ell^2 $}: In $\m{F}$, each red cluster $R_\ell$ is merged with $p|R_\ell|$ blue vertices from some cluster $B_i$. This expression counts the number of such blue-red pairs (recall $|R_\ell| = x_\ell)$. (One important thing to note is that since $S$ is a YES instance of the $\thrp$, we do not need to cut any red cluster $R_\ell$ in $\m{C}$ to form the {\fair} $\m{F}$.)
        
        \item \textbf{Reason behind the term $\frac{p^2}{2} \sum_{\ell = 1}^n x_\ell(T - x_\ell)$}: In $\m{F}$, each red cluster $R_\ell$ is merged with $p|R_\ell|$ blue vertices from some cluster $B_i$. Now, these $p|R_\ell|$ blue vertices of $B_i$ gets separated from other $pT - p|R_\ell|$ blue vertices of $B_i$. This expression counts the number of such blue-blue pairs. In the above expression, $1/2$ comes from the fact that we are counting a pair twice, once while considering the subset of $B_i$ that consists of $p|R_\ell|$ vertices and again while considering a subset of other $(pT - p|R_\ell|)$ blue vertices of $B_i$.
    \end{itemize}
    
    Hence, $(\m{C},\tau) $ is a YES instance of the $ \clof$.
\end{proof}

Now, we argue that our reduction maps a NO instance of the $\thrp$ to a NO instance of the $\clof$.

\begin{lemma}\label{lem:no-instance-p=2}
 For $p = 2$, if $S $ is a NO instance of the $\thrp$, then $(\m{C}, \tau) $ is also a NO instance of the $ \clof$.
\end{lemma}

\begin{proof}
    Let $\m{F}^*$ be any arbitrary closest clustering to $\m{C}$. We prove that $\dist(\m{C}, \m{F}^*) > \tau$.

    By the definition, $\dist(\m{C}, \m{F}^*)$ counts the total number of pairs $(u,v)$ that are together in $\m{C}$ but separated by $\m{F}^*$ and the number of pairs $(u,v)$ that are in different clusters in $\m{C}$ but together in $\m{F}^*$.

    To count a pair $(u,v)$, we use a charging scheme. For each pair $(u,v)$ we charge the vertex $u$, $1/2$ and the vertex $v$, $1/2$.

    Let us define $\costf{v}$ to be half times the number of vertices $u$ such that $u,v$ are together in $\m{C}$ but separated by $\m{F}^*$, or $u,v$ that are in different clusters in $\m{C}$ but together in $\m{F}^*$. More specifically, 

    \begin{align}
        \costf{v} := \frac{1}{2} | \{ u \mid &u,v \in C, C \in \m{C}\, \, \text{but}\, \, u \in F_a, v\in F_b, F_a, F_b \in \m{F}^*, a \neq b \n \\
        & \text{or} \, \, u,v \in F_a, F_a \in \m{F}^*, u \in C_i, v \in C_j, C_i, C_j \in \m{C}, i \neq j \} | \n 
    \end{align}

    For a cluster $F_a \in \m{F}^*$, let us define $\costf{F_a}$ (by abuse of notation) to be 

    \[
        \costf{F_a} := \sum_{v \in F_a} \costf{v}.
    \]
        
    Observe that 

    \begin{align}
        \costf{F_a} = &p\left( \sum_{\ell = 1}^n |R_{a,\ell}|\right)^2 + \frac{p^2}{2} \left( \sum_{\ell = 1}^n |R_{a,\ell}|\right) \left( T - \sum_{\ell = 1}^n |R_{a,\ell}|\right) \n \\
        &+ \frac{1}{2}\sum_{\ell_1 \neq \ell_2}|R_{a, \ell_1}||R_{a, \ell_2}| + \frac{1}{2}\sum_{\ell = 1}^n|R_{a, \ell}| \left(|R_{\ell}| -|R_{a, \ell}| \right) \label{equn:np-p=2-zero}
    \end{align}
    where $R_{a,\ell} = R_\ell \cap F_a$.

    \begin{itemize}
        \item \textbf{Reason behind the term $p\left( \sum_{\ell = 1}^n |R_{a,\ell}|\right)^2$}: $\left( \sum_{\ell = 1}^n |R_{a,\ell}|\right)$ is the number of red vertices in $F_a$. Since it is merged with $p\left( \sum_{\ell = 1}^n |R_{a,\ell}|\right)$ blue vertices, hence the number of blue-red pairs in $F_a$ is $p\left( \sum_{\ell = 1}^n |R_{a,\ell}|\right)^2$.
        \item \textbf{Reason behind the term $\frac{p^2}{2} \left( \sum_{\ell = 1}^n |R_{a,\ell}|\right) \left( T - \sum_{\ell = 1}^n |R_{a,\ell}|\right)$}: By \cref{clm:claim-np} and \cref{clm:np-all-partition-mop} we get that for all $F_a \in \m{F}^*$, $\blue{F_a} \subseteq B_i$ for some $B_i \in \m{C}$. Hence, the vertices in $\blue{F_a}$ are separated from $pT - |\blue{F_a}|$ vertices in $B_i$. These blue-blue pairs are counted in this expression
        \begin{align*}
            &\frac{1}{2} \left( p\sum_{\ell = 1}^n |R_{a,\ell}|\right) \left( pT - p\sum_{\ell = 1}^n |R_{a,\ell}|\right) \n \\
            = &\frac{p^2}{2} \left( \sum_{\ell = 1}^n |R_{a,\ell}|\right) \left( T - \sum_{\ell = 1}^n |R_{a,\ell}|\right)
        \end{align*}
        In the above expression, $1/2$ comes from the fact that for a pair $(u,v)$ we are charging $1/2$ for the vertices $u \in F_a$ while considering the cluster $F_a$ and again we will charge $1/2$ for the vertex $v$ while considering a cluster $F_b$ such that $v \in \blue{F_b}$, $\blue{F_b} \subseteq B_i$ and $F_b \neq F_a$.
        \item \textbf{Reason behind the term $\frac{1}{2}\sum_{\ell_1 \neq \ell_2}|R_{a, \ell_1}||R_{a, \ell_2}|$}: This counts the number of red-red pairs $(u,v)$ in $F_a$ such that $u \in R_{\ell_1}$ and $v \in R_{\ell_2}$.
        \item \textbf{Reason behind the term $\frac{1}{2}\sum_{\ell = 1}^n|R_{a, \ell}| \left(|R_{\ell}| -|R_{a, \ell}| \right)$}: The term counts the number of pairs $(u,v) \in R_\ell$ such that $u \in R_\ell \cap F_a$ and $v \in R_\ell \cap F_b$ for $F_b \neq F_a$. In the above expression, $1/2$ comes from the fact that for a pair $(u,v)$ we are charging $1/2$ for the vertices $u \in R_\ell \cap F_a$ while considering the cluster $F_a$ and again we will charge $1/2$ for the vertex $v$ while considering a cluster $F_b$ such that $v \in R_\ell \cap F_b$.
    \end{itemize}

    It is straightforward to observe

    \[
        \dist(\m{C}, \m{F}^*) \geq \sum_{F_a \in \m{F}^*} \costf{F_a}.
    \]

    Now we show that 

    \[
        \sum_{F_a \in \m{F}^*} \costf{F_a} > \tau.
    \]

    Let, $x_{a,\ell} = |R_{a,\ell}|$ and recall $x_\ell = |R_\ell|$.

    \begin{align}
        &\sum_{F_a \in \m{F}^*} \costf{F_a} \n \\
        = \, \, &p \sum_{F_a \in \m{F}^*} \left( \sum_{\ell = 1}^n x_{a,\ell} \right)^2 + \frac{p^2}{2} \sum_{F_a \in \m{F}^*}\left( \sum_{\ell = 1}^n x_{a,\ell}\left( T - \sum_{\ell = 1}^n x_{a,\ell}\right) \right) + \frac{1}{2}\sum_{F_a \in \m{F}^*}\sum_{\ell_1 \neq \ell_2} x_{a,\ell_1} x_{a,\ell_2} \n \\
        &+ \frac{1}{2}\sum_{F_a \in \m{F}^*}\sum_{\ell = 1}^n x_\ell (x_\ell - x_{a,\ell}) \n \\
        = \, \, &p \sum_{F_a \in \m{F}^*} \left( \sum_{\ell = 1}^n x_{a,\ell}^2 + \sum_{\ell_1 \neq \ell_2}x_{a,\ell_1} x_{a,\ell_2}\right) + \frac{p^2}{2} \sum_{F_a \in \m{F}^*}\left( \sum_{\ell = 1}^n x_{a,\ell}\left( T - \sum_{\ell = 1}^n x_{a,\ell}\right) \right) + \frac{1}{2}\sum_{F_a \in \m{F}^*}\sum_{\ell_1 \neq \ell_2} x_{a,\ell_1} x_{a,\ell_2} \n \\
        &+ \frac{1}{2}\sum_{\ell = 1}^n \sum_{F_{a_1} \neq F_{a_2}} x_{a_1, \ell}x_{a_2, \ell} \n \\
        = \, \, &p \sum_{F_a \in \m{F}^*} \sum_{\ell = 1}^n x_{a,\ell}^2 
         + \frac{p^2}{2} \sum_{F_a \in \m{F}^*}\left( \sum_{\ell = 1}^n x_{a,\ell}\left( T - x_{a,\ell}\right) \right) + \frac{1}{2}\sum_{\ell = 1}^n \sum_{F_{a_1} \neq F_{a_2}} x_{a_1, \ell}x_{a_2, \ell} \n \\
         &- \frac{p^2}{2} \sum_{F_a \in \m{F}^*}\sum_{\ell_1 \neq \ell_2} x_{a,\ell_1} x_{a,\ell_2} + \left(p + \frac{1}{2} \right)\sum_{F_a \in \m{F}^*}\sum_{\ell_1 \neq \ell_2} x_{a,\ell_1} x_{a,\ell_2} \n \\
         \geq \, \,  &p \sum_{F_a \in \m{F}^*} \sum_{\ell = 1}^n x_{a,\ell}^2 + \frac{p^2}{2} \sum_{F_a \in \m{F}^*}\left( \sum_{\ell = 1}^n x_{a,\ell}\left( T - x_{a,\ell}\right) \right) + \frac{1}{2}\sum_{\ell = 1}^n \sum_{F_{a_1} \neq F_{a_2}} x_{a_1, \ell}x_{a_2, \ell} \n \\ &(\text{as} \, \, \, p = 2) \label{equn:np-p=2-one} \\
         = \, \, &p \sum_{F_a \in \m{F}^*} \sum_{\ell = 1}^n x_{a,\ell}^2 + \frac{p^2}{2} \sum_{F_a \in \m{F}^*}\left( \sum_{\ell = 1}^n x_{a,\ell}\left( T - x_{\ell} + (x_{\ell} - x_{a, \ell})\right) \right) + \frac{1}{2}\sum_{\ell = 1}^n \sum_{F_{a_1} \neq F_{a_2}} x_{a_1, \ell}x_{a_2, \ell} \n \\
        = \, \, &p \sum_{F_a \in \m{F}^*} \sum_{\ell = 1}^n x_{a,\ell}^2 + \frac{p^2}{2} \sum_{F_a \in \m{F}^*}\sum_{\ell = 1}^n x_{a,\ell} (x_{\ell} - x_{a, \ell}) + \frac{p^2}{2} \sum_{F_a \in \m{F}^*} \sum_{\ell = 1}^n x_{a,\ell}\left( T - x_{\ell}\right) \n \\
        &+ \frac{1}{2}\sum_{\ell = 1}^n \sum_{F_{a_1} \neq F_{a_2}} x_{a_1, \ell}x_{a_2, \ell} \n \\
        = \, \, &p\sum_{\ell = 1}^n \left( \sum_{F_a \in \m{F}^*}  x_{a,\ell}^2 + \frac{p}{2}\sum_{F_a \in \m{F}^*} x_{a, \ell}(x_\ell - x_{a,\ell}) + \frac{1}{2p} \sum_{F_{a_1} \neq F_{a_2}} x_{a_1, \ell}x_{a_2, \ell}  \right) + \frac{p^2}{2} \sum_{\ell = 1}^n \left( T - x_{\ell}\right) \sum_{F_a \in \m{F}^*} x_{a,\ell}\n \\ 
        = \, \, &p\sum_{\ell = 1}^n \left( \sum_{F_a \in \m{F}^*} x_{a,\ell}^2 + \left(\frac{p}{2} + \frac{1}{2p} \right) \sum_{F_{a_1} \neq F_{a_2}} x_{a_1,\ell} x_{a_2, \ell}\right) + \frac{p^2}{2} \sum_{\ell = 1}^n x_\ell (T - x_\ell) \n \\
        > \, \, &p\sum_{\ell = 1}^n \left( \sum_{F_a \in \m{F}^*} x_{a,\ell}^2 +  \sum_{F_{a_1} \neq F_{a_2}} x_{a_1,\ell} x_{a_2, \ell}\right) + \frac{p^2}{2} \sum_{\ell = 1}^n x_\ell (T - x_\ell) \label{equn:np-p=2-two} \\
       = \, \, &p\sum_{\ell = 1}^n x_\ell^2 + \frac{p^2}{2} \sum_{\ell = 1}^n x_\ell (T - x_\ell) = \tau \n
    \end{align}

    The reason behind \cref{equn:np-p=2-two} is that $(p/2 + 1/2p) > 1$ for $p > 1$ and there must exist at least one $\ell$ and $F_{a_1}$, $F_{a_2} \in \m{F}^*$ such that $x_{a_1,\ell}, x_{a_2, \ell} \neq 0$. We prove this in the following claim.

    \begin{claim}
        There exists $\ell, a_1, a_2$ such that $x_{a_1,\ell}, x_{a_2, \ell} \neq 0$.
    \end{claim}

    \begin{proof}
        For the sake of contradiction, suppose for all $\ell$, there exists at most one $F_a \in \m{F}^*$ such that $x_{a,\ell} \neq 0$. 
        
        Since $\sum_{F_a \in \m{F}^*} x_{a,\ell} = x_\ell$, we have a cluster $F_a \in \m{F}^*$ such that $x_{a,\ell} = x_\ell$.
       
        Now we create partitions $S_i$ of $S$ in the following way: For each $B_i \in \m{C}$,

        \[
            S_i = \{ x_\ell \mid x_{a,\ell} = x_\ell \, \, \text{and} \, \, \blue{F_a} \subseteq B_i \}.
        \]

        Note by \cref{clm:claim-np} and \cref{clm:np-all-partition-mop} we get for each $F_a \in \m{F}^*$ there must exist a $B_i$ such that $\blue{F_a} \subseteq B_i$.

        Since there are $n/3$ clusters $B_i \in \m{C}$, the number of such subsets would be $n/3$. Clearly, the above subsets $S_i$'s create a partitioning of $S$, that is

        \begin{itemize}
            \item $S_a \cap S_b = \emptyset$ for $a \neq b$ (since, $B_i \cap B_j = \emptyset$, hence, if $\blue{F_a}\subseteq B_i$ then $\blue{F_a}\not\subseteq B_j$ for $j \neq i$);
            \item $\bigcup S_a = S$ (since $x_{a,\ell}$ cannot be zero for all $a$).  
        \end{itemize}

        Since $|B_i| = pT$ and all clusters $F_a \in \m{F}^*$ such that $\blue{F_a} \subseteq B_i$ are fair we have,

        \[
            \sum_{x_\ell \in S_i}x_\ell = T.
        \]

        Further, since each $x_\ell \in (T/4,T/2)$, we must have $|S_i| = 3$. However, since $S$ is a NO instance of $\thrp$, such a partitioning of $S$ does not exist, leading to a contradiction.

        Hence, we conclude that there must exist $\ell, a_1, a_2$ such that $x_{a_1,\ell}, x_{a_2, \ell} \neq 0$.
    \end{proof}

    Thus we finally get that $\dist(\m{C}, \m{F}^*) > \tau$, which implies $(\m{C},\tau) $ is a NO instance of the $ \clof$, for $p = 2$.
\end{proof}

Similar to the proof of \cref{lem:main-one}, we conclude the proof of \cref{lem:main-two}.

\begin{proof}[Proof of \cref{lem:main-two}]
    By~\cref{lem:yes-instance-p=2} and~\cref{lem:no-instance-p=2} we infer that for any $p = 2$, $S $ is a YES instance of the $ \thrp$ if and only if $(\m{C}, \tau) $ is a YES instance of the $ \clof$. It now immediately follows from our reduction that for any $p = 2$, $\clof$ is \textbf{NP}-hard because $\thrp$ is strongly $\npc$ (~\cref{thm:garey-johnson}).
\end{proof}

\begin{remark}\label{remark:np-hard-one}
    We would like to remark that the above \textbf{NP}-hardness proof can be extended for any fixed ratio (between the number of blue and red vertices) of $p/q$ where $1 < p/q \leq (1 + \sqrt{2})$ with modifications in the reduction mentioned in \cref{remark:np-hard}, with $\tau$ being set to $\tau = pq\sum_{\ell = 1}^n x_\ell^2 + \frac{p^2}{2} \sum_{\ell = 1}^n x_\ell(T - x_\ell)$. The reason behind this is that \cref{equn:np-p=2-one} holds for any ratio $ p\in [1 - \sqrt{2}, 1 + \sqrt{2}]$, and while expressing $\costf{F_a}$ in \cref{equn:np-p=2-zero} we use \cref{clm:claim-np} and \cref{clm:np-all-partition-mop}, both of which also hold for any $p, q > 1$.

\end{remark}

From \cref{lem:main-one} and \cref{lem:main-two} we conclude \cref{thm:main}.

\chapter{Closest Fair Clustering with more than Two Colors}\label{sec:closest-fair-with-more-than-two-colors}

In this chapter, we find an approximate closest fair clustering to given clustering $\m{C}$, when the vertex set $V$ is divided into multiple colored disjoint groups $\chi = \{c_1, \ldots, c_d\}$. We use the notation $\sigma(v)$ to denote the color assigned to $v$ from $\chi$.

\section{Approximate Closest Fair Clustering for Equi-Proportion Groups}
\label{sec1}

In this section, we provide an approximation algorithm to find a closest {\fair} when all the groups are of equal size. More specifically we prove \cref{thm:closest-fair-multicolor-balanced}, which we restate below.

\equalmulti*

First, let us handle the case when $|\chi|$ is a power of $2$. To do that, we provide an algorithm $\fptwo$, which produces an $O(|\chi|^{1.6})$-close fair clustering $\outfptwo$ to a clustering $\m{C}$ when $|\chi|$ is a power of $2$.

\subsection{Details of the Algorithm: Power-of-Two Colors} 

Let the input of the algorithm $\fptwo$ be a clustering $\m{C} = \{C_1, C_2, \ldots, C_m\}$, where each cluster $C_a \in \m{C}$ consists of points colored from a color set $\chi = \{c_1, c_2, \ldots, c_d\}$. We assume that the number of colors $|\chi|$ is a power of two.

The goal is to produce a fair clustering $\outfptwo$ such that, in each cluster $F_a \in \outfptwo$, the number of points of each color is equal, that is $c_p(F_a) = c_q(F_a)$ where $p \neq q$ and $c_r(F_a)$ denotes the number of points of color $c_r$ in $F_a$.

The algorithm proceeds in $\log |\chi|$ iterations, indexed by $i = 1$ to $\log |\chi|$. In each iteration, the color set $\chi = \{c_1, \ldots, c_{|\chi|}\}$ is partitioned into blocks of size $2^i$:
\[
B_j^i = \{c_{(j-1)\cdot 2^i + 1}, \ldots, c_{j \cdot 2^i}\}, \quad \text{for } j = 1, \ldots, |\chi|/2^i.
\]

\noindent (Note: This defines $|\chi| / 2^i$ blocks in total.)

Let, after iteration $i$ of the algorithm $\fptwo$ the input clustering $\m{C}$ is changed to a clustering $\m{N}^{i}$. The algorithm $\fptwo$ at each iteration \(i\) maintains an invariant that for every block \(B_j^i\) and for every cluster \(N^{i}_a \in \m{N}^{i}\), the number of points of each color within \(B_j^i\) is equal in \(N^{i}_a\). Formally, for all \(x, y \in \{1, \ldots, 2^i\}\) with \(x \ne y\), we have:
\[
c_{(j-1)\cdot 2^i + x}(N^{i}_a) = c_{(j-1)\cdot 2^i + y}(N^{i}_a),
\]
where \(c_r(N^{i}_a)\) denotes the number of points of color \(c_r\) in cluster \(N^{i}_a\).

Hence, after $\log |\chi|$ iterations when there is only one block, we have that for the clustering $\m{N}^{\log |\chi|} = \outfptwo$ in each cluster $F_a \in \outfptwo$, the number of points of each color is equal. 

Let us now describe the algorithm $\fptwo$ which maintains the above invariant. To describe the algorithm first we need to define some notations.

\begin{itemize}
    \item As described previously the clustering $\m{C}$ changes at each iteration $i$ of the algorithm $\fptwo$.

    Let, $\m{N}^i = \{ N_1^i, N_2^i, \ldots, N_z^i \}$ be the clustering formed at iteration $i$.

    Specifically, the clustering $\m{C}$ goes through the following transformations.

    \[
        \m{C} \rightarrow \m{N}^1 \rightarrow \m{N}^2 \rightarrow \cdots \rightarrow \m{N}^{\log |\chi|} = \outfptwo.
    \]

    \item $B_j^i(N^i_a)$: For any cluster $N^i_{a} \in \m{N}^i$, we have $B_j^i(N^i_a)$ denotes the set of points present in $N^i_a$ having color in the block $B_j^i$.

    More Specifically,

    \[
        B_j^i(N^i_a) = \{ v \in N^i_a \mid \sigma(v) \in B_j^i\}
    \]

    \item Let $T^j_a$ denote the surplus between $B_j^i$ and $B_{j+1}^i$ within cluster $N_a^i$. If $|B_j^i(N_a^i)| \geq |B_{j+1}^i(N_a^i)|$, then the surplus is a subset of $B_j^i(N_a^i)$ of size $|B_j^i(N_a^i)| - |B_{j+1}^i(N_a^i)|$ such that all colors $c_m, c_\ell \in B_j^i$ satisfy $c_m(T^j_a) = c_\ell(T^j_a)$ for all $c_m \neq c_\ell$. Otherwise, the surplus is defined analogously from $B_{j+1}^i(N_a^i)$.

\end{itemize}

Now we are ready to describe the algorithm $\fptwo$

\paragraph{Description of $\fptwo$:}

The algorithm proceeds in $\log |\chi|$ iterative phases to enforce fairness across pairs of buckets. The high-level structure of the algorithm is as follows:

\begin{enumerate}
    \item \textbf{Initialization:} Set the initial clustering $\m{N}^0 = \{N_1^0, N_2^0, \ldots, N_f^0\} = \mathcal{C}$ (input of $\fptwo$).
    
    \item \textbf{Iterative Refinement:} For each phase $i = 1$ to $\log |\chi|$:
    \begin{enumerate}
        \item Initialize $\m{N}^i$ as a copy of $\m{N}^{i-1}$.
        \item Iterate over $j = 1$ to $|\chi| / 2^i$ in steps of $2^i$:
        \begin{enumerate}
            \item For each cluster $N^i_a \in \m{N}^i$:
            \begin{itemize}
                \item Compute the surplus between the bucketed subsets $B_j^i(N^i_a)$ and $B_{j+1}^i(N^i_a)^i$, defined as $T^j_a$.
                \item Remove this surplus from $N^i_a$.
                \item Store the removed surplus in set $S_j$ if $|B_j^i(N^i_a)| \geq |B_{j+1}(N^i_a)^i|$, else store it in $S_{j+1}$.
            \end{itemize}
            \item Merge the stored surpluses using the subroutine \texttt{multi-GM}$(S_j, S_{j+1})$ to form new fair clusters.
            \item Add the resulting clusters of the multi-GM step back to $\m{N}^i$.
        \end{enumerate}
    \end{enumerate}
    
    \item \textbf{Output:} Return the final fair clustering $\m{N}^{\log |\chi|}$.
\end{enumerate}

\medskip

\textbf{Remarks.} The algorithm ensures local fairness between adjacent buckets $B_j^i$ and $B_{j+1}^i$ by removing and redistributing surplus elements. As the phases progress, the granularity of this balancing increases, resulting in a globally fair clustering. The \texttt{multi-GM} subroutine plays a key role in recombining the surplus elements in a way that preserves fairness across buckets.

\medskip

Let us now describe the $\greedymerge$ algorithm.

\paragraph{Cescription of $\greedymerge$:}

\texttt{multi-gm} is a procedure that takes as input two collections of point sets:
\begin{itemize}
    \item \texttt{Set1}: a set of point subsets whose elements have colors from a color block $B_j^i$,
    \item \texttt{Set2}: a set of point subsets whose elements have colors from the adjacent color block $B_{j+1}^i$.
\end{itemize}
Each block $B_j^i$ and $B_{j+1}^i$ contains the same number of colors. The goal is to iteratively pair subsets from \texttt{Set1} and \texttt{Set2} to form new subsets that are locally fair, meaning each resulting subset has an equal number of points from each color in the union of $B_j^i \cup B_{j+1}^i$. These fair subsets are returned as a new collection.

The algorithm proceeds greedily by:
\begin{itemize}
    \item Iteratively selecting one subset from each of \texttt{Set1} and \texttt{Set2}.
    \item Comparing their sizes and extracting a balanced subset from the larger set to match the size of the smaller one.
    \item Merging the two into a fair subset and adding it to the output collection.
    \item Updating or removing the subsets used and repeating this process until no more valid pairings are possible.
\end{itemize}

The final output is a collection of fair subsets, each containing an equal number of points of every color from $B_j^i \cup B_{j+1}^i$. We provide the pseudocode of the algorithms $\fptwo$ and $\greedymerge$ in \cref{algo:fair-power-of-two} and \cref{algo:greedymerge} respectively.

\subsection{Approximation Guarantee of $\fptwo$}

In this section, we analyze the algorithm $\fptwo$ by establishing \cref{lem:fair-power-of-two} stated below,

\begin{lemma}\label{lem:fair-power-of-two}
    Given a clustering $\m{C}$ as input, the algorithm $\fptwo$ computes a $O(|\chi|^{1.6})$-close $\fair$, where $|\chi|$ is a power of $2$.
\end{lemma}

\begin{proof}
    To prove this lemma, we need to define the following term
    \begin{itemize}
        \item $i$th fairness constraint of algorithm $\fptwo$: A clustering $\gencls$ is said to satisfy the $i$th fairness constraint of algorithm $\fptwo$ if, for every cluster $\gencl \in \gencls$ and for all pairs of distinct colors $c_k \ne c_\ell$ in the set $B^i_j$, the number of points of color $c_k$ in $\gencl$ equals the number of points of color $c_\ell$ in $\gencl$; that is,
        \[
            c_k(\gencl) = c_\ell(\gencl).
        \]
    \end{itemize}

    Now we prove the following claim
    \begin{claim}\label{clm:main}
        In the algorithm $\fptwo$, the clustering $\m{N}^i$ is $2$-close clustering to $\m{N}^{i-1}$ that satisfies the $i$th fairness constraint of $\fptwo$, where $\m{N}^0 = \m{C}$.
    \end{claim}

    We will prove \cref{clm:main} later. First, let us prove \cref{lem:fair-power-of-two} assuming \cref{clm:main}.

     To prove \cref{lem:fair-power-of-two}, we need to prove $\m{N}^{\log |\chi|}$, the output of the algorithm $\fptwo$ is $O(\card{\chi}^{1.6})$-close to $\m{C}$. Let, $\m{N}^*$ be the closest fair clustering to $\m{C}$.

    We will prove using mathematical induction that after any iteration $i$ of the algorithm $\fptwo$ we have 

    \begin{align}
        \dist(\m{C}, \m{N}^i) \leq (3^i - 1) \dist(\m{C}, \m{N}^*) \label{equn:analysis-fptwo-one}
    \end{align}

    For the base case, that is when $i = 1$ by \cref{clm:main} we have $\m{N}^1$ is a $2$-close clustering to $\m{C}$ that satisfies the $1$st fairness constraint of algorithm $\fptwo$. Since $\m{N}^*$ also satisfies the $1$st fairness constraint we have
    \[
        \dist(\m{C}, \m{N}^1) \leq 2 \dist(\m{C}, \m{N}^*) 
    \]

    Now, let us assume \cref{equn:analysis-fptwo-one} is true for $i = k-1$, that is
    \[
        \dist(\m{C}, \m{N}^{k - 1}) \leq (3^{k - 1} - 1) \dist(\m{C}, \m{N}^*)
    \]

    Now, we prove it for $i = k$.

    \begin{align}
        \dist(\m{N}^{k - 1}, \m{N}^k) &\leq 2 \dist(\m{N}^{k - 1}, \m{N}^*) \label{equn:analysis-fptwo-two} \\
        &\leq 2(\dist(\m{N}^{k - 1}, \m{C}) + \dist(\m{C}, \m{N}^*)) \n \\ &\text{(triangle inequality)} \nonumber \\
        &\leq 2((3^{k - 1} - 1)\dist(\m{C}, \m{N}^*) + \dist(\m{C}, \m{N}^*)) \n \\ &\text{(by IH)} \nonumber \\
        &= 2 \cdot 3^{k - 1} \dist(\m{C}, \m{N}^*) \label{equn:analysis-fptwo-two-(i)}
    \end{align}

    Here, \cref{equn:analysis-fptwo-two} is true because $\m{N}^*$ also satisfies the $i$th fairness constraint of the algorithm $\fptwo$ and due to \cref{clm:main}.

    Now, we have
    \begin{align}
        \dist(\m{C}, \m{N}^k) &\leq \dist(\m{C}, \m{N}^{k - 1}) + \dist(\m{N}^{k - 1}, \m{N}^k) \n \\
        &\text{(triangle inequality)} \nonumber \\
        &\leq (3^{k - 1} - 1) \dist(\m{C}, \m{N}^*) + 2 \cdot 3^{k - 1} \dist(\m{C}, \m{N}^*) \nonumber \\
        &\text{(by IH and \cref{equn:analysis-fptwo-two-(i)})} \nonumber \\
        &= (3^k - 1) \dist(\m{C}, \m{N}^*) \label{equn:analysis-fptwo-three}
    \end{align}

    Hence, now we can conclude for $i = \log |\chi|$, 
    \begin{align}
        \dist(\m{C}, \m{N}^{\log |\chi|}) &\leq (3^{\log |\chi|} - 1) \dist(\m{C}, \m{N}^*) \n \\
        &= O(|\chi|^{1.6}) \dist(\m{C}, \m{N}^*) \, \, (\log_23 = 1.6)
    \end{align}

    Thus the algorithm $\fptwo$ computes a $O(\chi^{1.6})$-close $\fair$ which completes the proof of \cref{lem:fair-power-of-two}.

    Now, let us prove \cref{clm:main}. To prove \cref{clm:main}, we establish two intermediate claims: \cref{clm:lower-bound-opt} and \cref{clm:upper-bound-algo}.

    Let $\m{N}^{i^*}$ denote the closest clustering to $\m{N}^{i - 1}$ that satisfies the $i$th fairness constraint of algorithm $\fptwo$.

    In \cref{clm:lower-bound-opt}, we derive a lower bound on the distance $\dist(\m{N}^{i - 1}, \m{N}^{i^*})$, and in \cref{clm:upper-bound-algo}, we provide an upper bound on the distance $\dist(\m{N}^{i - 1}, \m{N}^i)$, where $\m{N}^i$ is the clustering obtained after the $i$th iteration of algorithm $\fptwo$, starting from the initial clustering $\m{C}$.

    By comparing the bounds from \cref{clm:lower-bound-opt} and \cref{clm:upper-bound-algo}, we will complete the proof of \cref{clm:main}.

    To formalize \cref{clm:lower-bound-opt}, lets recall and introduce some notations:

    \begin{itemize}
        \item Recall, $B^{i - 1}_k(N_a^{i - 1})$ denotes the set of data points in the cluster $N_a^{i - 1} \in \m{N}^{i - 1}$ whose colors belong to the block $B^{i - 1}_k$.
    
        \item Recall, $\Sn$ denotes the surplus between $B_{2k - 1}^{i-1}$ and $B_{2k}^{i - 1}$ in cluster $N_a^{i-1}$. Specifically, 
        \begin{itemize}
            \item If $|B_{2k - 1}^{i-1}(N_a^{i-1})| \geq |B_{2k}^{i}(N_a^{i-1})|$, then
            \[
                \Sn \subseteq B_{2k - 1}^{i-1}(N_a^{i-1}), \quad |\Sn| = |B_{2k - 1}^{i-1}(N_a^{i-1})| - |B_{2k}^{i - 1}(N_a^{i-1})|,
            \]
            such that $\forall c_j \neq c_\ell \in B_{2k}^{i-1}$, we have $c_j(\Sn) = c_\ell(\Sn)$.
            \item Otherwise,
            \[
                \Sn \subseteq B_{j+1}^{i - 1}(N_a^{i-1}), \quad |\Sn| = |B_{j+1}^{i - 1}(N_a^{i-1})| - |B_j^{i-1}(N_a^{i-1})|,
            \]
            such that $\forall c_k \neq c_\ell \in B_{j+1}^{i - 1}$, we have $c_k(\Sn) = c_\ell(\Sn)$.
        \end{itemize}
        \item Let, $T_a$ denotes the union of the surpluses within the cluster $N_a^{i - 1}$, computed across all paired color blocks $(B_{2k - 1}^{i - 1}, B_{2k}^{i - 1})$ at iteration $i - 1$. More specifically,
        \begin{align*}
            T_a &= \bigcup_{k = 1}^{|\chi| / 2^{i}} \Sn
        \end{align*}
    \end{itemize}

\begin{claim}\label{clm:lower-bound-opt}
\begin{align}
\dist(\m{N}^{i - 1}, \m{N}^{i^*}) \geq 
\frac{1}{2}\sum_{N_a^{i - 1} \in \m{N}^{i - 1}}  
&|T_a| \cdot \left(\left|N_a^{i - 1}\right| - \left| T_a \right| \right) +  |T_a|^2 \nonumber
\end{align}
\end{claim}

\begin{proof}
    Consider a cluster $N_a^{i - 1} \in \m{N}^{i - 1}$. Suppose in $\m{N}^{i^*}$ the cluster $N_a^{i - 1}$ is partitioned into $X_1, X_2, \ldots, X_t$, more specifically, 

    \begin{itemize}
        \item For all $j \in [t]$, $X_j \subseteq N_{r_j}^{i^*}$ for some $N_{r_j}^{i^*} \in \m{N}^{i^*}$.
        \item For all $j \neq \ell \in [t]$, we have $N_{r_j}^{i^*} \neq N_{r_\ell}^{i^*}$.
        \item $\bigcup_{j \in [t]} X_j = N_a^{i - 1}$.
    \end{itemize}

    By abuse of notation, let us define $B^{i - 1}_{2k - 1}(X_j)$ and $B^{i - 1}_{2k}(X_j)$ be the set of points in $X_j$ that has a color from the blocks $B^{i - 1}_{2k - 1}$ and $B^{i - 1}_{2k}$ respectively. 

    WLOG assume, $|B^{i - 1}_{2k - 1}(N_a^{i - 1})| > |B^{i - 1}_{2k}(N_a^{i - 1})|$.

    Recall, the blocks are created in such a way that for two consecutive blocks $B^{i - 1}_{2k - 1}$ and $B^{i - 1}_{2k}$, we have 
    \[
        |B^{i - 1}_{2k - 1}| = |B^{i - 1}_{2k}|.
    \]

    Let's create arbitrary pairing of colors $(c,\hat{c})$ where $c \in B^{i - 1}_{2k - 1}$ and $\hat{c} \in B^{i - 1}_{2k}$, we call $\hat{c} \in B^{i - 1}_{2k}$ a color corresponding to $c \in B^{i - 1}_{2k - 1}$.

    \begin{itemize}
        \item \textbf{Surplus between two colors $c \in B^{i - 1}_{2k - 1}$ and $\hat{c} \in B^{i - 1}_{2k - 1}$ in a cluster $N_a^{i - 1}$} is defined as
        \[
            s^a(c,\hat{c}) \subseteq B^{i - 1}_{2k - 1}(N_a^{i - 1})
        \]
        of size $\max(0, c(N_a^{i - 1}) - \hat{c}(N_a^{i - 1}))$
        \item \textbf{Surplus between two colors $c \in B^{i - 1}_{2k - 1}$ and $\hat{c} \in B^{i - 1}_{2k}$ for a partition $X_j$} is defined as
        \[
            \sigma^j_c \subseteq B^{i - 1}_{2k - 1}(X_j)
        \]
        of size $\max(0, c(X_j) - \hat{c}(X_j))$
        It is straightforward to see that 
        \[
            \sum_{j = 1}^t |\sigma^j_c| \geq |s^a(c,\hat{c})|
        \]

        Let, us assume $y \in [t]$ be an index such that
    \[
        \sum_{j = 1}^{y - 1}|\sigma^j_c| < |s^a(c,\hat{c})| \leq \sum_{j = 1}^y |\sigma^j_c|
    \]

    Again assume, 
    \[
    \hat{\sigma^y_c} \subseteq \sigma^y_c
    \]

    such that,

    \[
        \sum_{j = 1}^{y - 1}|\sigma^j_c| + |\hat{\sigma^y_c}| = |s^a(c,\hat{c})|
    \]

    Let us now redefine the notation $\sigma^j_c$ for $j \in [t]$

    \[
    \sigma^j_c := 
        \begin{cases}
        \sigma^j_c & \text{if } j < y, \\
        \hat{\sigma^y_c} & \text{if } j = y, \\
        \emptyset & \text{if } j > y.
        \end{cases}
    \]

    That is, the sets $\sigma^j_c$ from $j = 1$ to $(y - 1)$ remains unchanged. We shrink the set $\sigma^y_c$ to include only as many elements as needed to make the summation $|s^a(c,\hat{c})|$. We ignore the sets $\sigma^j_c$ from $j = (y + 1)$ to $t$ completely. 

        \item \textbf{Surplus with respect to consecutive pair of blocks $B^{i - 1}_{2k - 1}$ and $B^{i - 1}_{2k}$ for a partition $X_j$} is defined as
        \[
            \Sx = \bigcup_{c \in B^{i - 1}_{2k - 1}} \sigma^j_c
        \]
        \item \textbf{Surplus with respect to a partition $X_j$} is defined as
        \[
            S_j = \bigcup_{k = 1}^{m_{i - 1}} S^k_j
        \]
        where $m_{i - 1}$ is the number of blocks created at iteration $(i - 1)$.
    \end{itemize}
    It is straightforward to see that
    \[
        \sum_{j = 1}^t |S_j| = T_a
    \]

    Now, since $\m{N}^{i^*}$ satisfies $i$th fairness constraint of $\fptwo$, hence in $N_{r_j}^{i^*} \in \m{N}^{i^*}$ we have,
    
    \[
        B_{2k - 1}^{i - 1}(N_{r_j}^{i^*}) = B_{2k}^{i - 1}(N_{r_j}^{i^*}). 
    \]

    Recall, $X_j \subseteq N_{r_j}^{i^*}$, hence there must exist at least $|\Sx|$ vertices having colors from the color block $B^{i - 1}_{2k}$ in $N_{r_j}^{i^*}$ that belongs to clusters other than $N_a^{i - 1}$. Let us denote this set of vertices by $M^k_j$. More specifically,

    \begin{itemize}
        \item $M^k_j \subseteq N_{r_j}^{i^*}$ such that following conditions are satisfied.
        \begin{enumerate}
            \item $M^k_j \cap N_a^{i - 1} = \emptyset$.
            \item $|M^k_j| = |\Sx|$.
            \item The vertices in $M^k_j$ have colors from the color block $B^{i - 1}_{2k}$.
        \end{enumerate}
    \end{itemize}

    Let,
    \[
        M_j = \bigcup_{k = 1}^{m_{i - 1}} M^k_j
    \]
    Note, since $|M^k_j| = |\Sx|$ we also have $|M_j| = |S_j|$.

    Let us also define
    \[
        M(N_a^{i - 1}) = \bigcup_{j = 1}^t M_j
    \]

    By the definition of $\dist(\m{N}^{i - 1}, \m{N}^{i^*})$ is the number of pairs $(u,v)$ that are separated in $\m{N}^{i - 1}$ and together in $\m{N}^{i^*}$ or viceversa. Formally, we say that $(u,v)$ are together in $\m{N}^{i^*}$ if there exists $N_k^{i^*} \in \m{N}^{i^*}$ such that $u,v \in N_k^{i^*}$ and we say $(u,v)$ are separated in $\m{N}^{i^*}$ if there exists $N_k^{i^*}, N_j^{i^*} \in \m{N}^{i^*}$ such that $k \neq j$, $u \in N_k^{i^*}$ and $v \in N_j^{i^*}$

    Now, to lower bound $\dist(\m{N}^{i - 1}, \m{N}^{i^*})$ let us define some costs.
    \begin{itemize}
        \item $\mcostone(N_a^{i - 1}):$ For a cluster $N_a^{i - 1} \in \m{N}^{i - 1}$, $\mcostone(N_a^{i - 1})$ denotes the number of pairs $(u,v)$ such that 
        \begin{enumerate}[(a)]
            \item $u \in (N_a^{i - 1} \setminus T_a)$ and $v \in T_a$ but separated in $\m{N}^{i^*}$  or,
            \item $u \in (N_a^{i - 1} \setminus T_a$ and $v \in M(N^{i - 1}_a)$.
        \end{enumerate}
        
        \item $\mcosttwo(N_a^{i - 1}):$ For a cluster $N_a^{i - 1} \in \m{N}^{i - 1}$, $\mcosttwo(N_a^{i - 1})$ denotes the number of pairs $(u,v)$ such that
        \begin{enumerate}[(a)]
            \item $u, v \in T_a$ but separated in $\m{N}^{i^*}$  or,
            \item $u \in T_a$ and $v \in M(N_a^{i - 1})$.
        \end{enumerate}
    \end{itemize}

    We can verify that the pairs counted in $\mcostone(N_a^{i - 1}, k)$, and $\mcosttwo(N_a^{i - 1}, k)$ are disjoint and thus we have.

    \begin{align}
        \dist(\m{N}^{i - 1}, \m{N}^{i^*}) \geq &\sum_{N_a^{i - 1} \in \m{N}^{i - 1}} \frac{1}{2} \mcostone(N_a^{i - 1}) \n \\ &+ \frac{1}{2}\mcosttwo(N_a^{i - 1}) \label{equ:clm-lower-bound-opt}
    \end{align}

    We multiply $\mcostone(N_a^{i - 1})$ and $\mcosttwo(N_a^{i - 1})$ with $1/2$ to avoid overcounting of pairs. In both the costs we count the pairs $(u,v)$ in $N_a^{i - 1}$ and $M(N_a^{i - 1})$. This pair $(u,v)$ may be counted twice because we take summation overall $N_a^{i - 1} \in \m{N}^{i - 1}$.

    Now, to prove \cref{clm:lower-bound-opt}, we prove the following

    \begin{enumerate}
        \item $\mcostone(N_a^{i - 1}) \geq |T_a| \left(\left|N_a^{i - 1}\right| - \left| T_a \right| \right)$.
        \item $\mcosttwo(N_a^{i - 1}) \geq \left|T_a \right|^2$.
    \end{enumerate}

    It is easy to see that combining \cref{equ:clm-lower-bound-opt} and the above statements will prove this \cref{clm:lower-bound-opt}. So, let us now prove the above statements.

    \begin{claim}
        $\mcostone(N_a^{i - 1}) \geq |T_a| \left(\left|N_a^{i - 1}\right| - |T_a| \right)$
    \end{claim}

    \begin{proof}
        Recall in $N^{i^*}$, the cluster $N_a^{i - 1}$ is partitioned into $X_1, \ldots, X_t$.

        Now, consider a partition $X_j$ of $N_a^{i-1}$ where $j \in [t]$. Let us count the number of pairs $(u,v)$ such that $u \in X_j \setminus S_j$ and $v \in S_\ell$ for $\ell \neq j$. The number of such pairs is

        \begin{align}
	   \left|X_j \setminus S_j\right| \sum_{\ell \neq j}|S_\ell| \label{eq:cost-one-eq-one}
        \end{align}

        Let us also count the number of pairs $(u,v)$ such that $ u \in X_j \setminus S_j$ and $ v \in M_j$. The number of such pairs is 

        \begin{align}
            &|X_j \setminus S_j| |M_j| \n \\
            = &|X_j \setminus S_j| |S_j| \label{eq:cost-one-eq-two} 
        \end{align}

        Now, combining \cref{eq:cost-one-eq-one} and \cref{eq:cost-one-eq-two} we get the number of pairs $(u,v)$ such that,

        \[
            (u,v) \in X_j \setminus S_j \times S_\ell \, \, \text{for $\ell \neq j$}
        \]

        or

        \[
            (u,v) \in X_j \setminus S_j \times M_j
        \]

        is

        \begin{align}
            &\left|X_j \setminus S_j\right| \sum_{\ell \neq j}|S_\ell| \n \\
            + &|X_j \setminus S_j| |S_\ell| \n \\
            = &\left|X_j \setminus S_j\right| |T_a| \label{eq:cost-one-eq-three}
        \end{align}

        Now from \cref{eq:cost-one-eq-three} we get,

        \begin{align}
            \mcostone(N_a^{i - 1}, k) &\geq \sum_{j = 1}^t \left|X_j \setminus S_j\right| |T_a| \n \\
            &= |T_a| \sum_{j = 1}^t \left|X_j \setminus S_j\right| \n \\
            &= \left|T_a\right| \left(\left|N_a^{i - 1}\right| - |T_a|\right) \n
        \end{align}
 
    \end{proof}
    
    \begin{claim}
        $\mcosttwo(N_a^{i - 1}) \geq \left|T_a \right|^2$
    \end{claim}

    \begin{proof}
        Consider a partition $X_j$ of $N_a^{i-1}$ where $j \in [t]$. Let us count the number of pairs $(u,v)$ such that $u \in S_j$ and $v \in S_\ell$ for $\ell \neq j$. The number of such pairs is

        \begin{align}
	   \left|S_j\right| \sum_{\ell \neq j}|S_\ell| \label{eq:cost-two-eq-one}
        \end{align}

        Let us also count the number of pairs $(u,v)$ such that $ u \in S_j$ and $ v \in M_j$. The number of such pairs is 

        \begin{align}
            &|S_j| |M_j| \n \\
            = &|S_j| |S_j| \label{eq:cost-two-eq-two} 
        \end{align}

        Now, combining \cref{eq:cost-two-eq-one} and \cref{eq:cost-two-eq-two} we get the number of pairs $(u,v)$ such that,

        \[
            (u,v) \in S_j \times S_\ell \, \, \text{for $\ell \neq j$}
        \]

        or

        \[
            (u,v) \in S_j \times M_j
        \]

        is

        \begin{align}
            &\left|S_j\right| \sum_{\ell \neq j}|S_\ell|
            + |S_j| |S_j| \n \\
            = &\left|S_j\right| |T_a| \label{eq:cost-two-eq-three}
        \end{align}

        Now from \cref{eq:cost-two-eq-three} we get,

        \begin{align}
            \mcosttwo(N_a^{i - 1}) &\geq \sum_{j = 1}^t \left|S_j\right| |T_a| \n \\
            &= |T_a| \sum_{j = 1}^t \left|S_j\right| \n \\
            &= \left|T_a\right| \left|T_a\right|\n \\
            &= \left|T_a\right|^2 \n
        \end{align}
    \end{proof}
    
\end{proof}

\begin{claim}\label{clm:upper-bound-algo}
    \begin{align}
        \dist(\m{N}^{i - 1}, \m{N}^i) \leq &\sum_{N_a^{i - 1} \in \m{N}^{i - 1}} \left|T_a\right| \left(\left|N_a^{i - 1}\right| - \left| T_a \right| \right) \n \\
        &+ \frac{1}{2} \left|T_a \right|^2 \n
    \end{align}
\end{claim}

\begin{proof}
    In the algorithm $\fptwo$, from each cluster $N_a^{i - 1} \in \m{N}^{i - 1}$ we cut the set $T_a$.

    Hence, the pairs $(u, v)$ s.t. $u \in N_a^{i - 1} \setminus T_a$ and $v \in T_a$ are counted in $\dist(\m{N}^{i - 1}, \m{N}^i)$. 

    The number of such pairs is
    \begin{align}
        \left|T_a\right| \left(\left|N_a^{i - 1}\right| - \left| T_a \right| \right) \label{eq:upper-bound-equn-one}
    \end{align}

    Again in~\cref{algo:greedymerge} the set $T_a$ can further get splitted into multiple subsets $R_1, R_2, \ldots, R_t$ (say). Each of these sets $R_i$ for $i \in [t]$ gets merged with $|R_i|$ points from a different cluster.

    Hence, the following pairs $(u,v)$ are counted in $\dist(\m{N}^{i - 1}, \m{N}^i)$ which satisfies 

    \begin{enumerate}
        \item $u \in R_i$ and $v$ belongs to the set that merged to $R_i$.
        \item $u \in R_i$ and $v \in R_j$ for $i \neq j$.
    \end{enumerate}

    To count such pairs $(u,v)$ we use a charging scheme, we charge $1/2$ for the vertex $u$ and $1/2$ for the vertex $v$. That is we define for a set $R_i$. 

    \begin{align}
        \pay{R_i} = \frac{1}{2} |\{ (u,v) \mid &u \in R_i \, \, \text{and $v \in$ the set merged to $R_i$} \n \\
        &\text{ or $u \in R_i$ and $v \in R_j$ for $i \neq j$}\} |\n
    \end{align}

    The total number of such pairs is
    \begin{align}
        \sum_{i = 1}^t \pay{R_i} &= \sum_{i = 1}^t \frac{1}{2} |R_i|^2 + \frac{1}{2} |R_i||(T_a \setminus R_i)| \n \\
        &= \frac{1}{2} |T_a| \sum_{i = 1}^t |R_i| \n \\
        &= \frac{1}{2} |T_a|^2 \label{eq:upper-bound-equn-two}
    \end{align}

    By \cref{eq:upper-bound-equn-one} and \cref{eq:upper-bound-equn-two} we get

    \begin{align}
        \dist(\m{N}^{i - 1}, \m{N}^i) \leq &\sum_{N_a^{i - 1} \in \m{N}^{i - 1}} \left|T_a\right| \left(\left|N_a^{i - 1}\right| - \left| T_a \right| \right) \n \\
        &+ \frac{1}{2} \left|T_a \right|^2 \n
    \end{align}
\end{proof}

It is straightforward to see that \cref{clm:lower-bound-opt} and \cref{clm:upper-bound-algo} proves \cref{clm:main}.

\end{proof}

\cref{lem:fair-power-of-two} proves \cref{thm:closest-fair-multicolor-balanced} when $|\chi|$ is a power of $2$. Now, we prove \cref{thm:closest-fair-multicolor-balanced} for any values of $|\chi|$. To do that, we need to describe the algorithm $\fequi$.

\subsection{Details of Algorithm: General Case}

To describe the algorithm $\fequi$ we need to use a subroutine $\fmulti$. In the subroutine $\fmulti$,
we are given a clustering \( \mathcal{I} = \{ I_1, I_2, \ldots, I_\eta \} \) over a point set \( V \). Each point \( v \in V \) is assigned a color from a finite set \( \zeta = \{ z_1, z_2, \ldots, z_r \} \). The total number of points of each color satisfies the global proportion:

\[
z_1(V) : z_2(V) : \ldots : z_r(V) = p_1 : p_2 : \ldots : p_r
\]

WLOG assume $p_1 > p_2 > \cdots > p_r$.

Furthermore, for each cluster \( I_i \in \mathcal{I} \), the number of points of each color \( z_j \) is a multiple of \( p_j \). Recall from \cref{sec:closest-fair-two-colors} such a clustering $\m{I}$ is called \emph{$p$-divisible clustering} (\texttt{pdc} for short).

The goal of $\fmulti$ is to construct a clustering \( \outmpf \) such that for each cluster \( F \in \outmpf \),
\[
z_1(F) : z_2(F) : \ldots : z_r(F) = p_1 : p_2 : \ldots : p_r
\]

\paragraph{Description of the subroutine $\fmulti$:}

The algorithm proceeds in \( \lceil \log_2 r \rceil \) iterations. At each iteration, we refine the current clustering to more closely satisfy the target color proportions. Let \( \mathcal{F}^0 := \mathcal{I} \) denote the initial clustering. The algorithm maintains a sequence of progressively refined clusterings:

\[
    \mathcal{I} = \mathcal{F}^0 \rightarrow \mathcal{F}^1 \rightarrow \mathcal{F}^2 \rightarrow \ldots \rightarrow \mathcal{F}^{\lceil \log_2 r \rceil} = \outmpf
\]

In each iteration \( t \), the algorithm enforces proportionality over small blocks of colors, where the block size increases with \( t \). As the iterations progress, these local balancing constraints compound, eventually ensuring that the final clustering \( \mathcal{F} \) satisfies the full global ratio \( p_1 : p_2 : \cdots : p_r \) within every cluster.

\paragraph{Description of the color blocks:}

We define a hierarchical merging of the color set \( \zeta = \{ z_1, z_2, \ldots, z_r \} \) over \( \lceil \log_2 r \rceil \) iterations. At each iteration \( t \), we maintain a partition of \( \zeta \) into blocks \( \{ B^t_1, B^t_2, \ldots, B^t_{m_t} \} \), where each block represents a group of colors whose proportions will be jointly enforced in that round.

\begin{itemize}
    \item \textbf{Iteration 0:} Initialize each color into its own singleton block:
    \[
        B^0_j = \{ z_j \} \quad \text{for } j = 1, 2, \ldots, r
    \]
    
    \item \textbf{Iteration \( t \geq 1 \):} Merge adjacent blocks from the previous iteration:
    \[
        B^t_i = B^{t-1}_{2i - 1} \cup B^{t-1}_{2i} \quad \text{for } i = 1, 2, \ldots, \left\lfloor \frac{m_{t-1}}{2} \right\rfloor
    \]
    \[
        \text{If } m_{t-1} \text{ is odd, then let } B^t_{\lceil m_{t-1}/2 \rceil} = B^{t-1}_{m_{t-1}} \text{ (unchanged)}
    \]
    
    \item Continue this process until the final iteration \( T = \lceil \log_2 r \rceil \), where a single block \( B^{\lceil \log_2 r \rceil}_1 = \zeta \) is formed.
\end{itemize}

At each level \( t \), we enforce the proportionality constraints jointly over the colors in each block \( B^t_i \) for all clusters in the current clustering \( \mathcal{F}^{t-1} \). This recursive construction guarantees that after \( \lceil \log_2 r \rceil \) rounds, the final clustering \( \mathcal{F}^{\lceil \log_2 r \rceil} \) satisfies the global target ratio \( p_1 : p_2 : \cdots : p_r \) in every cluster.

\paragraph{Balancing Rule:}

Consider a block \( B_i^t \) at iteration \( t \), defined as:

\[
    B_i^t = B_{2i - 1}^{t - 1} \cup B_{2i}^{t - 1}
\]

Let
\[
    B_{2i - 1}^{t - 1} = \{ z_{a_1}, z_{a_2}, \ldots, z_{a_s} \}, \quad 
    B_{2i}^{t - 1} = \{ z_{b_1}, z_{b_2}, \ldots, z_{b_u} \}
\]
where \( s \geq u \).

By construction, at iteration \( t-1 \), each cluster \( F^{t-1} \in \mathcal{F}^{t-1} \) satisfies:

\[
    z_{a_1}(F^{t-1}) : \ldots : z_{a_s}(F^{t-1}) = p_{a_1} : \ldots : p_{a_s}
\]
and
\[
    z_{b_1}(F^{t-1}) : \ldots : z_{b_u}(F^{t-1}) = p_{b_1} : \ldots : p_{b_u}
\]

Suppose the point counts in \( F^{t-1} \) satisfy:

\[
    z_{a_j}(F^{t-1}) = p_{a_j} \cdot x \quad \text{for } j \in [s]
\]
and
\[
    z_{b_k}(F^{t-1}) = p_{b_k} \cdot y 
\]
for $k \in [u]$.

Our goal is to construct a cluster \( F^t \in \mathcal{F}^t \) such that:

\begin{align}
    &z_{a_1}(F^t) : \ldots : z_{a_s}(F^t) : z_{b_1}(F^t) : \ldots : z_{b_u}(F^t) \n \\
    &= p_{a_1} : \ldots : p_{a_s} : p_{b_1} : \ldots : p_{b_u} \n
\end{align}

To achieve this,

\begin{itemize}
    \item If \( x > y \), merge \( p_{b_k} \cdot (x - y) \) points of color \( z_{b_k} \) into \( F^{t-1} \) for each \( k \in [u] \).
    \item If \( x < y \), cut \( p_{b_k} \cdot (y - x) \) points of color \( z_{b_k} \) from \( F^{t-1} \) for each \( k \in [u] \).
\end{itemize}

This ensures that the scaling factor for both sub-blocks becomes equal ($x$ in both the cases), aligning the entire block \( B_i^t \) to the correct global ratio structure. We provide the pseudocode of $\fmulti$ in \cref{algo:make-pdc-fair}.

Let us now describe the algorithm $\fequi$. The algorithm $\fequi$ takes a clustering $\m{C}$ as input. Each vertex of the clustering is colored from the colors $\chi = \{ c_1, c_2, \ldots, c_d\}$ where $|\chi|$ is not a power of two. 
    \paragraph{Description of the algorithm $\fequi$:} We divide $\chi$ into $\log |\chi|$ many disjoint color groups $G_1, G_2, \ldots, G_{\log |\chi|}$ such that the size of each group, $|G_j|$ where $j \in [\log |\chi|]$ is a power of two. We do it greedily according to the binary representation of $|\chi|$. Consider the binary representation of $|\chi|$. For each index $j \in [\log \lceil|\chi|\rceil]$, if the corresponding bit is 1, create a group of size $2^{j-1}$.

    We apply the algorithm $\fptwo$ to get an intermediate clustering $\m{I} = \{I_1, \ldots, I_\eta\}$ such that for each cluster $I_j \in \m{I}$ where $j \in [k]$ we have for any pair of colors $c_a, c_b \in G_\ell$ where $\ell \in [\log|\chi|]$, 
    
    \[c_a(I_j) = c_b(I_j)\]

    For a subset $S \subseteq V$, let us define $G_\ell(S)$ as the set of vertices $v \in S$ such that $v$ has a color in $G_\ell$. 
    Now, our goal is to get a clustering $\m{F} = \{ F_1, \ldots, F_\iota \}$ from $\m{C}$ such that for each cluster $F_k \in \m{F}$ we have $c_u(F_k) = c_v(F_k)$ where $c_u, c_v \in \chi$.

     Given the intermediate clustering \(\mathcal{I}\) as input, the algorithm \(\fmulti\) produces a final $\fair$ \(\mathcal{F} = \{ F_1, \ldots, F_{\iota} \}\) and thus for each cluster \( F_j \in \mathcal{F} \), the following proportion holds.

\begin{align*}
    &|G_1(F_j)| : |G_2(F_j)| : \ldots : |G_{\log |\chi|}(F_j)| \n \\
    &= |G_1| : |G_2| : \ldots : |G_{\log |\chi|}|
\end{align*}

To apply $\fmulti$, we consider a group of colors $G_\ell$ as a single metacolor $z_\ell$. Here the clustering $\m{I}$ is a $p$-divisible clustering because for any cluster $I_j \in \m{I}$ we have $|G_\ell(I_j)|$ is divisible by $|G_\ell|$.

Furthermore, the algorithm \(\fmulti\) ensures uniformity within each group: for every group \( G_{\ell} \) and for any pair of colors \( c_a, c_b \in G_{\ell} \) with \( \ell \in [\log |\chi|] \), it guarantees that \( c_a(F_j) = c_b(F_j) \). As a consequence, for any pair of colors \( c_u, c_v \in \chi \), we have \( c_u(F_j) = c_v(F_j) \), i.e., each color is equally represented within every cluster of \(\mathcal{F}\).

\subsection{Approximation Guarantee of $\fequi$}

To analyse $\fequi$, we first analyze the subroutine $\fmulti$ for which we prove the following lemma.


\begin{lemma} \label{lem:analyze-fmulti}
    The algorithm \(\fmulti\) outputs a clustering \(\mathcal{F}\) that is \( O( r^{2.81}) \)-close $\fair$ to the input clustering \(\mathcal{I} \), where $r$ is the number of colors.
\end{lemma}

\begin{proof}
    To prove this lemma, let us define the $t$th fairness constraint of $\fmulti$ algorithm.

    \begin{itemize}
        \item $t$th fairness constraint of $\fmulti$ : Let \( \mathcal{C} \) be a clustering and let \( \{ B_k^t \} \) denote the color blocks at the $t$th iteration in the $\fmulti$ algorithm. We say that \( \mathcal{C} \) satisfies the \( t \)th Fairness Constraint of $\fmulti$ if, for every cluster \( C_i \in \mathcal{C} \) and every block \( B_k^t \), the color counts satisfy:

\[
z_{a_1}(C_i) : z_{a_2}(C_i) : \cdots : z_{a_m}(C_i) = p_{a_1} : p_{a_2} : \cdots : p_{a_m}
\]
where $B_k^t = \{ z_{a_1}, \ldots, z_{a_m} \}$

    \end{itemize}

To prove \cref{lem:analyze-fmulti}, we need to prove the following claim

\begin{claim}\label{clm:each-step-4-approx}
    For all iterations $t$ of the algorithm $\fmulti$, $\m{F}^t$ is a $6$-close clustering to $\m{F}^{t - 1}$ that satisfies the $t$th fairness constraint of $\fmulti$. More specifically,
    \[
        \dist(\m{F}^{t - 1}, \m{F}^t) \leq 6 \dist(\m{F}^{t - 1}, \m{F}^{t^*}) 
    \]
    where $\m{F}^{t^*}$ is the closest clustering to $\m{F}^{t - 1}$ that satisfies $t$th fairness constraint of $\fmulti$.
\end{claim}

For now, let us assume \cref{clm:each-step-4-approx} and prove \cref{lem:analyze-fmulti}.

\paragraph{Proof of \cref{lem:analyze-fmulti}:} To prove \cref{lem:analyze-fmulti}, we need to prove $\m{F}^{\log r}$, the output of the algorithm $\fmulti$ is $r^{2.8}$-close to $\m{I}$. Let $\m{F}^*$ be the closest fair clustering to $\m{I}$.

    We will prove using mathematical induction that after any iteration $t$ of the algorithm $\fmulti$ we have 

    \begin{align}
        \dist(\m{I}, \m{F}^t) \leq (7^t - 1) \dist(\m{I}, \m{F}^*) \label{equn:analysis-fptwo-one-multi}
    \end{align}

    For the base case, that is when $t = 1$ by \cref{clm:each-step-4-approx} we have $\m{F}^1$ is a $6$-close clustering to $\m{I}$ that satisfies the $1$st fairness constraint of algorithm $\fmulti$. Since $\m{F}^*$ also satisfies the $1$st fairness constraint we have
    \[
        \dist(\m{D}, \m{F}^1) \leq 6\dist(\m{D}, \m{F}^*) 
    \]

    Now, let us assume \cref{equn:analysis-fptwo-one-multi} is true for $t = k-1$, that is
    \[
        \dist(\m{I}, \m{F}^{k - 1}) \leq (7^{k - 1} - 1) \dist(\m{I}, \m{F}^*)
    \]

    Now, we prove it for $t = k$.

    \begin{align}
        \dist(\m{F}^{k - 1}, \m{F}^k) &\leq 6 \dist(\m{F}^{k - 1}, \m{F}^*)\label{equn:analysis-fmulti-two} \\
        &\leq 6 (\dist(\m{F}^{k - 1}, \m{I}) + \dist(\m{I}, \m{F}^*)) \n \\ &(\text{by triangle inequality}) \n \\
        &\leq 6 \cdot (7^{k - 1} - 1) \dist(I,\m{F}^*) + 6 \dist(I,\m{F}^*) \n \\
        &(\text{by inductive hypothesis}) \n \\
        &= 6 \cdot 7^{k - 1} \dist(\m{I},\m{F}^*) \label{equn:analysis-fmulti-two-(i)}
    \end{align}

    Here, \cref{equn:analysis-fmulti-two} is true because $\m{F}^*$ also satisfies the $k$th fairness constraint of the algorithm $\fmulti$ and due to \cref{clm:each-step-4-approx}.

    Now, we have
    \begin{align}
        \dist(\m{I}, \m{F}^k) &\leq \dist(\m{I}, \m{F}^{k - 1}) + \dist(\m{F}^{k - 1}, \m{F}^k) \n \\
        &(\text{by triangle inequality})\n \\
        &\leq (7^{k - 1} - 1) \dist(\m{I}, \m{F}^*) + 6 \cdot 7^{k - 1} \dist(\m{I}, \m{F}^*)  \n \\
        & (\text{by inductive hypothesis and \cref{equn:analysis-fmulti-two-(i)}}) \n \\
        &= (7^k - 1) \dist(\m{I}, \m{F}^*) \label{equn:analysis-fmulti-three}
    \end{align}

    Hence, now we can conclude for $t = \log r$, 
    \begin{align}
        \dist(\m{I}, \m{F}^{\log r}) &\leq O(7^{\log r}) \dist(\m{I}, \m{F}^*) \n \\
        &= O(r^{2.8}) \dist(\m{I}, \m{F}^*) \, \, (\textbf{as} \log_27 = 2.8)
    \end{align}

    Thus the algorithm $\fmulti$ computes a $O(r^{2.8})$-close $\fair$ which completes the proof of \cref{lem:analyze-fmulti}.

    Now, let us prove \cref{clm:each-step-4-approx}. To prove \cref{clm:each-step-4-approx}, we establish four intermediate claims: \cref{clm:lower-bound-opt-multi-one}, \cref{clm:lower-bound-opt-multi-two}, \cref{clm:lower-bound-opt-new} and \cref{clm:upper-bound-algo-multi}.

    Let $\m{F}^{t^*}$ denote the closest clustering to $\m{F}^{t - 1}$ that satisfies the $t$th fairness constraint of algorithm $\fmulti$.

    In \cref{clm:lower-bound-opt-multi-one}, \cref{clm:lower-bound-opt-multi-two} and \cref{clm:lower-bound-opt-new} we derive lower bounds on the distance $\dist(\m{F}^{t - 1}, \m{F}^{t^*})$, and in \cref{clm:upper-bound-algo-multi}, we provide an upper bound on the distance $\dist(\m{F}^{t - 1}, \m{F}^t)$, where $\m{F}^t$ is the clustering obtained after the $t$th iteration of algorithm $\fmulti$, starting from the initial clustering $\m{I}$.

    By comparing the bounds from \cref{clm:lower-bound-opt-multi-one}, \cref{clm:lower-bound-opt-multi-two}, \cref{clm:lower-bound-opt-new} and \cref{clm:upper-bound-algo-multi}, we will complete the proof of \cref{clm:each-step-4-approx}.

    To state \cref{clm:lower-bound-opt-multi-one}, we define the surplus and deficit for a cluster \( F_a^{t - 1} \in \mathcal{F}^{t - 1} \).

    Recall \( \mathcal{F}^{t - 1} \) denote the clustering obtained after the \((t - 1)\)th iteration of the $\fmulti$ algorithm and \( \{ B^{t - 1}_1, B^{t - 1}_2, \ldots, B^{t - 1}_{m_{t - 1}} \} \) be the set of color blocks at this iteration. For a color \( c_u \in \chi \), recall, \( p_u \) denote its proportion in the vertex set \( V \), i.e.,
\[
c_1(V): c_2(V): \cdots : c_{|\chi|}(V) = p_1 : p_2 : \cdots : p_{|\chi|}.
\]

WLOG, we assume $p_1 > p_2 >\cdots > p_{|\chi|}$. Note that due to this assumption for two colors $c_u \in B^{t - 1}_{j}$ and $c_v \in B^{t - 1}_{k}$ where $k > j$ we have $p_v < p_u$.

We define the following for a cluster \( F_a^{t - 1} \in \mathcal{F}^{t - 1} \):

Recall the color blocks $B_{2k -1}^{t - 1}$ and $B_{2k}^{t - 1}$ are created in such a way that we have $|B^{t - 1}_{2k - 1}| \geq |B^{t - 1}_{2k}|$. Suppose $\tilde{B^{t - 1}_{2k - 1}} \subseteq B^{t - 1}_{2k - 1}$ such that $|\tilde{B^{t - 1}_{2k - 1}}| = |B^{t - 1}_{2k}|$.

We create an arbitrary pair of colors $(c_j, c_{\hat{j}})$ where $c_j \in \tilde{B^{t - 1}_{2k - 1}}$ and $c_{\hat{j}} \in B^{t - 1}_{2k}$. We say $c_{\hat{j}} \in B^{t - 1}_{2k}$ is a color corresponding to $c_j \in B^{t - 1}_{2k - 1}$. Note for the colors present in $B^{t - 1}_{2k - 1} \setminus \tilde{B^{t - 1}_{2k - 1}}$ we have no corresponding color.

Now we define the following terms

\begin{itemize}
    \item \textbf{Weight of a vertex $v$:} For $v \in V$, suppose it has a color $c_j \in B^{t - 1}_{2k - 1}$ and its corresponding color $c_{\hat{j}} \in B^{t - 1}_{2k}$.

    In this case, since $p_j > p_{\hat{j}}$ we define 
    \[
        w(v) = \frac{p_{\hat{j}}}{p_j}
    \]
    otherwise if $v$ has color $c_{\hat{j}} \in B_{2k}^{t - 1}$ and its corresponding color $c_j \in B_{2k - 1}^{t - 1}$ then we define
    \[
        w(v) = 1
    \]
    Note, $w(v) \leq 1$ for all $v \in V$.
    \item \textbf{Surplus w.r.t. two corresponding colors $c_j$ and $c_{\hat{j}}$ in a cluster $F_a^{t - 1} \in \m{F}^{t - 1}$} is defined as
    \[
        s^a(c_j, c_{\hat{j}}) = \max( 0, c_{\hat{j}}(F_a^{t - 1}) - \frac{p_{\hat{j}}}{p_j}c_j(F_a^{t - 1}))
    \]
    \item \textbf{Surplus w.r.t. consecutive blocks $B^{i - 1}_{2k - 1}$ and $B^{i - 1}_{2k}$ in a cluster $F_a^{t - 1} \in \m{F}^{t - 1}$} is defined as
    \[
        T^k_a = \sum_{c_j \in B^{i - 1}_{2k - 1}} s^a(c_j, c_{\hat{j}})
    \]
    \item \textbf{Surplus of a cluster $F_a^{t - 1} \in \m{F}^{t - 1}$} is defined as
    \[
        T_a = \sum_{k = 1}^{m_{t - 1}} T^k_a
    \]
    where $m_{t - 1}$ is the number of blocks at iteration $(t - 1)$.
    \item \textbf{deficit w.r.t. two corresponding colors $c_j$ and $c_{\hat{j}}$ in a cluster $F_a^{t - 1} \in \m{F}^{t - 1}$} is defined as
    \[
        d^a(c_j, c_{\hat{j}}) = \max (0, \frac{p_{\hat{j}}}{p_j}c_j(F_a^{t - 1}) - c_{\hat{j}}(F_a^{t - 1}))
    \]
    \item \textbf{deficit w.r.t. consecutive blocks $B^{i - 1}_{2k - 1}$ and $B^{i - 1}_{2k}$ in a cluster $F_a^{t - 1} \in \m{F}^{t - 1}$} is defined as
    \[
        D^k_a = \sum_{c_j \in B^{i - 1}_{2k - 1}} d^a(c_j, c_{\hat{j}})
    \] 
    \item \textbf{deficit of a cluster $F_a^{t - 1} \in \m{F}^{t - 1}$} is defined as
    \[
        D_a = \sum_{k = 1}^{m_{t - 1}} D^k_a
    \]
\end{itemize}

    Now, we can state \cref{clm:lower-bound-opt-multi-one}.

    \begin{claim}\label{clm:lower-bound-opt-multi-one}
        \begin{align}
        \dist(\m{F}^{t - 1}, \m{F}^{t^*}) \geq \frac{1}{4} &\sum_{F_a^{t - 1} \in \m{F}^{t - 1}}  T_a \left(\left|F_a^{t - 1}\right| - T_a \right) \n \\
        &+ D_a \left(\left|F_a^{t - 1}\right| -  T_a \right) \n
    \end{align}
    \end{claim}

    \begin{proof}
        Consider a cluster $F_a^{t - 1} \in \m{F}^{t - 1}$. Suppose in $\m{F}^{t^*}$ the cluster $F_a^{t - 1}$ is partitioned into $X_1, X_2, \ldots, X_s$, more specifically, 

    \begin{itemize}
        \item For all $j \in [t]$, $X_j \subseteq F_{r_j}^{t^*}$ for some $F_{r_j}^{t^*} \in \m{F}^{t^*}$.
        \item For all $j \neq \ell \in [t]$, we have $F_{r_j}^{t^*} \neq F_{r_\ell}^{t^*}$.
        \item $\bigcup_{j \in [s]} X_j = F_a^{t - 1}$.
    \end{itemize}

    According to the definition of $\dist(\m{F}^{t - 1}, \m{F}^{t^*})$, in $\dist(\m{F}^{t - 1}, \m{F}^{t^*})$ we count the number of pairs $(u,v)$ that are present in different clusters in the clustering $\m{F}^{t - 1}$ but in the same cluster in the clustering $\m{F}^{t^*}$ or present in the same cluster in the clustering $\m{F}^{t - 1}$ but present in different clusters in the clustering $\m{F}^{t^*}$.

    To give a lower bound on $\dist(\m{F}^{t - 1}, \m{F}^{t^*})$ we calculate the cost incurred by each cluster $F_a^{t - 1} \in \m{F}^{t - 1}$. Let us describe this formally,
    \begin{align}
        \pay{F_a^{t - 1}} = |&\{(u,v) \mid (u,v \in F_a^{t - 1} \land u \nsim_{\m{F}^{t^*}} v) \n \\
        & \lor (u \in F_a^{t - 1}, v \notin F_a^{t - 1} \land u \sim_{\m{F}^{t^*}} v) \}| \n
    \end{align}
    where $u \sim_{\m{F}^{t^*}} v$ denotes $u$ and $v$ belongs to the same cluster in the clustering $\m{F}^{t^*}$.

    It is straightforward to see that 
    \begin{align}
        \dist(\m{F}^{t - 1}, \m{F}^{t^*}) \geq \frac{1}{2} \sum_{F_a^{t - 1} \in \m{F}^{t - 1} }\pay{F_a^{t - 1}} \label{eq:main-multi}
    \end{align}

    In the above expression, we multiply by $1/2$ because we took summation overall $F_a^{t - 1} \in \m{F}^{t - 1} $. To count a pair $(u,v)$ where $u \in F_a^{t - 1}$ and $v \in F_b^{t - 1}$ for $a \neq b$ we charge $1/2$ when calculating $\pay{F_a^{t - 1}}$ and again $1/2$ when calculating $\pay{F_b^{t - 1}}$. 

    Now, we provide a lower bound on $\pay{F_a^{t - 1}}$. To provide this lower bound, we need to define some terms.

    \begin{itemize}
    \item \textbf{Weight of a color block:}
    \[
    w(B_j^{t - 1}) = \sum_{c_u \in B_j^{t - 1}} p_u.
    \]
    
    \item \textbf{Scaling factor of a color block in cluster \( F_a^{t - 1} \):}
    \[
    h(B_j^{t - 1}) = \frac{c_u(F_a^{t - 1})}{p_u} \quad \text{for any } c_u \in B_j^{t - 1},
    \]
    which is well-defined since
    \[
    \frac{c_v(F_a^{t - 1})}{p_v} = \frac{c_u(F_a^{t - 1})}{p_u} \quad \forall c_u \neq c_v \in B_j^{t - 1}.
    \]
    \end{itemize}

    Create two sets $\cb$ and $\mb$.
    \[
        \cb = \{ (B_i^{t - 1}, B_{i + 1}^{t - 1}) \mid h(B_{i + 1}^{t - 1}) > h(B_i^{t - 1})\}
    \]
    
    \[
        \mb = \{ (B_i^{t - 1}, B_{i + 1}^{t - 1}) \mid h(B_{i + 1}^{t - 1}) \leq h(B_i^{t - 1})\}
    \]
    Informally, the set $\cb$ contains a pair of color blocks $B_i^{t - 1}$ and $B_{i + 1}^{t + 1}$ if the scaling factor of $B_{i + 1}^{t + 1}$ is greater than the scaling factor of $B_{i}^{t + 1}$. In the algorithm $\fmulti$, we cut the surplus for these types of pairs of color blocks. For the pair of color blocks in $\mb$ we merge the deficit.

    Now, for the pairs of color blocks $(B_{i}^{t + 1}, B_{i + 1}^{t + 1}) \in \cb$ let us define some notations w.r.t. a partition $X_j$.

    \begin{itemize}
    \item Surplus w.r.t. two corresponding colors $c_j$ and $c_{\hat{j}}$ in a partition $X_\ell$ is defined as

    \[ 
        \sigma^\ell_{c_j} = c_{\hat{j}}(X_\ell) - \frac{p_{\hat{j}}}{p_j} c_j(X_\ell)
    \]
    
    Here, $(c_j, c_{\hat{j}}) \in B^{t - 1}_{2k - 1} \times B^{t - 1}_{2k}$ where $(B^{t - 1}_{2k - 1}, B^{t - 1}_{2k}) \in \cb$.

    It is straightforward to see that,
    \[
        s^a(c_j, c_{\hat{j}}) \geq \sum_{\ell = 1}^s \sigma^\ell_{c_j}
    \]
    Similar to the proof of \cref{lem:fair-power-of-two} we can redefine $\sigma^\ell_{c_j}$ in such a way that
    \[
        s^a(c_j, c_{\hat{j}}) = \sum_{\ell = 1}^s \sigma^\ell_{c_j}
    \]
    \item Surplus w.r.t. two consecutive blocks $B^{i - 1}_{2k - 1}$ and $B^{i - 1}_{2k}$ in a partition $X_\ell$ is defined as
    \[
        S^k_{\ell} = \sum_{c_j \in B^{t - 1}_{2k - 1}} \sigma^\ell_{c_j}
    \]
    \item Surplus of the partition $X_\ell$ is
    \[
        S_\ell = \sum_{B^{t - 1}_{2k - 1} \in \cb} S^k_\ell
    \]
    It is straightforward to see that 
    \[
        T_a = \sum_{\ell = 1}^s S_\ell.
    \]
\end{itemize}
    Since, $\m{F}^{t^*}$ satisfies the $t$th fairness constraint of the algorithm $\fmulti$ we have for a cluster $F_{r_\ell}^{t^*} \in \m{F}^{t^*}$
    \[
        \dfrac{c_u(F_{r_\ell}^{t^*})}{p_{u}}= \dfrac{c_v(F_{r_\ell}^{t^*})}{p_{v}}\, \,  \forall c_u \in B^{t - 1}_{2k - 1}, c_v \in B^{t - 1}_{2k}
    \]
    recall $X_\ell \subseteq F_{r_\ell}^{t^*} \in \m{F}^{t^*}$.

    Let, $M_\ell \subseteq  F_{r_\ell}^{t^*}\setminus F^{t-1}_{a} $

    To maintain the $t$th fairness constraint of the algorithm $\fmulti$ we must have
    \[
        |M_\ell| \geq \sum_{v \in M_\ell} w(v) \geq S_\ell
    \]
    Reasons behind the above inequalities
    \begin{itemize}
        \item \textbf{$1$st inequality:} $w(v) \leq 1$.
        \item \textbf{$2$nd inequality:} This follows from the requirement that the cluster $F^{t^*}_{r_\ell}$ must satisfy the $t$th fairness constraint of the algorithm $\fmulti$. That is, for every pair $(c_j, c_{\hat{j}}) \in B^{t - 1}_{2k - 1} \times B^{t - 1}_{2k}$, we must ensure that after adding the set $M_\ell$ to $X_\ell$ in the cluster $F^{t^*}_{r_\ell}$, the following holds:
\[
    \sum_{v \in C_j(X_\ell \cup M_\ell)} w(v) = \sum_{v \in C_{\hat{j}}(X_\ell \cup M_\ell)} w(v)
\]
where $C_r(S)$ denotes the set of vertices of color $c_r$ in a subset $S \subseteq V$. In other words, the total weight of the vertices of color $c_j$ and $c_{\hat{j}}$ in the updated cluster must be equal, for every such pair. The surplus $ S_\ell$ precisely captures the imbalance in these weights in $X_\ell$, and hence the total weight added must be at least $S_\ell$ to restore this balance.
    \end{itemize}
    Let, us define $\mcostone(F_a^{t - 1})$ be the number of pairs $(u,v)$ such that
    \begin{enumerate}
        \item Either $u \in M_\ell$ and $v \in X_\ell$
        \item or $u \in X_m$ and $v \in X_\ell$ for $m \neq \ell$.
    \end{enumerate}
    Hence,
    \begin{align}
        &\mcostone(F_a^{t - 1}) \n \\
        &\geq \frac{1}{2}\sum_{\ell = 1}^s\left(|M_\ell|(|X_\ell|) + \sum_{m \neq \ell}|X_m|(|X_\ell|)\right) \n \\
        &\geq \frac{1}{2} \sum_{\ell = 1}^s \left(S_\ell(|X_\ell| - S_\ell) + \sum_{m \neq \ell}S_m(|X_\ell| - S_\ell)\right) \n \\
        &\geq \frac{1}{2} \sum_{\ell = 1}^s T_a (|X_\ell| - S_\ell) \n \\
        &\geq \frac{1}{2} T_a (|F_a^{t - 1}| - T_a) \label{eq:cost-one-multi} 
    \end{align}

    Similarly, now for a pair of color blocks $(B_{2k - 1}^{t - 1}, B_{2k}^{t + 1}) \in \mb$ and for a partition $X_\ell$, let us define
    \begin{itemize}
        \item $\hat{M}_\ell \subseteq V \setminus F_a^{t - 1}$ that serves the deficit for the color blocks $B_{2k - 1}^{t - 1}$ and $B_{2k}^{t - 1}$ in the cluster $F_{r_\ell}^{t^*}$.
        \item $G_m = \bigcup_{(B_{2k-1}^{t - 1}, B_{2k}^{t - 1}) \in \mb} F_{r_m}^{t^*} \cap B_{2k - 1}^{t - 1}(F_a^{t - 1})$ : it denotes the part of the set $B_{2k - 1}^{t - 1}(F_a^{t - 1})$ that lies in the cluster $F_{r_m}^{t^*} \in \m{F}^{t^*}$ for $m \neq \ell$
    \end{itemize}

    It is straightforward to see that 
    \[
        |\hat{M}_\ell| + \sum_{m \neq \ell} |G_m| \geq D_a
    \]
    Let, us define $\mcosttwo(F_a^{t - 1})$ be the number of pairs $(u,v)$ such that
    \begin{enumerate}
        \item Either $u \in \hat{M}_\ell$ and $v \in X_\ell$
        \item or $u \in G_m$ and $v \in X_\ell$ for $m \neq \ell$.
    \end{enumerate}
    Hence,
    \begin{align}
        \mcosttwo(F_a^{t - 1}) &\geq \frac{1}{2}\sum_{\ell = 1}^s|\hat{M}_\ell||X_\ell| \n \\
        &+ \sum_{m \neq \ell}|G_m||X_\ell| \n \\
        &\geq \frac{1}{2}\sum_{j = 1}^s D_a (|X_\ell| - |S_\ell|) \n \\
        &\geq \frac{1}{2} D_a (|F_a^{t - 1}| - T_a) \label{eq:cost-two-multi} 
    \end{align}
    Since, in $\mcostone(F_a^{t - 1})$ and $\mcosttwo(F_a^{t - 1})$ we count disjoint pairs. Hence, 
    \begin{align}
        \pay{F_a^{t - 1}} \geq \mcostone(F_a^{t - 1}) + \mcosttwo(F_a^{t - 1}) \label{eq:pay-single-cluster}
    \end{align}

    Now. by \cref{eq:main-multi}, \cref{eq:cost-one-multi}, \cref{eq:cost-two-multi} and \cref{eq:pay-single-cluster} we conclude the proof of \cref{clm:lower-bound-opt-multi-one}.
    
\end{proof}

   \begin{claim}\label{clm:lower-bound-opt-multi-two}
        \begin{align}
        \dist(\m{F}^{t - 1}, \m{F}^{t^*}) \geq &\sum_{F_a^{t - 1} \in \m{F}^{t - 1}} \frac{1}{2}T_a^2  \n
    \end{align}
\end{claim}

\begin{proof}
    Similar to the proof of the previous claim, we define $S_\ell$ as the surplus of a partition $X_\ell$ and $M_\ell \subseteq V \setminus F_a^{t - 1}$ be the set of points that are merged to $X_\ell$ to satisfy the $t$th fairness constraint of $\fmulti$.

    Let, us define $\pay{F_a^{t - 1}, X_\ell}$ be the number of pairs $(u,v)$ s.t.
    \begin{itemize}
        \item $u \in M_\ell$, $v \in X_\ell$
        \item $u \in X_m$, $v \in X_\ell$ for all $m \neq \ell$.
    \end{itemize}
    Now,
    \begin{align*}
        \dist(\m{F}^{t - 1}, \m{F}^{t^*}) \geq \frac{1}{2} \sum_{F_a^{t - 1} \in \m{F}^{t - 1}} \sum_{\ell = 1}^s \pay{F_a^{t - 1}, X_\ell}
    \end{align*}

    We multiply by $1/2$ to avoid overcounting of pairs. Note we can overcount a pair in the following situations
    \begin{enumerate}[(i)]
        \item Since we take the sum overall $F_a^{t - 1} \in \m{F}^{t - 1}$, we can overcount a pair $(u,v)$ if $u \in F_b^{t - 1}$ and $v \in F_a^{t - 1}$ where $b \neq a$ when considering the cluster $F_b^{t - 1}$ in the summation.
        \item Since we take the sum over all partitions $X_\ell$ of a cluster $F_a^{t - 1}$, we can overcount a pair $(u,v)$ if $u \in X_m$ and $v \in X_\ell$ where $m \neq \ell$ when we consider the partition $X_m$ in the summation.
    \end{enumerate}
    Now, we only need to show
    \[
        \sum_{\ell = 1}^s \pay{F_a^{t - 1}, X_\ell} \geq T_a^2
    \]
    \begin{align}
        \sum_{\ell = 1}^s \pay{F_a^{t - 1}, X_\ell} & \geq \sum_{\ell = 1}^s (|M_\ell| |X_\ell| + \sum_{m \neq \ell}|X_m||X_\ell|) \n \\
        &\geq \sum_{\ell = 1}^s (|M_\ell| S_\ell + \sum_{m \neq \ell}S_m S_\ell) \n \\
        &\geq \sum_{\ell = 1}^s (S_\ell^2 + \sum_{m \neq \ell}S_m S_\ell) \n \\
        &\geq T_a^2 \n
    \end{align}
\end{proof}

\begin{claim}\label{clm:lower-bound-opt-new}
    \[\dist(\m{F}^{t - 1}, \m{F}^{t^*}) \geq \sum_{F_a^{t - 1} \in \m{F}^{t - 1}} \frac{1}{2} D_a^2 \]
\end{claim}

\begin{proof}
    For a pair of color blocks $(B_{2k - 1}^{t - 1}, B_{2k}^{t + 1}) \in \mb$ and for a partition $X_\ell$, let us define
    \begin{itemize}
        \item $\hat{M}_\ell \subseteq V \setminus F_a^{t - 1}$ that serves the deficit for the color blocks $B_{2k - 1}^{t - 1}$ and $B_{2k}^{t - 1}$ in the cluster $F_{r_\ell}^{t^*}$.
        \item $G_m = \bigcup_{(B_{2k-1}^{t - 1}, B_{2k}^{t - 1}) \in \mb} F_{r_m}^{t^*} \cap B_{2k - 1}^{t - 1}(F_a^{t - 1})$ : it denotes the part of the set $B_{2k - 1}^{t - 1}(F_a^{t - 1})$ that lies in the cluster $F_{r_m}^{t^*} \in \m{F}^{t^*}$ for $m \neq \ell$
    \end{itemize}

    It is straightforward to see that 
    \[
        |\hat{M}_\ell| + \sum_{m \neq \ell} |G_m| \geq D_a
    \]
    Let, us define $\pay{F_a^{t - 1}, X_\ell}$ be the number of pairs $(u,v)$ s.t.
    \begin{enumerate}
        \item Either $u \in \hat{M}_\ell$ and $v \in X_\ell$
        \item or $u \in G_m$ and $v \in X_\ell$ for $m \neq \ell$.
    \end{enumerate}
    \begin{align*}
        \dist(\m{F}^{t - 1}, \m{F}^{t^*}) \geq \frac{1}{2} \sum_{F_a^{t - 1} \in \m{F}^{t - 1}} \sum_{\ell = 1}^s \pay{F_a^{t - 1}, X_\ell}
    \end{align*}

    We multiply by $1/2$ to avoid overcounting of pairs. Note, we can overcount a pair in the following situations
    \begin{enumerate}[(i)]
        \item Since we take the sum overall $F_a^{t - 1} \in \m{F}^{t - 1}$, we can overcount a pair $(u,v)$ if $u \in F_b^{t - 1}$ and $v \in F_a^{t - 1}$ where $b \neq a$ when considering the cluster $F_b^{t - 1}$ in the summation.
        \item Since we take the sum over all partitions $X_\ell$ of a cluster $F_a^{t - 1}$, we can overcount a pair $(u,v)$ if $u \in G_m$ and $v \in X_\ell$ where $m \neq \ell$ when we consider the partition $X_m$ in the summation. Note $G_m \subseteq X_m$.
    \end{enumerate}
    Now, we only need to show
    \[
        \sum_{\ell = 1}^s \pay{F_a^{t - 1}, X_\ell} \geq D_a^2
    \]
    \begin{align}
        \sum_{\ell = 1}^s \pay{F_a^{t - 1}, X_\ell} &\geq \sum_{\ell = 1}^s\left(|\hat{M}_\ell||X_\ell| + \sum_{m \neq \ell}|G_m||X_\ell|\right) \n \\
        &\geq \sum_{\ell = 1}^s D_a |X_\ell| \n \\
        &\geq  D_a |F_a^{t - 1}| \n \\ 
        &\geq  D_a^2\, \, \text{(\textbf{as}$|F_a^{t - 1}| \geq D_a$)} \n
    \end{align}
\end{proof}

    \begin{claim}\label{clm:upper-bound-algo-multi}
        \begin{align}
        \dist(\m{F}^{t - 1}, \m{F}^t) \leq &\sum_{F_a^{t - 1} \in \m{F}^{t - 1}}  T_a\left(\left|F_a^{t - 1}\right| - T_a \right) \n \\
        &+ D_a \left(\left|F_a^{t - 1}\right| - D_a \right) \n \\
        &+ \frac{1}{2}D_a^2 + \frac{1}{2}T_a^2 \n
    \end{align}
    \end{claim}

    \begin{proof}
        For each cluster $F_a^{t - 1} \in \m{F}^{t - 1}$ we cut the surplus many vertices, $T_a$ and merge deficit many vertices $D_a$ to it.

        Hence, in $\dist(\m{F}^{t - 1}, \m{F}^t)$ for a cluster $F_a^{t - 1} \in \m{F}^{t - 1}$ we count the cost of cutting and merging to it, which is
        \[
            T_a (|F_a^{t - 1}| - T_a) + D_a (|F_a^{t - 1}| - D_a)
        \]
        Again for a cluster $F_a^{t - 1} \in \m{F}^{t - 1}$ the deficit many vertices that we merge can come from several other clusters. A trivial upper bound on this cost is given by $(1/2) \cdot D_a^2$.

        The surplus many vertices $T_a$, that we cut from $F_a^{t - 1} \in \m{F}^{t - 1}$ can further get divided. A trivial upper bound on the cost of dividing $T_a$ many points is $(1/2)\cdot T_a^2$.

        Hence we get,

        \begin{align}
        \dist(\m{F}^{t - 1}, \m{F}^t) \leq &\sum_{F_a^{t - 1} \in \m{F}^{t - 1}}  T_a\left(\left|F_a^{t - 1}\right| - T_a \right) \n \\
        &+ D_a \left(\left|F_a^{t - 1}\right| - D_a \right) \n \\
        &+ \frac{1}{2}D_a^2 + \frac{1}{2}T_a^2 \n
        \end{align}
    \end{proof}
    
    Now we complete the proof of \cref{clm:each-step-4-approx}.
    \begin{proof}[\textbf{Proof of \cref{clm:each-step-4-approx}}:]
        By \cref{clm:upper-bound-algo-multi} we get,
        \begin{align}
        \dist(\m{F}^{t - 1}, \m{F}^t) \leq &\sum_{F_a^{t - 1} \in \m{F}^{t - 1}}  T_a\left(\left|F_a^{t - 1}\right| - T_a \right) \n \\
        &+ D_a \left(\left|F_a^{t - 1}\right| - D_a \right) \n \\
        &+ \frac{1}{2}D_a^2 + \frac{1}{2}T_a^2 \label{eq:upper-bound-multi-zero}
    \end{align}
        By \cref{clm:lower-bound-opt-multi-one} we get
        \begin{align}
            &\sum_{F_a^{t - 1} \in \m{F}^{t - 1}}  T_a\left(\left|F_a^{t - 1}\right| - T_a \right) \n \\
        &+ D_a \left(\left|F_a^{t - 1}\right| - D_a \right) \leq 4 \dist(\m{F}^{t - 1}, \m{F}^{t^*}) \label{eq:upper-bound-multi-one}
        \end{align}
        By \cref{clm:lower-bound-opt-new} we get
        \begin{align}
            \sum_{F_a^{t - 1} \in \m{F}^{i - 1}} \frac{1}{2}D_a^2 \leq \dist(\m{F}^{t - 1}, \m{F}^{t^*}) \label{eq:upper-bound-multi-two}
        \end{align}
        By \cref{clm:lower-bound-opt-multi-two} we get
        \begin{align}
            \sum_{F_a^{t - 1} \in \m{F}^{i - 1}} \frac{1}{2}T_a^2 \leq \dist(\m{F}^{t - 1}, \m{F}^{t^*}) \label{eq:upper-bound-multi-three}
        \end{align}
        Now, by combining \cref{eq:upper-bound-multi-zero}, \cref{eq:upper-bound-multi-one}, \cref{eq:upper-bound-multi-two} and \cref{eq:upper-bound-multi-three} we get,
        \[
            \dist(F^{t - 1}, F^{t}) \leq 6 \dist(F^{t - 1}, F^{t^*})
        \]
    \end{proof}
    We have shown earlier that \cref{clm:each-step-4-approx} implies \cref{lem:analyze-fmulti}. Thus, we complete the proof of \cref{lem:analyze-fmulti}.
\end{proof}

Using \cref{lem:fair-power-of-two} and \cref{lem:analyze-fmulti}, we can prove \cref{thm:closest-fair-multicolor-balanced}. 

\begin{proof}[Proof of \cref{thm:closest-fair-multicolor-balanced}]
    To prove this theorem, we analyse the algorithm $\fequi$. The algorithm $\fequi$ takes a clustering $\m{C}$ as input, applies the algorithm $\fptwo$ as a subroutine to create an intermediate clustering $\m{I}$ and then applies the subroutine $\fmulti$ to get the final fair clustering $\m{F}$.

    By \cref{lem:analyze-fmulti} we get that, $\m{F}$ is $O(\log^{2.8} |\chi|)$-close to the clustering $\m{I}$. By \cref{lem:fair-power-of-two} we get that $\m{I}$ is $O(|\chi|^{1.6})$-close to the clustering $\m{C}$. By applying the triangle inequality to the two preceding results, we conclude that the output clustering $\m{F}$ produced by $\fequi$ is $O(|\chi|^{1.6} \log^{2.8} |\chi|)$-close to the input clustering $\m{C}$. 

    Let $\m{F}^*$ be the closest fair clustering to $\m{C}$. Hence, we get,

    \begin{align}
        \dist(\m{C}, \m{F}) &\leq \dist(\m{C}, \m{I}) + \dist(\m{I}, \m{F}) \, \, \text{(triangle inequality)} \n \\
        &\leq \dist(\m{C}, \m{I}) + O(\log^{2.8}|\chi|) \dist(\m{I}, \m{F}^*) \n \\
        &\leq O(|\chi|^{1.6}) \dist(\m{C}, \m{F}^*) + O(\log^{2.8}|\chi|) \dist(\m{I}, \m{F}^*) \n \\
        &\leq O(|\chi|^{1.6}) \dist(\m{C}, \m{F}^*) + O(\log^{2.8}|\chi|) (\dist(\m{C}, \m{I}) \n \\ &+ \dist(\m{C}, \m{F}^*)) \, \, \text{(triangle inequality)} \n \\
        &\leq O(|\chi|^{1.6} \log^{2.8}|\chi| + |\chi| + \log^{2.8}|\chi|) \dist(\m{C}, \m{F}^*) \n \\
        &\leq O(|\chi|^{1.6} \log^{2.8}|\chi|) \dist(\m{C}, \m{F}^*) \n
    \end{align}
    This completes the proof of \cref{thm:closest-fair-multicolor-balanced}.
\end{proof}

\section{Approximate Closest Fair Clustering for Arbitrary-Proportion Groups}
\label{sec2}

In this section, we extend our analysis from the equiproportion setting to the general case where color groups may appear in arbitrary proportions. In this setting, we provide a polynomial-time algorithm to compute an approximate closest fair clustering to a given input clustering $\m{C}$. More specifically, we prove \cref{thm:closest-fair-multicolor-general}, which we restate below.

\genmulti*

To prove the above theorem, we take a $2$-step approach similar to \cref{sec:closest-fair-two-colors}, which was constrained to only two colors.

\begin{enumerate}[(i)]
    \item First, we convert the clustering $\m{C}$ to a clustering $\m{M}$ such that for each cluster $M_i \in \m{M}$, $c_j(M_i)$ is divisible by $p_j$. Recall, we call such a clustering a $p$-divisible clustering.
    \item In the second step, we will provide the clustering $\m{M}$ as input to the algorithm $\fmulti$ and get a fair clustering $\m{F}$ as output. 
\end{enumerate}

Now, let us describe the $\fgen$ that computes an approximate closest fair clustering to a given input clustering $\m{C}$.
We first describe a subroutine $\pdca$ that converts a clustering $\m{C}$ on a vertex set $V$ to a $p$-divisible clustering $\m{M}$.

\subsection{Details of the Algorithm}

To describe the algorithm $\pdca$, let us define some terms.

\begin{itemize}
    \item $\sigma(C_k,c_j)$: Surplus of $C_k$ w.r.t. to the color $c_j$, formally,
    \[
        \sigma(C_k,c_j) \subseteq C_k
    \] 
    that contains $c_j(C_k) \mod p_j$ points of color $c_j$ if $p_j \nmid c_j(C_k)$, otherwise it contains $p_j$ points of color $c_j$.
    \item $\sigma_j = \sum_{C_k \in \m{C}} \sigma(C_k,c_j)$ (Note $\sigma_j$ must be a multiple of $p_j$)
    \item $\delta(C_k,c_j)$: Ceficit of $C_k$ w.r.t. to the color $c_j$, $\delta(C_k,c_j) \subseteq V \setminus C_k$ that contains $p_j - |\sigma(C_k,c_j)|$ points. 
    \item $\kappa^j(C_k)$: Cut cost of $C_k$ w.r.t. a color $c_j$, formally 
    \[
        \kappa^j(C_k) = |\sigma(C_k,c_j)| (|C_k| - |\sigma(C_k,c_j)|)
    \]
    \item $\mu^j(C_k)$: Merge cost of $C_k$ w.r.t. a color $c_j$, formally
    \[
        \mu^j(C_k) = |\delta(C_k, c_j)| (|C_k|)
    \]
\end{itemize}

Now, let's discuss the algorithm $\pdca$.

\textbf{Input:} A clustering $\mathcal{C}$ of data points, a set of colors $\chi = \{c_1, c_2, \ldots, c_d\}$, and a desired proportion vector $\mathbf{p} = (p_1, \ldots, p_d)$ such that $c_1(V):c_2(V):\cdots:c_d(V) = p_1:p_2:\cdots:p_d$.

\textbf{Goal:} Convert the input clustering $\mathcal{C}$ into a $p$-divisible clustering $\mathcal{M}$ where each cluster contains a multiple of $p_j$ points for each color $c_j$.

\textbf{Description of the subroutine $\pdca$:}

\begin{itemize}
    \item For each color $c_j$, prepare $\sigma_j / p_j$ empty \texttt{extra clusters} $\{P_1, P_2, \ldots, P_{\sigma_j/p_j} \}$
    \item For each cluster $C_i \in \m{C}$:
    \begin{itemize}
        \item If the number of surplus $c_j$-colored points in $C_i$ is small ($\leq p_j/2$), that is $|\sigma(C_i, c_j)| \leq p_j /2$ mark $C_i$ for cutting ($\cut$ set).
        \item Otherwise, mark it for merging ($\merge$ set).
    \end{itemize}
    \item While there are clusters in the $\cut$ set:
    \begin{itemize}
        \item cut the surplus $\sigma(C_k, c_j)$ from cluster $C_k$, that is $C_k = C_k \setminus \sigma(C_k, c_j)$ and try to distribute it:
        \begin{itemize}
            \item If there are clusters in $C_\ell \in \merge$, donate surplus to them using their deficits. That is
            \[
               C_\ell = C_\ell \cup T. 
            \]
            where
            \[
                T \subseteq \sigma(C_k,c_j)
            \]
            of size $\min(|\sigma(C_k,c_j)|, |\delta(C_\ell,c_j)|)$.
            \item Otherwise, distribute the surplus to a prepared cluster $P_m \in$\texttt{extra clusters}, ensuring that each such cluster is of size $p_j$. That is
            \[
                P_m = P_m \cup Q.
            \]
            where
            \[
                Q \subseteq \sigma(C_k,c_j)
            \]
            of size $\min(p_j, |\sigma(C_k,c_j)|, |\sigma(C_k,c_j)| - |P_m|)$.
        \end{itemize}
    \end{itemize}
\end{itemize}

\begin{itemize}
\item After all \texttt{CUT} clusters are processed, handle the remaining \texttt{MERGE} clusters:
    \begin{itemize}
        \item Pick a cluster with the minimum cut-merge cost difference and distribute its surplus, similar to the above step.
        \item Do it until all the deficits are satisfied. Note that in the previous step, we may pick a cluster multiple times. 
    \end{itemize}
\end{itemize}
We provide the pseudocode of the algorithm $\pdca$ in \cref{alg:p-divisible-clustering}. 

Now, we are ready to describe the algorithm $\fgen$. 

\paragraph{Description of the algorithm $\fgen$:} The algorithm $\fgen$ proceeds in two stages. First, we apply the algorithm $\pdca$ to the input clustering $\m{C}$ to obtain a $p$-divisible clustering $\m{M}$. Then, we apply the algorithm $\fmulti$ on $\m{M}$ to obtain the final fair clustering $\m{F}$.

\subsection{Approximation Guarantee of $\fgen$}

In this section, we analyze the algorithm~$\fgen$ presented earlier and show that it satisfies the approximation guarantee stated in \cref{thm:closest-fair-multicolor-general}. To analyze $\fgen$, first let us analyze the subroutine $\pdca$. More specifically, we prove the following lemma.

\begin{lemma}\label{lem:main-multiple-of-p}
    Given a clustering $\m{C}$, the algorithm $\pdca$ outputs a clustering $\m{M}$ s.t. $\m{M}$ is $O(|\chi|)$-close-$p$-divisible clustering to $\m{C}$.
\end{lemma}

\begin{proof}
    For a color $c_j \in \chi$, we created two sets $\cut$ and $\merge$ in the algorithm $\pdca$. Let us now define two cases.

\begin{itemize}

	\item Cut case for color $c_j$: $\cutcase$: 
        
	\[
		\text{If} \, \, \sum_{C_k \in \cut} |\sigma(C_k, c_j)| \geq \sum_{C_k \in \merge} |\mu(C_k, c_j)|
	\]
	\item Merge case for color $c_j$: $\mergecase$:
	\[
		\text{If} \, \, \sum_{C_k \in \cut} |\sigma(C_k, c_j)| < \sum_{C_k \in \merge} |\mu(C_k, c_j)|
	\]

\end{itemize}

For $\cutcase$, let us define some costs incurred by our algorithm $\pdca$

\begin{itemize}
     \item From each cluster $C_k \in \cut$, $\pdca$ cuts the surplus part $\sigma(C_k, c_j)$ from $C_k$. Hence, the cost paid for cutting these surplus points is the number of pairs $(u,v)$ such that $u \in \sigma(C_k, c_j)$ and $v \in (C_k \setminus \sigma(C_k, c_j))$. We denote this by:
     \begin{align}
         \mcostone(\m{M})^{c_j} =  \sum_{C_k \in \cut} |\sigma(C_k, c_j)| (|C_k| - |\sigma(C_k, c_j)|)\label{equn:cost-paid-one}
     \end{align}
    \item For each cluster $C_m \in \merge$, the algorithm $\pdca$  merges the deficit amount $|\delta(C_m, c_j)|$ to these clusters. Hence, the cost paid for merging the deficit to $C_m$ is the number of pairs $(u,v)$ such that $u \in \delta(C_m, c_j)$ and $v \in C_m$
    
     \begin{align}
         \mcosttwo(\m{M})^{c_j} =  \sum_{C_m \in \merge} |\delta(C_m, c_j)| |C_m| \label{equn:cost-paid-two}
     \end{align}
    
     \item The $\sigma(C_k, c_j)$ points that are cut from $C_k$ can get merged with the surplus of other clusters $\sigma(C_\ell, c_j)$ (say) which are also cut from a cluster $C_\ell \neq C_k$. We call the cost of merging $\sigma(C_k, c_j)$  with the surplus of other clusters as $\mcostthree(\m{M})^{c_j}$

     Let, $\sigma(C_k, c_j)$ gets further divided into multiple parts of size $\alpha_1, \ldots, \alpha_t$, so we get,
     \begin{align}
         \mcostthree(\m{M})^{c_j} \leq \sum_{C_k \in \m{C}} \dfrac{1}{2}\sum_{i = 1}^t\alpha_i (p_j - \alpha_i) \label{equn:cost-paid-three}
     \end{align}
    In the above expression we provide an upper bound on the number of pairs $(u,v)$ such that $u \in \sigma(C_k, c_x)$ and $v \in V \setminus C_k$ that are present together in a cluster in the clustering $\m{M}$. The above expression provides such an upper bound because of the fact that the surpluses which we cut from the clusters in the $\cut$ in our algorithm is used to fulfil the deficit of a cluster $C_m \in \merge$ where $m \neq k,\ell$ and we know $\delta(C_m, c_j) < p_j$. 
     
     \item These $\sigma(C_k, c_j)$ points from $C_k$ can also further be split into several parts $W_1, W_2, \ldots, W_t$ (say). These parts of $\sigma(C_k, c_j)$ points belong to different clusters in $\m{M}$ and thus would incur some cost. We call this cost as $\mcostfour(\m{M})^{c_j}$.
     \begin{align}
         \mcostfour(\m{M})^{c_j} =  \sum_{C_k \in \m{C}} \frac{1}{2} \sum_{s = 1}^t |W_s|(\sigma(C_k, c_j) - |W_s|)\label{equn:cost-paid-four}
     \end{align}
    In the above expression, we count the number of pairs $(u,v)$ such that $u,v \in \sigma(C_k, c_j)$ but present in different clusters in the clustering $\m{M}$.
\end{itemize}

For $\mergecase$, let us define some costs incurred by our algorithm $\pdca$.

\begin{itemize}
        \item In this case for a cluster $C_k \in \m{C}$, we may cut multiple subsets of size $p_j$ and a single subset of size $\sigma(C_k, c_j)$. Let us assume $Y_{k,z}$ denotes the $z$th such subset of the cluster $C_k$ and $y_{k,z}$ takes the value $1$ if we cut $z$th such subset from $C_k$. The cost of cutting $z$th such subset from $C_k$ is given as
     \begin{align}
         &\kappa_0(C_k) = |\sigma(C_k, c_j)| (|C_k| -  |\sigma(C_k, c_j)|) \n \\
         &(\text{cost of cutting the $0$th subset})\n \\
         &\kappa_z(C_k) = p_j (|C_k| -|\sigma(C_k, c_j)| - zp_j) \n \\
         &(\text{cost of cutting the $z$th subset for $z \geq 1$})\n 
    \end{align}
    Thus, we define 
    \begin{align}
         &\mcostfive(\m{M})^{c_j} = \sum_{C_k \in \m{C}} \sum_{z = 0}^t y_{k,z} \kappa_z(C_k) \label{equn:cost-paid-one-merge} \\
         &\text{where} \, \, \left( \text{t} = \frac{c_j(C_k) - |\sigma(C_k,c_j)|}{p_j} \right) \n
     \end{align}
     \item Suppose the algorithm $\pdca$ merges at a cluster $C_m \in \merge'$. Here, $\merge' \subseteq \merge$ denotes the set of clusters where the algorithm $\pdca$ has merged the deficit amount of points. More specifically, it is defined as
     \begin{align*}
         &C_m \in \merge' \iff \exists M_\ell \in \m{M} \n \\ 
         &\text{s.t.}\, \, C_m \subseteq M_\ell
     \end{align*}
        
    Then, the cost paid for merging the deficit to $C_m$ is
    \begin{align}
        \mcostsix(\m{M})^{c_j} = \sum_{C_m \in \merge'} \delta(C_m, c_j) |C_m| \label{equn:cost-paid-two-merge}
    \end{align}
    
     \item The $|Y_{k,z}|$ points that are cut from $C_k$ can get merged with the subsets $Y_{\ell,z'}$ of some other cluster $C_\ell$. We call the cost of merging a subset $Y_{k,z}$ of $C_k$ with the subset $Y_{\ell,z'}$ of another cluster $C_\ell$ as $\mcostseven(M)^{c_j}$.

     Let, $Y_{k,z}$ gets further divided into multiple parts of size $\alpha_1, \ldots, \alpha_t$, so we get, 
     \begin{align}
         \mcostseven(\m{M})^{c_j} \leq  \sum_{C_k \in \m{C}} \dfrac{1}{2}\sum_{i = 1}^t\alpha_i (p_j - \alpha_i) \label{equn:cost-paid-three-merge}
     \end{align}
     \item The $|Y_{k,z}|$ points that are cut from $C_k$ can also further be split into several parts $W_1, W_2, \ldots, W_t$ (say). These parts of $Y_{k,z}$ can belong to different clusters in $\m{M}$ (output of $\pdca$) and thus would incur some cost. We call this cost as $\mcosteight(M^{c_j})$.
     \begin{align}
         \mcosteight(\m{M})^{c_j} =  \sum_{C_k \in \m{C}} \frac{1}{2} \sum_{j = 1}^t |W_j|(|W_{i,z}| - |W_j|)
     \end{align}
\end{itemize}

There is a cost which can occur in both the $\cutcase$ and $\mergecase$.
\begin{itemize}
    \item Suppose for a cluster $C_k \in \m{C}$, deficit of $c_j$ and another color $c_r \in \chi$ is filled up by the subsets of some other clusters $C_\ell$ and $C_m$ in $\m{C}$ respectively such that $\ell \neq m$ then this would incur some cost which is the number of pairs $(u,v)$ such that $u \in \delta(C_k, c_j)$ and $v \in \delta(C_k, c_r)$.
     \begin{align}
         \mcostnine(\m{M})^{c_j} = \sum_{C_k \in D} |\delta(C_k, c_j)| |\delta(C_k, c_r)|
    \end{align}
\end{itemize}

Let us define $\pay(\m{M})^{c_j}$ as the number of pairs $(u,v)$ such that at least one of $u$ and $v$ is colored $c_j$ and the following conditions are true.
\begin{itemize}
    \item $u$ and $v$ are present in the same cluster in $\m{C}$ but in separate clusters in $\m{M}$.
    \item $u$ and $v$ are present in the separate clusters in $\m{C}$ but in the same cluster in $\m{M}$.
\end{itemize}

It is straightforward to see that,
\[
    \pay{\m{M}}^{c_j} \leq \sum_{i = 1}^9 \text{cost}_i(\m{M})^{c_j}
\]
and
\begin{align}
    \dist(\m{C}, \m{M}) = \sum_{c_j \in \chi} \pay{\m{M}}^{c_j} \label{eq:main-equation-multiple-of-p}
\end{align}
Now, to prove the above \cref{lem:main-multiple-of-p}, we take the help of the following claims from \cite{chakraborty2025towards}.

\begin{claim}\cite{chakraborty2025towards} \label{clm:one-chakraborty}
    $\mcostone(\m{M})^{c_j} + \mcosttwo(\m{M})^{c_j} + \mcostthree(\m{M})^{c_j} + \mcostfour(\m{M})^{c_j} \leq 3.5 \dist(\m{C}, \m{M}^*)$
\end{claim}

\begin{claim}\cite{chakraborty2025towards} \label{clm:two-chakraborty}
    $\mcostfive(\m{M})^{c_j} + \mcostsix(\m{M})^{c_j} + \mcostseven(\m{M})^{c_j} + \mcosteight(\m{M})^{c_j} \leq 3 \dist(\m{C}, \m{M}^*)$
\end{claim}

\begin{claim}\cite{chakraborty2025towards} \label{clm:three-chakraborty}
    $\mcostnine(\m{M})^{c_j} \leq \dist(\m{C}, \m{M}^*)$
\end{claim}

Now we complete the proof of \cref{lem:main-multiple-of-p}

    By \cref{clm:one-chakraborty}, \cref{clm:two-chakraborty}, \cref{clm:three-chakraborty} and \cref{eq:main-equation-multiple-of-p} we get
    \begin{align}
    \dist(\m{C}, \m{M}) &= \sum_{c_j \in \chi} 3.5 \dist(\m{C}, \m{M}^*) + 3 \dist(\m{C}, \m{M}^*) \n \\&+ \dist(\m{C}, \m{M}^*) \n \\
    &= \sum_{c_j \in \chi} 7.5 \dist(\m{C}, \m{M}^*) \n \\
    &= O(|\chi|) \dist(\m{C}, \m{M}^*) \n
\end{align}
\end{proof}

Now we are ready to prove \cref{thm:closest-fair-multicolor-general}.

\begin{proof}[Proof of \cref{thm:closest-fair-multicolor-general}]
To prove the theorem, we describe an algorithm $\fgen$ that proceeds in two stages. First, we apply the algorithm $\pdca$ to the input clustering $\m{C}$ to obtain a $p$-divisible clustering $\m{M}$ that is $O(|\chi|)$-close to $\m{C}$ (by \cref{lem:main-multiple-of-p}). Then, we apply the algorithm $\fmulti$ on $\m{M}$ to obtain the final fair clustering $\m{F}$, which is $O(|\chi|^{2.81})$-close to $\m{M}$ (by \cref{lem:analyze-fmulti}). By combining the guarantees from both steps, we conclude that $\m{F}$ is $O(|\chi|^{3.81})$-close $\fair$ to the input $\m{C}$.
    \begin{align}
        \dist(\m{C}, \m{F}) &\leq \dist(\m{C}, \m{M}) + \dist(\m{M}, \m{F}) \, \, \text{(triangle inequality)} \n \\
        &\leq O(|\chi|) \dist(\m{C}, \m{F}^*) + O(|\chi|^{2.81}) \dist(\m{M}, \m{F}^*) \n \\
        &\leq O(|\chi|) \dist(\m{C}, \m{F}^*) + O(|\chi|^{2.81}) (\dist(\m{C}, \m{M}) \n \\ &+ \dist(\m{C}, \m{F}^*)) \, \, \text{(triangle inequality)} \n \\
        &\leq O(|\chi| + |\chi|^{3.81} + |\chi|^{2.81}) \dist(\m{C}, \m{F}^*) \n \\
        &\leq O(|\chi|^{3.81}) \dist(\m{C}, \m{F}^*) \n
    \end{align}
    This completes the proof of \cref{thm:closest-fair-multicolor-general}.
\end{proof}
\chapter{Implications to Weighted Fair Correlation and Fair Consensus Clustering}\label{sec:implications-to-fair-correlation-and-fair-consensus-clustering}
Building on the algorithms developed in the previous sections that compute an $\alpha$-close fair clustering for a given (possibly unfair) clustering~$\m{D}$, we now show how these results can be leveraged to obtain approximation algorithms for the weighted fair correlation clustering and fair consensus clustering problems.

\section{Implication to weighted Fair Correlation Clustering}

In this section, we present an algorithm for computing an approximate weighted fair correlation clustering, assuming access to a procedure that can compute an $\alpha$-close fair clustering for any given (potentially unfair) clustering~$\m{D}$. We further provide a detailed analysis of the algorithm and establish its approximation guarantee. Specifically, we prove the following result.

\clfairtocor*

\begin{proof}
    Let us first describe the algorithm $\fcc$, which computes a fair clustering for a given weighted correlation clustering instance \( G \).

\begin{itemize}
    \item \textbf{Input:} Weighted Correlation clustering instance \( G \).
    \item \textbf{Output:} A fair clustering \( \mathcal{F} \).
    \item \textbf{Procedure:}
    \begin{enumerate}
        \item Compute a clustering \( \m{D} \) from the weighted insatnce \( G \) using a \( \rho \)-approximation algorithm to find an unfair correlation clustering on a weighted instance.
        \item Apply the closest fair clustering algorithm to \( \m{D} \) to obtain a fair clustering \( \mathcal{F} \) that is \( \alpha \)-close to \( \m{D} \).
        \item Return \( \mathcal{F} \).
    \end{enumerate}
\end{itemize}

Let us now prove that $\m{F}$ is $(\alpha + \rho + \alpha\rho)$ approximate fair correlation clustering of $G$.

We know for a weighted correlation clustering instance $G$, by definition for a clustering $\m{K}$ on $G$ we have
\begin{align*}
    \cost(\m{K}) = \sum_{u \sim_{\m{K}}v} w^-(u,v) + \sum_{u \nsim_{\m{K}}v} w^+(u,v)
\end{align*}
recall where $u \sim_{\m{K}}v$ denotes that $u$ and $v$ are together in the clustering $\m{K}$ and $u \nsim_{\m{K}}v$ denotes that they are separated in $\m{K}$.

Let $\optcorr$ be the optimal fair correlation clustering of $G$. We need to prove that

\begin{align*}
    \cost(\m{F}) \leq (\alpha + \rho + \alpha\rho) \cost(\optcorr).
\end{align*}

To show the above, we need to prove the following claims

\begin{claim}\label{clm:corr-clm-one}
    For any two arbitrary clusterings $\m{M}$ and $\m{K}$ on a vertex set $V$ we have
    \[
        \cost(\m{M}) \leq \dist(\m{M}, \m{K}) + \cost(\m{K})
    \]
\end{claim}

\begin{proof}
    We have
    \begin{align}
        \cost(\m{M}) &= \sum_{u \sim_{\m{M}} v} w^{-}(u,v) + \sum_{u \nsim_{\m{M}} v} w^+(u,v) \n \\
        &= \sum_{\substack{u \sim_{\m{M}} v \\\land u \sim_{\m{K}} v}} w^{-}(u,v) + \sum_{\substack{u \sim_{\m{M}} v \\ \land u \nsim_{\m{K}} v}} w^{-}(u,v) \n \\
        &+\sum_{\substack{u \nsim_{\m{M}} v \\ \land u \sim_{\m{K}} v}} w^+(u,v) + \sum_{\substack{u \nsim_{\m{M}} v \\ \land u \nsim_{\m{K}} v}} w^{+}(u,v) \n
    \end{align}

    Again, similarly, we have,
    \begin{align}
        \cost(\m{K}) 
        &= \sum_{\substack{u \sim_{\m{K}} v \\\land u \sim_{\m{M}} v}} w^{-}(u,v) + \sum_{\substack{u \sim_{\m{K}} v \\ \land u \nsim_{\m{M}} v}} w^{-}(u,v) \n \\
        &+\sum_{\substack{u \nsim_{\m{K}} v \\ \land u \sim_{\m{M}} v}} w^+(u,v) + \sum_{\substack{u \nsim_{\m{K}} v \\ \land u \nsim_{\m{M}} v}} w^{+}(u,v) \n
    \end{align}

    We also have,
    \begin{align}
        \dist(\m{M}, \m{K}) = \sum_{\substack{u \sim_{\m{M}} v \\\land u \nsim_{\m{K}} v}} 1 + \sum_{\substack{u \nsim_{\m{M}} v \\ \land u \sim_{\m{K}} v}} 1 \n
    \end{align}

    Hence,
    \begin{align}
        \cost(\m{K}) + \dist(\m{M}, \m{K}) &= \sum_{\substack{u \sim_{\m{M}} v \\\land u \nsim_{\m{K}} v}} (1 + w^+(u,v)) + \sum_{\substack{u \nsim_{\m{M}} v \\ \land u \sim_{\m{K}} v}} (1 + w^-(u,v)) \n \\
        &+ \sum_{\substack{u \sim_{\m{K}} v \\\land u \sim_{\m{M}} v}} w^{-}(u,v) + \sum_{\substack{u \nsim_{\m{K}} v \\ \land u \nsim_{\m{M}} v}} w^{+}(u,v) \n \\
        & \geq  \cost(\m{M}) \, \, \, \text{(since $w^+(u,v), w^-(u,v) \leq 1$)} \n
    \end{align}
\end{proof}

\begin{claim}\label{clm:corr-clm-two}
    For any two arbitrary clusterings $\m{M}$ and $\m{K}$ on a vertex set $V$ we have
    \[
        \dist(\m{M}, \m{K}) \leq \cost(\m{M}) + \cost(\m{K})
    \]
\end{claim}

\begin{proof}
    We have
    \begin{align}
        \dist(\m{M}, \m{K}) = \sum_{\substack{u \sim_{\m{M}} v \\\land u \nsim_{\m{K}} v}} 1 + \sum_{\substack{u \nsim_{\m{M}} v \\ \land u \sim_{\m{K}} v}} 1 \n
    \end{align}

    \begin{align}
        \cost(\m{M}) 
        &= \sum_{\substack{u \sim_{\m{M}} v \\\land u \sim_{\m{K}} v}} w^{-}(u,v) + \sum_{\substack{u \sim_{\m{M}} v \\ \land u \nsim_{\m{K}} v}} w^{-}(u,v) \n \\
        &+\sum_{\substack{u \nsim_{\m{M}} v \\ \land u \sim_{\m{K}} v}} w^+(u,v) + \sum_{\substack{u \nsim_{\m{M}} v \\ \land u \nsim_{\m{K}} v}} w^{+}(u,v) \n
    \end{align}
    and
    \begin{align}
        \cost(\m{K}) 
        &= \sum_{\substack{u \sim_{\m{K}} v \\\land u \sim_{\m{M}} v}} w^{-}(u,v) + \sum_{\substack{u \sim_{\m{K}} v \\ \land u \nsim_{\m{M}} v}} w^{-}(u,v) \n \\
        &+\sum_{\substack{u \nsim_{\m{K}} v \\ \land u \sim_{\m{M}} v}} w^+(u,v) + \sum_{\substack{u \nsim_{\m{K}} v \\ \land u \nsim_{\m{M}} v}} w^{+}(u,v) \n
    \end{align}

    Hence,
    \begin{align}
        \cost(\m{M}) + \cost(\m{K}) 
        &= \sum_{\substack{u \sim_{\m{K}} v \\\land u \sim_{\m{M}} v}} 2w^{-}(u,v) + \sum_{\substack{u \sim_{\m{K}} v \\ \land u \nsim_{\m{M}} v}} 1 \n \\
        &+\sum_{\substack{u \nsim_{\m{K}} v \\ \land u \sim_{\m{M}} v}} 1 + \sum_{\substack{u \nsim_{\m{K}} v \\ \land u \nsim_{\m{M}} v}} 2w^{+}(u,v) \n \\
        &\text{(since, $w^+(u,v) + w^-(u,v) = 1$)} \n \\
        & \geq \dist(\m{M}, \m{K}) \n
    \end{align}
    
\end{proof}

Now we are ready to prove \cref{thm:closest-fair-to-correlation}.

\begin{align}
    \cost(\m{F}) &\leq \cost(\m{D}) + \dist(\m{D}, \m{F}) \label{ineq:one} \\
    &\leq \rho \cost(\optcorr) + \dist(\m{D}, \m{F}) \label{ineq:two} \\
    &\leq \rho \cost(\optcorr) + \alpha \dist(\m{D}, \m{F}_{\m{D}}^*) \label{ineq:three} \\
    &\leq \rho \cost(\optcorr) + \alpha \dist(\m{D}, \optcorr) \label{ineq:four} \\
    &\leq \rho \cost(\optcorr) + \alpha (\cost(\m{D}) + \cost(\optcorr))\label{ineq:five} \\
    &\leq \rho \cost(\optcorr) + \alpha (\rho\cost(\optcorr) + \cost(\optcorr)) \label{ineq:six} \\
    &= (\alpha + \rho + \alpha \rho) \cost(\optcorr). \n 
\end{align} 

\begin{itemize}
    \item The inequality \cref{ineq:one} is by \cref{clm:corr-clm-one}.
    \item The inequality \cref{ineq:two} is since $\m{D}$ is $\rho$-approximate correlation clustering.
    \item The inequality \cref{ineq:three} is since $\m{F}$ is $\alpha$-approximate closest fair to $\m{D}$.
    \item The inequality \cref{ineq:four} is since $\optcorr$ is also a fair clustering.
    \item The inequality \cref{ineq:five} is by \cref{clm:corr-clm-two}.
    \item The inequality \cref{ineq:six} is again since $\m{D}$ is $\rho$-approximate correlation clustering.
\end{itemize}
\end{proof}


\section{Implication to Fair Consensus Clustering}

In this section, we present an algorithm for computing an approximate fair consensus clustering, assuming access to a procedure that can compute an $\alpha$-close fair clustering for any given (potentially unfair) clustering~$\m{D}$. We also provide a detailed analysis of the algorithm and establish its approximation guarantee. Specifically, we prove the following result. 

\clfairtocon*

\begin{proof}
    Let us consider the following algorithm: Suppose we are given $m$ clusterings $\inpset_1,\ldots,\inpset_m$. Then, for each $\inpset_i$, compute an $\alpha$-close $\fair$ $\m{F}_i$. Then, output the $\fair$ $\m{F}_k$ that minimizes the $\ell$-mean objective function, i.e.,
    \[
    \m{F}_k = \arg \min_{i} \left(\sum_{j=1}^m \left(\dist(\inpset_j, \m{F}_i)\right)^\ell \right)^{1/\ell}.
    \]

 It remains to argue that the above algorithm achieves $(\alpha + 2)$-approximation to the fair consensus clustering problem.

Let $\optcon$ be an (arbitrary) optimal fair consensus clustering, and let $\inpset_{i^*}$ be a closest (breaking ties arbitrarily) clustering among $\inpset_1,\inpset_2,\cdots,\inpset_m$, to $\optcon$. We emphasize that we consider $\inpset_{i^*}$ only for the sake of the analysis. Since $\optcon$ is a $\fair$, we get
\begin{align}
    \label{eq:closest-to-opt}
    \dist(\inpset_{i^*},\m{F}_{i^*}) & \le \alpha \cdot \dist(\inpset_{i^*},\optcon).
\end{align}

We next derive the following

\begin{align*}
    &\left(\sum_{j=1}^m \left(\dist(\inpset_j, \m{F}_{i^*})\right)^\ell \right)^{1/\ell}\\
    & \le \left(\sum_{j=1}^m \left(\dist(\inpset_j,\optcon) + \dist(\optcon ,\m{F}_{i^*}) \right)^\ell \right)^{1/\ell} &&\text{(By the triangle inequality)}\\
    & \le \left(\sum_{j=1}^m \left(\dist(\inpset_j,\optcon) + \dist(\optcon ,\inpset_{i^*}) + \dist(\inpset_{i^*} ,\m{F}_{i^*}) \right)^\ell \right)^{1/\ell} &&\text{(By the triangle inequality)}\\
    & \le \left(\sum_{j=1}^m \left(\dist(\inpset_j,\optcon) + \dist(\optcon ,\inpset_{i^*}) + \alpha \cdot \dist(\inpset_{i^*},\optcon) \right)^\ell \right)^{1/\ell}&&\text{(By \cref{eq:closest-to-opt})}\\
    & \le \left(\sum_{j=1}^m \left(\dist(\inpset_j,\optcon) + \dist(\inpset_j,\optcon) + \alpha \cdot \dist(\inpset_j,\optcon) \right)^\ell \right)^{1/\ell} &&\text{(Since $\inpset_{i^*}$ is closest to $\optcon$)}\\
    &=(\alpha + 2) \left(\sum_{j=1}^m \left(\dist(\inpset_j,\optcon)\right)^\ell \right)^{1/\ell}.
\end{align*}

Now, since our algorithm outputs $\m{F}_k$ that minimizes the $\ell$-mean objective function, the approximation guarantee follows, and that concludes the proof.
\end{proof}

\chapter{Fair Consensus Clustering in Streaming setting}\label{sec:streaming}

In this chapter, we present an algorithm $\stalgo$ for solving fair consensus clustering in a streaming setting. Recall in the fair consensus clustering problem, the input is the set of clusterings $\setofclusterings = \{C_1, \ldots, C_m\}$ on a set of vertices $V$ (with $|V| = n$) and our goal is to output a fair clustering $\m{F}$, such that

\[
     \left( \sum_{i = 1}^m \dist(\m{C}_i, \m{F})^\ell\right)^{1/\ell} \leq  \beta \left( \sum_{i = 1}^m \dist(\m{C}_i, \optcon)^\ell\right)^{1/\ell}
\]
where $\optcon$ is an optimal fair consensus clustering, the distance, $\dist(\cdot, \cdot)$ between two clusterings is the measured by the number of pairs $(u,v)$ that are together in one clustering but separated by the other, and $\beta \in \mathbb{R}^{>1}, \ell \in \mathbb{Z}^{>1}$ are some parameters. 

Let us now describe the streaming model. We consider an \emph{insertion-only} streaming model. The input is a sequence of triples $(p,j,b)$, where:
\begin{itemize}
  \item $p=(u,v)$ is an unordered pair of vertices in $V$,
  \item $j\in[m]$ identifies a clustering $\m{C}_j\in\setofclusterings$, and
  \item $b\in\{0,1\}$ indicates whether $u$ and $v$ lie in the same cluster in the clustering $\m{C}_j$ ($b=0$) or in different clusters ($b=1$).
\end{itemize}

Here, a triple $((u,v),j,b)$ refers an information about the clustering $\m{C}_j$. For each clustering $\m{C}_j$, we have $n^2$ such tuples, hence the length of the stream is $mn^2$. These triples $(p,j,b)$ can arrive in any arbitrary order. Let us name the above streaming model as $\pconstream$. Our main result is stated in the following theorem.

\streaming*

\section{Details of the algorithm: $\stalgo$}
    In this section, we present our streaming algorithm, $\stalgo$ for \emph{fair consensus clustering}. The algorithm relies on a subroutine, denoted by~$\fc$, which we describe first.

The subroutine~$\fc$ takes as input a set of clusterings~$\setofclusterings$ and outputs a set of candidate clusterings~$\canset$. The crucial guarantee is that~$\canset$ contains at least one fair clustering~$\m{F}$ that achieves a good approximation. The subroutine~$\fc$ assumes access to an $\alpha$-close fair clustering algorithm.

The candidate set~$\canset$ is constructed as follows. For each clustering~$\m{C}_i \in \setofclusterings$, we include its $\alpha$-close fair clustering~$\m{F}_{\m{C}_i}$ in~$\canset$. In addition, for every ordered triple
\[
    \m{T} = \{\m{C}_i, \m{C}_j, \m{C}_k\}
    \in \setofclusterings \times \setofclusterings \times \setofclusterings,
\]
we add to~$\canset$ a clustering~$\m{T}_{i,j,k}$ obtained by applying the subroutine~$\clsfitting$ to~$\m{T}$.

The pseudocodes of~$\fc$ and~$\clsfitting$ are given in
\cref{alg:streaming-fair} and~\cref{alg.cluster.fitting}, respectively.

\begin{algorithm}
\caption{$\fc(\setofclusterings)$}
\label{alg:streaming-fair}
\KwIn{A set $\setofclusterings$ of clusterings}
\KwOut{A set $\canset$ of candidate fair clusterings}
\BlankLine
Initialize an empty set $ \canset $\;
    \For{each $ \m{C}_i \in \setofclusterings  $}{
        $ \m{F}_i \gets $ a $ \approxFactorClsFair $-close fair clustering to $ \concls{i} $ \;
        $ \canset \gets \canset\cup \{\m{F}_i\} $ \;
    }
    
    \For{each $ \tripleconcls = \{\m{C}_i, \m{C}_j, \m{C}_k\} $ of $ \setofclusterings $}{
        $ \fairtriple \gets \clsfitting{ \tripleconcls } $\;
        $ \canset \gets \canset\cup \{\fairtriple\} $
    }
    \KwRet{$\canset$}
\end{algorithm}

\begin{algorithm}[htbp]
    \SetKwInOut{KwIn}{Input}
    \SetKwInOut{KwOut}{Output}
    \KwIn{Clusterings $ T = \{\concls{i}, \concls{j}, \concls{k}\} $}
    \KwOut{A fair clustering $ \fairtriple $}
        $ V(G)\gets V $

        $ E^{+}(G) \gets \{(a,b)| a,b \in V,\ a \text{ and } b \text{ are together in at least } 2 \text{ clusterings}\} $

        $ E^{-}(G) \gets \{(a,b)| a,b \in V,\ a \text{ and } b \text{ are separated in at least } 2 \text{ clusterings}\} $

        $ \fairtriple\gets $ a $ \approxFactorFairCorCls $-approximation fair correlation clustering of $ G = \left(V(G), E^{+}(G)\cup E^{-}(G)\right) $.

        \KwRet{$ \fairtriple $}
        \caption{$ \clsfitting{T} $}
    \label{alg.cluster.fitting}
\end{algorithm}
    
    Now we are ready to describe the algorithm $\stalgo$. We work in the insertion-only stream described previously.
The algorithm maintains two memory structures in parallel:
\begin{enumerate}
    \item \textbf{Sampled Store $1$ ($\mathsf{M}_1$):}
          Prior to the arrival of the stream, we independently sample 
          $4g \log m$ indices $j_1, \ldots, j_{4g \log m}$ from $[m]$ for some parameter $g$ we fix the value of it later. 
          For these sampled indices, all their corresponding triples that appear in the stream 
          are stored in $\mathsf{M}_1$. 

    \item \textbf{Sampled Store $2$ ($\mathsf{M}_2$):}
          Similarly, before the stream starts, we sample 
          $64\varepsilon^{-2}\log m$ indices 
          $k_1, \ldots, k_t$ from $[m]$, where $t = 64\varepsilon^{-2}\log m$ 
          and $0 < \varepsilon \leq 1$. 
          For these sampled indices, all their triples are stored in $\mathsf{M}_2$ as they appear in the stream.
\end{enumerate}

After the stream ends.
\begin{enumerate}
    \item We use union-find, to construct the clusterings $\m{C}_{j_1}, \ldots, \m{C}_{j_{\log m}}$ from its triples stored in $\mathsf{M}_1$. Then we apply a subroutine $\fc$ (\cref{alg:streaming-fair}) on these $\lceil\log m \rceil$ clusterings to get a candidate set of fair clusterings $\widetilde{\m{F}}$. The subroutine $\fc$ assumes access to a $\alpha$-close fair clustering algorithm.
    \item We use union-find, to construct the clusterings $\m{C}_{k_1}, \ldots, \m{C}_{k_t}$ from its triples stored in $\mathsf{M}_2$. Let, $\m{W} = \{\m{C}_{k_1}, \ldots, \m{C}_{k_t}\}$.

    We use $\m{W}$ to find the best fair clustering $\m{F} \in \widetilde{\m{F}}$ with respect to $\m{W}$, more specifically we find
    \[
       \m{F} = \argmin_{\m{F}' \in \canset} \sum_{\m{C}_i \in \m{W}}  \dist(\m{C}_i, \m{F}')
    \]
    \item Return $\m{F}$.
\end{enumerate}

We provide the pseudocode of $\stalgo$ in the appendix.

\section{Analysis of $\stalgo$}
\label{sec:analysis}

In this section, we analyze the approximation guarantee of the algorithm $\stalgo$, along with its time and space complexities.

\subsection{Approximation Guarantee}
\label{subsec:approx-guarantee}

We now establish the approximation guarantee of $\stalgo$ and thus prove \cref{thm:explicit.fair.consensus.clustering.randomized}. The \cref{thm:explicit.fair.consensus.clustering.randomized} is obtained by appropriately instantiating the parameters in the following more general lemma.

\begin{lemma}
\label{lem:fair.consensus.clustering.randomized}
    Suppose there exists an algorithm with running time $\runtimeClsFair{n}$ that, given a clustering over $n$ points, computes an $\approxFactorClsFair$--close fair clustering. 
    Further, assume there exists an $\approxFactorFairCorCls$--approximation algorithm for fair correlation clustering on a graph with $n$ vertices, running in time $\runtimeFairCorCls{n}$.
    
    Let $\eta > 0$, $0 < \beta < 1$, $c > 1$, $s > 1$, and $g > s$ be fixed parameters satisfying $sc - s - 2c > 0$. 
    Then, for any collection of $m$ clusterings over $n$ points, there exists a randomized algorithm that, with probability at least $1 - \frac{1}{m}$, runs in time
    \[
        O\!\left(
            n^{2}\log^{4} m
            + \runtimeFairCorCls{n}\log^{3} m
        \right),
    \]
    uses $O(n \log m)$ space, and outputs a
    $\left(\frac{1+\errCoreset}{1-\errCoreset} \cdot r \right)$--approximate fair consensus clustering, where
    {\small
    \begin{align}
        r = \max \Biggl\{ 
        &\approxFactorClsFair + 2 - (\approxFactorClsFair + 1)\beta, \nonumber \\
        &\approxFactorClsFair + 2 
        + (\approxFactorClsFair + 1)\frac{\beta(g - s) + s}{g(s - 1)}
        - \frac{2\eta}{c}, \nonumber \\
        &\approxFactorClsFair + 2 
        + (\approxFactorClsFair + 1)\frac{\beta(g(2c + s) - sc) + sc}{g(cs - s - 2c)}
        - 2\!\left(1 - \frac{1}{s} - \frac{1}{c}\right)\!\eta, \nonumber \\
        &1 + 3(\approxFactorFairCorCls + 1)\eta
        \Biggr\}. \nonumber
    \end{align}
    }
\end{lemma}

We now focus on proving \cref{lem:fair.consensus.clustering.randomized}. 
To this end, we first introduce the necessary notation. For any clustering~$\m{D}$, we define its objective value as
\[
    \obj{\inpconclss, \m{D}}
    \;=\;
    \left(
        \sum_{\m{C}_i \in \inpconclss}
        \dist(\m{C}_i, \m{D})^{\ell}
    \right)^{\frac{1}{\ell}},
\]
where~$\ell > 1$ is a fixed constant.

Fix an optimal fair clustering~$\optconcls$. For notational convenience, we define
\[
    \optconval := \obj{\inpconclss, \optconcls}
    \quad\text{and}\quad
    \avgconval := \frac{\optconval}{m}.
\]

For each clustering~$\concls{i}$, let~$\unalignedSet{i}$ denote the set of unordered vertex pairs~$(u,v) \in V \times V$ such that $u$ and $v$ are clustered together (respectively, separated) in~$\optconcls$, but are separated (respectively, clustered together) in~$\concls{i}$. 
By definition, this implies
\[
    \dist(\concls{i}, \optconcls)
    \;=\;
    \card{\unalignedSet{i}}.
\]

Let us also define
    \begin{align}
    S_{h} = \{\concls{j}| \card{\unalignedSet{h}\cap \unalignedSet{j}} > \eta\avgconval\}.\nonumber
\end{align}
for some parameter $\eta > 0$.

We now prove the following lemma.

\begin{lemma}
\label{lem:if.unaligned.is.small}
    If there exists a clustering $\concls{i}$ such that
    $\card{\unalignedSet{i}} \leq (1-\beta)\avgconval$,
    then
    \[
        \obj{\inpconclss, \fairconcls{i}}
        \;\leq\;
        \bigl(\approxFactorClsFair + 2 - (\approxFactorClsFair + 1)\beta\bigr)\optconval,
    \]
    where $\fairconcls{i}$ denotes an $\approxFactorClsFair$-approximate closest fair clustering to $\concls{i}$ returned by the closest fair clustering algorithm.
\end{lemma}

\begin{proof}
    By the triangle inequality, we have
    \begin{align}
        \obj{\inpconclss, \fairconcls{i}}
        &= \sum_{j=1}^{m} \dist(\concls{j}, \fairconcls{i}) \nonumber \\
        &\leq \sum_{j=1}^{m}
        \Bigl(
            \dist(\concls{j}, \optconcls)
            + \dist(\optconcls, \concls{i})
            + \dist(\concls{i}, \fairconcls{i})
        \Bigr) \nonumber \\
        &\leq \optconval
        + m \dist(\optconcls, \concls{i})
        + m \approxFactorClsFair \dist(\optconcls, \concls{i}) \nonumber \\
        &\leq \optconval
        + (1+\approxFactorClsFair)(1-\beta)m\avgconval \nonumber \\
        &= \bigl(\approxFactorClsFair + 2 - (\approxFactorClsFair + 1)\beta\bigr)\optconval. \nonumber
    \end{align}
    The second inequality follows from the fact that $\fairconcls{i}$ is an
    $\approxFactorClsFair$-close fair clustering to $\concls{i}$, and the final
    inequality uses the assumption
    $\dist(\optconcls, \concls{i}) = \card{\unalignedSet{i}} \leq (1-\beta)\avgconval$.
\end{proof}

Now suppose there exists a constant $g > 1$ such that at least $\frac{m}{g}$ clusterings
$\concls{i}$ satisfy $\card{\unalignedSet{i}} \leq (1-\beta)\avgconval$ for some $0 < \beta < 1$.
Then, with high probability, our algorithm samples at least one such clustering.
Let $\fairconcls$ be an $\approxFactorClsFair$-close fair clustering corresponding
to a sampled clustering of this type. By \cref{lem:if.unaligned.is.small}, we obtain
\[
    \obj{\inpconclss, \fairconcls}
    \;\leq\;
    \bigl(\approxFactorClsFair + 2 - (\approxFactorClsFair + 1)\beta\bigr)\optconval.
\]

Hence, in the remaining part of the analysis, we assume that the number of clusterings $ \concls{i} $ with $ \card{\unalignedSet{i}}\leq (1-\beta)\avgconval $ is less than $ \frac{m}{g} $. More specifically, we have the following ordering
\begin{align}
    \card{\unalignedSet{1}}\leq \card{\unalignedSet{2}}\leq \dots \card{\unalignedSet{t}} \leq (1-\beta)\avgconval < \card{\unalignedSet{t+1}}\leq \dots \leq \card{\unalignedSet{m}}, \label{eq.ordering.unalignedSet}
\end{align}
where $ t < \frac{m}{g} \leq t+1 $.

We prove the following lemmas under the assumption~\eqref{eq.ordering.unalignedSet}.

\begin{lemma}
\label{lem:extended.S1.small.Sl.large}
    Suppose that~\eqref{eq.ordering.unalignedSet} holds. For any
    $t < h \leq \frac{m}{s}$, where $s$ is a constant satisfying $s < g$, we have
    \[
        \card{\unalignedSet{h}}
        \;\leq\;
        \left(
            1 + \frac{\beta(g - s) + s}{g(s - 1)}
        \right)\avgconval.
    \]
\end{lemma}

\begin{proof}
    Since $\optconval = m \avgconval$ is the sum of all $\card{\unalignedSet{i}}$,
    we have
    \begin{align}
        m\avgconval
        &= \sum_{i=1}^{m} \card{\unalignedSet{i}} \nonumber \\
        &\geq \sum_{i=t+1}^{h-1} \card{\unalignedSet{i}}
        + \sum_{i=h}^{m} \card{\unalignedSet{i}} \nonumber \\
        &> (h - t - 1)(1 - \beta)\avgconval
        + (m - h + 1)\card{\unalignedSet{h}}, \nonumber
    \end{align}
    where the last inequality follows from~\eqref{eq.ordering.unalignedSet}.

    Rearranging terms yields
    \begin{align}
        \card{\unalignedSet{h}}
        &<
        \frac{m - (1 - \beta)(h - t - 1)}{m - h + 1}\avgconval \nonumber \\
        &=
        \left(
            1
            + \frac{\beta(h - 1)}{m - h + 1}
            + \frac{t(1 - \beta)}{m - h + 1}
        \right)\avgconval. \nonumber
    \end{align}

    Since $h \leq \frac{m}{s}$, we have
    \[
        \frac{\beta(h - 1)}{m - h + 1}
        = \frac{\beta}{\frac{m}{h - 1} - 1}
        \;\leq\;
        \frac{\beta}{s - 1}.
    \]
    Moreover, using $t < \frac{m}{g}$, we obtain
    \[
        \frac{t}{m - h + 1}
        <
        \frac{m/g}{m - m/s + 1}
        \;\leq\;
        \frac{s}{g(s - 1)}.
    \]
    Since $0 < \beta < 1$, it follows that
    \[
        \frac{t(1 - \beta)}{m - h + 1}
        <
        \frac{s(1 - \beta)}{g(s - 1)}.
    \]

    Combining the above bounds, we conclude that
    \begin{align}
        \card{\unalignedSet{h}}
        &<
        \left(
            1
            + \frac{\beta}{s - 1}
            + \frac{s(1 - \beta)}{g(s - 1)}
        \right)\avgconval \nonumber \\
        &=
        \left(
            1
            + \frac{\beta(g - s) + s}{g(s - 1)}
        \right)\avgconval. \nonumber
    \end{align}
\end{proof}

\begin{lemma}\label{lem:Sh.large.is.good}
    Suppose that~\eqref{eq.ordering.unalignedSet} holds. If there is a clustering $ \concls{h} $ with $ t< h\leq \frac{m}{s} $ satisfying $ \card{S_{h}} \geq \frac{m}{c} $ for some constant $c > 1$, then
    \begin{align}
        \obj{\inpconclss, \fairconcls{h}} \leq \left( \approxFactorClsFair + 2 + \left(\approxFactorClsFair+1\right)\dfrac{\beta(g-s)+s}{g(s-1)} - \dfrac{2 \eta}{c}\right)\optconval.\nonumber
    \end{align}
\end{lemma}
\begin{proof}
    As~\eqref{eq.ordering.unalignedSet} holds, applying~\cref{lem:extended.S1.small.Sl.large} with $ h\leq \frac{m}{s} $, we obtain $ \card{\unalignedSet{h} }\leq (1 + \frac{\beta(g-s)+s}{g(s-1)})\avgconval $.
    \begin{align}
        \obj{\inpconclss, \fairconcls{h} } &= \sum_{i=1}^{m} \dist(\concls{i}, \fairconcls{h}) \nonumber \\ 
        &\leq \sum_{i=1}^{m}\left(\dist(\concls{i},\concls{h}) + \dist(\concls{h}, \fairconcls{h})\right) \nonumber \\ &\leq \sum_{i=1}^{m}\left( \card{\unalignedSet{i} } + (\approxFactorClsFair+1 )\card{\unalignedSet{h} } - 2\card{\unalignedSet{i}\cap \unalignedSet{h} }\right) \nonumber \\
                                           &\leq \optconval + (\approxFactorClsFair+1 )\left(1+\dfrac{\beta(g-s)+s}{g(s-1)}\right)\optconval - \card{S_{h}} 2 \eta\avgconval \nonumber \\
                                           &\leq \left( \approxFactorClsFair + 2 + (\approxFactorClsFair+1 )\dfrac{\beta(g-s)+s}{g(s-1)} - \dfrac{2 \eta}{c} \right) \optconval. \nonumber
    \end{align}
\end{proof}

We define
\begin{align}
    \neigh{h} := \{\concls{j}| h< j \leq h + \dfrac{m}{c} + \dfrac{m}{s}:\ \card{\unalignedSet{j}\cap \unalignedSet{h} } \leq \eta\avgconval \}, \nonumber \\
    \farNeigh{h,k} := \{ \concls{j}| \card{\unalignedSet{j}\cap \unalignedSet{h} } \leq \eta\avgconval \text{ and } \card{\unalignedSet{j}\cap \unalignedSet{k} }\leq \eta\avgconval \}.\nonumber
\end{align}

\begin{lemma}\label{lem:far.neigh.small.is.good}
    Suppose that~\eqref{eq.ordering.unalignedSet} holds. If for some $ h\leq \frac{m}{s} $ with $ \card{S_{h}}< \frac{m}{c} $, there exists a clustering $ \concls{k}\in \neigh{h} $ such that $ \card{\farNeigh{h,k}} < \frac{m}{s} $, then
    { \scriptsize
    \begin{align}
        \obj{\inpconclss, \fairconcls{k} } \leq \left( \approxFactorClsFair + 2 + (\approxFactorClsFair + 1 )\dfrac{\beta(g(2c+s)-sc)+sc)}{g(cs-s-2c)} - 2\left( 1- \dfrac{1}{s} - \dfrac{1}{c}  \right)\eta \right) \optconval. \nonumber
    \end{align} 
    }
\end{lemma}
\begin{proof}
    With $ \concls{k}\in \neigh{h} $, by definition of $ \neigh{h} $, it follows that $ k\leq h + \frac{m}{c} + \frac{m}{s}\leq \frac{m}{sc/(2c+s)} $, where the last inequality is obtained by using the assumption $ h\leq \frac{m}{s} $. Applying~\cref{lem:extended.S1.small.Sl.large} with $ k \leq \frac{m}{sc/(2c+s)} $, we get $ \card{\unalignedSet{k}}\leq \left( 1+ \frac{\beta(g(2c+s)-sc)+sc}{g(sc-s-2c)} \right) \avgconval $.

    We now turn to give a lower bound for $ \card{S_{k}} $. Note that by definition of $ \farNeigh{h,k} $, for every clustering $ \concls{j}\in \inpconclss $, if $ \concls{j}\notin \farNeigh{h,k} $, then at least one of $ \card{\unalignedSet{j}\cap \unalignedSet{h} } > \eta\optconval $ or $ \card{\unalignedSet{j}\cap \unalignedSet{k} }> \eta\optconval $ must hold. In other words, it must be the case that $ \concls{j}\in S_{h} $ or $ \concls{j}\in S_{k} $ (or both) holds. Hence, $ \card{S_{h}} + \card{S_{k}} \geq \card{\inpconclss } - \card{\farNeigh{h,k}} $. Using the assumptions $ \card{S_{h}} < \frac{m}{c} $ and $ \card{\farNeigh{h,k} } < \frac{m}{s} $, we obtain $ \card{S_{k}}\geq m - \frac{m}{c} - \frac{m}{s} $.

    Finally, using $ \card{\unalignedSet{k} } \leq \left(1 +  \frac{\beta(g(2c+s)-sc)+sc}{g(sc-s-2c)}\right)\avgconval $ and $ \card{S_{k} }\geq m - \frac{m}{c} - \frac{m}{s} $, we have
    { \small
    \begin{align}
        &\>\>\>\>\>\obj{\inpconclss, \fairconcls{k} }  \nonumber \\
        &\leq \sum_{i=1}^{m}(\card{\unalignedSet{i} } + (1+\approxFactorClsFair )\card{\unalignedSet{k} } - 2\card{\unalignedSet{i}\cap \unalignedSet{k} }) \nonumber \\ 
                                           &\leq \optconval + (\approxFactorClsFair+1 )\left(1+ \dfrac{\beta(g(2c+s)-sc)+sc}{g(cs-s-2c)}\right)\optconval - \card{S_{k}}2 \eta \avgconval \nonumber \\ 
                                           &\leq \left(\approxFactorClsFair + 2 + (\approxFactorClsFair+1 ) \dfrac{\beta(g(2c+s)-sc)+sc}{g(cs-s-2c)} - 2 \left(1 - \dfrac{1}{s} - \dfrac{1}{c} \right) \eta \right) \optconval. \nonumber
    \end{align} 
    }
\end{proof}

\begin{lemma}\label{lem:cluster.fitting.helps}
    If there is a set $ T = \{\concls{i}, \concls{j}, \concls{k}\} $ such that, for all $ r\neq s \in \{i,j,k\} $, we have $ \card{\unalignedSet{r} \cap \unalignedSet{s}} \leq \eta\avgconval $, then $ \obj{\inpconclss, \fairtriple} \leq (1 + 3(\approxFactorFairCorCls+1)\eta)\optconval $.
\end{lemma}
\begin{proof}
    We define a set of bad pairs
    \begin{align}
        B = (\unalignedSet{i} \cap \unalignedSet{j}) \cup (\unalignedSet{j} \cap \unalignedSet{k}) \cup (\unalignedSet{k} \cap \unalignedSet{i}), \nonumber
    \end{align}
    that is, $ B $ contains all pairs of points $ (a,b) $ such that $ a,b $ are together (resp.\ separated) in a majority of clusterings in $ T $, while in $ \optconcls $, $ a $ and $ b $ are separated (resp.\ together). 

    Note that $ \card{B} \leq 3\eta\avgconval $, as $ \card{\unalignedSet{r} \cap \unalignedSet{s}} \leq \eta\avgconval $, for all $ r\neq s \in \{i,j,k\} $.

    Let $ G $ be a graph with vertex set $ V $ and the edge set $ E(G) = E^{+}(G)\cup E^{-}(G) $, where
    \begin{align}
    E^{+}(G) &= \left\{(a,b)\mid a,b\in V,\, 
        a \text{ and } b \text{ are together} \right. \nonumber\\
             &\quad \left. \text{in at least } 2 \text{ clusterings among } \concls{i}, \concls{j}, \concls{k}\right\}, \text{ and} \nonumber\\
    E^{-}(G) &= \left\{(a,b)\mid a,b\in V,\, 
        a \text{ and } b \text{ are separated} \right. \nonumber\\
             &\quad \left. \text{in at least } 2 \text{ clusterings among } \concls{i}, \concls{j}, \concls{k} \right\}.\nonumber
\end{align}
    Observe that for each pair of vertices $ (a,b) $, if $ (a,b)\notin B $, then $ a,b $ are together (resp.\ separated) in $ \optconcls $ if $ (a,b) $ is an edge in $ E^{+}(G) $ (resp.\ $ E^{-}(G) $). On the other hand, if $ (a,b)\in B $, then $ a,b $ are separated (resp.\ together) if $ (a,b) $ is an edge in $ E^{+}(G) $ (resp.\ $ E^{-}(G) $). Therefore, if for every pair $ (a,b)\in B $, we flip their labels (from ``$+$'' to  ``$-$'' and vice versa) we obtained a new graph $ G' $ such that, any two vertices in a same cluster in $ \optconcls $ are connected by an ``$+$'' edge, and any two vertices is different clusters in $ \optconcls $ are connected by a ``$-$'' edge. Note that, the correlation cost of a clustering $ \concls $ of $ G $ can be viewed as the number of pairs $ (a,b) $ whose labels need to be flipped so that, the edges inside every cluster of $ \concls $ are all ``$+$'' edges, and the edges between the clusters of $ \concls $ are all ``$-$'' edges. Hence, $ \optconcls $ has a  correlation cost $ \costcor{\optconcls} = \card{B} $.

    Let $ \fairtriple $ be the output of $ \clsfitting{\tripleconcls} $ (\cref{alg.cluster.fitting}). In other words, $ \fairtriple $ is a clustering obtained by running a $ \approxFactorFairCorCls $-approximation fair correlation clustering algorithm on $ G $. Then $ \costcor{\fairtriple}\leq \approxFactorFairCorCls\card{B} $, as the cost of an optimal fair correlation clustering is at most $ \card{B} $. Moreover, the number of pairs $ (a,b) $ that are clustered differently between $ \fairtriple $ and $ \optconcls $ is at most, every pair in $ B $ and the $ \approxFactorFairCorCls\card{B} $ pairs that are flipped by the approximation fair correlation clustering algorithm. As a result, we get

    \begin{align}
        \dist(\fairtriple, \optconcls) &\leq \card{B} + \approxFactorFairCorCls\card{B} \nonumber \\
                                          &\leq (1+\approxFactorFairCorCls)3 \eta\avgconval &( \text{as }\card{B}\leq 3 \eta\avgconval ). \nonumber
    \end{align}
    It remains to bound the objective value of $ \fairtriple $.
    \begin{align}
        \obj{\inpconclss, \fairtriple} &= \sum_{i=1}^{m}\dist(\concls{i}, \fairtriple) \nonumber \\
                                       &\leq \sum_{i=1}^{m}(\dist(\concls{i}, \optconcls) + \dist(\optconcls, \fairtriple) ) \nonumber \\
                                       &\leq \optconval + m(1+\approxFactorFairCorCls)3 \eta\avgconval \nonumber \\
                                       &= (1+3(\approxFactorFairCorCls+1) \eta)\optconval. \nonumber
    \end{align}
\end{proof}

\begin{lemma}\label{lem:cls.fitting.helps}
    Suppose that the algorithm $\stalgo$ sampled two clusterings $ \concls{h} $ and $ \concls{k} $ such that $ k \in \neigh{h} $ and $ \card{\farNeigh{h,k}} \geq \frac{m}{s} $. Then, with probability at least $ 1 - m^{-4} $, the algorithm $\stalgo$ also sampled a clustering $ \concls{\ell} $ such that, for $ \fairtriple = \clsfitting{\{\concls{h}, \concls{k}, \concls{\ell}\}} $, the following holds:
    \begin{align}
        \obj{\inpconclss, \fairtriple} \leq \left( 1 + 3(\approxFactorFairCorCls + 1)\eta \right)\optconval. \nonumber
    \end{align}
\end{lemma}
\begin{proof}
    Since $ \card{\farNeigh{h,k}}\geq \frac{m}{s} $, the probability that at least one clustering $ \concls{\ell} $ in $ \farNeigh{h,k} $ is sampled is at least
    \begin{align}
        1 - \left(1 - \dfrac{4g\log m}{m} \right)^{\frac{m}{s}} \geq 1 - e^{-4g\log m/s} = 1 - \dfrac{1}{m^{4g/s}} \geq 1 - m^{-4}. \nonumber
    \end{align}
    By definition of $ \neigh{h} $ and $ \farNeigh{h,k} $, for all $ r,s\in \{h,k,\ell\} $, we have $ \card{\unalignedSet{r}\cap \unalignedSet{s} }\leq \eta\avgconval $. Applying~\cref{lem:cluster.fitting.helps}, we obtain $ \obj{\inpconclss, \fairtriple }\leq (1+3(\approxFactorFairCorCls+1 ) \eta)\optconval $.
\end{proof}

Now, let us prove \cref{lem:fair.consensus.clustering.randomized}

\begin{proof}[Proof of~\cref{lem:fair.consensus.clustering.randomized}]
    If the number of clusterings $ \concls{i} $ with $ \card{\unalignedSet{i}}\leq (1-\beta)\avgconval $ is at least $ \frac{m}{g} $, then with probability at least $ 1 - m^{-4} $, our algorithm samples at least one such clustering. Using~\cref{lem:if.unaligned.is.small} and let $ \fairconcls $ be a $ \approxFactorClsFair $-close fair clustering to such sampled clustering, we have $ \obj{\inpconclss, \fairconcls}\leq (\approxFactorClsFair + 2 - (\approxFactorClsFair+1)\beta)\optconval $.

    It remains to consider the case when the number of such clusterings $ \concls{i} $ is less than $ \frac{m}{g} $. With probability at least $ 1 - m^{-4(1-s/g)} $, our algorithm samples at least one clustering $ \concls{h} $ with $ \frac{m}{g} < h\leq \frac{m}{s} $. If $ \card{S_{h}}\geq \frac{m}{c} $, then by~\cref{lem:Sh.large.is.good}, let $ \fairconcls{h} $ be a $ \approxFactorClsFair $-close fair clustering to $ \concls{h} $, we have
    \begin{align}
        \obj{\inpconclss, \fairconcls{h}} \leq \left( \approxFactorClsFair + 2 + \left(\approxFactorClsFair+1\right)\dfrac{\beta(g-s)+s}{g(s-1)} - \dfrac{2 \eta}{c}\right)\optconval. \nonumber
    \end{align}
    If $ \card{S_{h}} < \frac{m}{c} $, then $ \card{\neigh{h}}\geq \frac{m}{s} $. It follows that with probability at least $ 1 - m^{-4g/s} $, we sample a clustering $ \concls{k}\in \neigh{h} $. If $ \card{\farNeigh{h,k}} < \frac{m}{s} $, then by~\cref{lem:far.neigh.small.is.good}, let $ \fairconcls{k} $ be a $ \approxFactorClsFair $-close fair clustering to $ \concls{k} $, we have
    { \scriptsize
    \begin{align}
        \obj{\inpconclss, \fairconcls{k} } \leq \left( \approxFactorClsFair + 2 + (\approxFactorClsFair + 1 )\dfrac{\beta(g(2c+s)-sc)+sc)}{g(cs-s-2c)} - 2\left( 1- \dfrac{1}{s} - \dfrac{1}{c}  \right)\eta \right) \optconval. \nonumber
    \end{align} 
    }
    If $ \card{\farNeigh{h,k}} \geq \frac{m}{s} $, then by~\cref{lem:cls.fitting.helps}, with probability at least $ 1 - m^{-4} $, there exists a sampled clustering $ \concls{\ell} $ such that, for $ \fairtriple = \clsfitting{\{\concls{h}, \concls{k}, \concls{\ell}\}} $, we have
    \begin{align}
        \obj{\inpconclss, \fairtriple} \leq \left( 1 + 3(\approxFactorFairCorCls + 1)\eta \right)\optconval. \nonumber
    \end{align}

    Combining all the cases, with probability at least $ 1-m^{-3} $, there is a fair clustering $ \fairconcls\in \widetilde{\m{F}} $ such that $ \obj{\inpconclss, \fairconcls}\leq r\optconval $, where 
    { \small
    \begin{align}
        r = \max \Biggl\{ &(\approxFactorClsFair + 2 - (\approxFactorClsFair+1)\beta), \nonumber \\
                        &\left( \approxFactorClsFair + 2 + \left(\approxFactorClsFair+1\right)\dfrac{\beta(g-s)+s}{g(s-1)} - \dfrac{2 \eta}{c}\right), \nonumber \\
                        &\left( \approxFactorClsFair + 2 + (\approxFactorClsFair + 1 )\dfrac{\beta(g(2c+s)-sc)+sc)}{g(cs-s-2c)} - 2\left( 1- \dfrac{1}{s} - \dfrac{1}{c}  \right)\eta \right), \nonumber\\
                        &\left( 1 + 3(\approxFactorFairCorCls+1 )\eta \right) \Biggr\} .\nonumber
    \end{align} 
    }

    As the algorithm $\stalgo$ uses a set $ \m{W} $ of $ \Omega(\varepsilon^{-2}\log m ) $ input clusterings sampled uniformly at random as a evaluation set, according to a result developed by Indyk~\cite{indyk1999sublinear},~\cite[Theorem 31]{indyk2001high}, with probability at least $ 1- \frac{1}{m} $, our algorithm outputs a clustering $ \fairconcls' $ such that $ \obj{\inpconclss, \fairconcls'} \leq (1+\varepsilon)\obj{\inpconclss, \fairconcls} $. Therefore, with probability at least $ 1 - \frac{1}{m} $, our algorithm outputs a fair clustering $ \fairconcls' $ with $ \obj{\setofclusterings, \fairconcls'} \leq (1+\varepsilon)r\optconval $.
\end{proof}

\subsection{Time and Space Complexity}

Now we analyse the time and space complexity of the algorithm $\stalgo$.

\noindent\textbf{Time Complexity.}
In $\mathsf{M}_1$ we store $\log m$ many clusterings. Hence, there are $O(\log^3 m)$ many candidate clusterings in $\canset$. Computing a $ \approxFactorClsFair $-close fair clustering to each sampled clustering takes $\runtimeClsFair{n}\log m$ time and running the procedure $ \clsfitting $ on $ \log^{3} m $ triples of clusterings takes $ \runtimeFairCorCls{n}\log^{3} m $ time. To find the best clustering in $\canset$, we need to perform $O(\log^3 m)$ many checks. Now, in each check, we need to compute the distance between two clusterings, which takes $O(n^2)$ time, and since there are $O(\log m)$ clusterings in the $\m{W}$, the total time needed per check is $O(n^2 \log m)$. Hence, overall for all checks, the time needed is $O(n^2 \log^4 m)$. Hence, the overall time complexity is
\begin{align*}
    &O(n^2 \log^4 m + \runtimeClsFair{n} \log m + \runtimeFairCorCls{n} \log^3 m) \n \\
    &= O(n^2 \log^4 m + \runtimeFairCorCls{n} \log^3 m) \, \, \, \text{(since  $\runtimeFairCorCls{n} \geq \runtimeClsFair{n}$)}
\end{align*}

\noindent\textbf{Space Complexity.}
In $\mathsf{M}_1$, we store information of $\lceil \log m \rceil$ many clusterings, hence we require a space of $O(n \log m)$ in $\mathsf{M}_1$. In $\mathsf{M}_2$ also, we store $O(\log m)$ clusterings; hence, in $\mathsf{M}_2$, too we need $O(n \log m)$ space. Hence, overall, the space complexity is $O(n \log m)$.

\subsection{Proof of Main Theorem}

Now, we are ready to prove \cref{thm:explicit.fair.consensus.clustering.randomized}. To prove \cref{thm:explicit.fair.consensus.clustering.randomized} we need to use \cref{thm:closest-fair-to-correlation}. Recall \cref{thm:closest-fair-to-correlation} states that if there exists an algorithm to find $\alpha$-close $\fair$ to any given unfair clustering $\m{D}$, then given a correlation clustering instance $G$, there exists an algorithm to find an $(\alpha + \rho + \alpha \rho)$ - approximate fair correlation clustering, where $\rho$ is the state-of-the-art approximation factor for the correlation clustering problem.

\begin{proof}
    From the previous section, we get that the time and space complexity of $\stalgo$ are $O(n \log m)$ and $ O(n^2 + \runtimeFairCorCls{n} \log^3 m)$ respectively.
    
    We apply~\cref{lem:fair.consensus.clustering.randomized} and~\cref{thm:closest-fair-to-correlation} with $ \beta = \frac{1}{72(\approxFactorClsFair+1)(\approxFactorCorCls+1)} \in (0,1) $, $ \eta = \frac{1}{6(\approxFactorCorCls+1)} $, $ s=4 $, $ c=3 $, and $ g \geq \max\left(100\times \frac{4}{3} \times (\approxFactorClsFair+1)\frac{216(\approxFactorCorCls+1)}{23}, 100\times 6 \times (\approxFactorClsFair+1)\frac{72(\approxFactorCorCls+1)}{5}\right) $. Note that
    \begin{itemize}
        \item $ 1 + 3(\approxFactorFairCorCls+1)\eta = 1 + 3(\approxFactorClsFair+1)(\approxFactorCorCls+1)\eta = \frac{\approxFactorClsFair}{2} + 1.5 $.
        \item $ \approxFactorClsFair + 2 - (\approxFactorClsFair+1)\beta = \approxFactorClsFair + 2 - \frac{\approxFactorClsFair+1}{72(\approxFactorClsFair+1)(\approxFactorCorCls+1)} = \approxFactorClsFair + 2 - \frac{1}{72(\approxFactorCorCls+1)} < \approxFactorClsFair + 1.995 $.
        \item $ \approxFactorClsFair + 2 + (\approxFactorClsFair+1)\frac{\beta(g-s)+s}{g(s-1)} - \frac{2\eta}{c} = \approxFactorClsFair + 2 + (\approxFactorClsFair+1)\frac{\beta(g-4)+4}{3g} - \frac{2\eta}{3} = \approxFactorClsFair + 2 + \frac{1}{216(\approxFactorCorCls+1)} - \frac{4\beta(\approxFactorClsFair+1)}{3g} + \frac{4(\approxFactorClsFair+1)}{3g} - \frac{1}{9(\approxFactorCorCls+1)} = \approxFactorClsFair + 2 - \frac{23}{216(\approxFactorCorCls+1)} - \frac{4\beta(\approxFactorClsFair+1)}{3g} + \frac{4(\approxFactorClsFair+1)}{3g} \leq \approxFactorClsFair + 2 - \frac{23}{216(\approxFactorCorCls+1)} \frac{99}{100} < \approxFactorClsFair + 1.96 $, where in the first inequality, we drop the negative term $ -\frac{4\beta(\approxFactorClsFair+1)}{3g} $ and use the choice of $ g $ so that $ \frac{4(\approxFactorCorCls+1)}{3g} \leq \frac{23}{100\times 216 (\approxFactorCorCls+1)} $, and the last inequality is due to $ \approxFactorCorCls = 1.4371 $ .
        \item $ \approxFactorClsFair + 2 + (\approxFactorClsFair + 1 )\frac{\beta(g(2c+s)-sc)+sc)}{g(cs-s-2c)} - 2\left( 1- \frac{1}{s} - \frac{1}{c}  \right)\eta = \approxFactorClsFair + 2 + (\approxFactorClsFair+1)\frac{\beta(10g - 12)+12}{2g} - \frac{5\eta}{6} = \approxFactorClsFair + 2 + \frac{5}{72(\approxFactorCorCls+1)} - \frac{6\beta(\approxFactorClsFair+1)}{g} + \frac{6(\approxFactorClsFair+1)}{g} - \frac{5}{36(\approxFactorCorCls+1)} \leq \approxFactorClsFair + 2 - \frac{5}{72(\approxFactorCorCls+1)} - \frac{6\beta(\approxFactorClsFair+1)}{g} + \frac{6(\approxFactorClsFair+1)}{g} \leq \approxFactorClsFair + 2 - \frac{5}{72(\approxFactorCorCls+1)} \frac{99}{100} < \approxFactorClsFair + 1.98 $, where in the first inequality, we drop the negative term $ -\frac{6\beta(\approxFactorClsFair+1)}{g} $ and use the choice of $ g $ so that $ \frac{6(\approxFactorCorCls+1)}{g} \leq \frac{5}{100\times 72 (\approxFactorCorCls+1)} $, and the last inequality is due to $ \approxFactorCorCls = 1.4371 $.
    \end{itemize}
    Under this setting,~\cref{lem:fair.consensus.clustering.randomized} yields a streaming $ (\approxFactorClsFair+1.995) $-approximation fair consensus clustering algorithm using $ O(n\log m ) $ space, with a query time of $ O(n^{2}\log^{4}m + \runtimeClsFair{n}\log^{3}m) $, with probability at least $ 1 - \frac{1}{m} $.
\end{proof}

\chapter{Open Problems}\label{sec:future-directions}

In this chapter, we highlight three compelling open directions that naturally arise from our study of closest fair clustering. These questions aim to deepen our understanding of fairness in clustering under different structural and algorithmic constraints.

\section{Improving the Approximation Factor for Multiple Colors}

In \cref{thm:closest-fair-multicolor-balanced} and \cref{thm:closest-fair-multicolor-general}, we established algorithms for finding an $\alpha$-close fair clustering where the approximation factor $\alpha$ depends polynomially on the number of color groups $|\chi|$. While these results provide the first non-trivial guarantees for the multi-color setting, the dependence on $|\chi|$ can be undesirable—particularly when $|\chi|$ is large or comparable to the number of vertices $|V|$. 

An intriguing open direction is to design algorithms whose approximation factor is independent of $|\chi|$, that is, to obtain a \emph{constant-factor} approximation for the closest fair clustering problem. Achieving such a result would represent a significant conceptual and technical advancement, as it would eliminate the combinatorial dependence on the number of protected groups. 

Furthermore, as discussed in \cref{sec:implications-to-fair-correlation-and-fair-consensus-clustering}, a constant-factor approximation for the closest fair clustering problem would immediately imply constant-factor approximations for the fair correlation clustering and fair consensus clustering problems. This connection underscores the broader impact of improving the approximation factor in this setting.

\begin{question}
Given a clustering $\m{C}$ on a vertex set $V$, where $V$ is partitioned into disjoint color groups $\chi = \{c_1, \ldots, c_d\}$, can we design an algorithm that computes an $O(1)$-close fair clustering $\m{F}$ to $\m{C}$, independent of the number of colors $d$?
\end{question}

\section{Extending to Alternative Fairness Notions}

Several alternative definitions of fairness have been explored in the clustering literature, each motivated by different application scenarios and ethical considerations. One notable example is due to Ahmadian, Epasto, Kumar and Mahdian ~\cite{Ahmadian2020}, where each color $c_i$ is associated with a parameter $\sigma_i \in (0,1)$. In their formulation, a clustering $\m{F}$ is said to be \emph{fair} if for every cluster $F \in \m{F}$ we have $c_i(F) \leq \sigma_i |F|$ for all $i$, recall $c_i(F)$ denotes the number of vertices of color $c_i$ in $F$. This definition allows for flexible upper bounds on group representation rather than enforcing strict proportionality, thereby offering a more permissive and context-sensitive notion of fairness.

We refer to such definitions collectively as \emph{relaxed notions of fairness}. For a given clustering $\m{C}$, we define $\m{F}^*_{\m{C}}$ to be the \emph{closest relaxed fair clustering} if
\[
\dist(\m{C}, \m{F}^*_{\m{C}}) \leq \dist(\m{C}, \m{F})
\]
for all clusterings $\m{F}$ that satisfy any relaxed fairness criterion. 

While our results in previous sections address proportional fairness, it remains an intriguing question whether similar approximation guarantees can be achieved for these relaxed notions. This leads to the following open problem.

\begin{question}
Given a clustering $\m{C}$ on a vertex set $V$ partitioned into disjoint color groups $\chi = \{c_1, \ldots, c_d\}$, can we design an approximation algorithm to compute the closest relaxed fair clustering $\m{F}$ to $\m{C}$?
\end{question}

\section{Handling Overlapping Subgroups}

Fairness considerations in clustering often assume that the underlying demographic or protected groups form a disjoint partition of the dataset. However, in many real-world scenarios, individuals may belong to multiple groups simultaneously (for example, defined by intersections of gender, race, and age). This motivates the study of \emph{overlapping subgroup fairness}, where the protected groups $\chi = \{c_1, \ldots, c_d\}$ are not disjoint.

Ahmadian and Negahbani~\cite{ahmadian2023improved} studied this setting in the context of the fair correlation clustering problem. Their fairness definition is same as Ahmadian, Epasto, Kumar, and Mahdian~\cite{Ahmadian2020}, where each color $c_i$ is associated with a parameter $\sigma_i \in (0,1)$, and a clustering $\m{F}$ is said to be fair if for every cluster $F \in \m{F}$ we have $c_i(F) \leq \sigma_i |F|$ for all $i$. 

Ahmadian and Negahbani provided a \emph{bicriteria constant-factor approximation} for the fair correlation clustering problem under this definition. Here, “bicriteria” means that fairness constraints are allowed to be slightly violated in some clusters, i.e.,
\[
c_i(F) \leq (1 + \epsilon)\,\sigma_i |F|
\]
for some small $\epsilon \in (0,1)$. 

A natural open question is whether similar bicriteria or exact guarantees can be achieved for the \emph{closest fair clustering} problem under overlapping subgroup fairness. This would imply approximation for the fair consensus clustering problem.

\begin{question}
Given a clustering $\m{C}$ on a vertex set $V$, where the color groups $\chi = \{c_1, \ldots, c_d\}$ may overlap (i.e., each vertex may belong to multiple protected groups), can we develop an approximation algorithm to find the closest relaxed fair clustering $\m{F}$ to $\m{C}$?
\end{question}

\chapter{Appendix (Pseudocodes)}

\section{Pseudocode of the algorithm $\algog$}

\begin{algorithm}[H]
\DontPrintSemicolon
\caption{$\algog(\inpset)$}\label{alg:algo-for-general}
\KwData {Input set of clusters $\inpset$, in each $\inpcl{i}  \in \inpset$ the surplus of $\inpcl{i} $, $\surp{\inpcl{i} }$ is within $0 < \surp{\inpcl{i} } < p$}
\KwResult{A set of clusters $\out$, such that in each cluster $\outcl{i} \in \m{T}$ the number of blue vertices is a multiple of $p$.} 
   $\cut, \merge, \nc, \outg \gets \emptyset$
   
   \For{$\inpcl{i}  \in \inpset$}
   {
        \lIf{$ \surp{\inpcl{i}}=0 $}{
            add $ \inpcl{i}$ to the set $ \nc $\;
        } 
        \ElseIf{$\surp{\inpcl{i} } < p/2$}
        {
             add $\inpcl{i} $ to the set $\cut$ \;
        }
        \Else
        {
            add $\inpcl{i} $ to the set $\merge$ \;
        }
   }
   Sort the clusters in $\inpcl{k} \in \merge$ based on their $(\ccostf{\inpcl{k}} - \mcostf{\inpcl{k}})$ in non-increasing order \;
   
   \While{$\cut \neq \emptyset$ and $\merge \neq \emptyset$}{
        \For{$\inpcl{i} \in \cut$} {
            \For{$\inpcl{j} \in \merge$} {
                Let, $k = min(\surp{\inpcl{i}}, \defi{\inpcl{j}})$. \;
                
                cut a set of $k$ blue vertices from $\inpcl{i}$ and add to $\inpcl{j}$ \;
                \If{$k = \surp{\inpcl{i}}$}{
                    $\cut = \cut \setminus \{\inpcl{i}\}$. \;
                    $\nc = \nc \cup \inpcl{i}$ \;
                
                }
                \If{$k = \defi{\inpcl{j}}$}{
                    $\merge = \merge \setminus \{\inpcl{j}\}$. \; 
                    $\outg = \outg \cup \inpcl{j}$ \;
                    \label{step:cut-merge}
                }
            }
        }
   }
   \If{$\cut = \emptyset$} {
       \Return{$\outg \cup \algom(\nc, \merge)$} \;
    }
    \Else{
         \Return{$\outg \cup \nc \cup \algoc(\cut)$} \;
    }
\end{algorithm}

\section{Pseudocode of the algorithm $\algom$}

\begin{algorithm}
\DontPrintSemicolon
\caption{$\algom(\textsc{newcut},\merge')$}\label{alg:algo-for-merge}
\KwData{Two clustering $\textsc{newcut}$ and $\merge'$ such that for all clusters $C_i \in \nc$ we have $p \mid |\blue{C_i}|$ and for clusters $C_j \in \merge'$ we have $s(C_j)>p/2$ respectively. In $\merge'$, the clusters $C_j \in \merge'$ are sorted based on their $(\ccostf{C_j} - \mcostf{C_j})$}
\KwResult{A clustering $\out$, such that for all clusters $T_i \in \out$, $p \mid \blue{T_i}$.}
  $W\gets \sum_{C_j\in\merge'}\defi{C_j}$ \;
  \For{$C_\ell \in (\nc \cup \merge')$} {
        Initialize a variable $v_k$ to $0$\;
  }

  \While{$W \neq 0$} {
    Take the cluster $C_\ell \in (\nc \cup \merge')$ for which $\textit{cost}(W_{m,v_m})$ is minimum\;
    \tcp{ $\textit{cost}(W_{m,v_m}) = \kappa_{v_k}(C_\ell)$ if $v_k \geq 1$, for $v_k = 0$ $\textit{cost}(W_{m,v_m}) = \kappa_{0}(C_\ell) - \mcostf{C_\ell}$}
    \While{$\textit{size}(W_{m,v_m}) \neq 0$} {
        \tcp{ $\textit{size}(W_{m,v_m})$ is the size $v_k$th subset of $C_\ell$ }
        \For{$C_j \in \merge'$}{
         $\gamma = \min(\defi{C_j}, \textit{size}(W_{m,v_m}))$\;
         cut $\gamma$ many blue vertices from $C_\ell$ and merge to $C_j$\;
         \If{$\gamma = \defi{C_j}$}
         {
            $\textit{size}(W_{m,v_m}) = \textit{size}(W_{m,v_m}) - \gamma$\;
         }
         \If{$\gamma = \textit{size}(W_{m,v_m}))$}
         {
            $\defi{C_j} = \defi{C_j} - \gamma$ \;
            $W = W - p$ \;
            $v_{k} = v_{k} + 1$\;
         }
         
        }
    }
  }
  \For{$C_j \in \merge'$} {
    \If{$\defi{C_j} = 0$}{
    update $\merge' = \merge' \setminus \{C_j\}$
    }
  }
  \Return{$\out = \nc \cup \merge'$}
\end{algorithm}

\section{Pseudocode of the algorithm $\algoc$}

\begin{algorithm}[H]
\DontPrintSemicolon
\caption{$\algoc(\cut')$}\label{alg:algo-for-cut}
\KwData{Set of clusters $\cut'$, in each cluster $\gencl{i} \in \cut'$ the surplus of $\gencl{i}$, $\surp{\gencl{i}} \leq p/2$.}
\KwResult{Set of clusters $\outcut$, such that in each cluster $\outcl{i} \in \outcut$ the number of blue vertices is a multiple of $p$. }
ExtraSum = $\sum \surp{\inpcl{i} }$ \;

$\pcard = \frac{ExtraSum}{p}$ \;

Initialize $\pcard$ many sets $\pcl{1}, \pcl{2}, \ldots \pcl{\pcard}$ to $\emptyset$ \; 
\tcp{$\pcard$ many extra clusters}
Initialize $\pcard$ many variables $\ell_1, \ell_2, \ldots \ell_n$ to $p$ \;
\tcp{leftover space of extra clusters}
Boolean $flag = 0$ \;
\For{$\gencl{i} \in \cut'$} {
    \For{$j = 1$ to $n$} {
             Add $k = min ( \ell_j, \surp{\gencl{i} })$ many blue vertices from $\gencl{i} $ to the cluster $\pcl{j}$ \;
             
             $\ell_j = \ell_j - k$ \;
    }
}
\Return{$\cut' \cup \{\pcl{1}, \pcl{2}, \ldots, \pcl{\pcard}\}$}
\end{algorithm}

\section{Pseudocode of the algorithm $\algmf$}

\begin{algorithm}[H]
\DontPrintSemicolon
\caption{$\algmf(\out)$}\label{alg:algo-mf}
\KwData {a $ \mopqdef \ \out $.}
\KwResult{a fair clustering $ \fairset $.} 
    $ \tb, \tr, \fairset \gets \emptyset$ \;
    \For{$ \outcl{i} \in \out $}{
        \If{$ |\blue{\outcl{i}}| < \ratio|\red{\outcl{i}}| $}{
            add $ \outcl{i} $ to the set $ \tr $ \;
    } \ElseIf {$ |\blue{\outcl{i}}| > \ratio|\red{\outcl{i}}| $}{
            add $ \outcl{i} $ to the set $ \tb $ \;
        }
    }

        \For{$ \outcl{i} \in \tr $}{
            \For{$ \outcl{j} \in \tb $}{
                Let $ k = \min(|\red{\outcl{i}}|-\dfrac{|\blue{\outcl{i}}|}{\ratio}, \dfrac{|\blue{\outcl{j}}|}{\ratio} - |\red{\outcl{j}}|) $ \;
                Move $ k $ red vertices from $ \outcl{i} $ to $ \outcl{j} $ \;
                \If{$ k = |\red{\outcl{i}}|-\dfrac{|\blue{\outcl{i}}|}{\ratio} $}{
                    $ \tr = \tr\setminus \{\outcl{i}\} $\;
                    $ \fairset = \fairset \cup \{\outcl{i}\} $ \;
                }
                \If{$ k = \dfrac{|\blue{\outcl{j}}|}{\ratio} - |\red{\outcl{j}}| $}{
                    $ \tb = \tb\setminus \{\outcl{j}\} $ \;
                    $ \fairset = \fairset \cup \{\outcl{j}\} $ \;
                }
            }
        }
    \Return{$ \fairset $}
\end{algorithm}

\section{Pseudocode of the algorithm $\fptwo$}

\begin{algorithm}[H]
\DontPrintSemicolon
\KwIn{Clustering $\mathcal{C}$}
\KwOut{A fair clustering $\outfptwo$}
Let, $\m{N}^0 = \{ N^0_1, N^0_2, \ldots, N^0_\ell \} = \m{C}$. \;
\For{$i \gets 1$ \KwTo $\log |\chi|$}{
    $\m{N}^i = \m{N}^{i - 1}$ \;
    $j \gets 1$ \;
    \While{$j \neq |\chi| / 2^i$}{
        $ S_{j}, S_{j+1} \gets \emptyset $\;
        \ForEach{$N^{i}_a \in \mathcal{N}^{i}$}{
            $N^i_a \gets N^{i}_a \setminus T^j_a$\; 
            \If{$|B_j^i(N^{i}_a)| \geq |B_{j+1}^i(N^{i}_a)|$}{
                $S_j \gets S_{j} \cup T^j_a$\;
            }
            \Else{
                $S_{j+1} \gets S_{j+1} \cup T^j_a$\;
            }
        }
        $\m{N}^{i} = \m{N}^i \cup \texttt{multi-GM}(S_j, S_{j+1})$\;
        $j \gets j + 2^i$.
    }
}
\Return{$\m{N}^{\log |\chi|}$}
\caption{$\fptwo (\m{C})$}\label{algo:fair-power-of-two}
\end{algorithm}

\section{Pseudocode of the algorithm $\greedymerge$}

\begin{algorithm}[H]
\DontPrintSemicolon
\KwIn{Two sets \texttt{Set1}, \texttt{Set2} of subsets of $V$, from blocks $B_j^i$ and $B_{j+1}^i$ respectively}
\KwOut{A set \texttt{Fair} of fair merged vertex sets}

\texttt{Fair} $\gets \emptyset$\;

\While{\texttt{Set1} $\neq \emptyset$ \textbf{and} \texttt{Set2} $\neq \emptyset$}{
    Let $S_1 \in \texttt{Set1}$\;
    Let $S_2 \in \texttt{Set2}$\;

    \If{$|S_1| \geq |S_2|$}{
        Let $S \subseteq S_1$ such that $|S| = |S_2|$ and $S$ contains equal number of vertices of each color in $B_j^i$.\;
        \texttt{Fair} $\gets$ \texttt{Fair} $\cup \{ S \cup S_2 \}$\;
        $S_1 \gets S_1 \setminus S$\;
        \texttt{Set2} $\gets$ \texttt{Set2} $\setminus \{S_2\}$\;
        \If{$S_1 = \emptyset$}{
            \texttt{Set1} $\gets$ \texttt{Set1} $\setminus \{S_1\}$\;
        }
    }
    \Else{
        Let $S \subseteq S_2$ such that $|S| = |S_1|$ and $S$ contains equal number of vertices of each color in $B_{j+1}^i$\;
        \texttt{Fair} $\gets$ \texttt{Fair} $\cup \{ S \cup S_1 \}$\;
        $S_2 \gets S_2 \setminus S$\;
        \texttt{Set1} $\gets$ \texttt{Set1} $\setminus \{S_1\}$\;
        \If{$S_2 = \emptyset$}{
            \texttt{Set2} $\gets$ \texttt{Set2} $\setminus \{S_2\}$\;
        }
    }
}
\Return{\texttt{Fair}}\;
\caption{\texttt{multi-GM}$(\texttt{Set1}, \texttt{Set2})$} \label{algo:greedymerge}
\end{algorithm}

\section{Pseudocode of the algorithm $\fmulti$}

\begin{algorithm}[H]
\DontPrintSemicolon
\KwIn{Initial clustering \( \mathcal{I} = \mathcal{F}^0 \), color set \( \zeta = \{ z_1, \ldots, z_r \} \), target proportions \( (p_1 : \cdots : p_r) \) \\ For each cluster $I_j \in \m{I}$ we have $z_k(I_j)$ is a multiple of $p_k$.}
\KwOut{Clustering \( \mathcal{F} \) where each cluster satisfies the global color ratio}

Let \( T \gets \lceil \log_2 r \rceil \)\;
Initialize blocks \( B^0_j \gets \{ z_j \} \) for all \( j \in [r] \)\;
Set \( \mathcal{F}^0 \gets \mathcal{I} \)\;

\For(\tcp*[f]{Iterate through block levels}){$t \gets 1$ \KwTo $T$}{
    Merge adjacent blocks from level \( t - 1 \) to form \( \{ B^t_1, B^t_2, \ldots \} \)\;

    Let \( m_{t-1} \gets \text{number of blocks at iteration } t - 1 \)\;
    
    \If{$m_{t-1}$ is odd}{
        Copy the last block as-is: \( B^t_{\lceil \#B^{t-1} / 2 \rceil} \gets B^{t-1}_{m_{t-1}} \)\;
    }
    
    Initialize \( \mathcal{F}^t \gets \emptyset \)\;
    
    \ForEach(\tcp*[f]{Iterate through clusters}){$F \in \mathcal{F}^{t-1}$}{
        \ForEach(\tcp*[f]{Iterate through blocks}){$B^t_i = B^{t-1}_{2i-1} \cup B^{t-1}_{2i}$}{
            Let \( A = B^{t-1}_{2i-1} = \{ z_{a_1}, \ldots, z_{a_s} \} \)\;
            Let \( B = B^{t-1}_{2i} = \{ z_{b_1}, \ldots, z_{b_u} \} \)\;
            
            Compute scaling factors:\;
            \Indp
                \( x \gets \min_{j \in [s]} \left( \frac{z_{a_j}(F)}{p_{a_j}} \right) \)\;
                \( y \gets \min_{k \in [u]} \left( \frac{z_{b_k}(F)}{p_{b_k}} \right) \)\;
            \Indm
            
            \uIf(\tcp*[f]{Case 1: need to merge}){$x > y$}{
                \ForEach{$k \in [u]$}{
                    Merge \( p_{b_k} \cdot (x - y) \) vertices of color \( z_{b_k} \) into \( F \)\;
                }
            }
            \uElseIf(\tcp*[f]{Case 2: need to cut}){$x < y$}{
                \ForEach{$k \in [u]$}{
                    Cut \( p_{b_k} \cdot (y - x) \) vertices of color \( z_{b_k} \) from \( F \)\;
                }
            }
            
            Add updated cluster \( F \) to \( \mathcal{F}^t \)\;
        }
    }
}

\Return{\( \mathcal{F}^T \)}
\caption{$\fmulti$}
\label{algo:make-pdc-fair}
\end{algorithm}

\section{Pseudocode of the algorithm $\pdca$}

\begin{algorithm}[H]
\caption{$\pdca$}
\label{alg:p-divisible-clustering}
\KwIn{Clustering $\mathcal{C}$, colors $\chi = \{c_1, \ldots, c_t\}$, and ratios $p_1:p_2:\cdots:p_t$}
\KwOut{$p$-divisible clustering $\mathcal{M}$}

\ForEach{$c_j \in \chi$}{
    Create $\sigma_j / p_j$ empty clusters: $\texttt{extra\_clusters} = \{P_1, P_2, \ldots, P_{\sigma_j/p_j}\}$\;

    Initialize $\texttt{CUT} \gets \emptyset$, $\texttt{MERGE} \gets \emptyset$\;
    
    \ForEach{$C_i \in \mathcal{C}$}{
        \lIf{$|\sigma(C_i, c_j)| \leq p_j / 2$}{
            $\texttt{CUT} \gets \texttt{CUT} \cup \{C_i\}$
        }
        \lElse{
            $\texttt{MERGE} \gets \texttt{MERGE} \cup \{C_i\}$
        }
    }

    \While{$\texttt{CUT} \neq \emptyset$}{
        Pick and remove $C_k \in \texttt{CUT}$\;
        Remove surplus: $C_k \gets D_k \setminus \sigma(C_k, c_j)$\;

        \If{$\texttt{MERGE} \neq \emptyset$}{
            \While{$\sigma(C_k, c_j) \neq \emptyset$}{
                \ForEach{$C_\ell \in \texttt{MERGE}$}{
                    $T \gets \min(|\sigma(C_k, c_j)|, |\delta(C_\ell, c_j)|)$-sized subset of $\sigma(C_k, c_j)$\;
                    $C_\ell \gets C_\ell \cup T$\;
                    $\sigma(C_k, c_j) \gets \sigma(C_k, c_j) \setminus T$\;
                    \If{$c_j(C_\ell)$ is a multiple of $p_j$}{
                        $\texttt{MERGE} \gets \texttt{MERGE} \setminus \{C_\ell\}$\;
                    }
                }
            }
        }
        \Else{
            \While{$\sigma(C_k, c_j) \neq \emptyset$}{
                \ForEach{$P_m \in \texttt{extra\_clusters}$}{
                    $Q \gets$ subset of size $\min(p_j, |\sigma(C_k, c_j)|, p_j - |P_m|)$\;
                    $P_m \gets P_m \cup Q$\;
                    $\sigma(C_k, c_j) \gets \sigma(C_k, c_j) \setminus Q$\;
                    \If{$|P_m| = p_j$}{
                        $\texttt{extra\_clusters} \gets \texttt{extra\_clusters} \setminus \{P_m\}$\;
                    }
                }
            }
        }
    }

    \While{$\texttt{MERGE} \neq \emptyset$}{
        Pick $C_k \in \cut \cup \merge$ with minimum $\kappa^j(C_k) - \mu^j(D_k)$\;
        Remove surplus: $C_k \gets C_k \setminus \sigma(C_k, c_j)$\;
        \While{$\sigma(D_k, c_j) \neq \emptyset$}{
            \ForEach{$C_\ell \in \texttt{MERGE}$}{
                $T \gets \min(|\sigma(C_k, c_j)|, |\delta(C_\ell, c_j)|)$-sized subset\;
                $C_\ell \gets C_\ell \cup T$\;
                $\sigma(C_k, c_j) \gets \sigma(C_k, c_j) \setminus T$\;
                \If{$c_j(C_\ell)$ is a multiple of $p_j$}{
                    $\texttt{MERGE} \gets \texttt{MERGE} \setminus \{C_\ell\}$\;
                }
            }
        }
    }
}
\Return{$\mathcal{M}$ composed of updated $\mathcal{C}$ and filled $\texttt{extra\_clusters}$}\;
\end{algorithm}

\section{Pseudocode of $\stalgo$}

\begin{algorithm}[H]
\caption{$\stalgo$}
\label{alg:sone}
\DontPrintSemicolon
\SetKwInOut{KwIn}{Input}
\SetKwInOut{KwOut}{Output}
\SetKwFunction{BuildUF}{BuildClustersFromUF}
\SetKwFunction{FairCand}{\fc}
\SetKwFunction{ArgMin}{ArgMin}
\KwIn{Stream of triples $((u,v), j, b)$; parameters $n,m,\varepsilon \in (0,\tfrac15]$}
\KwOut{A fair consensus clustering $\m{F}$.}

\BlankLine
\textbf{Sampled indices and memory structures:}\;
Draw $s \gets \lceil \log m \rceil$ indices $J \gets \{j_1,\dots,j_s\} \subseteq [m]$ uniformly at random.\;
Let $\mathsf{M}_1 \leftarrow$ empty store (hash from index $\to$ list of triples).\;
Set $t \gets \lceil \varepsilon^{-2} \log m \rceil$ and draw $K \gets \{k_1,\dots,k_t\} \subseteq [m]$ uniformly at random.\;
Let $\mathsf{M}_2 \leftarrow$ empty store (hash from index $\to$ list of triples).\;

\BlankLine
\textbf{Single-pass streaming phase:}\;
\ForEach{triple $((u,v), j, b)$ arriving in the stream}{
    \If{$j \in J$}{append $((u,v), b)$ to $\mathsf{M}_1[j]$\;}
    \If{$j \in K$}{append $((u,v), b)$ to $\mathsf{M}_2[j]$\;}
}

\BlankLine
\textbf{Post-stream reconstruction:}\;
\ForEach{$j \in J$}{
    initialize a fresh union--find $\mathsf{UF}_j$ on $V$\;
    \ForEach{$((u,v), b) \in \mathsf{M}_1[j]$}{
        \If{$b=0$}{union $u$ and $v$ in $\mathsf{UF}_j$\;}
    }
    $\m{C}_j \gets$ \BuildUF{$\mathsf{UF}_j$}\tcp*{extract clusters from union--find}
}
Let $\mathcal{C}_J \gets \{\m{C}_{j_1},\dots,\m{C}_{j_s}\}$.\;

\BlankLine
\textbf{Generate fair candidate set:}\;
$\widetilde{\m{F}} \gets$ \FairCand{$\mathcal{C}_J$}\tcp*{uses $\alpha$-close fair clustering algorithm}

\BlankLine
\textbf{Build $\m{W}$ from $\mathsf{M}_2$:}\;
\ForEach{$k \in K$}{
    initialize a fresh union--find $\mathsf{UF}_k$ on $V$\;
    \ForEach{$((u,v), b) \in \mathsf{M}_2[k]$}{
        \If{$b=0$}{union $u$ and $v$ in $\mathsf{UF}_k$\;}
    }
    $\m{C}_k \gets$ \BuildUF{$\mathsf{UF}_k$}\;
}
$\m{W} \gets \{\m{C}_{k_1},\dots,\m{C}_{k_t}\}$.\;

\BlankLine
\textbf{Select the best fair clustering on the sampled store $2$:}\;
$\displaystyle \m{F} \gets \ArgMin_{\m{F}' \in \widetilde{\m{F}}} \ \sum_{\m{C}_i \in \m{W}} \dist(\m{C}_i, \m{F}')$\;

\Return $\m{F}$\;
\end{algorithm}

\printbibliography

\end{document}